\documentclass[12pt,twoside,titlepage,a4paper]{ethzthesis}

\usepackage{etex}

\usepackage[utf8]{inputenc}

\usepackage[main=english,ngerman]{babel}

\newcommand{\myTitle}{Design and Analysis of an Extreme-Scale,\\High-Performance, and Modular\\Agent-Based Simulation Platform\xspace }
\newcommand{\myPlainTitle}{Design and Analysis of an Extreme-Scale, High-Performance, and Modular Agent-Based Simulation Platform\xspace }

\newcommand{\myDissNumber}{30705}
\newcommand{\myName}{Lukas Johannes Breitwieser\xspace}

\newcommand{\myTime}{2024\xspace}

\newcommand{\myDOI}{10.5281/zenodo.14505960}

\usepackage{scrtime}

\newcommand{\thesisVersion}[0]{1.0}

\newcommand{\cognoLungInjury}[0]{cogno_mechanistic_2024,cogno_thesis,cogno_agent-based_2022,cogno_3d_2022}
\newcommand{\abmImmuneSystem}[0]{montealegre_agent-based_2012,shinde_review_2018,bauer2009,truszkowska_predicting_2022,perrin_agent-based_2006,pappalardo_agent_2018,folcik_using_2011,folcik_basic_2007,chiacchio_agent-based_2014,jacob_swarm-based_2011}
\newcommand{\abmChronicDiseases}[0]{li2016,montagna_agent-based_2017,nianogo_agent-based_2015,schryver_emulating_2015,mi_agent-based_2007,khademi_agent-based_2018,li_assessing_2014,archbold_agent-based_2024,giabbanelli_application_2021,morshed_systematic_2019,beheshti_comparing_2017,giabbanelli_using_2017,squires_long-term_2023}
\newcommand{\abmHealthcareSystemOpt}[0]{cassidy_mathematical_2019,squires_long-term_2023,cabrera_simulation_2012,ajmi_agent-based_2019}
\newcommand{\abmPharmacoDynamics}[0]{gao_developing_2017}
\newcommand{\abmCardiovasularRiskFactors}[0]{li2018}

\newcommand{\abmMedicine}[0]{\cognoLungInjury,\abmImmuneSystem,\abmChronicDiseases,\abmPharmacoDynamics,\abmCardiovasularRiskFactors}

\newcommand{\abmCancer}[0]{demontigny_2021,duswald_2024,\cognoLungInjury,cogno_biomedicin_2024,metzcar_review_2019,wang2015,olsen_multiscale_2013,wang_multi-scale_2013,an_agent-based_2015,wang_integrated_2015,wang_simulating_2015,poleszczuk_agent-based_2016,norton_multiscale_2019,heidary_double-edged_2020,rivera_agent-based_2022,van_genderen_agent-based_2024}
\newcommand{\abmNeuronGrowth}[0]{zublerdouglas2009framework,zublerdissertation,duswald2024calibrationstochasticagentbasedneuron,torben-nielsen_context-aware_2014,de_schutter_efficient_2023}
\newcommand{\abmCorticalLamination}[0]{abubacar_neuronal_growth,zublerdissertation}
\newcommand{\abmNervousSystem}[0]{\abmNeuronGrowth,\abmCorticalLamination,pennisi2015,caffrey_silico_2014,avin_agent-based_2021}
\newcommand{\abmMorphoGenesis}[0]{glen2019,tang_phenotypic_2011,lambert_bayesian_2018,walpole_agent-based_2015,camacho-gomez_3d_2022,dalle_nogare_netlogo_2020,bonabeau_classical_1997,thorne_agent-based_2007}
\newcommand{\abmBiofilms}[0]{kreft2001,wilmoth_microfluidics_2018,li_agent-based_2024,latif_multiscale_2018,sweeney_agent-based_2019,nagarajan_agent-based_2022,koshy-chenthittayil_agent_2021}
\newcommand{\abmBiologyOther}[0]{soheilypour_agent-based_2018,an_agent-based_2009,griffin_agent-based_2006,gorochowski_agent-based_2016}
\newcommand{\abmBiology}[0]{\abmCancer,\abmNervousSystem,\abmMorphoGenesis,\abmBiofilms}

\newcommand{\abmCOVID}[0]{hesam_2024,ozik_citycovid_2021,yin_data_2021,kou_multi-scale_2021,faucher_agent-based_2022,pescarmona_agent-based_2021,cattaneo_agent-based_2022,kerr_covasim_2021,truszkowska_exploring_2023,hoertel_facing_2020,kumaresan_fitting_2023,ogden_mathematical_2024,topirceanu_impact_2023}
\newcommand{\abmInfluenza}[0]{laskowski_agent-based_2011,depasse_does_2017,karimi_effect_2015,kumar_policies_2013}
\newcommand{\abmMalaria}[0]{gharakhanlou_spatial_2020,pizzitutti_validated_2015,modu_agent-based_2023,bomblies_agent-based_2014,smith_agent-based_2018}
\newcommand{\abmHIV}[0]{teweldemedhin_agent-based_2004,brookmeyer_combination_2014,anderle_modelling_2024}
\newcommand{\abmPublicHealthInterventions}[0]{yin_data_2021,kou_multi-scale_2021,faucher_agent-based_2022,pescarmona_agent-based_2021,getz_agent-based_2019,cattaneo_agent-based_2022,kerr_covasim_2021,hoertel_facing_2020,topirceanu_impact_2023,liu_role_2015}
\newcommand{\abmEpidemicForecasting}[0]{tabataba_epidemic_2017,kumaresan_fitting_2023,venkatramanan_using_2018}
\newcommand{\abmEpidemicPreparedness}[0]{ogden_mathematical_2024,woodul_agent-based_2023,marini_enhancing_2020}
\newcommand{\abmUrbanAreas}[0]{macal_chisim_2018,hackl_epidemic_2019}
\newcommand{\abmVaccinationHesitancy}[0]{martono_agent-based_2024,buttenheim_provider_2013}
\newcommand{\abmEpidemiologyOther}[0]{hunter_taxonomy_2017,miksch2019,estill2020,tracy2018}
\newcommand{\abmEpidemiology}[0]{\abmCOVID,\abmInfluenza,\abmHIV,woodul_agent-based_2023,venkatramanan_using_2018,tabataba_epidemic_2017,\abmUrbanAreas,\abmVaccinationHesitancy,\abmEpidemiologyOther}

\newcommand{\abmSocialSciences}[0]{retzlaff2021,axelrod1981,arthur1994,axelrod1997,hegselmann2002,fischbach2021}
\newcommand{\abmFinanceAndEconomics}[0]{axtell2024,palmer_artificial_1994}
\newcommand{\abmAgriculture}[0]{huber2018,kremmydas_review_2018,yamashita_development_2018,maes_agent-based_2017,schreinemachers_agent-based_2011,barbuto_improving_2019,bert_agent_2011,coronese_agrilove_2023,beckers_modelling_2018}
\newcommand{\abmTransport}[0]{li2021,huang2022,nguyen2021,tzouras2023,kim_estimating_2022,cunha_development_2022,he_microscopic_2020,rindsfuser_agent-based_2007,wang_prediction_2021,tumer_distributed_2007}
\newcommand{\abmEcology}[0]{grimm2005,grimm2013,deangelis2005,zhang2020,grimm_ten_1999,deangelis_individual-based_2018,deangelis_individual-based_2014,judson_rise_1994,uchmanski_individual-based_1996,mclane_role_2011,karsai_bottom-up_2016,heckbert_agent-based_2010}
\newcommand{\abmEnergy}[0]{tian_agent-based_2020,robinson_determinants_2015,sachs_agent-based_2019,moglia_agent-based_2018,rai_agent-based_2016}
\newcommand{\abmMarketing}[0]{north_multiscale_2010,romero_two_2023,schramm_agent-based_2010,negahban2014,rand_agentbased_2021,garcia_validating_2007,rand_agent-based_2011}
\newcommand{\abmCrime}[0]{malleson_analysis_2012,zhu_agent-based_2021,malleson_crime_2010,groff_state_2019}
\newcommand{\abmOther}[0]{huber2018,boone2017,castro2020,stieler2022,heppenstall2021,bauer2009,\abmMarketing,\abmEnergy,\abmCrime}

\newcommand{\mechanisticModels}[0]{baker_mechanistic_2018,craver_when_2006,curtis_mechanistic_1991,sedghi_taxonomy_2021,zhang_chapter_2010,wang_evolution_2023,muldbak_digital_2022}

\newcommand{\alphafold}[0]{jumper_highly_2021,yang_alphafold2_2023,varadi_alphafold_2022,senior_improved_2020}
\newcommand{\surrogateModels}[0]{angione_using_2022,cheng_review_2024,ganti_design_2020,garzon_machine_2022,cai_surrogate_2021,cozad_learning_2014,agarwal_machine-learning-based_2020,luo_review_2023,mai_machine_2021,weber_technical_2020,lu_efficient_2019,bocquet_surrogate_2023,zahura_training_2020,prina_machine_2024,zhu_building_2022}

\newcommand{\physicsInformed}[0]{batuwatta-gamage_physics-informed_2022,karniadakis_physics-informed_2021,sharma_review_2023,kashinath_physics-informed_2021,karimpouli_physics_2020,zhao_parameter_2022,nghiem_physics-informed_2023,pateras_taxonomic_2023,qian_lift_2020}
\newcommand{\ddmOverview}[0]{sarker_ai-based_2022,casalino_ai-driven_2021,noe_machine_2020,bernetti_data-driven_2020,noauthor_rise_2021,ourmazd_science_2020,baker_mechanistic_2018}

\newcommand{\bdmUsage}[0]{duswald_2024,demontigny_2021,\cognoLungInjury,duswald2024calibrationstochasticagentbasedneuron,hesam_2024,demontigny_2023,demetriades_interrogating_2022,gazeli_interrogating_2022,bdm-workshop,abubacar_neuronal_growth,jennings_cryo,jennings_cryo2,hiv_malawi}
 
\newcounter{dummy}

\usepackage{csquotes}  \usepackage[T1]{fontenc}
\usepackage{xspace}
\usepackage{textcomp}
\usepackage{mparhack}
\usepackage{relsize}

\usepackage[numbers,sort&compress]{natbib}
\setcitestyle{numbers,sort&compress,open={[},close={]},citesep={,},notesep={, }
}

\let\origcite\cite \def\cite#1{\unskip~\origcite{#1}}

\usepackage{amsmath}
\usepackage{mathtools}
\usepackage{isomath}

\usepackage{tabularx}
\usepackage{ltablex}
\usepackage{floatrow}
\usepackage{caption}
\usepackage{subcaption}
\usepackage{blindtext}

\usepackage{graphicx}
\usepackage{rotating}
\usepackage{tikz}
\usepackage{pgfplots}
\pgfplotsset{compat=newest}
\usetikzlibrary{shapes.geometric}

\newcommand*\circled[1]{
  \tikz[baseline=(char.base)]{
    \node[shape=circle,draw,inner sep=1pt,font=\footnotesize,minimum size=0.8\baselineskip] (char) {\figureversion{lining}#1};
  }
}

\usepackage{cleveref}

\PassOptionsToPackage{printonlyused}{acronym}
\usepackage{acronym}

\usepackage{enumitem}

\definecolor{chapter-color}{cmyk}{1, 0.50, 0, 0.25}
\definecolor{link-color}{cmyk}{1, 0.50, 0, 0.25}
\definecolor{cite-color}{cmyk}{0, 0.7, 0.9, 0.2}

\usepackage{bookmark}

\let\chapterNumber\undefined \newfont{\chapterNumber}{eurb10 scaled 5500}

\usepackage{siunitx}
\DeclareSIUnit\au{a.u.}
\let\u=\SI

\usepackage{microtype}
\usepackage{textcase}
\usepackage[scaled=0.85]{beramono}
\DeclareRobustCommand{\spacedallcaps}[1]{\textls[160]{\scshape\MakeTextUppercase{#1}}}
\DeclareRobustCommand{\spacedlowsmallcaps}[1]{\textls[80]{\scshape\MakeTextLowercase{#1}}}

\usepackage{pifont}

\newcommand{\xmark}{\ding{55}}

\usepackage{flafter}

\usepackage[frozencache,cachedir=.]{minted}
\newmintedfile[inputcppcode]{cpp}{
  linenos,
  xleftmargin=\parindent, fontsize=\scriptsize,
  baselinestretch=1,
  frame=lines,
  framerule=1pt
}
\newminted{cpp}{
  linenos,
  xleftmargin=\parindent, fontsize=\scriptsize,
  baselinestretch=1,
  frame=lines,
  framerule=1pt
}

\newif\ifgrammarlyexport
\grammarlyexportfalse

\ifgrammarlyexport
  \usepackage[none]{hyphenat}
  \usepackage{microtype}
  \DisableLigatures{encoding = *, family = * }
\fi

 \usepackage[final]{pdfpages}

\usepackage{float}
\usepackage{listings}
\usepackage[ruled, linesnumbered]{algorithm2e}

\newfloat{lstfloat}{htbp}{lop}
\floatname{lstfloat}{Listing}

\usepackage{booktabs}
\usepackage{multirow}
\usepackage{adjustbox}  \usepackage{rotating}

\newcommand{\bdm}{BioDynaMo}
\newcommand{\ta}{TeraAgent}
\newcommand{\etal}{et al.}
\newcommand{\cpp}{C\texttt{++}}

\newcommand{\systemA}{System A (Table~\ref{tab:systems})}
\newcommand{\systemB}{System B (Table~\ref{tab:systems})}
\newcommand{\systemC}{System C (Table~\ref{tab:systems})}
\newcommand{\stref}[1]{\ref{ap:st#1}}

\newcommand{\result}[1]{#1}
\definecolor{dora}{rgb}{0.80, 0.60, 0.00}

\newcommand{\jgl}[1]{\textcolor{dora}{JGL: #1}}
\newcommand{\agy}[1]{\textcolor{red}{#1}}
\newcommand{\agycomment}[1]{\textcolor{red}{\textbf{[AGY:}#1\textbf{]}}}

\newcommand{\ESDefinition}{100 billion} \newcommand{\ESMaxAgentsApprox}{500 billion}
\newcommand{\ESMaxAgentsApproxDE}{500 Milliarden}
\newcommand{\ESMaxAgentsExact}{501.51 billion}
\newcommand{\ESMaxAgentsInTrillion}{half a trillion}
\newcommand{\ESNumCPUs}{84'096}
\newcommand{\ESNumNodes}{438}
\newcommand{\ESMemoryConsumptionInTB}{92}
\newcommand{\ESRuntimePerSecond}{147}
\newcommand{\ESFactorImprovementSOTA}{84}

\newif\ifsubmission
\submissionfalse

\ifsubmission
  \renewcommmand{\agy}[1]{#1}
  \renewcommand{\agycomment}[1]{}
  
  \renewcommand{\jgl}[1]{}
\fi

\newif\ifversion
\versionfalse

\newif\ifthesis
\thesistrue
\newcommand{\ThesisPaper}[2]{\ifthesis#1\else#2\fi}

\begin{document}

\raggedbottom \selectlanguage{english}
\pagenumbering{roman}

\cleardoublepage\setcounter{page}{1}

\clearpage{}\begin{titlepage}
\begin{center}
        \ifversion
          \InputIfFileExists{preamble/version.tex}{}{}
        \fi
        \large
        \begingroup
            \spacedlowsmallcaps{Diss. ETH No. \myDissNumber}
        \endgroup

        \hfill

        \vfill

        \begingroup
\spacedallcaps{\large \textbf{\myTitle}}
\endgroup

        \vfill

        \begingroup
            A thesis submitted to attain the degree of\\
            \vspace{0.5em}
            \spacedlowsmallcaps{Doctor of Sciences}\\
(Dr.\ sc.\ ETH Zurich)
        \endgroup

        \vfill

        \begingroup
            presented by\\
            \vspace{0.5em}
            \spacedlowsmallcaps{\Large \myName}\\
            \vspace{0.5em}
            Diplom-Ingenieur, Graz University of Technology \\
            born on 26.04.1987
\endgroup

        \vfill

        \begingroup
            accepted on the recommendation of\\
            \vspace{0.5em}
            Prof.\ Dr.\ Onur Mutlu, examiner\\
            Dr.\ Fons Rademakers, co-examiner\\
            Prof.\ Dr.\ Can Alkan, co-examiner\\
            Dr.\ Arnau Montagud, co-examiner\\
            Dr.\ Mohammad Sadrosadati, co-examiner
        \endgroup

        \vfill

        \myTime 

        \vfill
    \end{center}
\end{titlepage}
\clearpage{}
\clearpage{}\thispagestyle{empty}

\hfill

\vfill

\noindent\myName: \textit{\myPlainTitle,} \textcopyright\ \myTime

\bigskip

\noindent\spacedlowsmallcaps{Latest version at}: \href{https://doi.org/\myDOI}{\myDOI}

\clearpage{}
\cleardoublepage\clearpage{}\thispagestyle{empty}
\refstepcounter{dummy}

\vspace*{3cm}

\begin{center}
    To my wife Olena, \\
    and to my parents Monika and Fritz
\end{center}

\medskip
\clearpage{}
\cleardoublepage\clearpage{}\pdfbookmark[1]{Abstract}{Abstract}
\begingroup
\let\clearpage\relax
\let\cleardoublepage\relax
\let\cleardoublepage\relax

\chapter*{Abstract}

Agent-based modeling is an indispensable tool for studying complex systems in
  biology, medicine, sociology, economics, and other fields.
However, existing simulation platforms exhibit two major problems: 1)
  performance: they do not always take full advantage of modern hardware
  platforms, which leads to low performance and 2) modularity: they often have a
  field-specific software design.
First, the low performance of many agent-based simulation platforms has at
  least four undesirable consequences: i) It prevents simulations that can model
  large numbers of agents or complex agent behaviors, which is necessary in
  modeling large-scale and complex systems, e.g., in biology and epidemiology.
ii) It increases the development time of agent-based simulations, which are performed iteratively, leading to much longer latencies in performing such studies.
iii) It limits the capability to explore the parameter space or sensitivity analyses, which may lead to suboptimal or even incomplete simulation results.
iv) It increases the monetary cost required for computing power.
Second, platforms with an inflexible software design make it challenging to
  implement use cases in different domains.
Modelers who do not find a simulation platform that can be easily extended
  without modifying the platform's internals may start developing their own
  simulation tool to satisfy their modeling needs.
This situation not only wastes resources in reimplementing already existing
  functionality but may also lead to compromises due to the complexity of
  developing a simulator and the often limited development resources.

This dissertation presents a novel simulation platform called \bdm{} and its major improvement \ta{} that
  alleviate the performance and modularity problems via three major works.

First, we lay the platform's foundation by carefully defining the abstractions
  and interfaces, setting up the required software infrastructure, and
  implementing a multitude of features for agent-based modeling.
We demonstrate \bdm{}'s functionality and modularity with three use cases in
  neuroscience, epidemiology, and oncology.
We validate these models with experimental data or an analytical solution,
  which also demonstrates the correctness of the \bdm{} implementation.
These models show that in \bdm{} additional functionality can be added easily,
  and \bdm{}'s out-of-the-box features allow for concise model definitions in the
  range of 128--181 lines of C\texttt{++} code.

Second, we extend the \bdm{} platform by performing a rigorous performance
  analysis of agent-based simulation and identifying three key performance
  challenges for shared-memory parallelism, for which we present solutions.
1) To maximize parallelization, we present an optimized grid to search for neighbors and parallelize the merging of thread-local results.
2) We reduce the memory access latency with a non-uniform memory access aware agent iterator, agent sorting with a space-filling curve, and a custom heap memory allocator.
3) We present a mechanism to omit the collision force calculation under certain conditions.
Our solutions result in a up to three orders of magnitude speedup over the
  state-of-the-art, and the ability to simulate 1.72 billion agents on a single
  server.

Third, we introduce a distributed simulation engine called \ta{} that allows scaling out
  the computation of one simulation to multiple servers.
Distributed execution requires the exchange of agent information between
  servers. We identify such information exchanges as the key bottleneck that prevents the
  distributed engine from scaling out efficiently, for which we present two main
  solutions.
1) We add a tailored serialization mechanism to avoid unnecessary work.
2) We extend the agent serialization mechanism with delta encoding to reduce the amount of data transfer.
We choose delta encoding to exploit the iterative nature of agent-based
  simulations.
Our solutions enable \ta{} to 1) simulate \ESMaxAgentsApprox{} agents (a
  \ESFactorImprovementSOTA{}$\times$ improvement over the state-of-the-art), 2)
  scale to \ESNumCPUs{} CPU cores, 3) significantly reduce the simulation time
  (e.g., \ta{} simulates an iteration of 800 million agents in \SI{0.6}{\second}
  instead of \SI{5}{\second}), and 4) significantly increase the visualization
  performance by 39$\times$.

Since its publication and open source release, researchers have used \bdm{} to
  study radiotherapy of lung cancer, vascular tumor growth, invasion of Gliomas
  into surrounding tissue, the formation of retinal mosaics in the eye, freezing
  and thawing of cells, neuronal geometries, the formation of the cortical layers
  in the cerebral cortex, the spread of viruses on a country-scale, and more.
PhysicsWorld named the radiotherapy simulation based on \bdm{} as one of the
  top 10 breakthroughs in physics in 2024.

\endgroup

\cleardoublepage 

\begingroup
\let\clearpage\relax
\let\cleardoublepage\relax
\let\cleardoublepage\relax

\begin{otherlanguage}{ngerman}
	\pdfbookmark[1]{Zusammenfassung}{Zusammenfassung}
	\chapter*{Zusammenfassung}

Die agentenbasierte Modellierung ist ein unverzichtbares Instrument zur
	  Untersuchung komplexer Systeme in Biologie, Medizin, Soziologie, Wirtschaft und
	  anderen Bereichen.
	Die bestehenden Simulationsplattformen weisen jedoch zwei Hauptprobleme auf: 1)
	  Leistung: Sie nutzen die Vorteile moderner Hardwareplattformen nicht immer voll
	  aus, was zu einer geringen Leistung führt, und 2) Modularität: Sie haben oft
	  ein feldspezifisches Softwaredesign.
	Erstens: Die geringe Leistung vieler agentenbasierter Simulationsplattformen
	  hat mindestens vier unerwünschte Folgen: i) Sie verhindert die Entwicklung von Simulationen, die
	  eine große Anzahl von Agenten oder komplexe Verhaltensweisen
	  von Agenten modellieren können, was bei der Modellierung großer und komplexer
	  Systeme, zum Beispiel in der Biologie und Epidemiologie, von großer Relevanz ist.
	ii) Es erhöht die Entwicklungszeit von agentenbasierten Simulationen, die iterativ durchgeführt werden, was zu viel längeren Wartezeiten bei der Durchführung solcher Studien führen kann.
	iii) Sie schränkt die Möglichkeit ein, den Parameterraum zu erkunden oder Sensitivitätsanalysen durchzuführen, was zu suboptimalen oder sogar unvollständigen Simulationsergebnissen führen kann.
	iv) Sie erhöht die Kosten für die erforderliche Rechenleistung.
Zweitens erschweren Plattformen mit einem unflexiblen Softwaredesign die
	  Implementierung von Simulationen in verschiedenen Anwendungsbereichen.
	Modellierer, die keine Simulationsplattform finden, die leicht erweitert werden
	  kann, ohne die Interna der Plattform zu modifizieren, beginnen eventuell damit,
	  ihr eigenes Simulationswerkzeug zu entwickeln, um ihre
	  Modellierungsanforderungen zu erfüllen.
	Diese Situation vergeudet nicht nur Ressourcen bei der Neuimplementierung
	  bereits vorhandener Funktionalität, sondern kann aufgrund der Komplexität der
	  Entwicklung eines Simulators und der oft begrenzten Entwicklungsressourcen auch
	  zu Kompromissen führen.

  In dieser Dissertation wird eine neuartige Simulationsplattform namens \bdm{} und ihre deutliche Verbesserung \ta{}
	  vorgestellt, die die Leistungs- und Modularitätsprobleme durch drei
	  Hauptarbeiten mildert.

	Erstens, legen wir das Fundament der Plattform, indem wir die Abstraktionen und
	  Schnittstellen sorgfältig definieren, die erforderlichen
	  Software-Infrastruktur bereitstellen und eine Vielzahl von Funktionen
	  für agentenbasierte Modellierung implementieren.
	Wir demonstrieren die Funktionalität und Modularität von \bdm{} anhand von drei
	  Anwendungsfällen in der Neurowissenschaft, der Epidemiologie und der Onkologie.
	Wir validieren diese Modelle mit experimentellen Daten oder einer analytischen
	  Lösung, was auch die Korrektheit der \bdm{}-Implementierung demonstriert.
	Diese Modelle zeigen, dass in \bdm{} zusätzliche Funktionalitäten leicht
	  hinzugefügt werden können, und das die verfügbaren Funktionen von \bdm{}
    kurze und prägnante Modelldefinitionen im Bereich von 128--181 C\texttt{++} Code Zeilen
	  ermöglichen.

	Zweitens erweitern wir die \bdm{}-Plattform, indem wir eine rigorose
	  Leistungsanalyse der agentenbasierten Simulationen durchführen und drei
	  zentrale Leistungsherausforderungen für Shared-Memory-Parallelität identifizieren, für die wir
	  Lösungen präsentieren.
	1) Um die Parallelisierung zu maximieren, präsentieren wir ein optimiertes Gitter für die Suche nach Nachbarn und die Parallelisierung der Zusammenführung von thread-lokalen Ergebnissen.
	2) Wir reduzieren die Speicherzugriffslatenz mit einem Agenten-Iterator für Systeme mit ungleichmäßigem Speicherzugriff, der Verwendung einer raumfüllenden Kurve zur Sortierung von Agenten und einem benutzerdefinierten Heap-Speicher-Allokator.
	3) Wir stellen einen Mechanismus vor, der die Berechnung der Kollisionskräften unter bestimmten Bedingungen überflüssig macht.
	Unsere Lösungen führen zu einer Beschleunigung von bis zu drei Größenordnungen gegenüber dem Stand
	  der Wissenschaft und ermöglichen es 1,72 Milliarden Agenten auf einem einzigen
	  Server zu simulieren.

  Drittens fügen wir eine verteilte Simulations-Engine namens \ta{} hinzu, die es \bdm{}
	  ermöglicht, die Berechnung einer Simulation auf mehrere Server zu
	  verteilen.
	Die verteilte Ausführung erfordert den Austausch von Agenteninformationen
	  zwischen Servern. Wir identifizieren diesen Informationsaustausch als den wichtigsten Engpass, der
	  eine effiziente Skalierung des verteilten Systems
	  verhindert, wofür wir zwei Hauptlösungen vorstellen.
1) Wir fügen einen maßgeschneiderten Serialisierungsmechanismus hinzu, um unnötige Arbeit zu vermeiden.
2) Wir erweitern den Agenten-Serialisierungsmechanismus mit Delta-Kodierung, um die Menge der Datenübertragung zu reduzieren.
	Wir wählen die Delta-Kodierung, um die iterative Natur von agentenbasierten
	  Simulationen zu nutzen.
Unsere Lösungen ermöglichen \ta{} 1) \ESMaxAgentsApproxDE{}
	  Agenten zu simulieren (eine \ESFactorImprovementSOTA{}-fache Verbesserung
	  gegenüber dem Stand der Wissenschaft), 2) auf \ESNumCPUs{} CPU-Kernen zu skalieren, 3) die Simulationszeit erheblich
	  zu reduzieren (z.B.
	simuliert \ta{} eine Iteration von 800 Millionen Agenten in \SI{0.6}{\second}
	anstelle von \SI{5}{\second}), und 4) die Visualisierungsleistung
	um das 39-fache zu steigern.

  \bdm{} wurde seit der ersten Publikation und Open-Source-Veröffentlichung verwendet um 
	  die folgenden Prozesse zu simulieren: Strahlentherapie von Lungenkrebs, vaskuläres
	  Tumorwachstum, Invasion von Gliomen in das umgebende Gewebe, die Bildung von
	  Netzhautmosaiken im Auge, das Einfrieren und Auftauen von Zellen, neuronale
	  Geometrien, die Bildung der Schichten in der Großhirnrinde, die
	  landesweite Ausbreitung von Viren und vieles mehr.
	Die Fachzeitschrift PhysicsWorld bezeichnete die auf \bdm{} basierende Strahlentherapiesimulation
	  als eine der Top 10 Durchbrüche in der Physik im Jahr 2024.

\end{otherlanguage}

\endgroup

\vfill
\clearpage{}
\cleardoublepage\clearpage{}\pdfbookmark[1]{Acknowledgments}{acknowledgments}

\bigskip

\begingroup
\let\clearpage\relax
\let\cleardoublepage\relax
\let\cleardoublepage\relax
\chapter*{Acknowledgments}

\def\thanks#1{\begingroup
	\leftskip1em
	\noindent #1
	\par
	\endgroup
}

This work would not have been possible without the help of many individuals.
First and foremost, I would like to thank my Ph.D.
advisor, Onur Mutlu (ETH Zurich), and my second advisor, Fons Rademakers (CERN).

I am deeply grateful for Onur's supervision, feedback, and mindset of always
  aiming for the highest standards, which strongly influenced my development as a
  researcher.
Onur's constant encouragement to keep refining and improving in all aspects of
  my research was a key factor in the successful publication of my work.
I am especially grateful for the time and effort Onur dedicated to enhancing my
  writing to improve clarity and impact.

I want to thank Fons for giving me the creative freedom to lead the technical
  aspects of the BioDynaMo project while keeping the door open for any
  discussions.
I am deeply grateful for Fons' feedback and insights on growing and driving
  open-source projects based on his experiences founding and leading CERN's main
  data analysis platform, ROOT.

I want to express my gratitude to my co-examiners, Dr.
Mohammad Sadrosadati, Dr.
Arnau Montagud, and Prof.
Can Alkan, for their valuable time and insightful comments, which improved this
  dissertation.

I also want to thank all my co-authors Ahmad Hesam, Jean de Montigny, Vasileios
  Vavourakis, Alexandros Iosif, Jack Jennings, Marcus Kaiser, Marco Manca,
  Alberto Di Meglio, Zaid Al-Ars, Roman Bauer, Juan G\'{o}mez Luna, Abdullah
  Giray Yaglikci, Mohammad Sadrosadati, Tobias Duswald, Thomas Thorne, Barbara
  Wohlmuth, Frank P. Pijpers, Peter Hofstee, Sanja Bojic, Alex Sharp, Fons Rademakers, and Onur
  Mutlu for their collaboration, hard work, and valuable feedback.

I am very grateful to the SAFARI Research Group, CERN openlab, the CERN
  Knowledge Transfer Fund, the CERN Medical Application Office, the ETH Future
  Computing Laboratory, the BioPIM project, and the SAFARI Research Group's
  industrial partners including Huawei, Intel, Microsoft, and VMWare for funding
  and believing in my research.

The development of the distributed simulation engine was only possible with
  access to a supercomputer.
I, therefore, welcome the opportunity to thank SURF for the access to the Dutch
  National Supercomputer Snellius and Ahmad Hesam and Zaid Al-Ars for their help
  in applying for the grant.

Thanks to Axel Naumann from the ROOT Team and Vassil Vassilev from Princeton's
  Compiler Research Group for their help and discussions regarding the \cpp{}
  interpreter cling.

I also want to thank the CERN data center administrators Luca Azori, Joaquim
  Santos, Krzysztof Mastyna, and Guillermo Moreno, who swiftly resolved problems
  with our computing infrastructure.

I also want to thank all the anonymous reviewers whose feedback has helped
  strengthen my work and all other researchers and students who contributed to
  the \bdm{} platform.

My acknowledgments would not be complete without thanking Kristina Gunne, Tracy
  Ewen, Tulasi Blake, and Christian Rossi, for their timely help in all
  administrative matters and navigating the sometimes overwhelming bureaucracy of
  CERN, ETH, and international collaboration.

I owe a huge thanks to Ahmad Hesam and Tobias Duswald.
We spent countless late nights together, often being the last ones in the
  building.
Your company and friendship made these intense times much better.

I consider myself fortunate to have crossed paths with an incredible number of
  amazing individuals at CERN and ETH Zurich.
There are too many to mention individually, so I'd like to express my
  collective gratitude for their friendship, countless shared laughs, and
  unforgettable experiences outside the office.

I want to express my gratitude to my family for their endless support.
My parents, Monika and Fritz, have always encouraged me to have a positive
  attitude, a growth mindset, and to persevere.
My sisters, Judith and Magdalena, and grandparents bring so much joy into my
  life.
I'm also thankful for my warm and loving relationship with my parents-in-law,
  Vitalii and Lidiia.

Above all, I am most grateful for my wife, Olena, who has shown me endless
  love, support, patience, understanding, and encouragement.
She has always been there for me, and I appreciate her more than words can
  express.

\endgroup
\clearpage{}
\cleardoublepage\clearpage{}\phantomsection
\addcontentsline{toc}{chapter}{Contents}
\tableofcontents
\newpage
\listoffigures
\newpage
\listoftables

\setcounter{tocdepth}{3} \setcounter{secnumdepth}{3} \clearpage{}

\cleardoublepage\pagenumbering{arabic}\setstretch{1.3}
\clearpage{}\chapter{Introduction}
\label{ch:introduction}

Computer simulations have become integral to science in finding answers to
  questions that arise in complex systems.
Simulations are particularly useful for studying models where analytical
  solutions become impossible, or where experimentation is ``economically
  infeasible, ethically inappropriate, or ecologically dubious''
  \cite{weinzierl_pillars_2021}.
Joshua Epstein writes that models---and thus subsequently simulations---are not
  only used to make predictions, but also to ``explain, guide data collection,
  discover new questions, educate'' \cite{epstein_why_2008}, and 12 other
  reasons.
Due to computer simulations' importance in generating new insights, many
  researchers see simulation as a third pillar of science, complementing theory
  and experimentation \cite{weinzierl_pillars_2021}.

An important class of simulation is agent-based modeling (ABM), also called
  individual-based modeling.
Since its beginnings \cite{schelling_segration_1971}, ABM has come a long way
  and is now used in biology \cite{\abmBiology}, medicine \cite{\abmMedicine}, 
  epidemiology \cite{\abmEpidemiology}, social sciences \cite{\abmSocialSciences}, 
  finance and economics \cite{\abmFinanceAndEconomics},
  ecology \cite{\abmEcology}, transport \cite{\abmTransport}, 
  and more \cite{\abmOther}.
Agents or individuals are entities, which have attributes and a set of rules
  that govern their behavior and interactions.
Agents interact only with their local neighborhood.
Depending on the use case, these abstract entities can represent a person, a
  cell, a neurite segment, or any other individual that interacts locally.

Using three examples, we want to demonstrate that ABM is a versatile approach
  to model dynamic systems.
First, Craig Reynolds used ABM to model the swarm dynamics of birds
  \cite{reynolds_flocks_1987}.
In this example, agents are birds with a position and velocity.
Birds avoid collisions with each other, try to stay close to each other, and
  align their heading with their neighbors.
This simple model is sufficient to replicate realistic swarm movements.

Second, a different use case is cell sorting of two cell types
  (Section~\ref{sec:platform:soma-clustering}), which are initially randomly
  distributed in space (Figure~\ref{fig:soma-clustering}).
In this model, the agent is a spherical cell with a position, diameter, and
  cell type.
These cells secrete a substance and move toward high substance concentration
  (chemotaxis).
Over time, the two cell types separate and form clusters of the same type.

Third, ABM can be used in microeconomics to model the dynamics of a single-good
  market \cite{axtell2024}.
This model comprises two agent types: buyers and sellers.
Buyers have a maximum price they are willing to pay and a quantity they need to
  buy.
Sellers have a cost associated with producing the good and quantities they can
  supply.
A transaction happens if the buyer's offer exceeds the seller's price.
Agents adjust their prices, based on their preferences and the current market
  situation.
The price of the traded good emerges from the trades of the individual agents
  and may move toward an equilibrium in which supply and demand are balanced.

\section{Problem Discussion}

We believe there are three major problems in ABM that need to be addressed to
  realize its full potential.
These issues are 1) performance, scalability, and efficiency; 2) inflexible
  software design; and 3) software quality and reproducibility.

\paragraph{Performance, Scalability and Efficiency.}
Existing simulation tools are not efficient and performant enough (for
  extreme-scale simulations with \ESDefinition{} agents).
Since the slowdown of Moore's law \cite{moores-law} and end of Dennard scaling
  \cite{dennard_design_1974}, computing hardware has become increasingly parallel
  and heterogeneous.
Although the computational capacity kept growing at a high rate---which can be
  observed impressively on the TOP500 list \cite{top500perf}---without an
  efficient, parallelized, and distributed simulation engine, this processing
  power cannot be used for agent-based simulation.
In addition to parallelization and distribution, performance depends
  significantly on optimizations specific to the agent-based workload.
Many agent-based simulations are memory-bound
  \cite{breitwieser_biodynamo_2023}.
Therefore, their performance depends substantially on the memory layout of
  agents.
In the distributed setting, object serialization and data transfers are two
  important overheads that must be addressed (Section~\ref{ch:dse}).

The effects of low performance are exacerbated for ABM, because the simulation
  has to be executed many times for at least two reasons.
First, agent-based models are developed iteratively.
In several iteration cycles, the model is refined to match the studied
  phenomena, which requires at least one model execution per iteration cycle.
Second, parameter optimizations and sensitivity analyses might require a large
  number of model executions with different parameters, which can reach tens of
  millions \cite{duswald2024calibrationstochasticagentbasedneuron}.

\paragraph{Inflexible Software Design.}
An inflexible software design makes it hard for modelers to adapt and implement
  different simulations, increases the maintenance effort and software
  complexity, causes a steeper learning curve, and reduces the development speed.
An inflexible design may be caused by too much specialization to a specific
  domain or a lack of development resources to introduce well-defined interfaces
  and components.

In the context of ABM, an inflexible software design leads to further
  challenges as modelers decide to develop their own simulation tool if their
  model necessitates substantial modifications to the underlying simulation
  engine to function correctly.
We hypothesize that these modelers avoid the steep learning curve of
  understanding a perhaps large code base and the changes required to adapt it.
It appears easier to start from scratch, focusing only on the functionality
  needed for the specific model.
However, modelers often underestimate the effort required to develop an entire
  (agent-based) simulator.
The development requires domain knowledge and competence in numerical methods,
  software architecture, parallel and distributed computing, high-performance
  computing, and verification.
Lack of development time, funding, or workforce makes it hard to focus on all
  functional and non-functional requirements.
This limitation leads to further issues and negatively impacts software
  performance, modularity, quality, and reproducibility.

We identify inflexible software design as the root cause of these challenges
  and aim for a modular software design for the agent-based simulation platform
  presented in this dissertation.

\paragraph{Software Quality and Reproducibility.}

Software quality compromises lead to many issues, including reproducibility of
  results, increased maintenance effort, slow development speed, software
  crashes, and poor user experiences.
Achieving high software quality is not a one-time task, but a continuous,
  time-consuming, and labor-intensive process that demands constant attention.
To achieve high software quality, a project requires three key elements: 1) the
  necessary infrastructure, such as the ability to execute automated tests on all
  supported systems \cite{github_actions}, 2) established processes like
  test-driven development \cite{beck2000extreme}, and 3) a development team with
  the right mindset to adhere to best practices, improve them, and refrain from
  taking shortcuts.
Software quality is closely related to the reproducibility of results, a key
  concern in science.
The literature contains several articles that address the issue of scientific
  software falling short to reproduce results \cite{gewaltig_current_2014,
	  schutter_collaborative_2013, neylon_changing_2012, baxter_scientific_2006,
	  schutter_why_2008, nature_swpr_2018, medley_guidelines_2016}.
Hidden errors and inadequate testing can compromise the simulation results and
  invalidate the derived scientific insights.
In the past, this has even led to the retraction of published scientific
  manuscripts \cite{miller_scientists_2006}.

\section{Our Approach}

To address the aforementioned issues, we design and implement a novel
  agent-based simulation platform from the ground up.
We particularly focus on performance, scalability, modularity, reproducibility,
  and software quality.

We approach our solution in three main steps.
First, we lay the foundation of the project by developing all necessary
  infrastructure, defining interfaces and abstraction layers, and implementing a
  rich set of agent-based features.
These features are divided into (1) low-level functionality, which is
  transparent to the user; (2) high-level agent-based functionality, which is
  needed in models across many domains; and (3) domain specific model building
  blocks.
Second, we build a highly-efficient simulation engine using shared-memory
  parallelism on one server.
We perform a detailed performance analysis and develop solutions to increase
  the parallel fraction of the agent-based algorithm, improve the memory layout
  and data access patterns, and exploit domain knowledge to avoid unnecessary
  work.
Third, we add a distributed simulation engine, which allows the execution of
  one simulation on multiple servers and thus enables extreme-scale simulations
  with \ESMaxAgentsInTrillion{} agents.
This final engine also improves the serialization performance with a tailored mechanism
and reduces the required data transfers.

\section{Thesis Statement}

The thesis of this dissertation in computer science is:

\begin{quotation}
\noindent
	An agent-based simulation platform
	that is designed from the ground up to be high-performance, scalable, and modular
	can
	\begin{enumerate}
		\item significantly reduce the simulation runtime, thereby enabling larger and more
		      complex simulations, faster iterative development, and more extensive parameter
		      exploration and
		\item significantly improve adoption thereby enabling advances in many different domains.
	\end{enumerate}
\end{quotation}

\section{Contributions}
\label{ch:introduction:contributions}

This dissertation makes the following major contributions.

\begin{enumerate}
	\item We present a novel modular agent-based simulation platform called \bdm{}.
	      We provide a rich set of agent-based features commonly used in models across
	        different domains (Chapter~\ref{ch:platform}).
	      Low-level features (Section~\ref{sec:low-level-features}) are hidden from the
	        user and include parallelization, hardware acceleration, visualization,
	        web-based interface, and more.
	      High-level features (Section~\ref{sec:high-level-features}) are exposed to the
	        users and allow them to generate agent populations, perform statistical
	        analysis, simulate processes on different temporal scales, and more.
	      Model building blocks (Section~\ref{sec:model-building-blocks}) provide
	        functionality for a specific domain, for example, agent definitions for cells
	        and neurons.
	      We demonstrate this functionality and \bdm{}'s modular software design with
	        three simple yet representative simulations in the field of neuroscience
	        (Section~\ref{sec:pyramidal-cell}), oncology
	        (Section~\ref{sec:oncology-use-case}), and epidemiology
	        (Section~\ref{sec:epidemiology}).
	      These models show that additional functionality can be added easily, and
	        \bdm{}'s out-of-the-box features allow for concise model definitions with only
	        128 (Listing~\ref{lst:pc}), 154, and 181 lines of C\texttt{++} code.
	      \bdm{}'s functionality, modularity and flexibility are further examplified
	        through its application in simulating (vascular) tumor growth
	        \cite{duswald_2024, demontigny_2021}, (radiation-induced) lung fibrosis
	        \cite{cogno_mechanistic_2024, cogno_3d_2022, cogno_agent-based_2022,
            cogno_thesis} (named a top 10 breakthrough in physics in 2024 \cite{physicsworld-top10}), 
            formation of retinal cells during early development
	        \cite{demontigny_2023}, neuronal growth
	        \cite{duswald2024calibrationstochasticagentbasedneuron,
		        abubacar_neuronal_growth}, socioeconomic phenomena for the whole Dutch
	        population \cite{hesam_2024}, the spread of HIV in Malawi (Africa)
	        \cite{hiv_malawi}, freezing and thawing of tissue \cite{jennings_cryo,
		        jennings_cryo2}, and more
	        \cite{demetriades_interrogating_2022,gazeli_interrogating_2022, bdm-workshop}.
	      The platform's widespread use emphasizes its value for the broad agent-based
	        modeling community that spans many scientific domains.

  \item We present a high-performance and efficient simulation engine for \bdm{} using shared-memory parallelism.
	      The evaluation shows that \bdm{}'s simulation engine is 23$\times$ faster than
	        state-of-the-art serial simulators \cite{netlogo, zubler_simulating_2013} using
	        only one CPU core and three orders of magnitude faster than state of the art using
	        72 CPU cores (Section~\ref{sec:eval:comparison-with-cx3d-netlogo}).
	      We demonstrate that parallel efficiency with 72 physical cores and
	        hyperthreading enabled is 91.7\% (Section~\ref{ch:se:eval:scalability}).
	      The single-node engine can simulate 1.72 billion agents on a single server
	        (Section~\ref{sec:eval:comparison-with-biocellion}).
	      Building on our work, Duswald \etal{}
	        \cite{duswald2024calibrationstochasticagentbasedneuron} demonstrated that the
	        engine's efficiency enables high-throughput computing in which they calibrated
	        a neuron model using 50 million individual simulations.

  \item We present \ta{}, a distributed simulation engine for
	      extreme-scale simulations.
        The presented distributed engine is capable of 1) simulating \emph{\ESMaxAgentsExact{}
            agents} (Section~\ref{sec:de:eval:extreme-scale}), 2) scales to \ESNumCPUs{} CPU
	        cores (Section~\ref{sec:de:eval:scalability}), 3) significantly reduces the
          simulation time over the shared-memory parallelized \bdm{} version (e.g., \ta{}
	        simulates an iteration of 800 million agents in \SI{0.6}{\second} instead of
	        \SI{5}{\second}) (Section~\ref{sec:de:eval:scalability}), and (4) significantly
          increases the in-situ visualization performance by 39$\times$ over \bdm{}
	      (Section~\ref{sec:de:eval:interoperability}).

	\item We present six performance optimizations for the
        single-node engine \bdm{} that maximize the parallel part of the
	      agent-based algorithm (Section~\ref{sec:opt:maximize-parallelization}), improve the memory layout (Section~\ref{sec:opt:memory-layout}),
	      and avoid unnecessary work (Section~\ref{sec:static-agents}).
	      Together, these improvements speed up the simulation runtime between
	        33.1$\times$ and 524$\times$ (median: 159$\times$)
	        (Section~\ref{sec:se:eval:optimization-overview}).

  \item We present two performance improvements for the distributed simulation engine \ta{}, addressing the significant overhead of exchanging agents between different processes.
	      These exchanges are necessary because the agents are distributed across
	        different processes and are not all available within the same memory space as
	        in \bdm{}.
	      For each exchange, the selected agents and all their attributes are packed in a
	        contiguous buffer (i.e., serialization) before they can be sent to another
	        process.
	      The receiving process has to unpack the buffer before the agents can be used
	        (i.e., deserialization).
	      In our first optimization, we develop a tailored serialization mechanism
	        (Section~\ref{sec:design:serialization}).
	      This mechanism serializes the agents up to 296$\times$ faster (median
	        110$\times$).
	      It also improves deserialization performance with a maximum observed speedup of
	        73$\times$ (median 37$\times$) (Section~\ref{sec:de:eval:serialization}).
	      Second, we reduce the amount of data transferred by exploiting the iterative
	        nature of ABM.
	      We use delta encoding to transfer a compressed difference
	        (Section~\ref{sec:design:delta-encoding}), which reduces the data volume by up
	        to 3.5$\times$ (Section~\ref{sec:de:eval:delta-encoding}).

	\item This dissertation describes the development approach to achieve
	      high software quality.
	      Over 600 automated tests were developed to ensure the correctness of our
	        open-source simulation platform (Section~\ref{ch:platform:software-quality}).

	\item This dissertation provides a rigorous performance analysis of the
	      simulation engines and exclusive insights into the agent-based workload
	      characteristics (Section~\ref{sec:se:evaluation}).
	      These insights can be used to improve the performance of other agent-based
	        simulators.

\end{enumerate}

\section{Dissertation Outline}

This dissertation comprises 7 chapters.
Chapter~\ref{ch:background} provides background on agent-based simulation.
Chapter~\ref{ch:related-work} provides an overview of related work.
Chapter~\ref{ch:platform} presents the \bdm{} platform, details its user-facing
  features, and introduces three use cases in the domain of neuroscience,
  epidemiology, and oncology (based on \cite{breitwieser-bdm}).
Chapter~\ref{ch:engine} goes into depth about the performance-related aspects
  of the simulation engine, its optimizations for shared-memory parallelism, and
  performance evaluation (based on \cite{breitwieser_biodynamo_2023}).
Chapter~\ref{ch:dse} presents the design of the distributed simulation engine \ta{}
  that enables the execution of \emph{one} simulation on multiple servers.
Chapter~\ref{ch:conclusion-future-work} summarizes our findings and examines
  possible future research directions.
The appendix shows 
  (\ref{app:other-works} and \ref{app:contribution}) a complete list of the author's contributions 
  , (\ref{app:complex-simulations}) selected simulations without the author's
  involvement that demonstrate that the
  functionality provided by the \bdm{} platform can be used to model dynamic
  systems whose complexity exceeds the presented use cases in this dissertation,
  (\ref{app:table}) a list of agents, events, and operations,
  and (\ref{app:tutorials}) all supplementary tutorials.
\clearpage{}

\cleardoublepage \clearpage{}\chapter{Background}
\label{ch:background}

This section provides a deep dive into agent-based modeling, its underlying
  principles, comparison with other modeling paradigms, application domains, and
  challenges and limitations.

\section{What is Agent-Based Modeling?}

Although the term "agent-based modeling" lacks a precise definition in the
  literature, Macal outlines four characteristics of agent-based models that are
  commonly observed \cite{macal_everything_2016}.
Agent-based models involve (1) individuals with diverse states who (2)
  autonomously follow behavioral rules, (3) interact with each other or their
  immediate environment, and (4) adjust their behavior to achieve specific
  objectives \cite{macal_everything_2016}.
Of these four attributes, only the first two seem essential for an agent-based
  model, while the other two are optional.
Other works emphasize the importance of emergent behavior (i.e., behavior on
  the macro-level that occurs through the interactions on the micro-level) in
  agent-based models, which requires agent interaction \cite{bankes2002,
	  bandini2009}.

\subsection{Agents and Behaviors}

The term "agent" is an abstract placeholder for the individual autonomous
  entities within a complex system.
There are almost no limits to what an agent could represent.
This dissertation describes simulations in which an agent is a person in an
  infectious disease scenario (Section~\ref{sec:epidemiology}), a subcellular
  structure to simulate the development of a neuron
  (Section~\ref{sec:pyramidal-cell}), a cell to simulate cancer development
  (Section~\ref{sec:oncology-use-case}), or a blood vessel segment to simulate
  the interplay between tumor and vascular system \cite{duswald_2024}.
Further examples from the literature include vehicles in traffic simulation
  \cite{\abmTransport}, animals in an ecosystem \cite{\abmEcology}, or traders in
  a financial market \cite{axtell2024}.
Simulations can also contain multiple agent types.

Behaviors define the actions of individual agents depending on their state and
  (perhaps) their environment.
Examples of behaviors in epidemiological simulations are infection, disease
  progression, and recovery.
In biological models, behaviors could be movement, secretion, chemotaxis, and
  cell proliferation.

Behaviors might affect 1) its agent's state, 2) neighboring agents, or 3) the
  environment.
For instance, movement only changes the individual's position.
Meanwhile, secretion alters the substance concentration of the extracellular
  space, affecting the environment.
In an immunological model, a cytotoxic T-cell might cause a neighboring
  cancerous cell to undergo apoptosis.
If a behavior falls into category one or two is sometimes only
  implementation-dependent.
The infection behavior in an epidemiological model might be formulated as ``the
  agent infects \emph{itself} if an infected agent is nearby and the agent is
  susceptible''.
Alternatively, the same result can be achieved with ``if the agent is infected,
  it infects \emph{nearby susceptible agents}''.
Although the simulation outcome will be identical, the first option is
  favorable from a performance perspective, because in simulations where
  behaviors modify neighboring agents, additional thread synchronization is
  required to avoid race conditions, which degrades performance.

Agents and behaviors can be classified into different groups.
Here, we will explore several general categories: deterministic versus
  stochastic, reactive versus proactive and learning \cite{bandini2009,
	  axtell2024}.

Deterministic behaviors generate the same output for the same input.
For instance, a cell's movement to higher substance concentrations (i.e.,
  chemotaxis) is deterministic, while random agent movement (e.g., Brownian
  motion) is stochastic.
Stochastic (i.e., random) behaviors use pseudo-random numbers in their
  implementation leading to different output for repeated executions with the
  same input.

Bandini defines reactive agents as those that respond to the environment they
  perceive \cite{bandini2009}.
In contrast, proactive or cognitive agents use a mental model of the
  environment to guide their actions.
Axtell and Farmer call agents with a mental model as learning agents
  \cite{axtell2024}.

\subsection{Environment and Interactions}
Agents operate within an environment that facilitates their interactions.
Most simulations in this dissertation are based on a 3D space environment,
  which allows agents to search for neighbors within a specific distance for
  interaction.
Environments can also take different forms, such as graph-based environments,
  where agents are represented as graph vertices and can interact with connected
  agents through edges.
A graph-based environment can effectively model social interactions.
In financial simulations, an environment could take the form of a virtual stock
  exchange \cite{axtell2024}.

Environments often have boundary conditions that determine the characteristics
  near the simulation borders.
In three-dimensional simulations, the space may 1) expand to include agents
  that have moved beyond the previous iterations' simulation space, 2) be closed,
  repelling agents that attempt to escape, 3) be periodic (e.g., a torus), where
  agents exiting on one side re-enter on the opposite side, or 4) be absorbing,
  removing agents that cross the border from the simulation.

Agents interact with the environment in various ways.
Here, we give three examples.
First, there are interactions between agents themselves.
For example, cells in contact exert mechanical forces on each other.
These forces represent a symmetric interaction.
Asymmetric interactions can be seen in predator-prey relationships in ecology
  or between immune cells and pathogens.
Second, we have interactions between agents and the environment.
In financial simulations, agents might place orders at a virtual stock exchange
  (the environment) and then observe the price with some delay.
Third, there are hybrid interactions involving both agents and the environment.
For instance, in our body's circulatory system, cells may release a substance
  into the extracellular matrix.
This substance is then diffused by the environment and sensed by a distant
  cell.

\subsection{Emergence}
\label{ch:intro:emergence}

In many agent-based models, emergent phenomena \cite{heath2010} are a key
  aspect.
Bankes attaches central importance to this phenomenon but states that the term
  is often ill-defined and lacks a general algorithm to test if emergence
  happened in a simulation \cite{bankes2002}.
Emergent phenomena occur due to the interactions of individual agents with each
  other and the environment.
This emergence occurs at a higher level (i.e., macro-level) than the behavior
  of the individual agent (i.e., micro-level).
It follows the principle of ``more is different'' as described by Anderson
  \cite{anderson1972}.
Colloquially, emergence is often expressed as ``the whole is greater than the
  sum of its parts``.

Emergence can manifest in various forms, much like the diverse range of agents
  and behaviors resulting from different use cases.
Examples of emergent behavior include swarm dynamics of a flock of birds
  \cite{reynolds_flocks_1987}, hospitalization patterns during a pandemic
  \cite{hesam_2024}, traffic jams \cite{scerri2020}, wealth distribution
  \cite{epstein1996growing}, and many more.

\section{Where is Agent-Based Modeling Used?}
Different domains have influenced agent-based modeling, each with its own
  terminology.
In ecology and biology, researchers often refer to ABM as individual-based
  modeling
  \cite{an2021,grimm2013,deangelis2005,kreft2001,metzcar_review_2019,railsback_agent-based_2006,getz_agent-based_2019,thorne2007abm}.
According to \citep{macal_everything_2016}, the term multi-agent systems (MAS)
  \cite{lesser_cooperative_1978,uhrmacher2018,dorri2018,marietto_requirements_2003}
  is commonly used interchangeably with ABM.
This term is prevalent in computer science, particularly in swarm robotics and
  distributed artificial intelligence \cite{macal_everything_2016, panait2005}.
Penait and Luke state that MAS represents one of the two categories in
  distributed artificial intelligence, with the other being distributed
  problem-solving \cite{panait2005}.
In economics, the term agent-based computational economics (ACE) is sometimes
  used \cite{axtell2024}.

\subsection{A Brief History of Agent-Based Modeling}

\begin{sidewaysfigure}
\includegraphics[width=\textwidth]{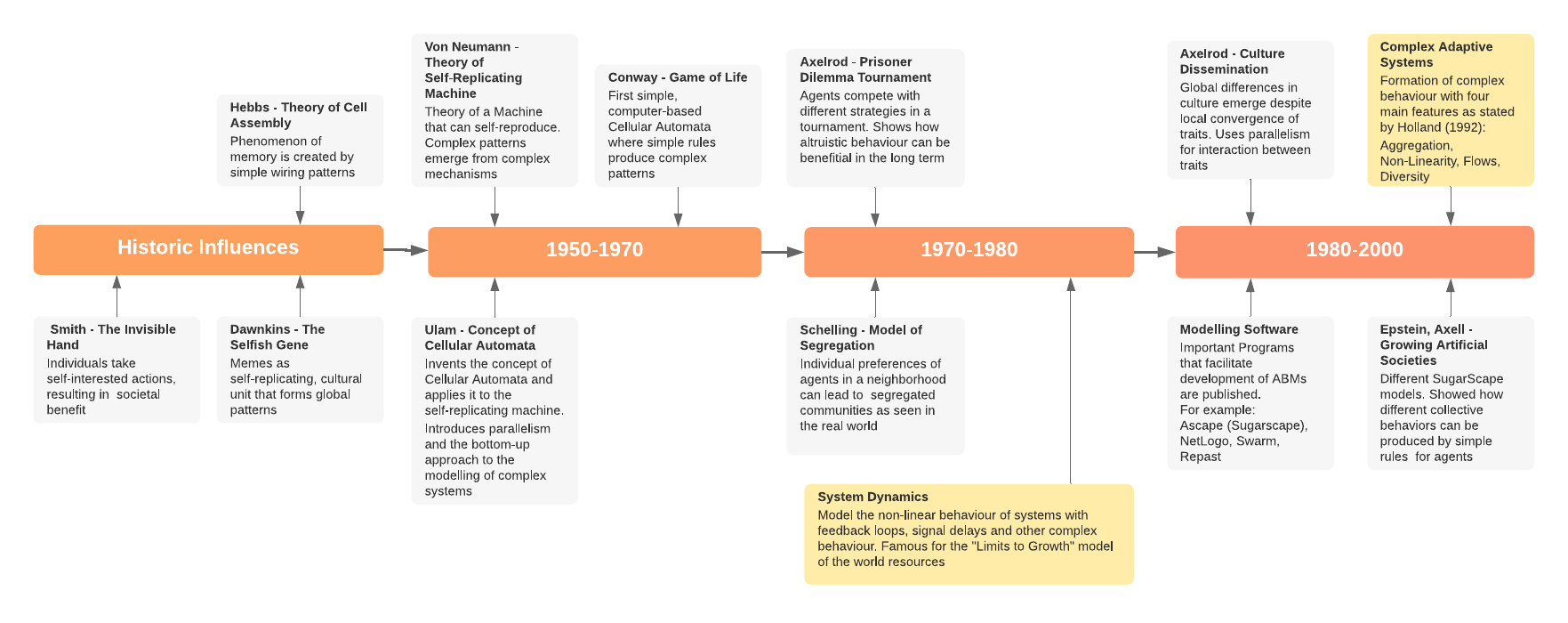}
\caption{Timeline part one of two of agent-based modeling from the social sciences perspective
		from Carl Orge Retzlaff \etal{} \cite{retzlaff2021}.
		Figure taken from \cite{retzlaff2021}, split in two parts, and used under the
		  \href{https://opensource.org/license/MIT}{MIT license}.
	}
	\label{fig:background:timeline1}
\end{sidewaysfigure}

\begin{sidewaysfigure}
\includegraphics[width=\textwidth]{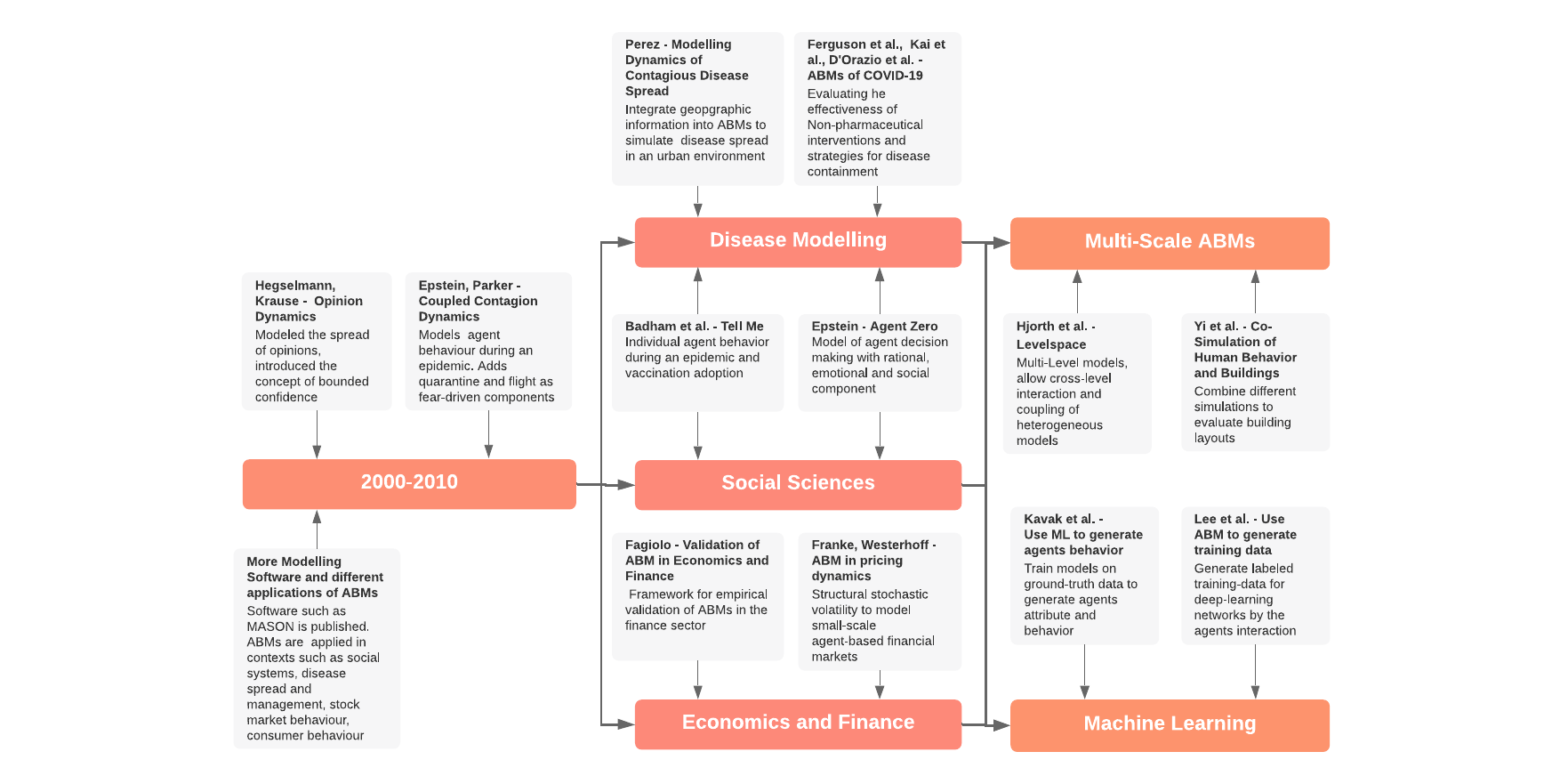}
\caption{Timeline part two of two of agent-based modeling from the social sciences perspective
		from Carl Orge Retzlaff \etal{} \cite{retzlaff2021}.
		Figure taken from \cite{retzlaff2021} split in two parts and used under the
		  \href{https://opensource.org/license/MIT}{MIT license}.
	}
	\label{fig:background:timeline2}
\end{sidewaysfigure}

The cellular automaton is a precursor to agent-based modeling as we know it
  today \cite{heath2009, liederkerke2015}.
Cellular automata are computer models where individual agents occupy a 2D grid
  (or lattice) and interact with neighboring grid cells.
Von Neumann and Moore are commonly used neighborhood patterns in these models
  \cite{neumann1966}.
Von Neumann was building upon the early work of Stanislaw Ulam \cite{ulam1952},
  which dates back to the 1950s.
Perhaps the best-known cellular automatons are Conway's ``Game of Life''
  \cite{gardner1970}, and Schellings ``Model of Segregation''
  \cite{schelling_segration_1971}, which may be seen as one of the first
  agent-based models \cite{macal_everything_2016}.
Seminal ABM works studied the ``Evolution of cooperation'' \cite{axelrod1981},
  flocking behavior \cite{reynolds_flocks_1987}, and ``the growth of artificial
  societies'' \cite{epstein1996growing}.
Figure~\ref{fig:background:timeline1} and \ref{fig:background:timeline2} show a
  timeline of ABM from the social sciences perspective \cite{retzlaff2021}.

The application areas of ABM are incredibly diverse, which we would like to
  describe in more detail below.
This overview is not meant to be exhaustive, but rather to outline the systems
  and problems that can be effectively represented using an agent-based approach.


\subsection{Biological Cell and Tissue Models}
\label{ch:background:biology}

ABM is extensively used in biology to model cancer \cite{\abmCancer}, the
  nervous system \cite{\abmNervousSystem}, morphogenesis \cite{\abmMorphoGenesis}
  biofilms \cite{\abmBiofilms}, and many more \cite{\abmBiologyOther}.
Simulations in these fields are often multi-scale to simulate processes that
  occur on different temporal or spatial scales, for example, cell division and
  diffusion.
Development and validation of these models often include laboratory experiments
  that have an impact on each other \cite{thorne_combining_2007}.
The laboratory results are used to improve the model, which may necessitate
  further validation and, consequently, additional laboratory experiments if the
  data is unavailable.

An important subgroup is models to study cancer development \cite{\abmCancer}
  which investigate the pathogenesis, invasion, interaction with the vascular
  system, and responses to treatment.
One advantage of the agent-based approach is its ability to integrate multiple
  data sources such as imaging or gene expression \cite{cogno_biomedicin_2024}
  and consider various interactions.
Agents are typical cells, subcellular structures, or larger tissue components
  like a vessel segment.
Cells may incorporate a state machine to distinguish between hypoxic,
  proliferative, quiescent, or necrotic conditions \cite{duswald_2024}.
Spatial environments are typically used in agent-based cancer models, and the
  mechanical properties of the tissue are modeled using pairwise forces between
  agents.
Metzgar categorizes the implementation of these models as on-lattice and
  off-lattice \cite{metzcar_review_2019}.
On-lattice methods include the already mentioned cellular automaton and the
  cellular Potts model (CPM) \cite{graner_simulation_1992} extension, in which
  agents can occupy multiple grid positions.
Off-lattice methods eliminate the constraints of discrete grid elements,
  allowing the agent to occupy any position in space.

In neuroscience, researchers use ABM to model the growth of neurons
  \cite{\abmNeuronGrowth}, the lamination of the cerebral cortex
  \cite{\abmCorticalLamination}, or neurodegenerative diseases
  \cite{pennisi2015}.
Neurons can be modeled as a tree of neurite segment agents, which elongate,
  retract, branch, bifurcate, form synapses, and grow toward attracting chemical
  cues \cite{zublerdouglas2009framework}.

\subsection{Epidemiology}
\label{ch:background:epidemiology}

Agent-based modeling allows for more realistic simulations of epidemiological
  questions and provides insights into the causal mechanisms
  \cite{\abmEpidemiology}, which provides benefits over compartmental models
  \cite{kermack_1927,alser_covidhunter_2022}.
Epidemiological applications include modeling the dynamics of infectious
  disease spread (such as COVID-19 \cite{\abmCOVID}, influenza
  \cite{\abmInfluenza}, malaria \cite{\abmMalaria}, and HIV \cite{\abmHIV}),
  assessing public health interventions \cite{\abmPublicHealthInterventions},
  epidemic forecasting \cite{\abmEpidemicForecasting}, appraising pandemic
  preparedness and response \cite{\abmEpidemicPreparedness}, understanding the
  dynamics in urban areas \cite{\abmUrbanAreas}, and more
  \cite{\abmEpidemiologyOther}.

ABM allows epidemiological models to generate heterogeneous populations of
  agents (i.e., people) that may consider personal risk factors, the settlement
  structure of a region, movement patterns, and personal preferences (e.g.,
  vaccination hesitancy \cite{\abmVaccinationHesitancy}).
Recent work, for example, incorporates data from the Dutch National Statistics
  Bureau to more accurately model the spread of COVID-19 and their corresponding
  hospitalizations \cite{hesam_2024}.
Another model from Ozik \etal{} provides actionable results and guided
  decision-making during the COVID-19 pandemic \cite{ozik_citycovid_2021}.
This work also reported the detection of health inequalities in the population,
  which can be a valuable insight to improve the future healthcare system.

\subsection{Medicine}
The application of agent-based modeling in medicine is multifaceted.
Besides use cases that overlap with biology
  (Section~\ref{ch:background:biology}) and epidemiology
  (Section~\ref{ch:background:epidemiology}), ABM is used for treatment planning
  \cite{\cognoLungInjury}, chronic disease management \cite{\abmChronicDiseases},
  healthcare system optimizations \cite{\abmHealthcareSystemOpt}, immune system
  response \cite{\abmImmuneSystem}, pharmacodynamics \cite{\abmPharmacoDynamics},
  and others.

ABM's strength lies in its ability to personalize treatment planning by
  enabling mechanistic models that consider a patient's individual attributes.
This personalized medicine approach may improve decision-making over
  phenomenological models that depend on clinical experience and possibly limited
  patient data.
Cogno et al., for example, laid the groundwork for one such mechanistic model
  in their simulation of radiation-induced lung injury \cite{\cognoLungInjury}
  using \bdm{} (Figure~\ref{fig:app:gallery:lung-tissue}).

Chronic diseases, including diabetes, obesity, cardiovascular diseases, and
  multimorbidities, negatively affect the patient's life and the healthcare
  system as a whole \cite{li2016}.
For patients, these diseases often lead to long-term physical discomfort,
  reduced quality of life, and reduced life expectancy.
Healthcare systems are under substantial financial strain due to the high
  prevalence of these diseases.
Agent-based models are used to study and improve the management and prevention
  of chronic diseases \cite{\abmChronicDiseases} to improve the patient's
  well-being and the cost-effectiveness of medical interventions.

Agent-based models are also used to optimize the processes in a healthcare
  system, such as a hospital emergency department.
These models help to study policies to reduce patients' waiting and admission
  time, cost of care, and undesirable treatment outcomes
  \cite{\abmHealthcareSystemOpt}.

\subsection{Social Sciences}
The ABM paradigm aligns well with the study of human interactions and the
  dynamics of a society \cite{retzlaff2021}.
Researchers studied cooperation \cite{axelrod1981}, income inequality
  \cite{epstein1996growing}, reasoning \cite{arthur1994}, culture dissemination
  \cite{axelrod1997}, and opinion dynamics \cite{hegselmann2002}.
For example, Axelrod studied the question ``Under what conditions will
  cooperation emerge in a world of egoists without central authority?'' 
  \cite{axelrod1984}.
Together with Hamilton, they invited experts to submit a program to compete in
  a tournament on the prisoner's dilemma \cite{poundstone2011prisoner}.
The prisoner's dilemma is a game theoretical problem with two players, each of
  whom can choose between two different actions: cooperation or defection.
Table~\ref{tab:background:prisoners-dilema} shows the rewards for the four
  possible states.
Defection would lead to the highest expected reward without knowing the other
  player's choice.
For cooperation to emerge, the prisoner's dilemma has to be played repeatedly.

This research maps naturally to agent-based modeling for four main reasons.
First, the different strategies are translated into agent behaviors, which can
  involve complex, non-linear actions.
Second, the focus lies on the \emph{emergence} (of cooperation), which is a key
  feature of ABM (Section~\ref{ch:intro:emergence}).
Third, Axelrod was also interested in evolutionary dynamics, which is another
  key strength of ABM.
Fourth, the agent-based model can help to understand the conditions that lead
  to cooperation or the robustness of these conditions.

In Axelrod's tournament, each strategy was evaluated against every other one.
The strategy with the highest reward turned out to be the simplest one: the
  strategy began with cooperation and then mirrored the opponent's previous
  choice in the current round (i.e., tit-for-tat) \cite{axelrod1981}.
Axelrod writes that this strategy could also be observed in trench warfare
  during World War 1: ``[u]nits actually violated orders from their own high
  commands in order to achieve tacit cooperation with each other [i.e., the
		  antagonist]'' \cite{axelrod1984}.

\begin{table}[h!]
	\centering
	\begin{tabular}{@{}rrr@{}}
		\toprule
		\textbf{Player 1 / Player 2} & \textbf{Cooperate} & \textbf{Defect} \\
		\midrule
		\textbf{Cooperate}           & \textsc{(3, 3)}    & \textsc{(0, 5)} \\
		\textbf{Defect}              & \textsc{(5, 0)}    & \textsc{(1, 1)} \\
		\bottomrule
	\end{tabular}
	\caption{Reward matrix for the prisoner's dilemma based on \cite{axelrod1981}.
		Tuples represent the reward for (Player 1, Player 2).
	}
	\label{tab:background:prisoners-dilema}
\end{table}

\subsection{Finance and Economics}
Robert Axtell and Doyle Farmer provide a thorough overview of the applications
  of ABM in finance and economics \cite{axtell2024}.
Research topics include market models, micro- and macroeconomics, stock
  markets, banking regulations, systemic risk modeling, housing markets, and more
  \cite{axtell2024}.
Studying the origin of price changes, agent-based financial market models were
  able to qualitatively and quantitatively reproduce the following statistical
  properties of financial markets: ``autocorrelation of price returns,
  autocorrelation of volatility, and the marginal distribution of price changes''
  \cite{axtell2024}.

\subsection{Other Fields}
The use of ABM is not restricted to the mentioned areas only; it is applied
  across various fields, ranging from agriculture \cite{\abmAgriculture} to
  zoology \cite{boone2017}, and numerous other fields
  \cite{\abmOther,\abmTransport,\abmEcology}.

\section{Why Agent-Based Modeling?}
The ABM paradigm was developed in response to the growing need for more
  flexibility and realism, driven by advancements in computing power.
Researchers sought to liberate themselves from artificial constraints and
  assumptions necessary to keep the model mathematically tractable, which
  includes ``linearity, homogeneity, normality, and stationarity"
  \cite{bankes2002, kaniyamattam2022}.
These assumptions are not limited to mathematical aspects but can also be based
  on domain-specific factors.
Axtell and Farmer provide two examples from economic models: perfect arbitrage
  and rational behavior \cite{axtell2024}.
Arbitrage allows investors to exploit price differences on different market
  places to make a profit.
Perfect arbitrage describes a situation in which the trades of the investor are
  risk-free \cite{shleifer_limits_1997}.
In economics, rational behavior describes the actions of an individual (i.e.,
  homo oeconomicus), who acts in self-interest and makes consistent choices
  \cite{vriend_rational_1996}.
These two idealized assumptions are incorporated into economic models.
ABM allows economists to deviate from these assumptions and create models that
  are more realistic \cite{shleifer_inefficient_2000}.

Agent-based models offer the ability to incorporate heterogeneity, local
  variations, nonlinearity, and the assimilation of various data sources
  \cite{hesam_2024}.
Unlike AI-based models, agent-based models, which are rooted in fundamental
  principles, produce simulation results that are easier to explain
  \cite{cogno_biomedicin_2024,baker_mechanistic_2018}.
Agent-based modeling is particularly suitable when the developmental aspect is
  a crucial component of the model because the simulation 1) evolves over time,
  2) allows for heterogeneous agents and behaviors, 3) supports feedback loops
  through agent and environment interactions, and 4) excels in capturing emergent
  behavior.
Examples include cancer growth studies \cite{\abmCancer} and development of the
  cerebral cortex \cite{\abmNervousSystem}.

\subsection{A Comparison to Other Modeling Paradigms}

\subsubsection{Equation-Based Modeling}
\label{ch:background:ebm}

In equation-based modeling (EBM) \cite{parunak1998}, a system's observables are
  described with a set of (differential) equations.
EBM is well-suited to describe physical systems like heat transfer,
  electromagnetism, or fluid dynamics.
However, solving or analyzing specific characteristics using the equations may
  necessitate properties like differentiability or smoothness.

To illustrate the difference between ABM and equation-based modeling
  \cite{parunak1998}, we compare a simple epidemiological model that divides the
  population into three groups: susceptible, infected, and recovered (SIR)
  \cite{kermack_1927}.
This model can be defined with three differential equations that depict the
  change over time: $dS/dt = - \beta S I / N$, $dI/dt = \beta S I / N - \gamma
	  I$, and $dR/dt = \gamma I$.
$S$, $I$, and $R$ are the number of susceptible, infected, and recovered individuals, $N$ is the
total number of individuals, $\beta$ is the mean transmission rate, and $\gamma$ is the recovery rate.

These equations can be solved analytically \cite{harko2014} or numerically.
An agent-based implementation with behaviors for movement, infection, and
  recovery is shown in Section~\ref{sec:epidemiology}.
From these three behaviors, the same observables emerge through agent
  interactions.
While the equation-based model is computationally (significantly) more
  affordable than an agent-based model, it is also quite rigid.
Although variations with different compartments exist for incorporating
  properties of various infectious diseases, the agent-based model can be
  extended in almost limitless ways.
The following list provides three examples to enhance realism of an
  epidemiological model using ABM.
\begin{itemize}
	\item Movement patterns could encompass activities such as going to work, school, shopping, and socializing with friends and family.
	\item The initial agent population could be generated to reflect the country's settlement structure.
	\item Risk factors that influence disease progression and hospitalization could be added to the agents.
\end{itemize}
These enhancements may enable scientists to examine various intervention
  methods and determine when hospitals become overwhelmed.
For detailed agent-based epidemiological models, readers are referred to
  \cite{hesam_2024, ozik_citycovid_2021}.

\subsubsection{System Dynamics}
\begin{sidewaysfigure}
\includegraphics[width=0.64\textwidth]{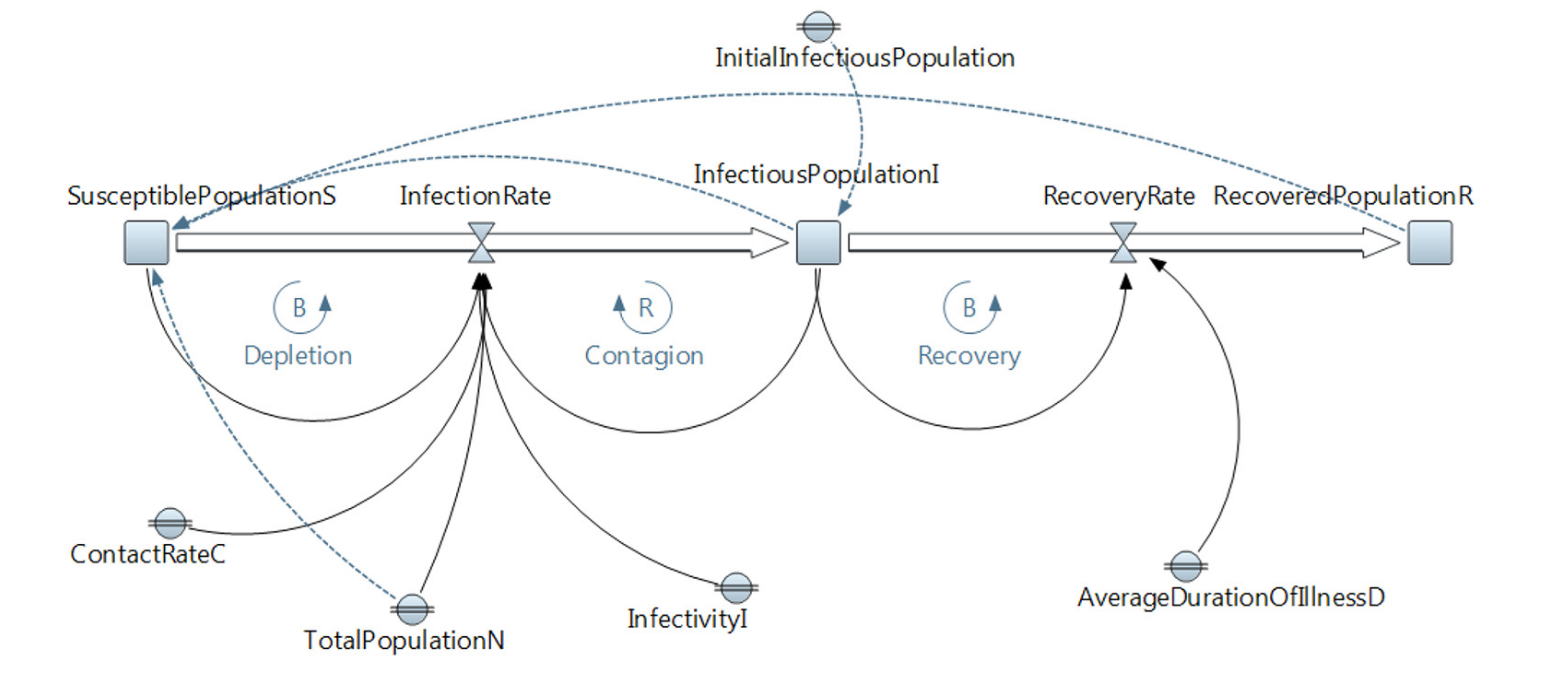}
	\includegraphics[width=0.34\textwidth]{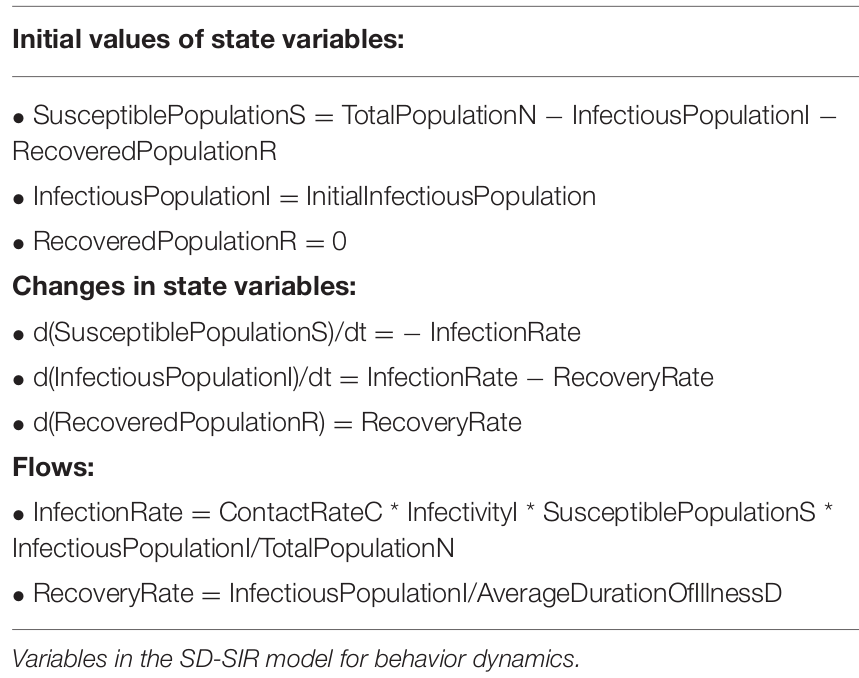}
\caption{
		Implementation of the SIR model using system dynamics
		from Pietro Cipresso \etal{} \cite{cipresso2015}.
		Figure taken from \cite{cipresso2015} without modification and used under
		  \href{https://creativecommons.org/licenses/by/4.0/}{CC BY 4.0}.
	}
	\label{fig:background:sd}
\end{sidewaysfigure}

System Dynamics (SD) is another top-down approach to describe the dynamics of a
  complex system \cite{forrester1961}, initially developed for inventory control
  \cite{forrester_beginning_1995}.
The model is often represented as a flow chart that contains the following key
  elements: stocks (i.e., the observables), flows between the stocks, and
  feedback loops.
Figure~\ref{fig:background:sd} shows the previously described SIR model
  (Section~\ref{ch:background:ebm}) implemented in SD \cite{cipresso2015}, which
  contains three stocks (susceptible, infected, and recovered), two flows
  (infection rate, and recovery rate), and three feedback loops (contagion,
  depletion, and recovery).

SD is well-suited for examining the high-level dynamics of a complex system
  that does not require any heterogeneity among its components.
One of the best-known SD models is ``The Limits to Growth'' from Meadows
  \etal{} \cite{meadows2013}, which studies population and resource dynamics.

Despite the significant difference between SD and ABM (top-down versus
  bottom-up, deterministic versus stochastic), Macal showed that the set of SD
  models is a subset of ABM models, which he calls the ``Agency Theorem''
  \cite{macal_agent-based_2010}.
Thus all well-formed SD models can be translated into an equivalent agent-based
  model \cite{macal_agent-based_2010}.

\subsubsection{Discrete-Event Simulation}
Discrete-event simulation (DES) \cite{robinson2005, hollocks2006, fishman2001}
  is a modeling method in which the simulation is assessed at specific time
  points triggered by an event.
Events may be random variables that are sampled from a probability
  distribution.
The event-driven nature of DES is in contrast to ABM where time evolves the
  simulation state \cite{cassandras2021}.
DES is commonly used in operations research to model and optimize queuing
  systems \cite{jain_analysis_2020}.
DES is, for example, used in hospitals \cite{forbus2022} to answer questions
  about mean waiting time, probability of waiting (longer than X minutes), or
  resource utilization \cite{fishman2001}.
These metrics are determined by aggregating the results of multiple simulation
  runs.

DES and ABM differ in several ways: top-down versus bottom-up approach,
  centralized versus decentralized control, passive versus active entities, and
  the presence of queues versus their absence \cite{siebers_discrete-event_2010}.

\subsubsection{Monte Carlo Simulation}
\begin{figure}[h]!
\includegraphics[width=0.6\textwidth]{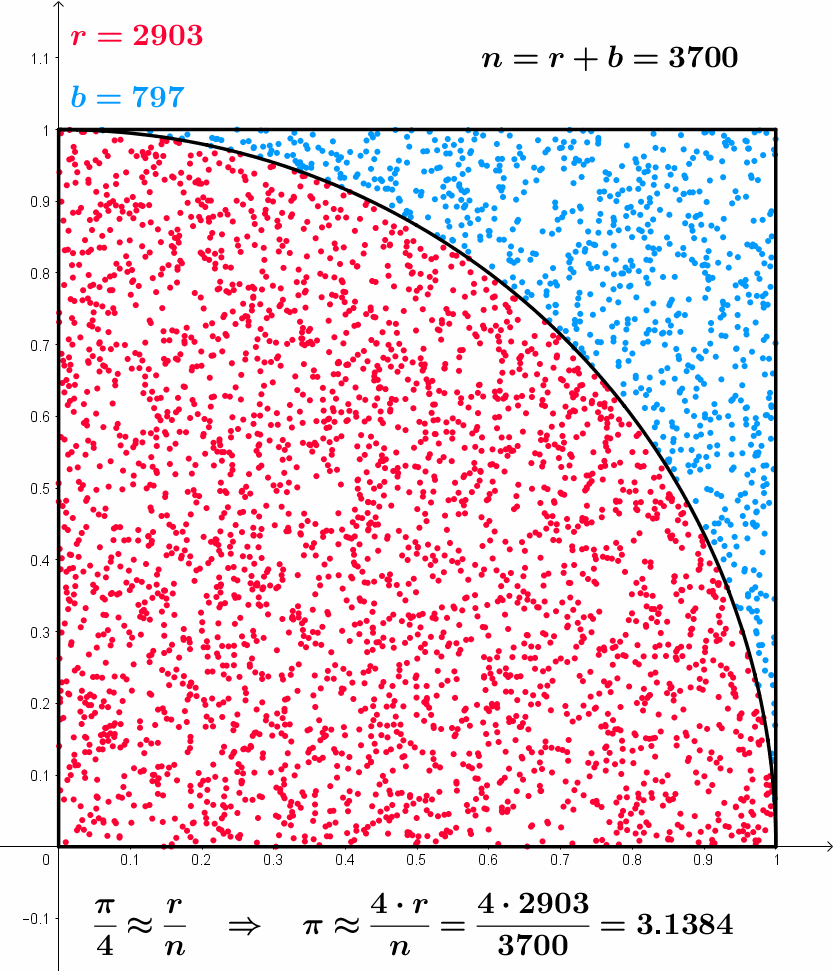}
\caption{
		Estimation of $\pi$ using the Monte Carlo method.
		Figure taken from \cite{pi-monte-carlo} and used under
		  \href{https://creativecommons.org/licenses/by/4.0/}{CC BY 4.0}.
		One frame was extracted from the animation.
	}
	\label{fig:background:monte-carlo}
\end{figure}

The Monte Carlo method is a generic way of using random numbers as input to a
  problem for simulation, estimation, or optimization \cite{kroese2014}.
This method can, for example, be used to estimate the value of $\pi$
  \cite{kalos2008}.
This estimation uses the fact that the area of the circle segment divided by
  the square is $\frac{\pi}{4}$.
Thus, by estimating the two areas, one can estimate $\pi$
  (Figure~\ref{fig:background:monte-carlo}).
The Monte Carlo method achieves this by: 1) defining points in 2D space as the
  input, 2) sampling points from a uniform distribution, 3) placing them on the
  square, and 4) aggregating the results: counting the points within distance of
  1 from the origin as well as the total number of points.
The estimation for $\pi$ is then $\frac{4\times r}{r + b}$, where $r$ and $b$
  correspond to the number of red and blue points in
  Figure~\ref{fig:background:monte-carlo}.
This simple yet powerful method is widely used in physics, chemistry,
  mathematics, statistics, economics, and finance \cite{kroese2014}.

Based on the example of calculating $\pi$ mentioned earlier, it's evident that
  MCM and ABM are fundamentally different.
While MCM is centered on random sampling for estimating quantities, quantifying
  uncertainties, or optimizing functions \cite{kroese2014}, it cannot simulate
  interactions between individual entities and emergent behavior like ABM can.

\subsubsection{Data-Driven Modeling}

Traditional simulation methods, as described in
  Section~\ref{ch:background:ebm}--\ref{fig:background:monte-carlo} for example,
  can be classified as mechanistic models.
Mechanistic models are built on causal relationships by combining fundamental
  physical, chemical, biological, or domain-specific principles
  \cite{\mechanisticModels}.
Consequently, mechanistic models lead to interpretable results and can be used
  to make predictions once validated.
However, mechanistic models are often computationally expensive, require deep
  domain knowledge to build, and pose challenges in integrating multiple sources
  of large unstructured data \cite{baker_mechanistic_2018}.

Data-driven modeling (DDM) \cite{\ddmOverview} addresses these challenges by
  using machine learning and artificial intelligence techniques to train models
  that can make predictions about complex systems based on the patterns extracted
  from the data.
Training typically necessitates access to large datasets.
Data-driven models extract statistical \emph{correlations} from these datasets
  rather than the causal relationships used in mechanistic models
  \cite{baker_mechanistic_2018}.
Once the model is trained, inference (i.e., ``simulation'' ) can be orders of
  magnitude faster than a simulation with a mechanistic model
  \cite{khattak_fast_2022,raikwar_transformers_2024}.
This execution speed improvement is utilized for constructing surrogate models,
  i.e., models that serve as substitutes for costly mechanistic models in
  parameter optimization and sensitivity analysis \cite{\surrogateModels}.

However, DDM are not only used to approximate mechanistic models, but can also
  significantly outperform their capabilities.
This potential has been impressively demonstrated by AlphaFold
  \cite{\alphafold}, a machine-learning approach to predict the 3D shape of
  proteins based on their amino acid sequence, which is a long-standing problem
  in biology.
AlphaFold won the CASP13 \cite{kryshtafovych_critical_2019} and CASP14
  \cite{kryshtafovych_critical_2021} protein folding prediction challenge with a
  large margin, which is considered a scientific breakthrough
  \cite{alquraishi_alphafold_2019}.

Baker \etal{} \cite{baker_mechanistic_2018} point out that the strengths and
  weaknesses of mechanistic and data-driven modeling are inverted and advocate
  for a combination of the two approaches \cite{zhang_combining_2020}.
This combination can be achieved in at least two ways.
First, by using physics-informed machine learning techniques that incorporate
  mechanistic elements into the training process to achieve better results
  \cite{\physicsInformed}.
Second, by utilizing data-driven methods to extract relevant patterns from a
  data set, which are then used as part of a mechanistic model
  \cite{baker_mechanistic_2018}.

\subsubsection{Hybrid Approach}
The transition between the presented modeling approaches is often fluid
  \cite{an2021}.
Many models often employ a hybrid approach to harness the advantages of various
  modeling paradigms.
Tissue models in biology often contain scalar or vector fields described as
  ordinary/partial differential equations, for example, to simulate substance
  diffusion \cite{duswald_2024} or heat transfer \cite{jennings_cryo,
	  jennings_cryo2}.
To simulate radiation-induced lung tissue damage, Nicolo Cogno \etal{}
  integrate an agent-based model with Monte Carlo simulation and equation-based
  substance diffusion \cite{cogno_mechanistic_2024, cogno_thesis}.
In this model, the agent-based model is used to simulate the tissue, while the
  Monte Carlo simulation provides the radioactive dose for each cell.
There are also works that combine ABM with SD \cite{howick2024} and DES
  \cite{dubiel2005}.

\section{How are Agent-Based Models Developed?}
Agent-based models are developed iteratively \cite{thorne_combining_2007,
	  demontigny_2023}.
This approach allows starting with a simple, easier-to-understand model and
  gradually adding complexity as needed.
In the beginning, modelers have to be clear about their modeling goal.
Do they want to predict a certain observable of a complex system, or are they
  more interested in explaining a specific mechanism \cite{epstein_why_2008}?

From this goal, the agent-based model definition can be derived by making
  simplifying assumptions of the complex system under investigation
  \cite{roberts_mathematical_2016} in a step by step manner.
There are five steps.

\paragraph{Step 1: Entity Definition.}
The modeler must define the main entities of the agent-based model: agents,
  behaviors, and the environment.
The definition of an agent includes its attributes and functionality.
To return to the epidemiological SIR model presented earlier
  (Section~\ref{ch:background:ebm}), agents can be defined as simple persons with
  a position in three-dimensional space and an attribute that determines the
  person's state (susceptible, infected, or recovered).
This agent type's functionality includes changing the position and
  transitioning between the three SIR states.
Afterward, the modeler must define the behaviors of these agents and how they
  interact.
A simple model could define the following behaviors for movement, infection,
  and recovery: 1) move randomly, 2) become infected with probability $p_i$, if
  there is an infected agent within distance $d$, and 3) recover with probability
  $p_r$.
This model would use a spatial environment, which supports agents' interaction
  if they are within distance $d$.
This model is explained in more depth in Section~\ref{sec:epidemiology}.

Once the modeler has defined the main components, the simulation's initial
  state must determined, by answering questions like: How are the agents
  initially distributed in space?
How many infected agents are there in the beginning and how are the
  distributed?

\paragraph{Step 2: Model Implementation.}
The so far pen-and-paper model is cast into code in the implementation stage.
To do that quickly and effectively, a modular simulation engine is needed that
  supports extension and modification without altering its internals.
Ideally, the simulation platform already contains the necessary functionality
  and model building blocks, such that the modeler only has to select and connect
  the required pieces.
Developing automated unit and integration tests is another important aspect.
Testing helps to achieve a high-quality implementation, and helps to save
  development time.
Finding a software error in a stochastic multi-million agent simulation is
  considerably more time-consuming than ensuring correctness by writing automated
  tests.

\paragraph{Step 3: Parameter Calibration.}
After the simulation is implemented, the modeler has to determine the values
  for all parameters that have been added to the model as placeholders.
Some parameters can be extracted from the literature , while others have to be
  determined using calibration and optimization.
In the SIR model, for example, the recovery probability for a specific disease
  may be extracted from the literature \cite{who_measles}, while the infection
  radius has to be determined through parameter calibration.
To do that, the simulation is repeatedly executed with different parameter
  values, with the goal of minimizing an objective function.
In the SIR model, the objective function may be defined as the mean squared
  error of the simulation result compared to the EBM model result.
An optimization algorithm (e.g., particle swarm optimization
  \cite{kennedy_1995}) determines the parameters for the next simulation
  execution based on the previous simulation results.

\paragraph{Step 4: Model Validation.}
The model is validated with data separately and independently from the
  parameter calibration stage.
This stage verifies that the model and its parameters accurately reflect the
  complex system under consideration.
The validation can be quantitative or qualitative, depending on the use case
  and may include sensitivity analyses to investigate the uncertainty of the
  simulation results based on the input parameters.

\paragraph{Step 5: Model Usage.}
The model is used for its intended purpose either to ``predict, explain, guide
  data collection, discover new questions'' \cite{epstein_why_2008} and more.
Additional validation may be needed to validate predictions beyond the
  observables (e.g., the number of susceptible, infected, and recovered agents)
  used in the previous steps.

If the simulation results in steps two to five are not satisfactory, the
  modeler returns to the first step and reevaluates the assumptions and
  simplifications based on the insights of the current development iteration.

\section{Challenges of Agent-Based Modeling}
Although ABM has been successfully used to model complex systems in many
  domains, this modeling paradigm also faces several challenges, which we cover
  in this section.

\paragraph{Performance and Scalability.}
For many agent-based models, the computational cost of simulating a vast number
  (e.g., millions or billions) of individual entities, exacerbated by the need to
  execute a simulation many times, is a key limitation \cite{hesam_2024,
	  duswald_2024, duswald2024calibrationstochasticagentbasedneuron,
	  abubacar_neuronal_growth, xing2015}.
Therefore, modelers are restricted by the number of agents they can simulate or
  by the complexity of the agent behaviors.

\paragraph{Parameter Calibration.}
The number of model parameters in ABM can become large quickly as more details
  are incorporated into the model.
Modelers face the challenge to determine realistic values for each parameter,
  either from the literature or through parameter calibration.
Besides the vast parameter space, calibration is made even more difficult by
  the non-determinism of many models, their non-linear nature, and limited amount
  of available data.
Even with available data, the simulation engine's performance constrains the
  number of model executions during parameter calibration, potentially resulting
  in suboptimal simulation outcomes \cite{estill2020}.

\paragraph{Validation.}
Validation is an important step to ensure that the model is in agreement with
  the real-world system before it can be trusted and used.
Beharathy and Silverman write that ABM modeling ``ha[s] often been criticized
  for relying extensively on informal, subjective and qualitative validation
  procedures'' \cite{bharathy_validating_2010}.
Several validation techniques exist, but unavailable data, the challenge of
  describing an agent-based model formally \cite{pullum_techniques_2012}, or the
  absence of a standard approach makes the process difficult \cite{tracy2018}.
Validation techniques include 1) face validation that relies on domain experts
  to determine if the model results matches the real-world system qualitatively
  \cite{bharathy_validating_2010}, 2) cross-validation which compares the results
  with the output of other models \cite{bharathy_validating_2010}, 3) extreme
  value analysis to investigate the consistency or limitations of the model
  \cite{bharathy_validating_2010} and 4) comparison with a mathematical model
  \cite{pullum_techniques_2012}.

\paragraph{No Standard Approach.}
ABM's versatility can also be seen as challenging, as there is often no
  standard approach to a modeling task, apart from the high-level development
  steps presented in the previous section.

Therefore, finding the right model complexity is difficult \cite{tracy2018,
	  bonabeau2002} and makes the development time-consuming.
Sensitivity analyses and approximate Bayesian computation might help to detect
  inputs that have little impact on the observables of the system and, therefore,
  indicate potential for model simplification \cite{tracy2018}.

In addition, emergent behavior, which is often the desired outcome of modeling,
  is a broad concept and thus difficult to manage.
Predicting emergent behavior is challenging, if at all possible, and there is
  no universal test to ascertain its occurrence in a simulation
  \cite{helbing_social_2012, bankes2002}.

\paragraph{Proper Simulation Platforms.}
Modelers rely on a simulation platform to handle the complex computational
  tasks.
Ideally, these platforms should be high-performing, user-friendly, easy to
  learn, modular, flexible, well-documented, freely available, robust,
  cross-platform compatible, interoperable, high-quality, and provide all
  necessary building blocks.
However, in reality, the available simulation platforms prioritize different
  requirements, resulting in shortcomings in other areas that complicate the
  modeler's work.
Given the diversity of the ABM field, flexibility and modularity are essential
  platform properties that are often overlooked.
This shortcoming, in turn, increases development time and hinders innovation in
  ABM.

\paragraph{Challenges Addressed.}
In this dissertation, we address the platform and performance challenges.
We create and implement a simulation platform that provides a comprehensive
  range of agent-based functionality, is flexible and modular for easy addition
  of new features, seamlessly integrates with third-party software, and focuses
  on high-quality.
We address the performance issue by utilizing parallel and distributed
  computing, addressing the memory bottleneck of today's computing systems,
  leveraging ABM-specific characteristics to avoid unnecessary work, and reducing
  data transfers between distributed processes.

\clearpage{}

\cleardoublepage \clearpage{}\chapter{Related Work}
\label{ch:related-work}

In this chapter, we will outline how our contributions (Section~\ref{ch:introduction:contributions}) 
  compare to the existing work in the field.
We will examine the key features of \bdm{}/\ta{} in relation to other agent-based
  simulators (Section~\ref{ch:related:other-simulators}), differentiate between prior work on optimization and our
  shared-memory (Section~\ref{ch:related:shared-memory}) and distributed optimizations (Section~\ref{ch:related:distributed}), provide context
  for our performance evaluation (Section~\ref{ch:related:performance-evaluation}), and also include a comparison with simulation
  tools beyond the agent-based domain (Section~\ref{ch:related:outside}).

\section{Other Agent-Based Simulators}
\label{ch:related:other-simulators}

Several ABM platforms have been published demonstrating the importance of
  agent-based models in complex systems research \citep{netlogo,
	  emonet_agentcell:_2005, zublerdouglas2009framework, koene_netmorph:_2009,
	  richmond_high_2010, collier2011repasthpc, lardon_idynomics:_2011,
	  rudge_computational_2012, mirams_chaste:_2013,
	  torben-nielsen_context-aware_2014, kang_biocellion:_2014,
	  cytowski_large-scale_2014, matyjaszkiewicz_bsim_2017,
	  ghaffarizadeh_physicell:_2018}.
In this section, we compare \bdm{}'s/\ta{}'s most crucial system properties with prior
  work.

\subsection{Extreme-Scale Model Support}

To our knowledge, \ta{} is the only simulation platform capable of simulating
  \ESMaxAgentsExact{} agents.
The largest reported agent populations in the literature is from Jon Parker
  \cite{parker_2007}, who simulated a specialized epidemiological model with up to 6 billion
  agents and Biocellion \cite{biocellion}, a distributed tool in the tissue
  modeling domain, with 1.72 billion agents.
Other distributed platforms exist \cite{repasthpc, chaste, dmason,
	  cytowski_large-scale_2015, richmond_high_2010}, but have not shown
  simulations on this extreme scale.

The NeuroMaC neuroscientific simulation platform
  \citep{torben-nielsen_context-aware_2014} claims to be scalable, but the
  authors do not present performance data and present simulations with only 100
  neurons.
Therefore, \bdm{}'s ability to simulate large-scale neural development, which
  we demonstrate in Chapters~\ref{ch:platform} and \ref{ch:engine}, is, to our
  knowledge, unrivaled.

\subsection{General-Purpose Application Area Support}
Many ABM platforms focus on a specific application area, for example: bacterial colonies
  \citep{emonet_agentcell:_2005, matyjaszkiewicz_bsim_2017,
	  rudge_computational_2012, lardon_idynomics:_2011}, cell colonies
  \citep{kang_biocellion:_2014, mirams_chaste:_2013, cytowski_large-scale_2014},
  and neural development \citep{zublerdouglas2009framework, koene_netmorph:_2009,
	  torben-nielsen_context-aware_2014}.
Pronounced specialization of an ABM platform may prevent its capacity to adapt
  to different use cases or simulation scenarios.
In contrast, \bdm{} can be adapted to many different fields due to its
  modularity and extensibility as we show in Chapter~\ref{ch:platform} 
  (and as also shown by various other works \cite{cogno_mechanistic_2024, cogno_agent-based_2022, cogno_3d_2022, cogno_thesis, duswald_2024, duswald2024calibrationstochasticagentbasedneuron, hesam_2024, demontigny_2021, demontigny_2023, jennings_cryo, jennings_cryo2, demetriades_interrogating_2022, gazeli_interrogating_2022}).
For example, \bdm{} has been used to study 
  (radiation-induced) lung injury \cite{cogno_mechanistic_2024, cogno_agent-based_2022, cogno_3d_2022, cogno_thesis}, 
  vascular tumor growth \cite{duswald_2024},
  gliomas \cite{demontigny_2021},
  country-scale COVID-19 hospitalizations \cite{hesam_2024}, 
  growth of pyramidal cells \cite{duswald2024calibrationstochasticagentbasedneuron},
  the formation of retinal mosaics \cite{demontigny_2023},
  cancer cell treatment with helium plasma jet \cite{gazeli_interrogating_2022},
  cancer drug pharmacodynamics \cite{demetriades_interrogating_2022},
  freezing and thawing of tissue \cite{jennings_cryo, jennings_cryo2}, 
  spread of HIV in Malawi \cite{hiv_malawi}, 
  development of the cerebral cortex \cite{abubacar_neuronal_growth}, 
  and other works that are currently in progress.

\subsection{Quality Assurance}
Automated software testing is the foundation of a modern development workflow.
Unfortunately, several simulation tools \citep{zublerdouglas2009framework,
	  rudge_computational_2012, lardon_idynomics:_2011, koene_netmorph:_2009,
	  torben-nielsen_context-aware_2014, cytowski_large-scale_2014} omit these tests.
Mirams \etal{} \cite{mirams_chaste:_2013} recognizes this shortcoming and
describes a rigorous development workflow in their paper.
\ta{} has over 600 automated tests (see Section~\ref{ch:platform:software-quality}) that are continuously executed on all
  supported operating systems to ensure high code quality.
The open-source code base \cite{bdm-github, breitwieser_supporting_2021, breitwieser_supporting_2022}, tutorials (see Appendix~\ref{app:tutorials}), and documentation not only help
  users get started (quickly) with modeling tasks, but also enable validation by external examiners.

\section{Shared-Memory Performance Optimizations}
\label{ch:related:shared-memory}

In order to achieve high-performance and efficiency, we improve the performance of the shared-memory parallelized simulation engine with a series of optimizations. This section compares our memory layout optimization, improved neighbor search algorithm, and utilization of domain-specific knowledge with prior work.

\subsection{Memory Layout Optimization}
Data movement between main memory and processor cores is a fundamental
  bottleneck in today's computing systems \cite{mutlu_processing_2019}.
Recent research in computer architecture explores new approaches to address
  this bottleneck, such as processing-in-memory, i.e., placing compute capability
  closer to the data~\cite{gomez-luna_benchmarking_2021, ahn_pim-enabled_2015}.
Agent-based simulation tools are also negatively impacted by the data movement
  bottleneck (Section~\ref{ch:dse:eval:gpm}).
We address this problem in software with a better memory layout that leads to
  more efficient bandwidth utilization and data reuse in processor caches.
Space-filling curves \cite{morton1966computer, hilbert_ueber_1891,
	  peano_sur_1890} can improve the memory layout by aligning objects that are
  close in 3D space.
Therefore, these curves are frequently used to optimize geometric data
  structures \cite{unibn, asano_space-filling_1997} and molecular dynamics
  simulations \cite{grime_highly_2014, anderson_general_2008,
	  nakano_scalable_1999}.
To our knowledge, none of the other agent-based simulation frameworks (e.g.,
  \cite{physicell, biocellion, mason, chaste, netlogo,
	  zublerdouglas2009framework, agentcell, bsim2, netmorph, idynomics,
	  torben-nielsen_context-aware_2014, morpheus, cytowski_large-scale_2015, dmason,
	  repast-simphony, repast-hpc, richmond_high_2010, CompuCell3D}) use
  space-filling curves to improve the cache hit rate and minimize the amount of
  remote DRAM accesses.
We introduce this proven technique to agent-based modeling and present a
  mechanism to determine the Morton order of a non-cubic grid in linear time.

\subsection{Neighbor Search}
Agent-based simulation platforms use various neighbor search algorithms:
  Delaunay triangulation \cite{zublerdouglas2009framework}, octree
  \cite{hauri_self-construction_2013}, and grid-based approaches \cite{netlogo,
	  richmond_high_2010, biocellion}.
Grids are also commonly used on GPUs \cite{hoetzlein_fast_2014, unibn,
	  gross_fast_2019, aaby_efficient_2010, dsouza}.
Depending on the dataset and specific search query ([fixed-]radius neighbor
  search or k-nearest neighbors), the literature recommends different algorithms
  \cite{vermeulen_comparative_2017, unibn}.
Our contribution is the efficient implementation and integration of the
  uniform grid into the simulation engine (Section~\ref{sec:grid}) and the insights into the
  performance differences for the agent-based workload (Section~\ref{sec:eval:environment}).

\subsection{Omitting Unnecessary Work}
In Section~\ref{sec:static-agents} we present a mechanism to omit the collision
  force calculation for static agents safely.
The neuroscience simulation tool NeuroMaC
  \cite{torben-nielsen_context-aware_2014} goes one step further and exclusively
  supports models where only the growth front of a neuron can change.
This is probably a good optimization for this line of research but is too
  restrictive for \bdm{}'s goal of becoming a general-purpose platform.
Related work can also be found in the traffic simulation domain, in which
  Andelfinger et al.
\cite{andelfinger_fast-forwarding_2018} proposes a
mechanism to
skip iterations of independent agents and fast-forward them to the next
interaction time.

\section{Distributed Performance Optimizations}
\label{ch:related:distributed}

In the distributed simulation engine, in which agents are distributed among multiple processes, information exchange is a key bottleneck. We address that challenge by optimizing serialization (i.e., packing agents into a contiguous buffer that can be transmitted) and reducing the amount of data transferred with delta encoding. The following section discusses related work in these two areas.

\subsection{Serialization}
There is a wide range of serialization libraries.
Chapter~\ref{ch:dse} compares the performance against ROOT IO \cite{brun_root_1997},
  which according to Blomer \cite{blomer_quantitative_2018} outperforms
  Protobuf \cite{protobuf}, HDF5 \cite{hdf5}, Parquet \cite{parquet}, and Avro
  \cite{avro}.
MPI \cite{mpi} also provides functionality to define derived data types, but targets use cases with regular patterns, for example, the last element of each row in a matrix. \ta{}'s agents are allocated on the heap with irregular offsets between them and, therefore, cannot use MPI's solution.

\subsection{Delta Encoding}
Delta encoding \cite{hunt_algorithm_1976} is a widely used concept to minimize the amount of data that is
  stored or transferred, which we apply to aura updates of the agent-based
  workload (Section~\ref{sec:design:delta-encoding}).
Other applications include backups \cite{burns1997efficient}, file revision
  systems such as git \cite{git}, network protocols \cite{rfc1144,
	  mogul2002delta}, cache and memory compression \cite{pekhimenko_base-delta-immediate_2012, pekhimenko_linearly_2013, lee_adaptive-latency_2015, vijaykumar_case_2015}, and more \cite{macdonald2000file, delta_encoding_vms}.
We did not find explicit mention of this concept in the literature to accelerate the distributed
  execution of agent-based simulations.

\section{Performance Evaluation}
\label{ch:related:performance-evaluation}

To our knowledge, this dissertation presents the most comprehensive performance
  analysis to date of an agent-based simulation platform.
Existing platforms report only limited performance results, including
  simulation execution times and occasionally scalability analyses
  \cite{zublerdouglas2009framework, chaste, biocellion, repasthpc, mason,
	  physicell}.
Performance data can also be found in model papers
  \cite{strazdins_performance_2011, murphy_simulating_2016} and in works that
  focus on hardware accelerators \cite{xiao_survey_2019}.
We improve upon these works by providing an in-depth analysis of each
  performance-relevant component.
Efforts in the direction of a standard agent-based benchmark have been made by
  Moreno et al.
\cite{moreno_designing_2019} and
Rousset et al. \cite{rousset_survey_2016}.
However, these synthetic benchmarks fall short of representing a realistic
  range of agent-based simulations by over-simplifying memory access patterns and
  assuming that agents always move randomly.
Compared to these, our benchmark simulations cover a broader spectrum of
  performance relevant simulation metrics (see
  Table~\ref{tab:sim-characteristics}).
Chapter~\ref{ch:engine} and~\ref{ch:dse} in this thesis provide detailed performance 
  analyses of the shared-memory and distributed simulation engine and the 
  different use cases presented in Chapter~\ref{ch:platform}.

\section{Comparisons With Simulators Outside the Agent-Based Modeling Field}
\label{ch:related:outside}

Particle-based simulations, i.e., simulations made of discrete particles that interact with each other , such as molecular dynamics (MD)
  \cite{phillips_scalable_2005, thompson_lammps_2022, vasp}, astrophysics (AP)
  \cite{spiridon_n-body_2013}, or computational fluid dynamic (CFD)
  \cite{jasak_openfoam_2009} simulations often face similar computational
  challenges to improve the performance of large-scale simulations.
Some agent-based models, for example cell sorting (Section~\ref{sec:platform:soma-clustering}), may also be seen as a particle-based method. 
LAMMPS \cite{thompson_lammps_2022}, for example, also uses a grid-based
  structure to determine neighbors.
While LAMMPS stores neighbor lists for each atom, which according to Thompson
  et al.
\cite{thompson_lammps_2022} ``consumes the most memory of any data structure in LAMMPS'' \bdm{} does not need these lists and therefore saves memory.
\bdm{} improves over LAMMPS and VASP \cite{vasp} by sorting agents using a
  space-filling curve (Section~\ref{sec:load-balancing}) and using a custom
  memory allocator (Section~\ref{sec:mem-allocator}) to reduce the memory access
  latency.
A NUMA-aware thread allocation mechanism, as the one used in \bdm{}
  (Section~\ref{sec:numa-iteration}), is not needed in LAMMPS or VASP because
  both tools support distributed parallelism with MPI.
In this work, we identify several computational challenges in ABM, which we
  tackle by using methods inspired by MD, AP, and CFD.
The main difference between ABM and other particle-based applications is that
  the computations can vary significantly from each other in terms of arithmetic
  intensity, the number of considered neighbors, data access patterns, and more,
  thus posing diverse computational challenges.

\clearpage{}

\cleardoublepage \clearpage{}\chapter{The \bdm{} Platform}
\label{ch:platform}

\section{Introduction}

Agent-based simulation (ABS) is a powerful tool assisting life scientists in
  better understanding complex biological systems.
In silico simulation is an inexpensive and efficient way to rapidly test hypotheses
  about the (patho)physiology of cellular populations, tissues, organs, or
  entire organisms \citep{Yankeelov2016, ji2017mathematical}.

However, the effectiveness of such computer simulations for scientific research
  is often limited, in part because of two reasons.
First, after the slowing down of Moore's law \citep{moores-law} and Dennard
  scaling \citep{dennard_design_1974}, hardware has become increasingly parallel
  and heterogeneous.
Most ABS platforms do not take full advantage of these hardware enhancements.
The resulting limited computational power forces life scientists to
  compromise either on the resolution of the model or on simulation size \citep{thorne2007abm}.
Second, existing ABS platforms have often been developed with a specific use case
  in mind.
This makes it challenging to implement the desired model, even if it deviates only
  slightly from its original purpose.

To help researchers tackle these two major challenges, 
we propose a novel open-source platform for biology dynamics modeling, \bdm{}.
We alleviate both of these problems by emphasizing performance and
  modularity.
\bdm{} features a high-performance simulation engine that is fully 
  parallelized to utilize multi-core CPUs and able to offload computation to hardware accelerators (e.g. a GPU).
The software comprises a set of fundamental biological functions and a
  flexible design that adapts to specific user requirements.
Currently, \bdm{} implements the neurite model and mechanical forces presented in \cite{zublerdouglas2009framework},
  but these components can easily be extended, modified, or replaced.
Hence, \bdm{} is well-suited for simulating a wide range of biological processes
  in tissue modeling and beyond.

\bdm{} provides by design five system properties:
\begin{itemize}
\item \textbf{Agent-based.}
        The \bdm{} project is established to support an agent-based modeling approach
          which allows one to simulate a wide range of developmental biological processes.
        A characteristic property of agent-based simulations is the absence of a
          central organizational unit that orchestrates the behavior of all agents.
        Quite to the contrary, each agent is an autonomous entity that
          individually determines its actions based on 
its current state, behavior, and the surrounding environment.

  \item \textbf{General purpose.}
\bdm{} is developed to become a general-purpose tool for agent-based
          simulation.
To simulate models from various fields, \bdm{}'s software
          design is extensible and modular.

  \item \textbf{Large scale.}
        Biological systems contain a large number of agents.
        The cerebral cortex, for example, comprises approximately 16 billion neurons
          \citep{azevedo_equal_2009}.
        Biologists should not be limited by the number of agents within a simulation.
        Consequently, \bdm{} is designed to take full advantage of modern hardware and use
          memory efficiently to scale up simulations.

  \item \textbf{Easily programmable.
        }
The success of an ABS platform depends, among other things, on how quickly a
          scientist, not necessarily an expert in computer science or high-performance programming,
          can translate an idea into a simulation.
        This characteristic can be broken down into four key requirements that \bdm{} is designed to fullfil:

        First, \bdm{} provides a wide range of common functionalities such as visualization, plotting,
          parameter parsing, backups, etc.
        Second, \bdm{} provides simulation primitives that minimize the
          programming effort necessary to build a use case.
        Third, as outlined in item ``General purpose", \bdm{} has a modular and
          extensible design.
Fourth, \bdm{} provides a coherent API and hides implementation details
          that are irrelevant for a computational model (e.g., details such as parallelization strategy,
          synchronization, load balancing, or hardware optimizations).

  \item \textbf{Quality assured.}
        \bdm{} establishes a rigorous, test-driven development process to foster
          correctness, maintainability of the codebase, and reproducibility of results.

\end{itemize}

The main contribution of this chapter is an open-source, high-performance, and modular
  simulation platform for agent-based simulations.
We provide the following evidence to support this claim:
(i) We detail the user-facing features of \bdm{} that enable users to build
  a simulation based on predefined building blocks and to define a model
  tailored to their needs.
(ii) We present three basic use cases in the field of neuroscience, oncology, and
  epidemiology to demonstrate \bdm{}'s capabilities and modularity.
(iii) We show that \bdm{} can produce biologically-meaningful simulation results
  by validating these use cases against experimental data or an analytical solution.
(iv) We present performance data on different systems and scale each use case to 
  one billion agents to demonstrate \bdm{}'s performance.

\section{Design Overview}

\subsection{Simulation Concepts}
\label{sec:simulation-concepts}

\begin{figure}[htbp]
\centering
\includegraphics[width=0.9\linewidth]{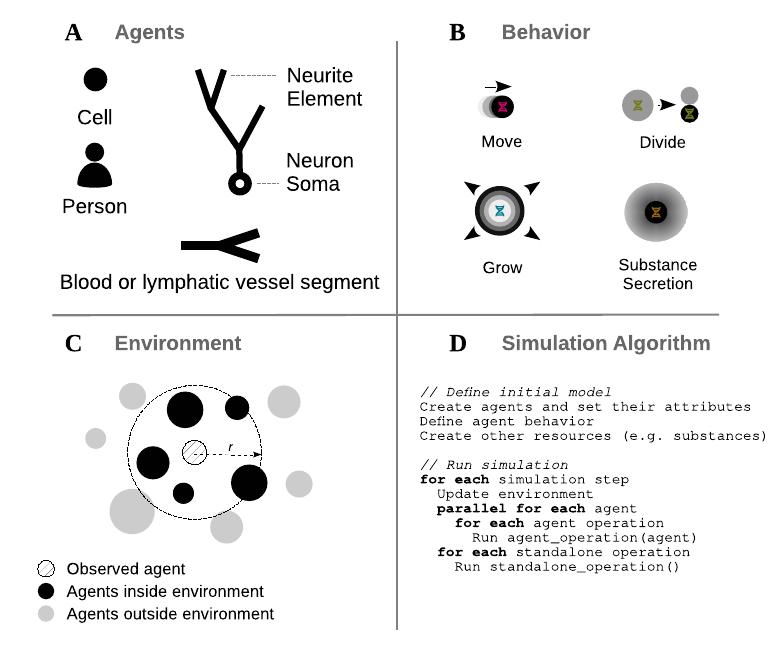}
\caption{{\bf Simulation concepts.}
Agents (A) have their own geometry, behavior (B), and environment (C). 
    (B) Agent behavior
    is defined in separate classes, which are inserted into agents.
The update of an agent
    is based on its current state and its surrounding environment. (C) The
    environment is determined by radius $r$ and contains other agents or extracellular
    substances.
    The simulation algorithm (D) can be divided into two main parts: the definition of the
    initial model and the execution of the simulation. 
}
    \label{fig:simulation-concepts}
\end{figure}

\bdm{} is implemented in the C++ programming language and supports simulations that follow an agent-based approach.
Figure~\ref{fig:simulation-concepts} gives an overview of \bdm{}'s main concepts, 
  Figure~\ref{fig:bdm-layers} presents the abstraction layers,
  while Figure~\ref{fig:sw-design} illustrates its object-oriented design.

An agent (Figure~\ref{fig:simulation-concepts}A) has a 3D geometry, behavior, and environment.
There is a broad spectrum of entities that can be modeled as an agent.
In the results section, we show examples where an agent represents a subcellular structure
(neuroscience use case), a cell (oncology use case), or a
person (epidemiology use case).
Figure~\ref{fig:simulation-concepts}B shows example agent behaviors such as growth factor secretion, chemotaxis, or cell division.
Behaviors can be activated or inhibited.
\bdm{} achieves this by attaching them to or removing them from the corresponding
  agent.

The \textit{Environment} is the vicinity that the agent can
  interact with (Figure~\ref{fig:simulation-concepts}C).
It comprises agents and other resources like chemical substances in the
  extracellular matrix.
Surrounding agents are, for example, needed to calculate mechanical
  interactions among agent pairs.

Currently, the user defines a simulation programmatically in C++ 
  (see Figure~\ref{fig:simulation-concepts}D and Section~\ref{sec:code}).
C++ is a great choice in terms of execution speed, efficiency, and
  interoperability with high-performance computing libraries, but
  is harder to program in due to its lower level of abstraction (versus higher
  level languages like Java or Python).
There are two main steps involved: initialization and execution.
During initialization, the user places agents in space, sets
  their attributes, and defines their behavior.
In the execution step, the simulation engine evaluates the defined model in the
  simulated physical environment by executing a series of operations.
We distinguish between agent operations and standalone operations (Figure~\ref{fig:simulation-concepts}D).
At a high level, an agent operation is a function that: (i)
  alters the state of an agent and potentially also its
  environment, (ii) creates a new agent, or (iii) removes an
  agent from the simulation.
Examples for agent operations are: execute all behaviors and calculate mechanical forces.
The simulation engine executes agent operations for each agent for each
  time step.
Alternatively, standalone operations perform a specific task during one time step and are 
  therefore only invoked once.
Examples include the update of substance diffusion and the export of visualization data.

\subsection{\bdm{} Features}
\label{sec:features}

\bdm{} is a simulation platform that can be used to develop
  agent-based simulations in various computational biology fields (e.g. neuroscience,
  oncology, epidemiology, etc.).
Although agent-based models in these different fields may intrinsically
  vary, there is a set of functionalities and definitions that they have in
  common.

\begin{figure}[tbp]
\centering
\includegraphics[width=\linewidth]{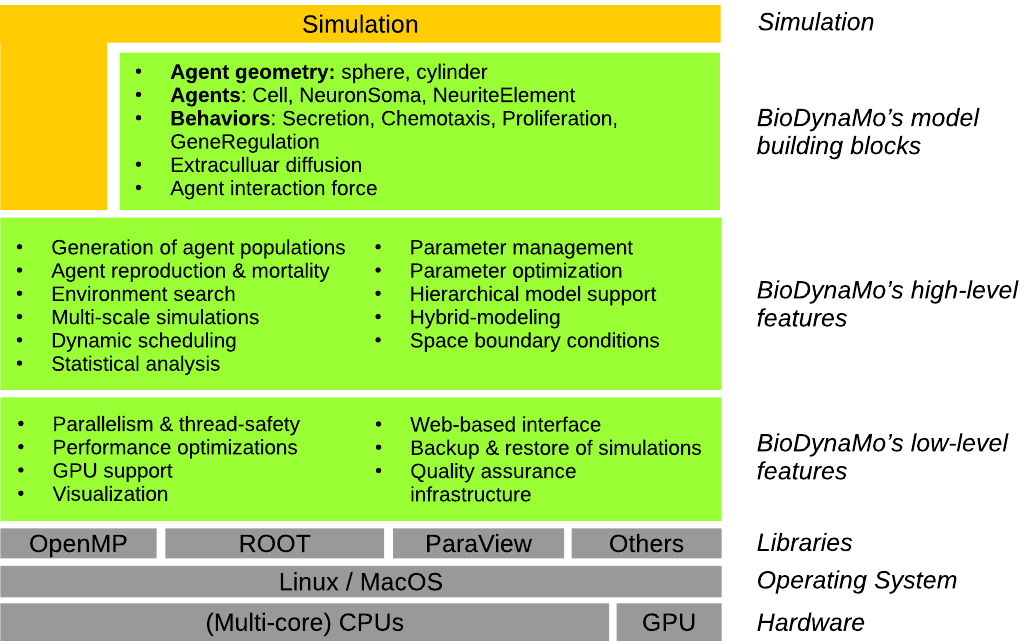}
\caption{{\bf \bdm{}'s layered architecture.}
\bdm{} is predominantly executed on multi-core CPUs, is able to offload 
    computations to the GPU, and supports Linux and MacOS operating systems.
    \bdm{} provides a rich set of low- and high-level features commonly required in
    agent-based models. 
    On top of these generic features, \bdm{} offers model building blocks
    to simplify the development of a simulation.
Even if \bdm{} does not provide the required building blocks, users still benefit from all 
    generic agent-based features (illustrated by the vertical extension of the ``Simulation" layer). 
}
    \label{fig:bdm-layers}
\end{figure}

\begin{sidewaysfigure}[p]
\centering
\includegraphics[width=\textwidth]{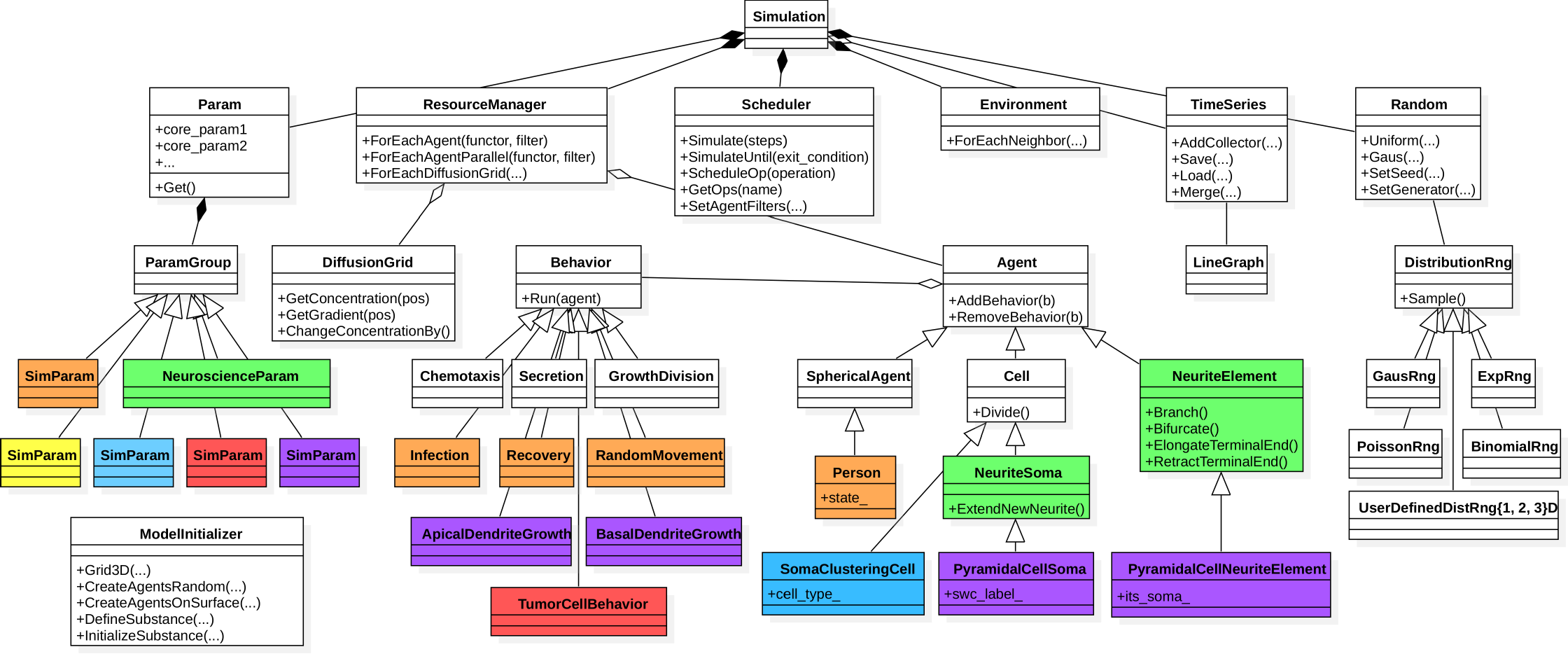}
\caption{{\bf Software design and modularity.}
    Overview of selected classes and functions that are important from the users' perspective.
    Classes in white (\bdm{} core) and green (\bdm{}'s neuroscience module) are part of the current \bdm{} installation.
    The remaining classes illustrate how we extended \bdm{} to implement the use cases and benchmarks presented in this chapter
    (purple: neuroscience use case, red: oncology use case, orange: epidemiology use case, blue: soma clustering benchmark, yellow: cell proliferation benchmark).
    A complete list of \bdm{} classes can be found at \href{https://biodynamo.org/api}{https://biodynamo.org/api}. 
    }
    \label{fig:sw-design}
\end{sidewaysfigure}

These commonalities can be divided into low- and high-level agent-based features
  and are an integral part of \bdm{}.
\bdm{} also provides model building blocks to accelerate the development of agent-based models.
The described layered architecture is shown in Figure~\ref{fig:bdm-layers}.

\section{Low-Level Features}
\label{sec:low-level-features}

The low-level features (Figure~\ref{fig:bdm-layers}) form the foundation of \bdm{} and provide crucial functionality responsible for 
high-performance execution and ease-of use.
These features are mostly hidden from the user and require control only in exceptional situations.

\subsection{Parallelism and Thread Safety}
\bdm{} exploits the inherent parallelism of agent-based models in which agents
  update themselves based on their current state, behavior, and local
  environment.
\bdm{}'s implementation uses OpenMP
  (\href{https://www.openmp.org/}{https://www.openmp.org/}) compiler directives
  to parallelize the loop over all agents
  (Figure~\ref{fig:simulation-concepts}D).
In addition to parallelizing the execution of agent operations for each agent,
  standalone operations like updating the diffusion grid and visualization are
  parallelized separately (Figure~\ref{fig:simulation-concepts}D).

\begin{figure}
	\centering
	\includegraphics[width=0.75\linewidth]{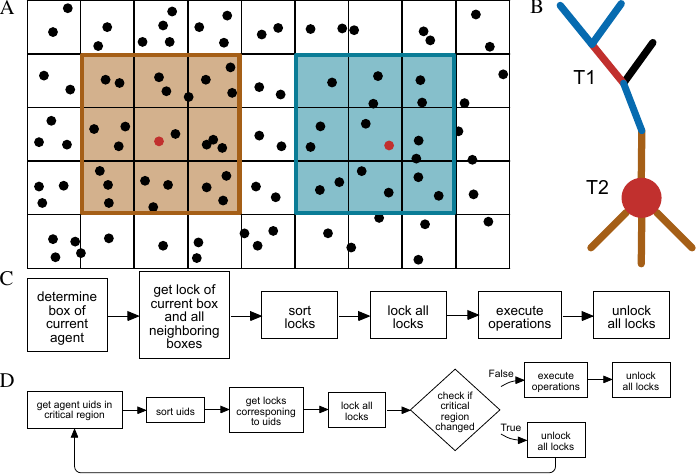}
	\caption{{\bf Thread safety in \bdm{}.}
		A: Currently updated agents in red and their local environment defined by the
		  uniform grid in ocher and blue.
		B: Neuron model in which two different threads (T1 and T2) update the agents in
		  red.
		Neighboring agents in ocher and blue might be modified during the update.
		C: Flow chart of the automatic thread safety mechanism, which ensures that the
		  local environments in A do not overlap.
		D: Flow chart of the user-defined thread safety mechanism, which ensures that
		  agents that might be modified (i.e., critical region) do not overlap.
	}
	\label{fig:thread-safety-design}
\end{figure}

Synchronization between threads is only needed if agents modify their
  environment.
In this case, two agents (handled by two different threads) might attempt to
  update the same resource in the local environment.
This scenario occurs in the neuroscience use case in which neurite elements
  modify neighboring segments.
Therefore, \bdm{} provides built-in synchronization mechanisms to ensure that
  even if two threads try to modify the same agent or resource, data is not
  corrupted.
There are two thread safety mechanisms to protect agents from data corruption:
  automatic and user-defined (Figure~\ref{fig:thread-safety-design}).
Automatic thread safety uses the environment to prevent two threads from
  processing agents with overlapping local environments (A,C).
This mechanism can be enabled with a single parameter, but might be too
  restrictive for some use cases.
User-defined thread safety on the other hand offers more fine-grained control
  over which agents must not be processed at the same time, but likely requires
  additional input from the user (B,D).

Other resources that are modified by agents (e.g., the \texttt{DiffusionGrid}
  to simulate extracellular diffusion) need their own protection mechanism.
This feature is used in the soma clustering benchmark where cells secrete a
  substance into the extracellular matrix.

For typical \bdm{} simulations users do not need to control parallel execution
  and thread synchronization.
This holds particularly true for all use cases and benchmarks presented in this
  chapter.
Only for more advanced uses, like adding a new environment algorithm or adding
  a shared resource that is \emph{not} an agent, users have to consider parallel
  execution.

\begin{figure}
	\centering
	\includegraphics[width=\linewidth]{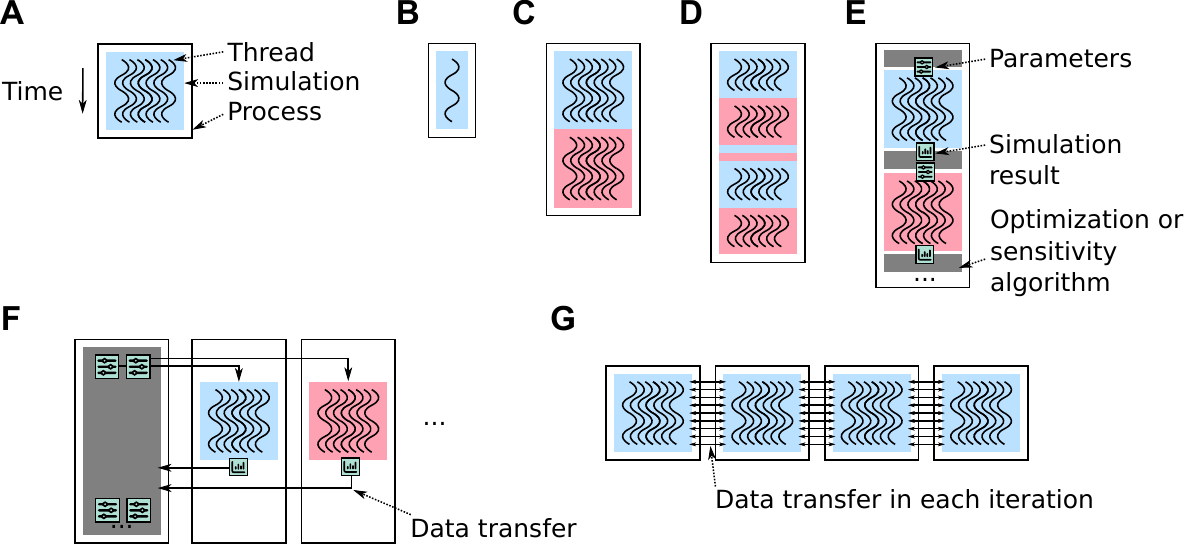}
	\caption{{\bf Parallelism modes.}
		A: Multithreaded (default) execution.
		B: Serial execution.
		C--G can also be executed with one thread in each simulation block.
		C: Multiple simulations in the same process.
		D: Multiple simulations in the same process with alternating execution and
		  potential exchange of information.
		E: Optimization or sensitivity analysis in the same process.
		F: Optimization or sensitivity analysis with multiple processes.
		G: Distributed execution (one simulation divided onto multiple servers).
	}
	\label{fig:parallelism-modes}
\end{figure}

Figure~\ref{fig:parallelism-modes} provides an overview of \bdm{}'s parallel
  execution modes.
\circled{A} shows the default mode of the single-node simulation engine.
All available CPU threads of the machine (i.e. laptop, workstation, or server)
  are utilized.
The user can change the number of threads used.
The thread number can be increased to oversubscribe the machine, or reduced
  down to one if the simulation is very small and the overheads outweigh the
  benefits of parallel execution \circled{B}.
This mode is used for example in the performance evaluation in
  Chapter~\ref{sec:eval:comparison-with-cx3d-netlogo} to compare \bdm{} with our
  serial baselines.
\circled{C}--\circled{G} can also be executed with one thread per simulation block.
\circled{C} shows that within the same process, multiple
simulations can be executed.
Simulations do not have to run until completion before the next simulation can
  be started.
It is possible to switch between simulations and exchange information\circled{D}~.
However, there can only be one active simulation at the same time within a
  process.
In other words, within the same process, multiple simulations cannot be updated
  in parallel.
This mode was used in \cite{demontigny_2021}.
\circled{E} uses the ability to execute multiple simulations in one process for
optimization or sensitivity analysis with one process.
This mode was used in the epidemiology use case to find the right simulation
  parameters for measles and seasonal influenza (Section~\ref{sec:epidemiology}).
\circled{F}
Hesam \etal{} \cite{hesam_2024} extends upon this solution by removing the
  restriction of one process.
The authors develop an MPI-based solution in which the master rank is running
  the optimization or sensitivity algorithm.
The algorithm generates new parameter sets, which are transferred to a free
  worker node.
The worker executes the simulation using the parameters received and returns
  the result.
Hesam \etal{} uses a modular design that allows for the easy replacement of the
  algorithm.
This solution is part of the \bdm{} installation.
Duswald \etal{} \cite{duswald2024calibrationstochasticagentbasedneuron}
  demonstrates that \bdm{} integrates well with third-party software that
  externally provides the algorithm and worker management.
Lastly, \circled{G} shows the distributed simulation engine which partitions
  \emph{one} simulation and executes it using multiple processes.
This important functionality is described in detail in Chapter~\ref{ch:dse}.

\subsection{Visualization}

\par
\bdm{} currently uses ParaView \citep{paraview} as a visualization engine.
There are two visualization modes, which we refer to as live mode and
  export mode.
With live mode, the simulation can be visualized during runtime, whereas with
  export mode, the visualization state is exported to file and can only be
  visualized post-simulation.
Live mode is a convenient approach to debug a simulation visually while it is
  executed. However, this can slow down the simulation considerably if used
  continuously.
In export mode, the visualization state can be loaded by the visualization
  package for post-simulation processing (slicing, clipping, rendering,
  animating, etc.).
\bdm{} can visualize substance concentrations and gradients (see
  Figure~\ref{fig:soma-clustering}), and the geometry of the supported agents.

Furthermore, it is possible to export any agent's data members. This
  information can then be used as input to ParaView filters, e.g., to highlight 
  elements based on a specific property.
The export of additional data members was used in Figure~\ref{fig:soma-clustering},
  for example, to color cells by their cell type.

\subsection{\bdm{} Notebooks}
\label{sec:notebooks}

\par
\begin{figure}
\centering
	\includegraphics[width=\linewidth]{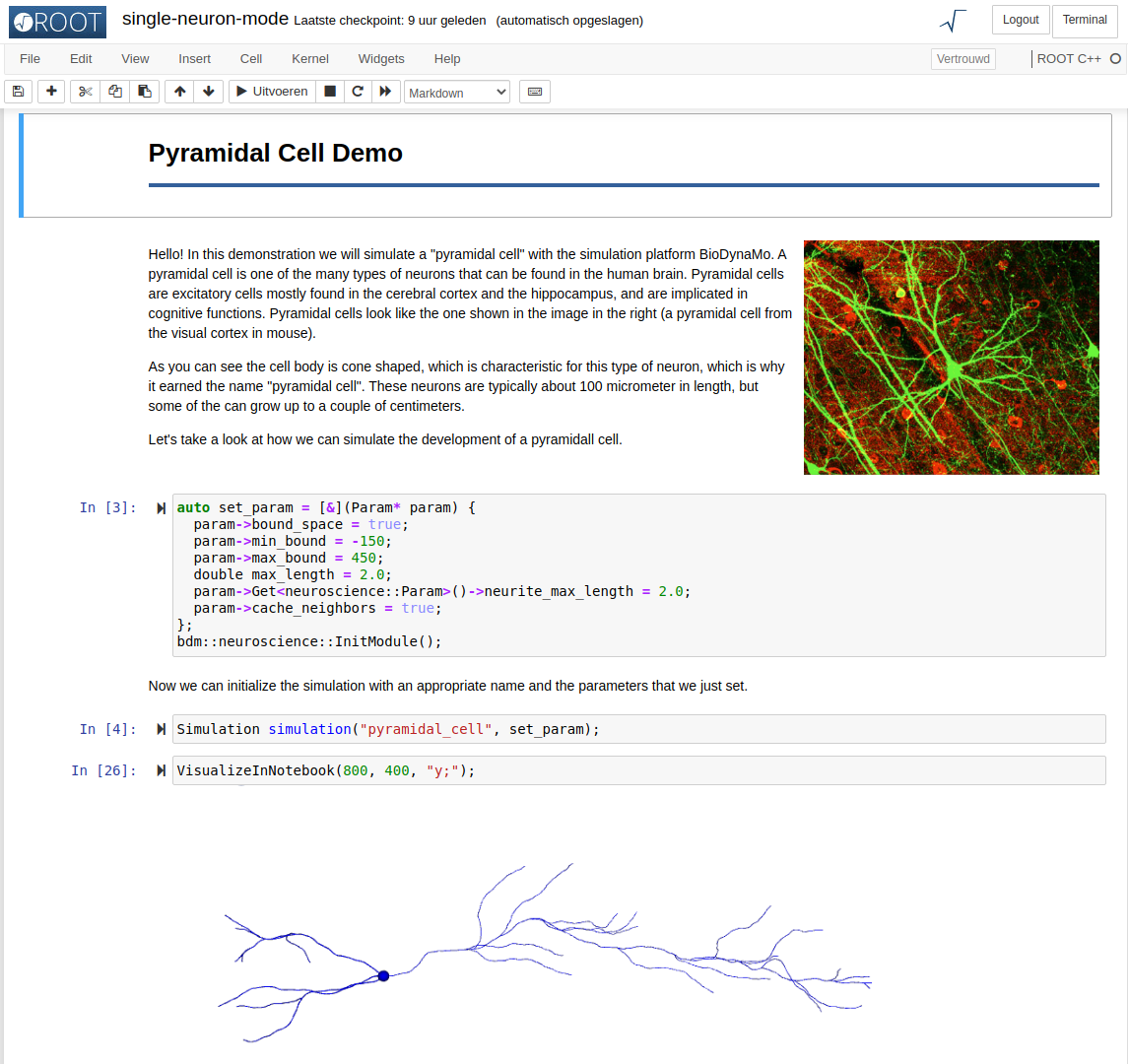}
\caption{{\bf \bdm{} notebook.} A convenient web interface to create
  and run simulations in a step-by-step manner. The inlining of text and media
  makes it possible to provide extra information. 
  A few intermediary blocks have been removed to fit the final 
  simulation output on the screenshot.
  }
  \label{fig:notebook}
\end{figure}

 Jupyter notebooks \citep{kluyver2016jupyter} is a widely used web application to quickly prototype
or demonstrate features of a software library.
With notebooks, it is possible to easily create a website with inline code
snippets that can be executed on-the-fly.
ROOT expands these notebooks by offering a C++ backend in addition to the
default Python backend.
This allows us to provide a web interface to easily and quickly get started with
\bdm{}.
Users do not need to install any software packages; a recent web browser is enough. 
It is also a convenient tool to interactively go through a demo or tutorial,
which opens up possibilities to use \bdm{} for educational purposes.
\bdm{} is the first agent-based simulation platform written in C++ that offers
  such an interface.
\bdm{} notebooks have already been successfully used to demonstrate and teach
about pyramidal cell growth and were well-received by high-school
students and teachers during CERN's official teachers and students programs.
Figure~\ref{fig:notebook} shows an example of how a \bdm{} notebook looks.
This example gives a brief introduction to pyramidal cells and
follows up with a step-by-step explanation of how to simulate their growth with
\bdm{}.
Interactive visualizations in the browser give users quick feedback about the
  simulation status.
Lastly, tutorials written as \bdm{} notebooks can be executed as part of our 
  continuous integration pipeline and ensure that documentation stays in 
  sync with the codebase.
In Supplementary Tutorial \stref{01}---\stref{15} we use this feature to explain \bdm{} to new users.

\subsection{Full Web-Based Development Environment}
\label{sec:p1:ll-gitpod}

\begin{figure}
\centering
	\includegraphics[width=\linewidth]{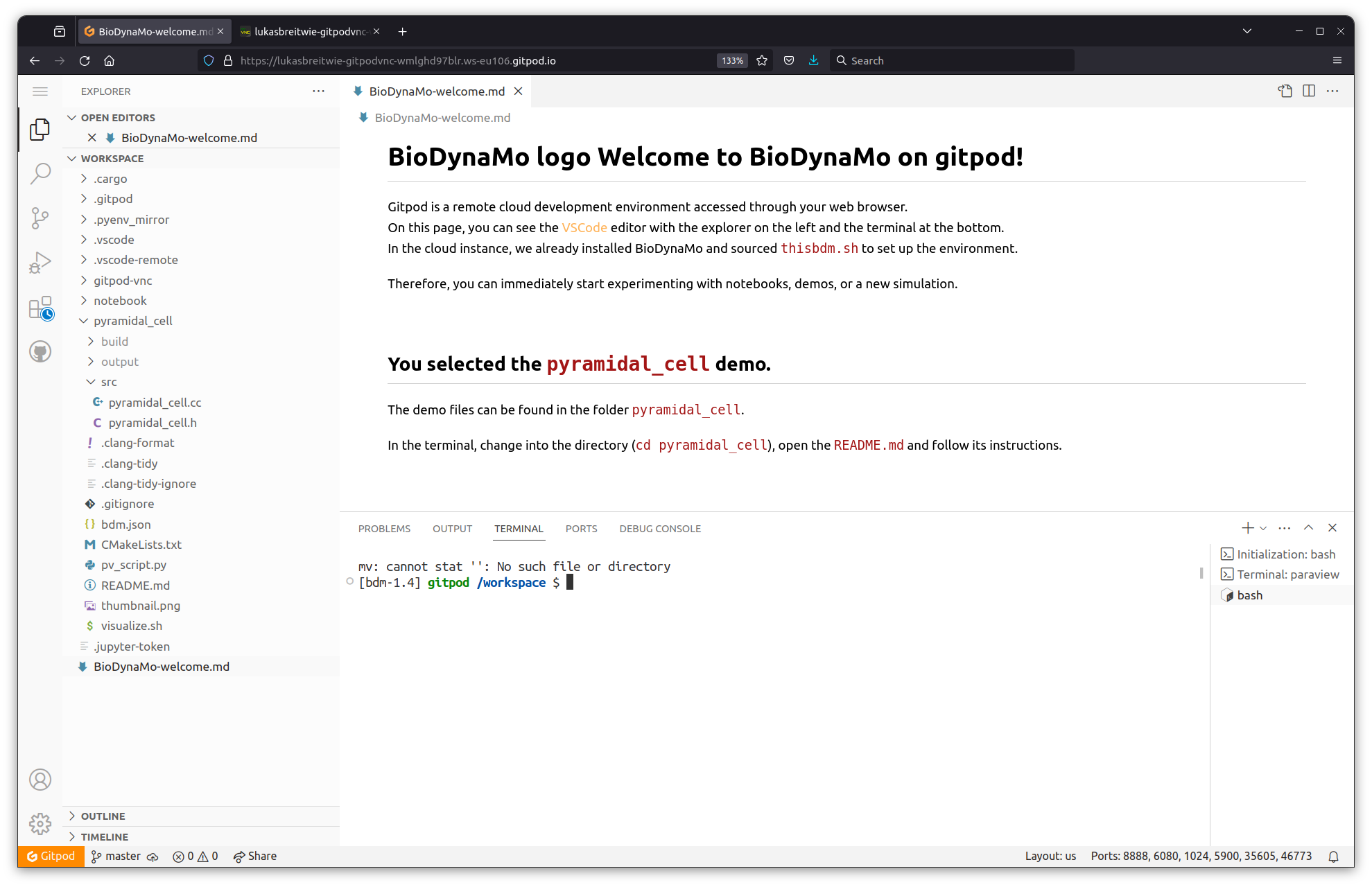}
	\includegraphics[width=\linewidth]{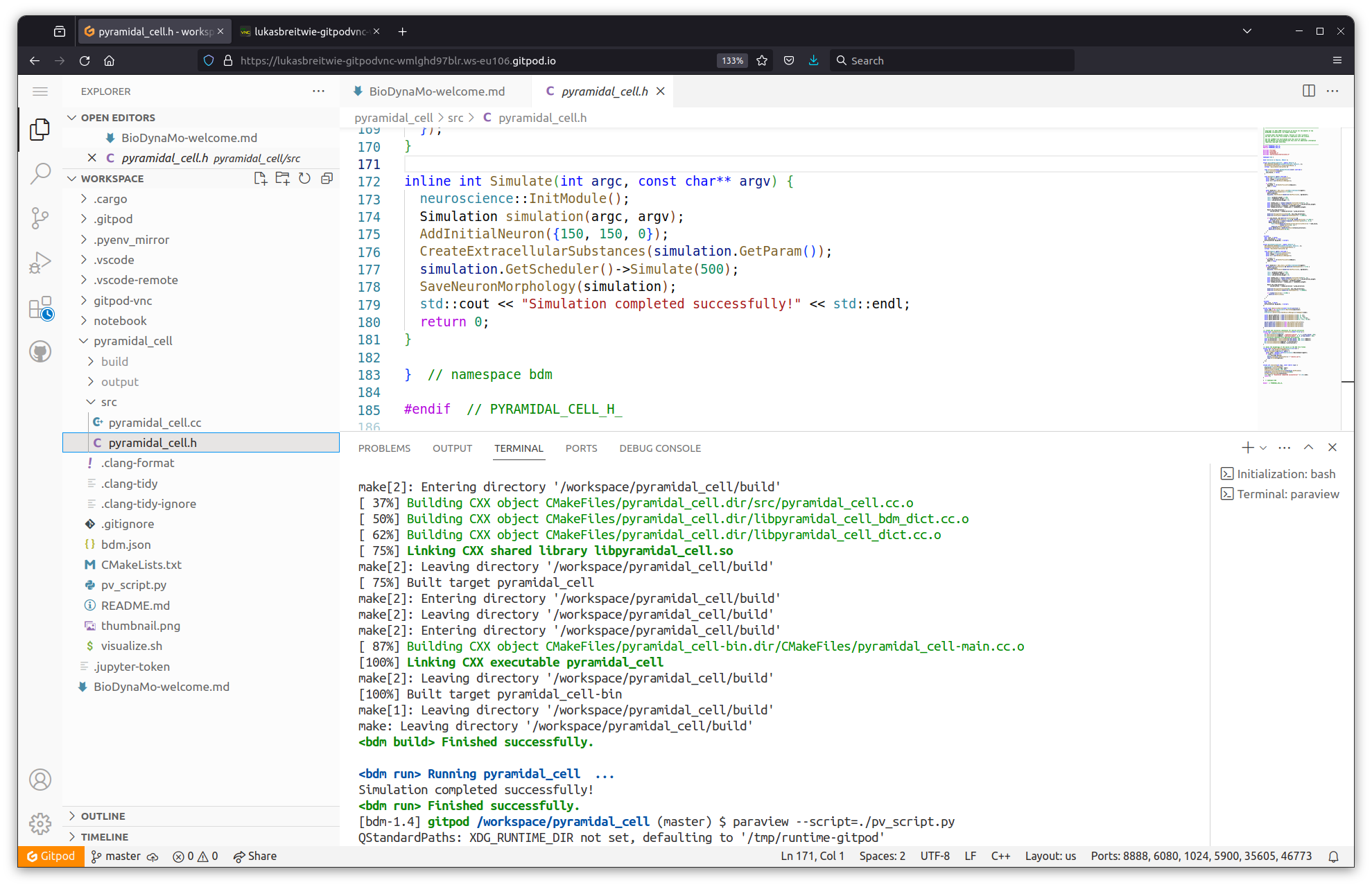}
\caption{{\bf Web-browser-based development environment.}
  Top: Welcome screen with user instructions.
  Bottom: Code pane and terminal with successful simulation execution.
	}
	\label{fig:p1:gitpod}
\end{figure}

For advanced simulations that contain more code than manageable in a notebook
  or require the full \bdm{} feature set, we provide a full web-based development
  environment, based on the Gitpod service \cite{gitpod}.
To that extent, we package \bdm{} in a docker image with Gitpod's base image
  and customize Gitpod's development environment.
If the user clicks on a demo link on the \bdm{} website, Gitpod creates a new
  instance based on this image.
The user interacts with this instance via a browser-based VSCode \cite{vscode}
  development environment.
Figure~\ref{fig:p1:gitpod} shows the welcome screen after clicking on the
  pyramidal cell growth demo link, which contains user instructions on how to
  proceed.
The image below shows the opened code pane and the terminal with a successful
  execution of this demo simulation.

\begin{figure}
\centering
	\includegraphics[width=\linewidth]{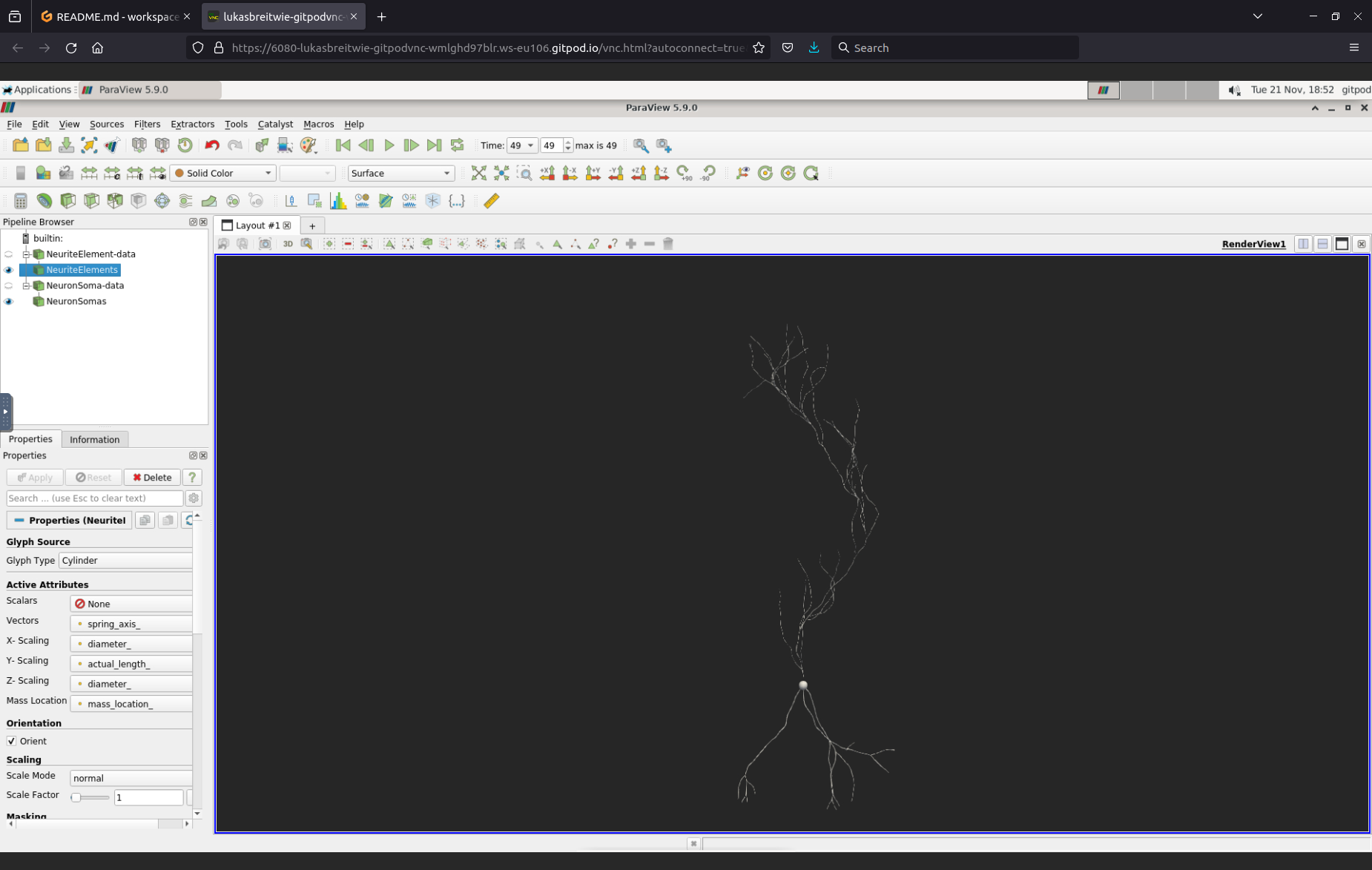}
\caption{{\bf Full ParaView instance in the web browser using a VNC connection.}}
	\label{fig:p1:gitpod-vis}
\end{figure}

Gitpod also supports GUI applications.
Executing ParaView in VSCode's terminal will open a new browser tab showing the
  full desktop version of \bdm{}'s main visualization tool in a VNC window
  (Figure~\ref{fig:p1:gitpod-vis}).
The ParaView command is exactly the same as if one would work on a local
  laptop or workstation (see last terminal line in Figure~\ref{fig:p1:gitpod}).

\subsection{Backup and Restore}
\par \bdm{} uses ROOT \cite{brun_root_1997} to integrate the backup and
restore functionality transparently. This allows system failures to occur without losing
valuable simulation data. Without any user intervention, all simulation data can
be persisted to disk as system-independent binary files, called ROOT files, and
restored into memory after a failure occurs. The ROOT file format is
well-established and is the primary format for storing large quantities
(exabyte) of data in high-energy physics experiments, such as CERN. To enable 
the backup and restore feature in BioDynaMo, one must simply specify the file 
name of the backup file. Additionally, one can set the interval at which a backup
is performed. A low interval value ensures a low amount of data loss whenever a
failure occurs, but also increases the incurred overhead for creating the backup
files. The advised backup interval depends on the duration of the simulation.

\subsection{Software Quality Assurance}
\label{ch:platform:software-quality}

\begin{figure}[h!]
\includegraphics[width=\linewidth]{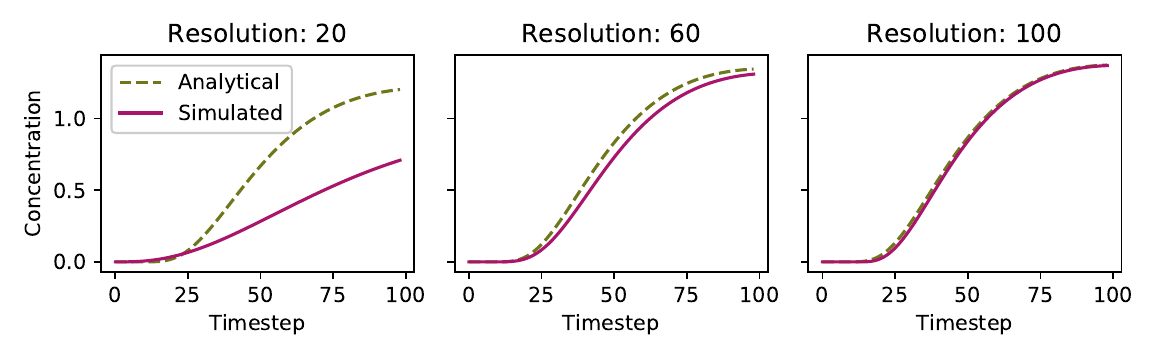}
\caption{{\bf Diffusion convergence test.}
    The simulated diffusion results converge towards the analytical solution as we
      increase the resolution of the diffusion grid. 
    The resolution represents the number of grid points in the
      diffusion grid along each dimension of the simulation space.
    We use an instantaneous point source at the origin and measure the concentration 
      change over time at $\sqrt{1000}$ micron away from the point source. 
  }
  \label{fig:diffusion-convergence}
\end{figure}

Compromising on software quality can have severe consequences that can
  culminate in the retraction of published manuscripts
  \citep{miller_scientists_2006}.
Therefore, we put tremendous effort into establishing a rigorous development
  workflow that follows industry best practices.
Test-driven development---a practice from agile development
  \citep{beck2000extreme}---is at the core of our solution.
\bdm{} has over 600 tests distributed among unit, convergence, system, and
  installation tests.
We monitor test coverage of our unit tests with the tool kcov \citep{kcov}, and currently
  cover 79.8\% lines of code.
Figure~\ref{fig:diffusion-convergence}, for example, shows the convergence test
  for extracellular diffusion. By comparing the computed results with the analytical 
  solution we observe an increasing accuracy of the diffusion solver when we increase
  the diffusion grid's resolution.
For each change to our repository (\href{https://github.com/BioDynaMo/biodynamo}{https://github.com/BioDynaMo/biodynamo}), GitHub Actions (\href{https://github.com/features/actions}{https://github.com/features/actions})
  executes the entire test suite and, upon success, updates the documentation on our website.
Installation tests are executed on each supported operating system and ensure
  that all demo simulations run on a default system.

\paragraph{Coding Style Guide.}
A coding style guide is a set of guidelines and best practices that improve
  the readability and maintainability of a codebase.
Consistency helps new developers and users quickly understand \bdm{}'s code
  base.
We are using \texttt{clang-format}, \texttt{clang-tidy}, and \texttt{cpplint}
  to monitor compliance with our style guide.
These tools are integrated into our build system and GitHub Actions to provide quick
  feedback to developers and code reviewers.

\section{High-Level Features}
\label{sec:high-level-features}

The high-level layer (Figure~\ref{fig:bdm-layers}) provides general functionality which is commonly required in agent-based models 
  across many fields.

\subsection{Generation of Agent Populations}
The first step in an agent-based model is to specify the starting condition of the simulation.
Therefore, \bdm{} provides functionality to create agent populations with specific properties.
Class \texttt{ModelInitializer} provides several functions to create agents in 3D space 
  (Figure~\ref{fig:generation-agent-populations}) and to 
  initialize extracellular substances (Figure~\ref{fig:sw-design} and Section~\ref{ch:platform:diffusion}). 
Furthermore, to initialize the attributes of an agent population, researchers can use one of 
  the many predefined random number generators that draw samples
  from a specific distribution (uniform, exponential, gaussian, binomial, etc.) 
  or define their own one.
These features are demonstrated in Supplementary Tutorial~\stref{01}, \stref{02} and \stref{08}.

\begin{figure}[tbp]
  \centering
\begin{subfigure}{\linewidth}
    \centering
    \begin{minipage}{0.4\textwidth}
    \begin{cppcode}
auto create_agent = [](const Real3& position) {
  auto* agent = new SphericalAgent(position);
  agent->SetDiameter(10);
  return agent;
};
uint64_t num_agents = 300;
    \end{cppcode}
    \end{minipage}
    \caption{Common code}
\end{subfigure}

  \begin{subfigure}{0.7\linewidth}
    \centering
    \includegraphics[width=0.55\textwidth]{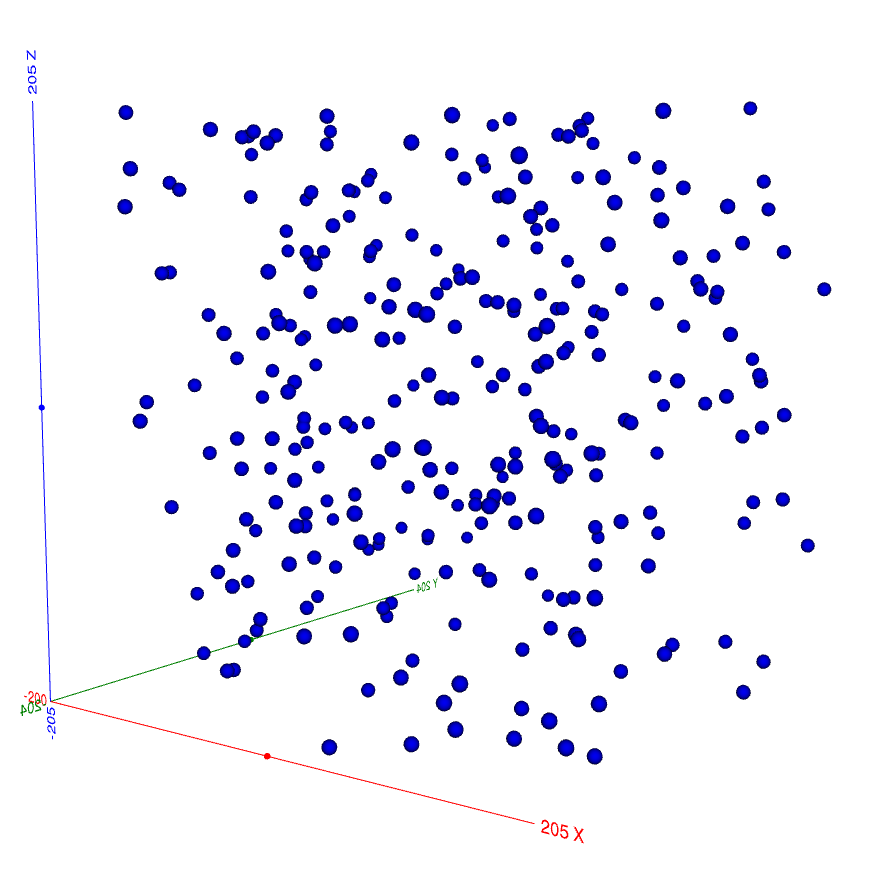}
    \begin{cppcode}
ModelInitializer::CreateAgentsRandom(-200, 200, num_agents, create_agent);
    \end{cppcode}
    \caption{Create agents randomly inside a 3D cube.}
  \end{subfigure}
  
  \begin{subfigure}{0.7\linewidth}
    \centering
    \includegraphics[width=0.55\textwidth]{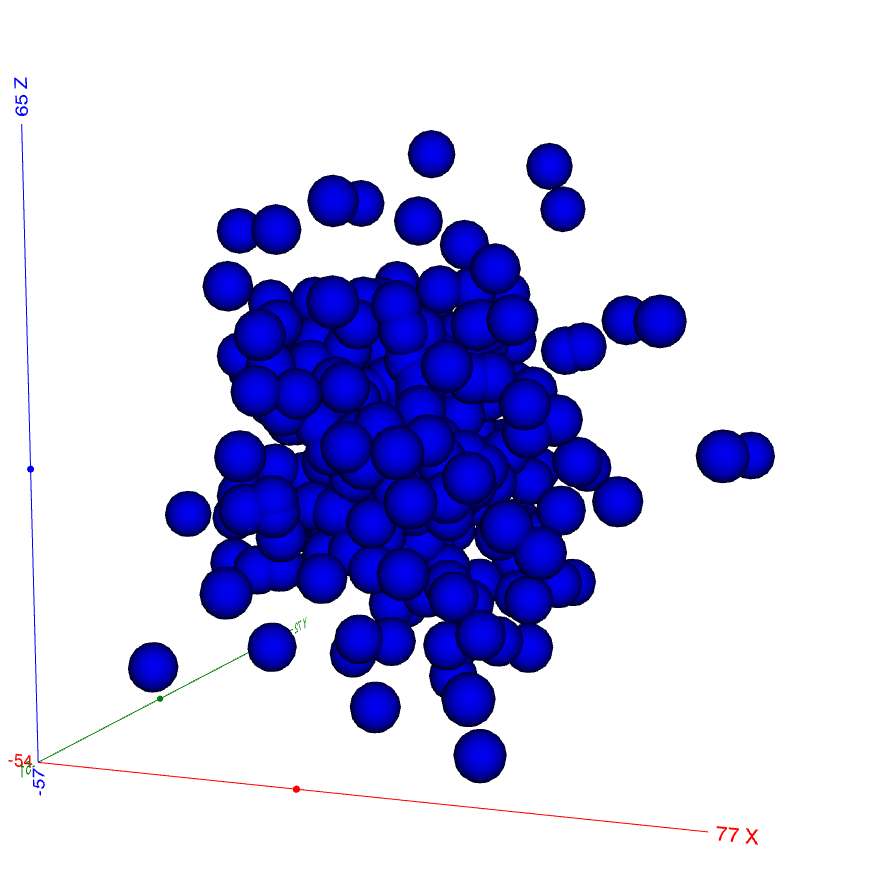}
    \begin{cppcode}
auto rng = simulation.GetRandom()->GetGausRng(0, 20);
ModelInitializer::CreateAgentsRandom(-200, 200, num_agents, create_agent, &rng);
    \end{cppcode}
    \caption{Create agents randomly inside a 3D cube using a gaussian distribution.}
  \end{subfigure}
\end{figure}

\begin{figure}[tbp]
  \ContinuedFloat
  \centering
\begin{subfigure}{0.7\linewidth}
    \centering
    \includegraphics[width=0.55\textwidth]{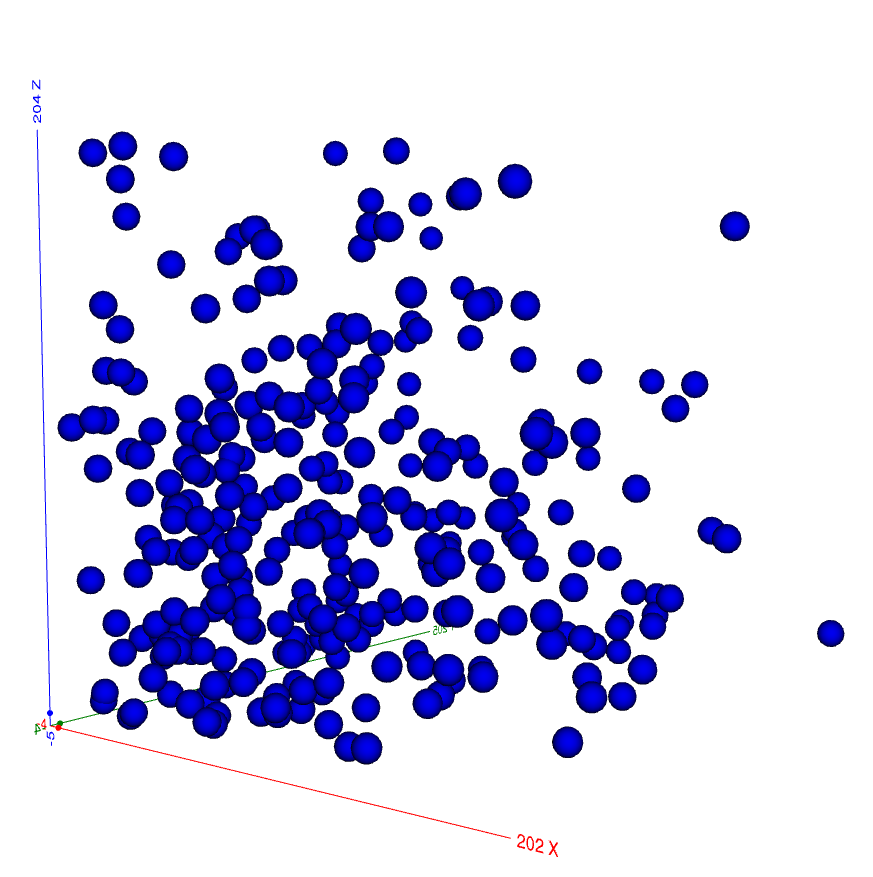}
    \begin{cppcode}
auto rng = simulation.GetRandom()->GetExpRng(100);
ModelInitializer::CreateAgentsRandom(-200, 200, num_agents, create_agent, &rng);
    \end{cppcode}
     \caption{Create agents randomly inside a 3D cube using an exponential distribution.}
\end{subfigure}

  \begin{subfigure}{\linewidth}
    \centering
    \includegraphics[width=0.45\textwidth]{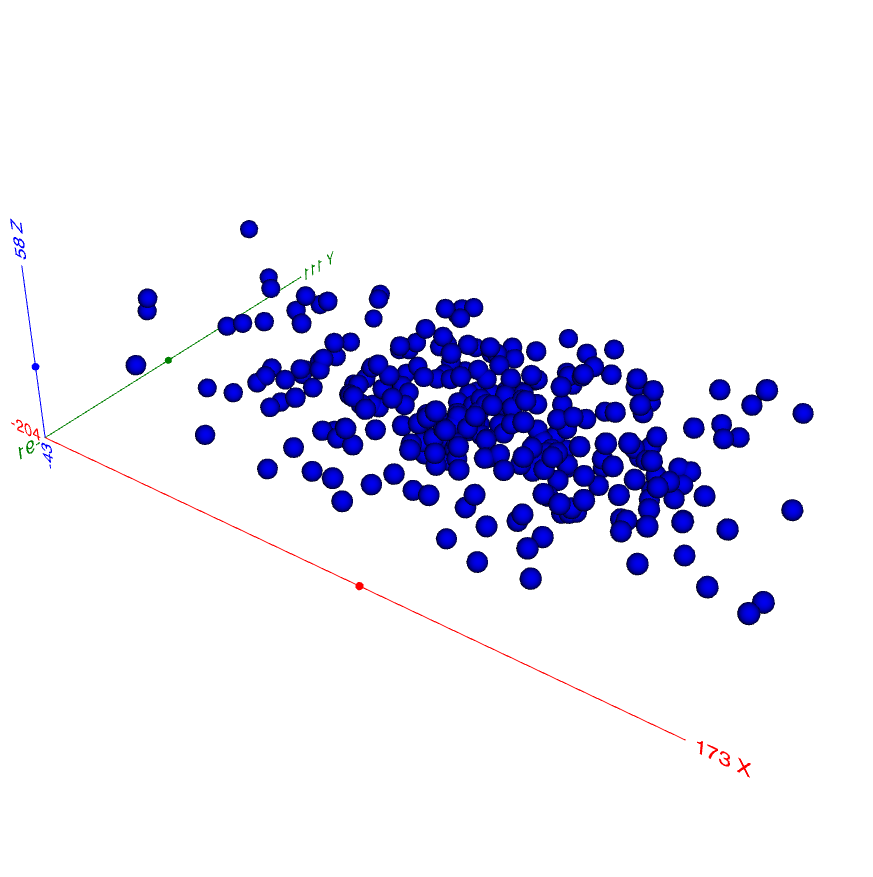}
    \\
    \begin{cppcode}
auto gaus3d = [](const real_t* x, const real_t* params) {
  auto mx = params[0];
  auto my = params[2];
  auto mz = params[4];
  auto sx = params[1];
  auto sy = params[3];
  auto sz = params[5];
  auto ret = (1.0 / (sx * sy * sz * std::pow(2.0 * Math::kPi, 3.0 / 2.0))) *
             std::exp(-std::pow(x[0] - mx, 2.0) / std::pow(sx, 2.0) - std::pow(x[1] - my, 2.0) / std::pow(sy, 2.0) -
                      std::pow(x[2] - mz, 2.0) / std::pow(sz, 2.0));
  return ret;
};
auto* random = simulation.GetRandom();
auto rng = random->GetUserDefinedDistRng3D(gaus3d, {0, 100, 0, 50, 0, 20}, -200, 200, -200, 200, -200, 200);
ModelInitializer::CreateAgentsRandom(-200, 200, num_agents, create_agent, &rng);
    \end{cppcode}
    \caption{Create agents randomly inside a 3D cube using a 3D gaussian distribution.}
\end{subfigure}
\end{figure}

\begin{figure}[tbp]
  \ContinuedFloat
  \centering
\begin{subfigure}{0.7\linewidth}
    \centering
    \includegraphics[width=0.55\textwidth]{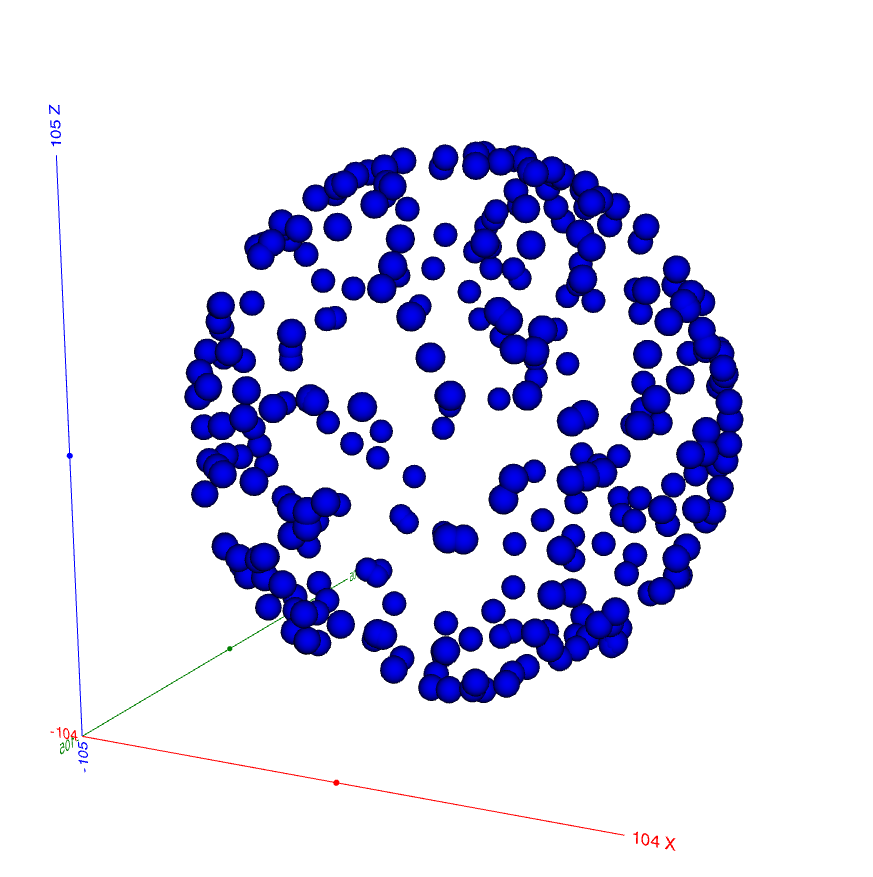}
    \begin{cppcode}
ModelInitializer::CreateAgentsOnSphereRndm({0, 0, 0}, 100, num_agents, 
                                           create_agent);
    \end{cppcode}
    \caption{Create agents randomly on a sphere}
\end{subfigure}

  \begin{subfigure}{0.7\linewidth}
    \centering
    \includegraphics[width=0.55\textwidth]{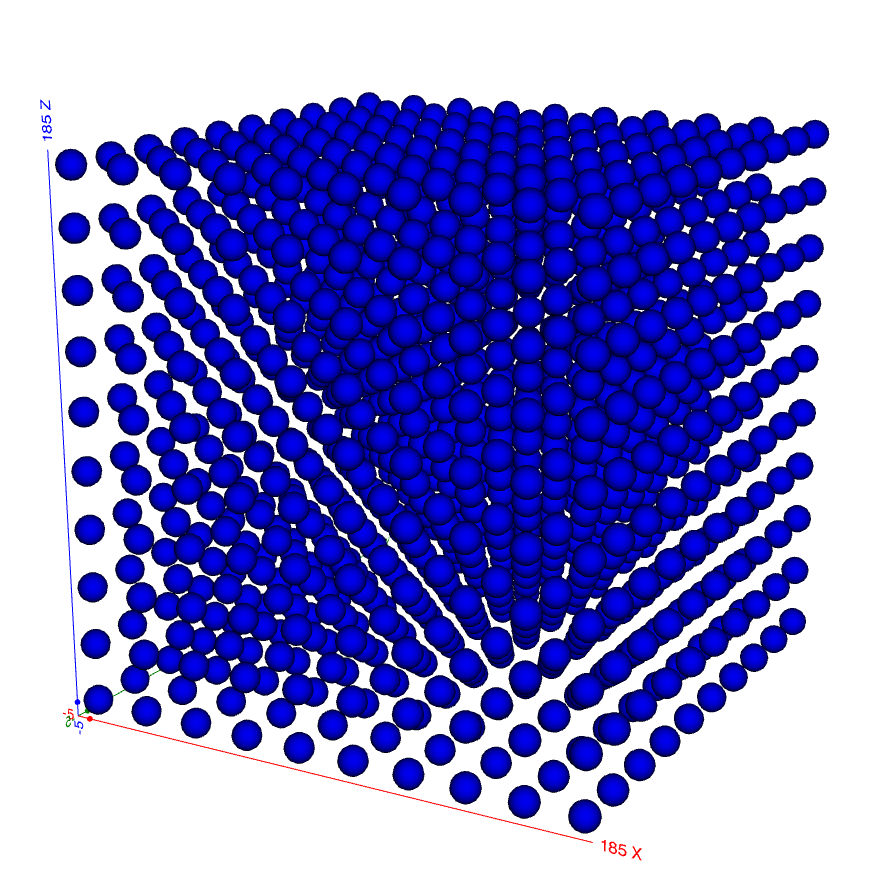}
    \\
    \begin{cppcode}
uint64_t agents_per_dim = 10;
real_t space_between_agents = 20;
ModelInitializer::Grid3D(agents_per_dim, space_between_agents, 
                         create_agent);
    \end{cppcode}
    \caption{Create agents on a 3D grid.}
\end{subfigure}
\end{figure}

\begin{figure}[tbp]
  \ContinuedFloat
  \centering
\begin{subfigure}{0.7\linewidth}
    \centering
    \includegraphics[width=0.55\textwidth]{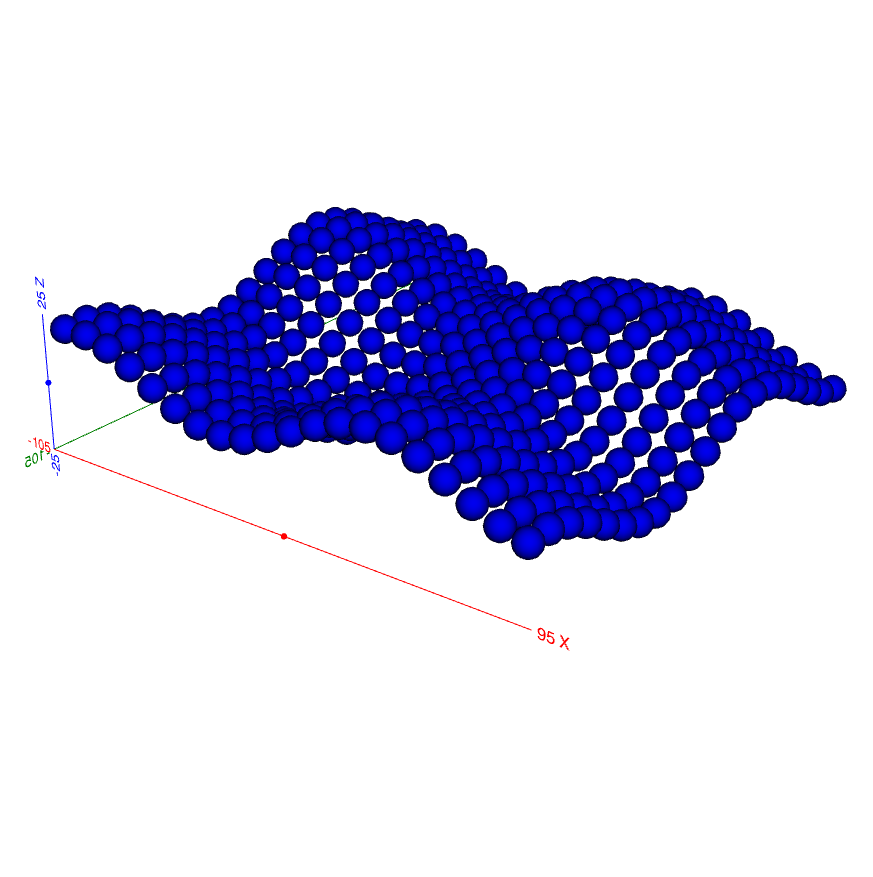}
    \begin{cppcode}
auto f = [](const real_t* x, const real_t* params) {
    return 10 * std::sin(x[0] / 20.) + 10 * std::sin(x[1] / 20.0);
};
ModelInitializer::CreateAgentsOnSurface(f, {}, -100, 100, 10, -100, 100, 10, 
                                        create_agent);
    \end{cppcode}
     \caption{Create agents on a surface.}
\end{subfigure}

  \begin{subfigure}{0.7\linewidth}
    \centering
    \includegraphics[width=0.55\textwidth]{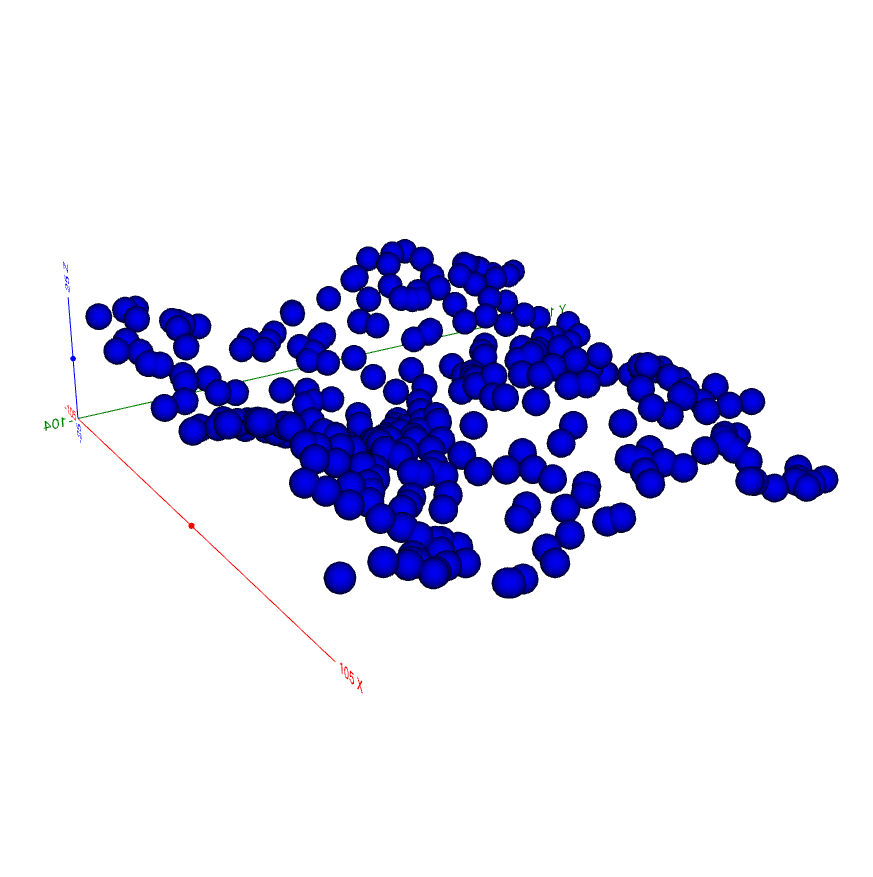}
    \\
    \begin{cppcode}
auto f = [](const real_t* x, const real_t* params) {
    return 10 * std::sin(x[0] / 20.) + 10 * std::sin(x[1] / 20.0);
};
ModelInitializer::CreateAgentsOnSurfaceRndm(f, {}, -100, 100, -100, 100, num_agents, 
                                            create_agent);
    \end{cppcode}
    \caption{Create agents on a surface randomly.}
\end{subfigure}
    \caption{{\bf Generation of agent populations in 3D space.} The full tutorial can be found at \ref{ap:st01}.
    }
   \label{fig:generation-agent-populations}
\end{figure}

\subsection{Agent Reproduction and Mortality} 
The addition and removal of agents during the execution of a simulation is an integral part of agent-based simulations.
Therefore, \bdm{} provides a framework to create new agents during a simulation and initialize their attributes.
By default, agents that are created in iteration \texttt{i} will be visible to other agents in iteration \texttt{i + 1}.
The removal of agents is handled identically.
The handling of when new or removed agents become visible to the simulation is encapsulated in the execution context. 
Therefore, a user could provide another implementation where agents are visible immediately. 

Besides adding and removing agents, a second major part is to provide a generic way to initialize 
  the attributes of an agent. 
To this end, \bdm{} simplifies the regulation of behaviors if new agents are created.
The user can control whether a behavior will be copied to a new agent or
  removed from the existing agent, based on the underlying process (e.g. cell division).
Similarly, agents and behaviors have a function \texttt{Initialize} which can be overridden by user-defined agents to initialize additional attributes.
These concepts are shown in Figure~\ref{platform:fig:event-sequence} and demonstrated in Supplementary Tutorial~\stref{03}--\stref{05}. 

\begin{sidewaysfigure}
\centering
  \includegraphics[width=0.9\linewidth]{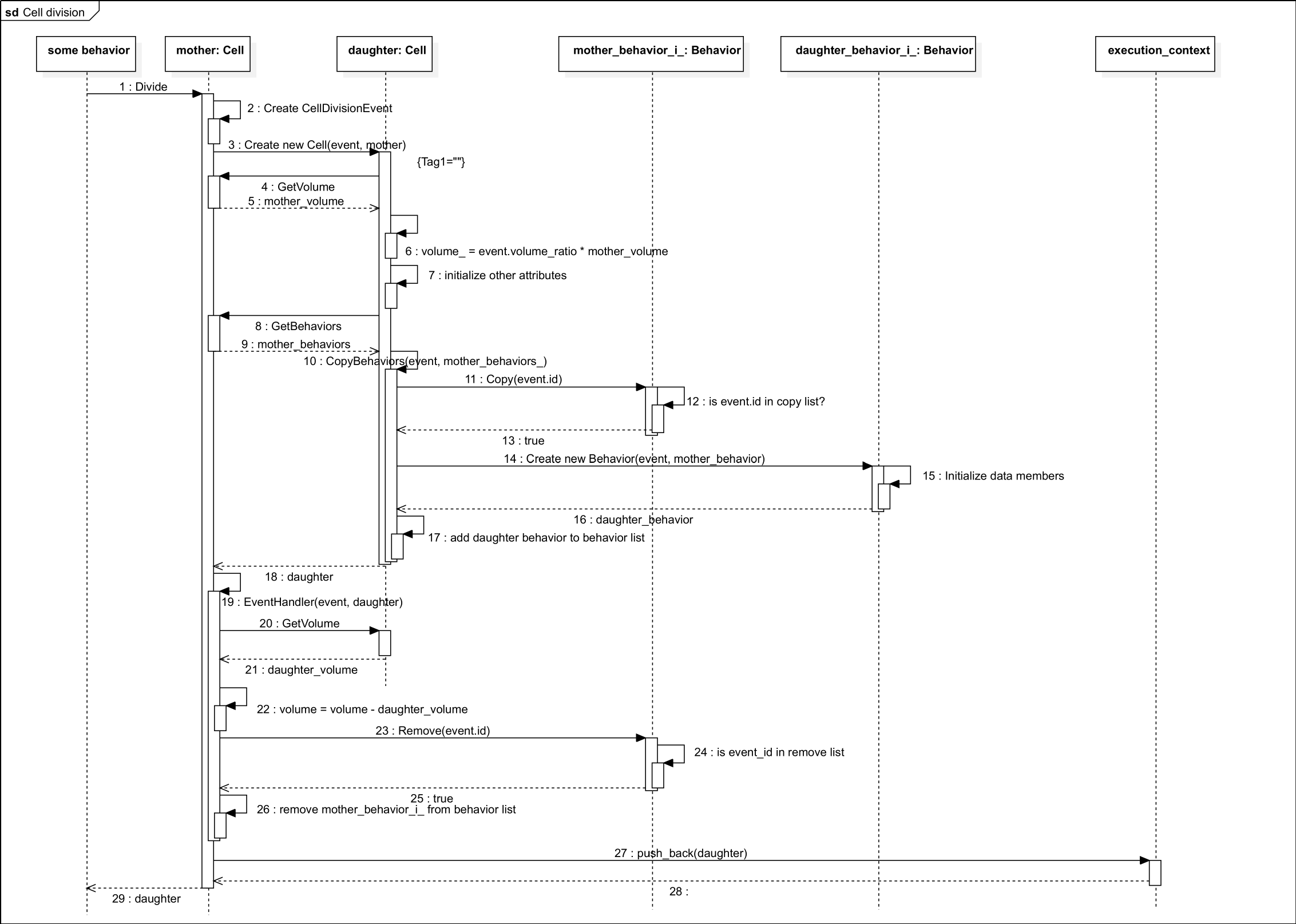}
  \caption{{\bf Event Sequence Diagram.}}
  \label{platform:fig:event-sequence}
\end{sidewaysfigure}

\subsection{Environment Search}
To determine the agents in the local environment (neighbors), \bdm{} uses an 
environment search algorithm (Supplementary Tutorial~\stref{06}). 
\bdm{}'s default environment algorithm is based on a uniform grid implementation.
The implementation divides the total simulation space into uniform boxes of
  the same size and assigns agents to boxes based on the center of
  mass of the agent.
Hence, the agents in the environment can be obtained by
  iterating over the assigned box and all its surrounding boxes
  (27 boxes in total).
The box size is chosen by the user or determined automatically based on the largest agent 
  in the simulation to
  ensure all mechanical interactions are taken into account. 
Alternatively, \bdm{} provides an octree and kd-tree environment implementation.
The interface is kept generic enough to support non-euclidean space environment definitions.

\subsection{Multi-Scale Simulations}
A biological simulation has to account for dynamic mechanisms that range from
  milliseconds to weeks (e.g. physical forces, reaction-diffusion processes, gene
  regulatory dynamics, etc.).
\bdm{} supports processes at different time scales by providing 
  a parameter to specify the time interval between two time steps and
  an execution frequency for each operation (Supplementary Tutorial~\stref{07}).
An execution frequency of one means that the corresponding operation is
  executed every time step.
In contrast, a frequency of three would mean that the
  operation is executed every three time steps.
This mechanism allows \bdm{} to simulate e.g. substance diffusion 
  and neurite growth in the same model.

\subsection{Statistical Analysis}
Statistical analysis plays a fundamental role in generating new insights from simulation data.
\bdm{} builds upon the rich features of CERN's primary data analysis framework ROOT \citep{brun_root_1997},
  which provides an extensive mathematical, histogram, graphing, and fitting library. 
\bdm{} complements this functionality by providing an easy mechanism to collect simulation data
  over time and a simplified API targeted to the agent-based use case.
These capabilities are demonstrated in Supplementary Tutorial~\stref{08}--\stref{11}.

\subsection{Hierarchical Model Support}
\cite{railsback2006} describe an agent-based model in which large agents have to be executed before smaller agents.
\bdm{} supports these hierarchical models with several functions in the 
   \texttt{ResourceManager}, \texttt{Scheduler}, and \texttt{Operation} class. 
The described order can be implemented in \bdm{} by adding three lines of code as demonstrated in Supplementary Tutorial~\stref{12}.
Additionally, it is possible to execute a different set of operations for large and small agents.

\subsection{Hybrid Modeling Support}
Some models benefit from the combination of multiple simulation methodologies---e.g. the combination of 
  an agent-based and continuum-based model.
\bdm{}'s flexible build system supports hybrid modeling capabilities and was demonstrated by \cite{demontigny_2021} to investigate cancer development.

\subsection{Dynamic Scheduling}
In \bdm{} the code that will be executed is controlled by behaviors and operations.
Behaviors can be attached to individual agents and thus allow very fine-grained control.
Agent operations are usually executed for all agents (if no agent filters are specified).
Both behaviors and operations can be created, added, removed, and destroyed during a simulation.
This feature gives the user maximum flexibility to change the executed simulation code over time
(Supplementary Tutorial~\stref{13}). 

\subsection{Parameter Management}
\bdm{} simplifies the definition of simulation parameters by liberating the user from the burden to write code to parse parameter files
  or command line arguments.

\subsection{Parameter Optimization} 
The epidemiology use case presented in the Section~\ref{sec:epidemiology} demonstrates how to define an experiment
  comprised of multiple input parameters, and a user-defined error function.
This experiment definition is used in conjunction with the optimlib library (\href{https://www.kthohr.com/optimlib.html}{https://www.kthohr.com/optimlib.html}) to 
  determine model parameters that match the ground truth.

\subsection{Space Boundary Conditions}
\label{sec:space-boundary}
\bdm{} support three boundary conditions: (i) open, where the simulation space grows to encapsulate all agents in the simulation, (ii) closed, where artificial walls prevent agents from exiting the simulation space, and (iii) toroidal, where agents that leave the space on one side, will enter on the opposite side.

\section{Model Building Blocks}
\label{sec:model-building-blocks}

Currently, \bdm{}'s building blocks (Figure~\ref{fig:bdm-layers}) belong to the (neural) tissue modeling domain.
Similar to the biological model presented in \citep{zublerdouglas2009framework}, \bdm{}
  supports spherical and cylindrical agent geometries, 
  mechanical interactions between agents, 
  and diffusion of extracellular substances.

With these features, researchers can simulate cell body dynamics, 
  neural growth, and gene regulatory networks.

Simulations to study the development of (neural) tissue are only
  one example of how \bdm{} could be used in the future.
By designing \bdm{} in a modular and extensible way, we laid the foundation to
  create new building blocks easily (Figure~\ref{fig:sw-design} and Section~\ref{sec:epidemiology}).
Supplementary Tutorial~\stref{15} for example demonstrates how to replace the
default mechanical force implementation with a user-defined one.

Table~\ref{tab:biological-features} lists the agents, behaviors, and operations that the \bdm{} \texttt{v1.0} installation contains.

\subsection{Mechanical Forces}
Growing realistic cell and tissue morphologies requires the consideration of
  mechanical interactions between agents.
Therefore, \bdm{} examines if two agents collide with each other at
  every timestep.
To find all possible collisions, it is sufficient to evaluate neighbors in the
  environment.
Whenever two agents (e.g. a cell body or a neurite element) overlap,
  a collision occurs.
If a collision is detected, the engine calculates the mechanical forces that act
  on them.

The mechanical force calculation between spheres and cylinders follows the same approach
as the implementation in Cortex3D \citep{zublerdouglas2009framework}.
Both in \bdm{} and Cortex3D, the magnitude of the force is computed based on
\citep{pattana2006division} and comprises a repulsive and
  attractive component:
\begin{flalign}
  \label{eq:force}
\begin{aligned}
    F_N = k\delta-\gamma\sqrt{r\delta}
  \end{aligned}
\end{flalign}
where $\delta$ indicates the spatial overlap between the two elements, and $r$ denotes a combined
measure of the two radii:
\begin{flalign}
  \label{eq:force-radius}
\begin{aligned}
    r = \frac{r_1 r_2}{r_1+r_2}
  \end{aligned}
\end{flalign}

where the radii denote the radii of the interacting spheres or cylinder.

Eq~\ref{eq:force} comprises the effects of the structural tension from the
  pressure between the colliding membrane segments, and the attractive force due
  to the cell adhesion molecules.
The magnitudes of these two force components depend upon the modifiable parameters $k$ and $\gamma$.
In the current form, as in Cortex3D, these are set to $2$ and $1$, respectively.
After the forces have been determined, the agents change their 3D
  location depending on the force resulting from all the mechanical interactions
  with neighbors.
More details about the implementation of the mechanical force, including the force
  between neighboring neurite elements, can be found in \citep{zublerdouglas2009framework}.

\subsection{Extracellular Diffusion}
\label{ch:platform:diffusion}

Signaling molecules, which differentiate and regulate cells, reach their
  destination through diffusion \citep{gurdon_morphogen_2001}.
A well-studied example of this process, called morphogen gradients, is the
  determination of vein positions in the wing of Drosophila
  \citep{bosch_dpp_2017}.

\bdm{} solves the partial differential equations that model the diffusion of
extracellular substances (Fick's second law) with the discrete central difference scheme \citep{smith1985numerical}.
A grid is imposed on the simulation space, and at each timestep, the concentration value of each grid point is updated according to

\begin{flalign}
  \label{eq:centraldiff}
\begin{aligned}
    u_{i,j,k}^{n+1} = u_{i,j,k}^n \times (1- \mu * \Delta t)
		&+ \frac{\nu \Delta t}{\Delta x^2}(u_{i+1,j,k}^n - 2 u_{i,j,k}^n + u_{i-1,j,k}^n) \\
		&+ \frac{\nu \Delta t}{\Delta y^2}(u_{i,j+1,k}^n - 2 u_{i,j,k}^n + u_{i,j-1,k}^n) \\
		&+ \frac{\nu \Delta t}{\Delta z^2}(u_{i,j,k+1}^n - 2 u_{i,j,k}^n + u_{i,j,k-1}^n) \,,
	\end{aligned}
\end{flalign}
where $u_{i,j,k}^n$ is the concentration value on grid point $(i,j,k)$ at
timestep $n$, $\nu$ is the diffusion coefficient, $\mu$ is the decay constant,
$ \Delta t$ is the duration of one timestep, and $\Delta x$, $\Delta y$, and
$\Delta z$ are the distances between grid points in the x, y, and z direction,
respectively.
The distances between the grid points are inversely proportional to the
resolution and determine the accuracy of the solver. 

In \bdm{}, it is possible to define the diffusion behavior at the simulation
  boundaries.
In the default implementation, which we use for our examples in the result
  section, substances diffuse out of the simulation space.

\bdm{} provides predefined substance initializers (e.g., Gaussian) and accepts 
  user-defined functions for arbitrary distributions to determine the initial 
  concentration values.
We used this functionality, for example, in the pyramidal cell growth simulation.

\subsection{Systems Biology Markup Language Integration}

Systems biology markup language (SBML) is a well-established standard to
  describe chemical reaction networks to model metabolism, or cell signaling
  \citep{sbml}.
The BioModels database contains a collection of more than 9000 models in this
  format \citep{biomodels}.

\begin{figure}[h]
  \centering
\includegraphics[width=0.8\linewidth]{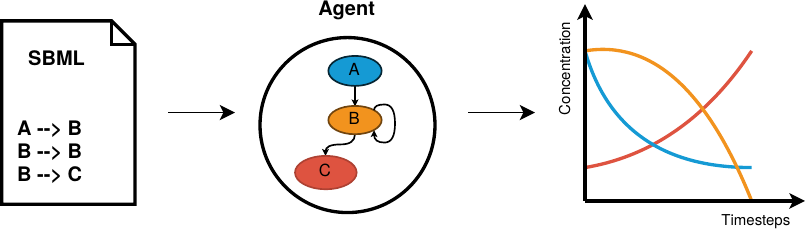}
\caption{{\bf SBML integration.}
    Chemical reaction networks defined in SBML format can be loaded into \bdm{} and
      assigned to any agent.
    The reaction equations are solved for each timestep.
  }
  \label{fig:sbml-integration}
\end{figure}

Therefore, \bdm{} provides the possibility to simulate chemical reaction networks
  described as SBML \cite{sbml} models (Figure~\ref{fig:sbml-integration}).
\bdm{} uses libroadrunner \cite{libroadrunner} to solve the reaction equations,
  which features various deterministic and stochastic solvers.
The intracellular concentration of substances can serve as a control mechanism 
  for agent behaviors displacement, division, branching, etc. 
Furthermore, it is
  feasible to couple the intracellular domain with the extracellular matrix by
  exocytosis and endocytosis of substances and extracellular diffusion.

\section{Exemplary Use Cases}

We demonstrate \bdm{}'s capacity to simulate disparate problems in 
  systems biology with simple yet representative use cases in neuroscience, 
  oncology, and epidemiology.
Since \bdm{} does not contain any epidemiological building blocks, this use case
  indicates how easy it is to implement a model based solely on features from the high- and low-level layer (Figure~\ref{fig:bdm-layers}).

For each use case we present the implemented model, validate the simulation results against verified experimental or analytical data, and report performance data for different problem sizes on multiple hardware configurations.
Furthermore, we provide pseudocode for all agent behaviors and a table with 
  model parameters.

\subsection{Neuroscience Use Case}
\label{sec:pyramidal-cell}

This example illustrates the use of \bdm{} to model neurite growth of pyramidal
  cells using chemical cues.
Initially, a pyramidal cell, composed of a 10 $\mu m$ cell body, three 0.5 $\mu m$ long
  basal dendrites, and one 0.5 $\mu m$ long apical dendrite (all of them considered 
  here as agents), is created in 3D space (L37--L51\footnote{Line numbers in Section~\ref{sec:pyramidal-cell} correspond to the code example in Listing~\ref{lst:pc}.}).
Furthermore, two artificial growth factors were initialized, following a Gaussian
  distribution along the z-axis (L54--L65).
The distribution of these growth factors guided dendrite growth and remained
  unchanged during the simulation.

\lstset{
  language=C++,
  basicstyle=\tiny\ttfamily,
  numbers=left,
  numberstyle=\tiny,
  frame=tb,
  columns=fullflexible,
  showstringspaces=false
}
\lstset{numbers=left,xleftmargin=3em,framexleftmargin=3em}

\begin{figure}[tbp]
  \centering
  \includegraphics[width=0.9\linewidth]{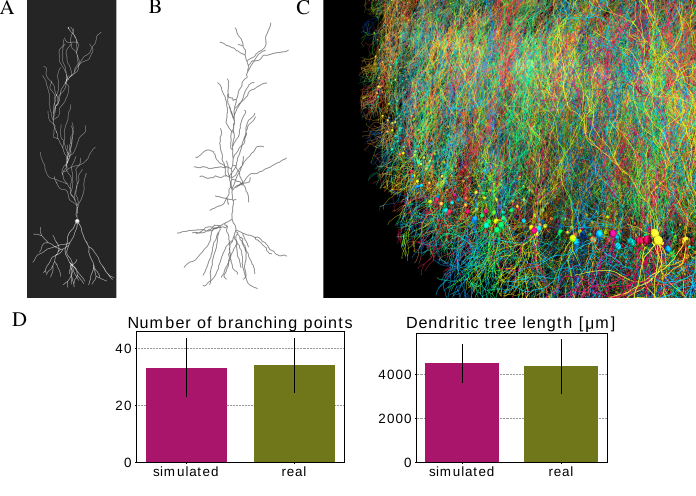}
  \caption{{\bf Pyramidal cell simulation.}
  (A) Example pyramidal cell simulated with \bdm{}.
  (B) Real neuron (R67nr67b-CEL1.CNG) taken from \citep{mellstrom_specific_2016} and 
    visualized with \href{https://neuroinformatics.nl/HBP/morphology-viewer/}{https://neuroinformatics.nl/HBP/morphology-viewer/}
  (C) Large-scale simulation.  
      The model started with 5000 initial pyramidal cell bodies and contained
      9 million agents after simulating 500 iterations.
      The simulation execution time was 35 seconds on a server with 72 CPU cores.
  (D) Morphology comparison between simulated neurons and experimental data from
    \citep{mellstrom_specific_2016}. Error bars represent the standard deviation.
  (A,C) A video is available in the Supplementary Information.
}
    \label{fig:pyramidal-cell}
\end{figure}

\begin{lstfloat}
\begin{minipage}{0.5\textwidth}
\inputcppcode[fontsize=\tiny]{neuroscience-example.cc}
\end{minipage}
\caption{\textbf{Pyramidal cell growth simulation code}. This example shows the 
  required C++ code to simulate the growth of a single pyramidal cell as shown in Figure~\ref{fig:pyramidal-cell}A.
All classes and functions which are not defined in this example are provided by \bdm{}.
Only the body of the two behavior's \texttt{Run} methods has been ommited, but 
  are provided in Supplementary File~S3 (\nameref{si:code}) and Algorithm~\ref{algo:dendrite_growth_module}.
}
\label{lst:pc}
\end{lstfloat}
 
Dendritic development was dictated by a behavior defining growth
  direction, speed, and branching behavior for apical and basal dendrites (L12--L35, see also Algorithm~\ref{algo:dendrite_growth_module}).
\begin{algorithm}[p]
  \caption{Apical and basal dendrite growth.}
  \label{algo:dendrite_growth_module}

\SetKwInOut{Input}{input}
  \SetKwData{Neurite}{neurite}
  \SetKwData{GrowthFactor}{growth\_factor}
  \SetKwData{Diameter}{diameter}
  \SetKwData{Position}{pos}
  \SetKwData{OldDirection}{old\_direction}
  \SetKwData{Direction}{direction}
  \SetKwData{GfGradient}{gradient}
  \SetKwData{DiamThreshold}{diameter\_threshold}
  \SetKwData{DiamThresholdTwo}{diameter\_threshold\_two}
  \SetKwData{GrowthSpeed}{growth\_speed}
  \SetKwData{BranchingProba}{branching\_probability}
  \SetKwData{OldDirectionWeight}{old\_direction\_weight}
  \SetKwData{RandomDirectionWeight}{randomness\_weight}
  \SetKwData{GfGradientWeight}{gradient\_weight}
  \SetKwData{Shrinkage}{shrinkage}
  \SetKwData{CanBranch}{can\_branch}
  \SetKwData{BranchingDirection}{branching\_direction}
  \SetKwFunction{GetDiameter}{GetDiameter}
  \SetKwFunction{SetDiameter}{SetDiameter}
  \SetKwFunction{GetPosition}{GetPosition}
  \SetKwFunction{GetDirection}{GetDirection}
  \SetKwFunction{GetGradient}{GetNormalizedGradient}
  \SetKwFunction{RandomUniform}{RandomUniform}
  \SetKwFunction{RandomUniformThree}{RandomUniform3}
  \SetKwFunction{Extend}{Extend}
  \SetKwFunction{IsTerminal}{IsTerminal}
  \SetKwFunction{IsApical}{IsApical}
  \SetKwFunction{Branch}{Branch}
  \SetKwFunction{Bifurcate}{Bifurcate}
  \SetKwFunction{CalcBranchingDirection}{CalculateBranchingDirection}

  \Input{\Neurite, \GrowthFactor, \DiamThreshold, \DiamThresholdTwo, \GrowthSpeed, \BranchingProba,
         \OldDirectionWeight, \RandomDirectionWeight, \GfGradientWeight, \Shrinkage, \CanBranch}

  \Diameter $\leftarrow$ \Neurite.\GetDiameter{}\;
  \If{$\Diameter > \DiamThreshold$}{
\OldDirection $\leftarrow$ \Neurite.\GetDirection{}\;
    \Position $\leftarrow$ \Neurite.\GetPosition{}\;
    \GfGradient $\leftarrow$ \GrowthFactor.\GetGradient{\Position}\;
    \Direction $\leftarrow$ $\OldDirection \times \OldDirectionWeight + \GfGradient \times \GfGradientWeight + \RandomUniformThree{-1, 1} \times \RandomDirectionWeight$\;
    \Neurite.\Extend{\GrowthSpeed, \Direction}\;
    \Neurite.\SetDiameter{\Diameter - \Shrinkage}\;
    \If{\Neurite.\IsApical{}} {
      \If{$\CanBranch$ {\bf and} \Neurite.\IsTerminal{} {\bf and}  $\Diameter < \DiamThresholdTwo$ {\bf and} $\RandomUniform{0, 1} < \BranchingProba$} {
        \BranchingDirection $\leftarrow$ \CalcBranchingDirection{\Neurite}\;
        \Neurite.\Branch{\BranchingDirection}\;
      }
    }
    \ElseIf {$\RandomUniform{0, 1} < \BranchingProba$} {
      \Neurite.\Bifurcate{}\;
    }
  }
\end{algorithm}
 At each step of the simulation, the dendritic growth direction depended on
  the local chemical growth factor gradient, the dendrite's previous direction,
  and a randomly chosen direction.
In addition, the dendrite's diameter tapered as it grew (shrinkage), until it
  reached a specified diameter, preventing it from growing any further.
The weight of each element on the direction varied between apical and basal
  dendrites.

These simple rules gave rise to a straight long apical dendrite with a simple
  branching pattern and more dispersed basal dendrites, as shown in
  Figure~\ref{fig:pyramidal-cell}A, similar to what can be observed in real
  pyramidal cell morphologies as shown in Figure~\ref{fig:pyramidal-cell}B or \cite{spruston2008pyramidal} (Figure~1A CA1).
Using our growth model, we were able to generate a large number of various
  realistic pyramidal cell morphologies.
We used a publicly available database of real pyramidal cells coming from
  \citep{mellstrom_specific_2016} for comparison and parameter tuning.
Table~\ref{tab:dendrites_parameters} shows the determined parameters.
Two measures were used to compare our simulated neurons and the 69 neurons
  composing the real morphologies database: the average number of branching points,
  and the average length of dendritic trees.
No significant differences were observed between our simulated neurons and the
  real ones ($p < 0.001$ using a T-test for two independent samples).
These results are shown in Figure~\ref{fig:pyramidal-cell}D.
The model specification of the pyramidal cell growth simulation consists of 127 lines of C++ code
  (Listing~\ref{lst:pc}).

\begin{figure}[tbp]
  \centering
  \includegraphics[width=\linewidth]{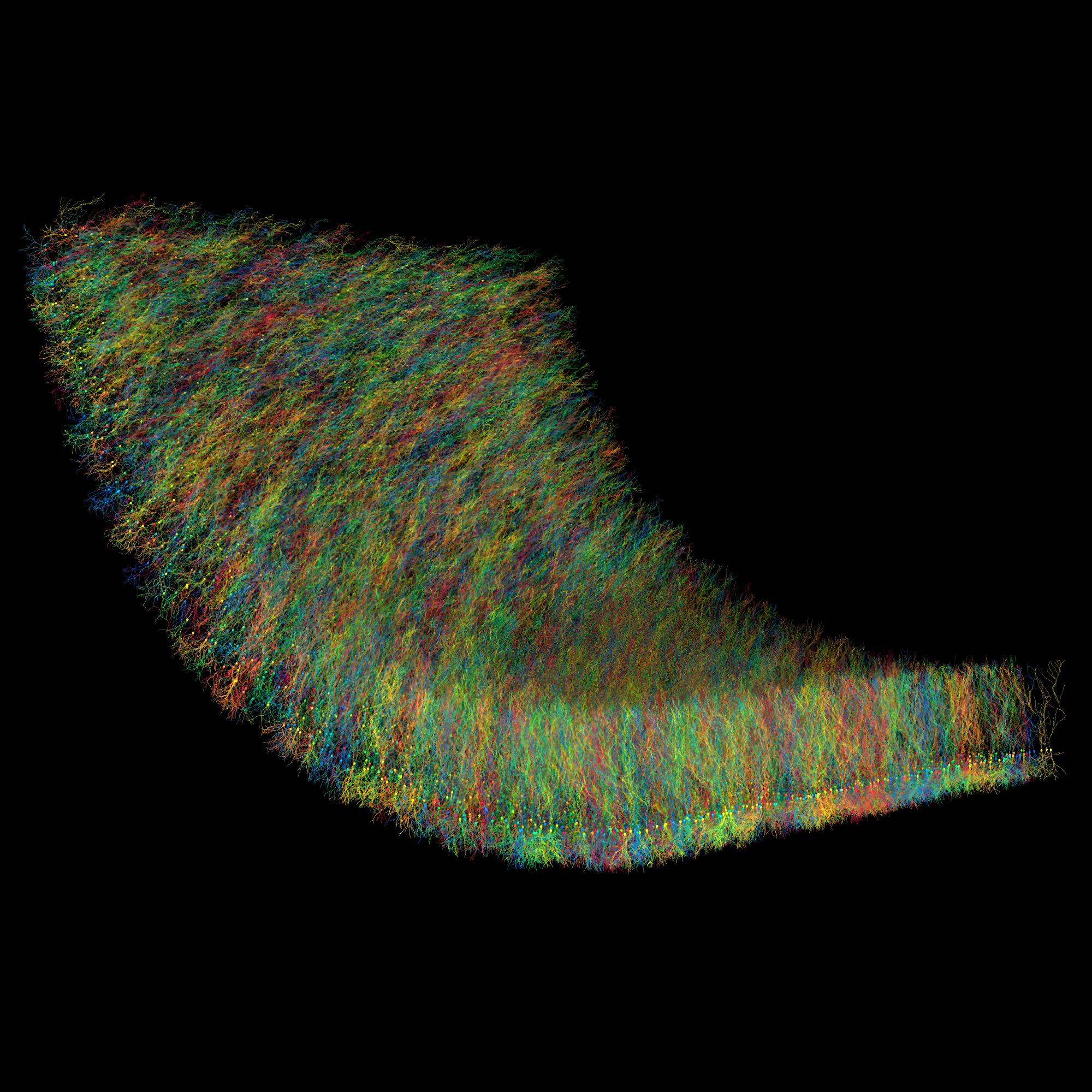}
  \caption{{\bf Large-scale pyramidal cell simulation.}
      The model started with 5000 initial pyramidal cell bodies and contained
      9 million agents after simulating 500 iterations.
      The simulation execution time was 35 seconds on a server with 72 CPU cores.
      A video is available in the Supplementary Information.}
    \label{fig:pyramidal-cell-large-scale}
\end{figure}

\begin{figure}[tbp]
  \centering
  \includegraphics[width=\linewidth]{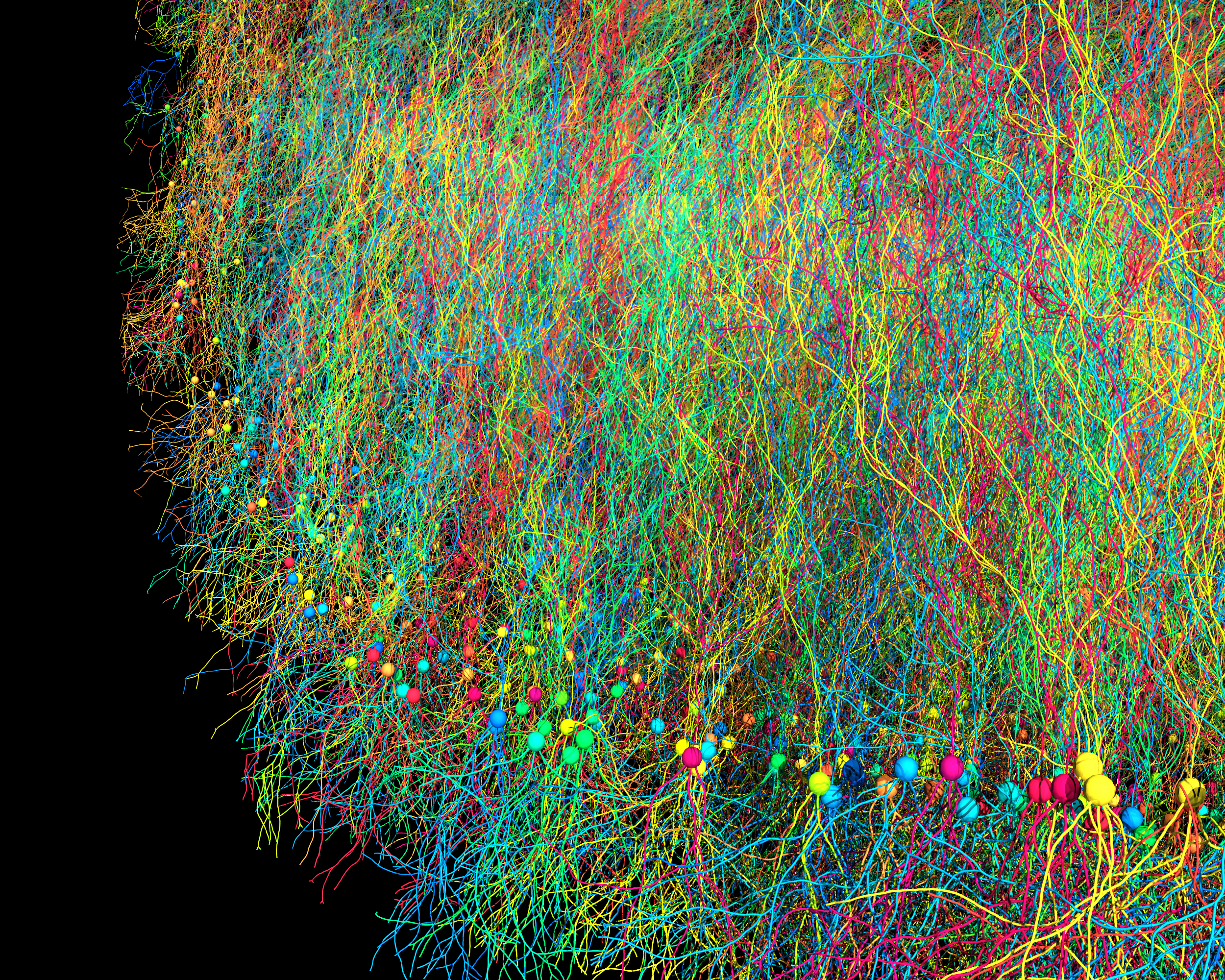}
  \caption{{\bf Large-scale pyramidal cell simulation.}
      Detailed view of the simulation shown in Figure~\ref{fig:pyramidal-cell-large-scale}.}
    \label{fig:pyramidal-cell-large-scale-detail}
\end{figure}

\begin{table}[p]
\caption{{\bf Model parameters for the pyramidal cell growth simulation.}}
\begin{tabular}{@{}lll@{}}
    \toprule
    \textbf{Parameter} & \textbf{Apical dendrite} & \textbf{Basal dendrite} \\
    \midrule
    {Diameter threshold}     & $0.575$    & $0.75$ \\
    {Diameter threshold two} & $0.55$     & $ $ \\
    {Old direction weight}   & $4$        & $6$ \\
    {Gradient weight}        & $0.06$     & $0.03$ \\
    {Randomness weight}      & $0.3$      & $0.4$ \\
    {Growth speed}           & $100$      & $50$ \\
    {Shrinkage}              & $0.00071$  & $0.00085$ \\
    {Branching probability}  & $0.038$    & $0.006$ \\
    \bottomrule
  \end{tabular}
\label{tab:dendrites_parameters}
\end{table}
 
Figure~\ref{fig:pyramidal-cell}C,~\ref{fig:pyramidal-cell-large-scale}, and \ref{fig:pyramidal-cell-large-scale-detail} show a large-scale simulation incorporating
  5000 neurons similar to the one described above and demonstrates the potential of
  \bdm{} for developmental, anatomical, and connectivity studies in the brain.
This simulation contained 9 million agents.

 \newpage
\subsection{Oncology Use Case}
\label{sec:oncology-use-case}

In this section, we present a tumor spheroid simulation to replicate
  in vitro experiments from \cite{Gongetal2015invitroMCF7}.
Tumor spheroid experiments are typically employed to investigate the
  pathophysiology of cancer and are also being used for pre-clinical drug screening
  \citep{Nunes_et_al:2019}.
Here we considered three in vitro test cases using a breast adenocarcinoma
  MCF-7 cell line \citep{Gongetal2015invitroMCF7} with different initial cell
  populations ($2000$, $4000$, and $8000$ MCF-7 cells).
Our goal was to simulate the growth of this mono cell culture embedded in a
  collagenous (extracellular) matrix.

The fundamental cellular mechanisms modeled here include cell growth,
  cell duplication, cell migration, and cell apoptosis.
All these processes are implemented in the class \texttt{TumorCellBehavior} (see Algorithm~\ref{algo:tumor_growth_module}).
The cell growth rate was derived from the published data \citep{Sutherland3998}, while cell migration
  (cell movement speed), cell survival, and apoptosis were fine-tuned after
  trial and error testing (see Table~\ref{tab:tumor-spheroids}).
Since the in vitro study considered the same agarose gel matrix composition among the experiments,
  the \bdm{} model assumes identical parameters for the cell--matrix interactions in the simulations.
Considering the homogeneous ECM properties, tumor cell migration was
  modeled as Brownian motion.

\begin{algorithm}[p]
  \caption{Cancer cell behavior.}
  \label{algo:tumor_growth_module}

  \SetKwInOut{Input}{input}
  \SetKwData{Cell}{cell}
  \SetKwData{Age}{age}
  \SetKwData{MaxDiameter}{max\_diameter}
  \SetKwData{RandomVector}{random\_vector}
  \SetKwData{Brownian}{brownian}
  \SetKwData{DisplacementRate}{displacement\_rate}

  \SetKwData{GrowthSpeed}{growth\_speed}
  \SetKwData{DivisionProbability}{division\_probability}
  \SetKwData{DeathProbability}{death\_probability}
  \SetKwData{MinimumCellAge}{minimum\_cell\_age}
  \SetKwFunction{IncreaseVolume}{IncreaseVolume}
  \SetKwFunction{GetDiameter}{GetDiameter}
  \SetKwFunction{RandomUniform}{RandomUniform}
  \SetKwFunction{RandomUniformThree}{RandomUniform3}
  \SetKwFunction{LTwoNorm}{L2Norm}
  \SetKwFunction{RemoveFromSimulation}{RemoveFromSimulation}
  \SetKwFunction{UpdatePosition}{UpdatePosition}
  \SetKwFunction{Divide}{Divide}

  \Input{\Cell, \MinimumCellAge, \DeathProbability, \DisplacementRate,
         \GrowthSpeed, \DivisionProbability}

  \RandomVector $\leftarrow$ \RandomUniformThree{-1, 1}\;
  \Brownian $\leftarrow$ $ \RandomVector \div \RandomVector.\LTwoNorm{}$\;
  \Cell.\UpdatePosition{$\Brownian \times \DisplacementRate$}\;

  \If{$\Age >= \MinimumCellAge$ {\bf and} $\RandomUniform{0, 1} < \DeathProbability$}{
    \Cell.\RemoveFromSimulation{}\;
    \Return\;
  }
  \Age $\leftarrow \Age + 1$\;

  \uIf{$\Cell.\GetDiameter < \MaxDiameter$} {
    \Cell.\IncreaseVolume{\GrowthSpeed}\;
  }
  \ElseIf{$\RandomUniform{0, 1} < \DivisionProbability$} {
      \Cell.\Divide{}\;
  }
\end{algorithm}

\begin{table}[p]
\caption{{\bf Model parameters for the tumor spheroid growth simulations.}}
\begin{tabular}{@{}llll@{}}
    \toprule
    \textbf{Parameter [dimensions]} & \textbf{2000} & \textbf{4000} & \textbf{8000} \\
    & \multicolumn{3}{c}{\bf cells/well} \\
    \midrule
    {Cell growth rate [$\mu$m\textsuperscript{3}/h]}      & $42.0$   & $35.0$   & $29.9$   \\
    {Minimum cell age to apoptosis [h]}                   & $87$     & $87$     & $87$     \\
    {Division probability}                                & $0.0215$ & $0.0215$ & $0.0215$ \\
    {Cell death probability}                              & $0.033$  & $0.033$  & $0.033$  \\
    {Maximum cell speed [$\mu$m/h]}                       & $1.0$    & $0.9$    & $0.2$    \\
    {Cell--ECM adherence}     & $1.8$    & $1.8$    & $1.8$    \\
    {\hspace{3mm} coefficient [dimensionless]} & & & \\
    {Random cell movement} & $0.005$  & $0.005$  & $0.0005$ \\
    {\hspace{3mm} = displacement rate [$\mu$m/h]} & & & \\
    \bottomrule
  \end{tabular}
\label{tab:tumor-spheroids}
\end{table}
 
\begin{figure}[!t]
  \centering
\includegraphics[width=0.9\linewidth]{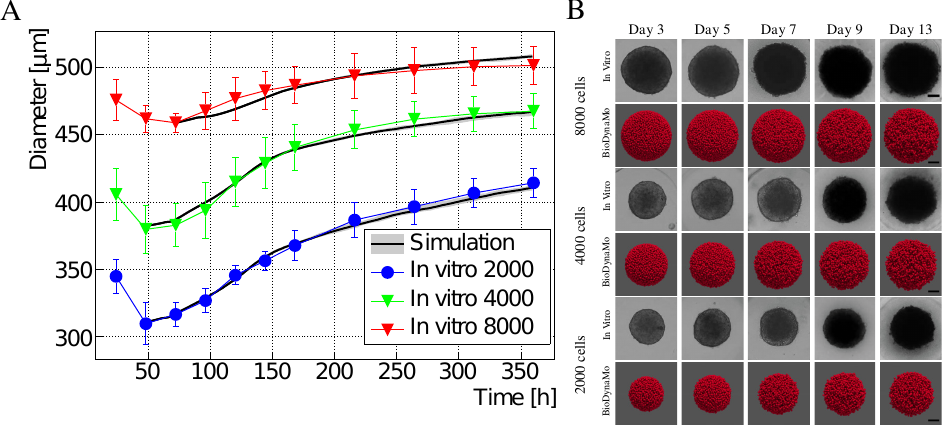}
\caption{{\bf Comparison between in vitro
        MCF-7 tumor spheroid experiments and our in silico simulations using
          BioDynaMo.
      }
    (A) Human breast adenocarcinoma tumor spheroid (MCF-7 cell line) development
      during a 15 day period, where different initial cell populations were
      considered (see Fig 3 in \citep{Gongetal2015invitroMCF7}).
    Error bars denote standard deviation to the experimental data.
    The mean of the in silico results is shown as a solid black line with
    a grey band depicting minimum and maximum observed value.
    (B) shows a qualitative comparison between the microscopy images and simulation
      snapshots.
    Scale bars correspond to 100$\mu$m.
    A video is available in the Supplementary Information.
  }
  \label{fig:tumor-spheroids}
\end{figure}

The in vitro experiments from
  \cite{Gongetal2015invitroMCF7} and the simulations using
  \bdm{} are depicted in Figure~\ref{fig:tumor-spheroids}.
Each line plot in Figure~\ref{fig:tumor-spheroids}A compares the mean diameter
  between the experiments and the simulations
  over time, which demonstrates the validity and accuracy of \bdm{}.
The diameter of the spheroids in the simulations were deducted from the volume
  of the convex hull that enclosed all cancer cells.
The in vitro experiments used microscopy imaging to measure the spheroid's
  diameters \citep{Gongetal2015invitroMCF7}.
Figure~\ref{fig:tumor-spheroids}B compares snapshots of the
  simulated tumor spheroids (bottom row) against microscopy images of in vitro
  spheroids (top row) at different time points.
The spheroid's morphologies between the in vitro experiments and the
  \bdm{} simulations are in excellent agreement.

Model specification required 154 lines of C++ code.

\subsection{Epidemiology Use Case}
\label{sec:epidemiology}

This section presents an agent-based model that describes the spreading of infectious diseases between humans.
The model divides the population into three groups: susceptible, infected, and recovered 
  (SIR) agents.
We compare our simulation results with the solution of the original SIR model from 
  \cite{kermack_1927}, which used the following three differential equations to describe the model dynamics:
$dS/dt = - \beta S I / N$, 
$dI/dt = \beta S I / N - \gamma I$, and
$dR/dt = \gamma I$.
$S$, $I$, and $R$ are the number of susceptible, infected, and recovered individuals, $N$ is the
  total number of individuals, $\beta$ is the mean transmission rate, and $\gamma$ is the recovery rate.

\begin{figure}
  \centering
\includegraphics[width=0.95\linewidth]{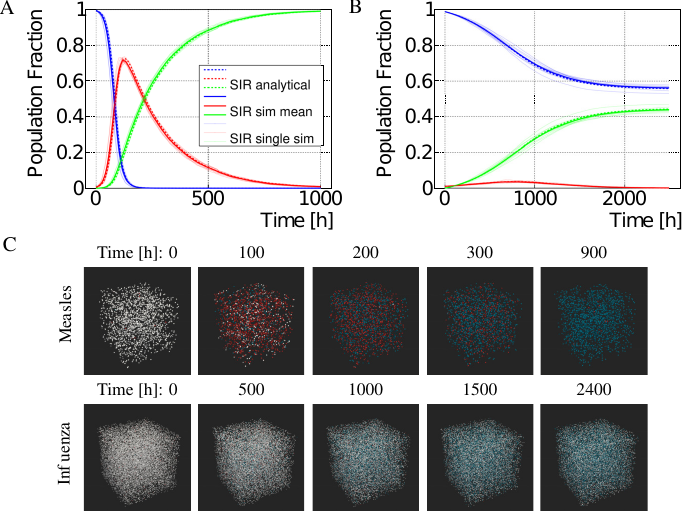}
\caption{{\bf Measles and seasonal influenza SIR model results.}
    (A,B) Comparison between agent-based (solid lines) and analytical (dashed lines) model for
    measles (A) and seasonal influenza (B). The agent-based simulation was repeated ten times. The 
    individual simulation results are shown as thin solid lines. The bold solid line represents the mean
    from all simulations. The legend is shared between the two plots.
    (C) Visualization of the measles and influenza model for different time steps in 3D space.
    Susceptible persons are shown in white, infected persons in red, and recovered persons in blue.
}
\label{fig:epidemiology}
\end{figure}

For our agent-based implementation (Figure~\ref{fig:epidemiology}C) we created a new agent
  (representing a person) that encompasses three new behaviors
  (see Figure~\ref{fig:sw-design}).
A susceptible agent became infected with the infection probability if an infected agent was within the infection radius (Algorithm~\ref{algo:infection}).
An infected agent recovered with the recovery probability at every time step (Algorithm~\ref{algo:recovery}).
All agents moved randomly in space with toroidal boundary condition. The absolute distance an agent may travel in every time step was limited (Algorithm~\ref{algo:random-movement}).

\begin{algorithm}[p]
  \caption{Infection behavior.}
  \label{algo:infection}

  \SetKwInOut{Input}{input}
  \SetKwData{Person}{person}
  \SetKwData{Environment}{environment}
  \SetKwData{InfectionRadius}{infection\_radius}
  \SetKwData{InfectionProbability}{infection\_probability}
  \SetKwData{Neighbors}{neighbors}
  \SetKwData{Neighbor}{neighbor}
  \SetKwData{Susceptible}{susceptible}
  \SetKwData{Infected}{infected}
  
  \SetKwFunction{RandomUniform}{RandomUniform}
  \SetKwFunction{GetState}{GetState}
  \SetKwFunction{SetState}{SetState}
  \SetKwFunction{GetNeighbors}{GetNeighbors}

  \Input{\Person, \Environment, \InfectionProbability, \InfectionRadius}

  \If{$\Person.\GetState{} == \Susceptible$ {\bf and} $\RandomUniform{0,1} < \InfectionProbability$}{
    \Neighbors $\leftarrow \Environment.\GetNeighbors{\InfectionRadius}$\;
    \For{{\bf each} \Neighbor in \Neighbors}{
      \If{$\Neighbor.\GetState{} == \Infected$}{
        \Person.\SetState{\Infected}\;
      }
    }
  }
\end{algorithm}

\begin{algorithm}[p]
  \caption{Recovery behavior.}
  \label{algo:recovery}

  \SetKwInOut{Input}{input}
  \SetKwData{Person}{person}
  \SetKwData{RecoveryProbability}{recovery\_probability}
  \SetKwData{Recovered}{recovered}
  \SetKwData{Infected}{infected}
  
  \SetKwFunction{RandomUniform}{RandomUniform}
  \SetKwFunction{GetState}{GetState}
  \SetKwFunction{SetState}{SetState}

  \Input{\Person, \RecoveryProbability}

  \If{$\Person.\GetState{} == \Infected$ {\bf and} $\RandomUniform{0,1} < \RecoveryProbability$}{
    \Person.\SetState{\Recovered}\;
  }
\end{algorithm}

\begin{algorithm}[p]
  \caption{Random movement behavior.}
  \label{algo:random-movement}

  \SetKwInOut{Input}{input}
  \SetKwData{Person}{person}
  \SetKwData{Speed}{speed}
  \SetKwData{MaxBound}{max\_bound}
  \SetKwData{Position}{position}
  \SetKwData{RandomMovement}{movement}
  \SetKwData{NewPosition}{new\_position}
  \SetKwData{Element}{el}
  
  \SetKwFunction{RandomUniform}{RandomUniform}
  \SetKwFunction{GetPosition}{GetPosition}
  \SetKwFunction{SetPosition}{SetPosition}
  \SetKwFunction{Fmod}{fMod}
  \SetKwFunction{LTwoNorm}{L2Norm}

  \Input{\Person, \Speed,  \MaxBound}
  
  \Position $\leftarrow$ \Person.\GetPosition{}\;
  \RandomMovement $\leftarrow$ \RandomUniformThree{-1, 1}.\LTwoNorm{}\;
  \NewPosition $\leftarrow \Position + \RandomMovement \times \Speed$\;
  \For{{\bf each} \Element in \NewPosition}{
    \Element $\leftarrow \Fmod{\Element, \MaxBound}$\;
    \If{$\Element < 0$} {
      \Element $\leftarrow \MaxBound + \Element$\;
    }
  }
  \Person.\SetPosition{\NewPosition}\;
\end{algorithm}

We selected two infectious diseases with different characteristics to verify our model: measles and seasonal influenza.
We obtained values for the basic reproduction number $R_0$ and recovery duration $T_R$ from the literature  
(Measles: $R_0 = 12.9$, $T_R = 8$ days \citep{guerra_2017, who_measles}, Influenza: 
  $R_0 = 1.3$, $T_R = 4.1$ days \citep{chowell_2008})
  and determined the parameters $\beta$ and $\gamma$ for the analytical model, based on $R_0 = \beta / \gamma$  and $\gamma = 1 / T_R$.  
For the agent-based model we set the recovery probability to $\gamma$, and determined the remaining parameters (infection radius, infection probability, and maximum movement in one time step)  using particle swarm optimization \citep{kennedy_1995} (see Table~\ref{tab:epidemiology}).
Figure~\ref{fig:epidemiology} shows that the agent-based model is in excellent agreement with the equation-based approach from \citep{kermack_1927} for measles and influenza.

\begin{table}[tbp]
  \centering
  \caption{{\bf Model parameters for the epidemiological use case.}}
\begin{tabular}{@{}llll@{}}
    \toprule
    \textbf{Parameter [dimension]}          & \textbf{Measles} & \textbf{Seasonal Influenza}\\
    \midrule
    {$\beta$ (analytical solution)}          & $0.06719$           & $0.01321$   \\
    {$\gamma$ (analytical solution)}         & $0.00521$           & $0.01016$   \\
    {Time step interval [h]}                 & $1$                 & $1$     \\
    {Number of time steps}                   & $1000$              & $2500$     \\
    {Cubic simulation space length [m]}      & $100$               & $215$ \\
    {Initial number of susceptible persons}  & $2000$              & $20000$  \\
    {Initial number of infected persons}     & $20$                & $200$  \\
    {Infection radius [m]}                   & $3.24179$           & $3.2123$  \\
    {Infection probability}                  & $0.28510$           & $0.04980$  \\
    {Recovery probability}                   & $0.00521$           & $0.01016$  \\
    {Max movement per time step [m]}         & $5.78594$           & $4.2942$  \\
    \bottomrule
  \end{tabular}
\label{tab:epidemiology}
\end{table}
 
Model specification required 181 lines of C++ code.

\subsection{Code Examples} 
\label{sec:code}

Figure~\ref{fig:sw-design} depicts in an abstract way that \bdm{}'s software design
  is open for extension.
With the four code examples in Listing~\ref{lst:pc} to \ref{lst:behavior}---taken directly from the 
  presented use cases and benchmarks---we want to emphasize how little 
  code is required to add new functionality.

\lstset{
  language=C++,
  basicstyle=\ttfamily,
  numbers=left,
frame=tb,
  columns=fullflexible,
  showstringspaces=false
}
\lstset{numbers=left,xleftmargin=3em,framexleftmargin=3em}

\begin{lstfloat}
\begin{lstlisting}
struct SimParam : public ParamGroup {
  BDM_PARAM_GROUP_HEADER(SimParam, 1);
  uint64_t cells_per_dim = 30;
  uint64_t iterations = 100;
};
\end{lstlisting}
\caption{Additional simulation parameters for the cell growth and division benchmark.}
\label{lst:sim-param}
\end{lstfloat}

\begin{lstfloat}
\begin{lstlisting}
/// Possible states.
enum State { kSusceptible, kInfected, kRecovered };

class Person : public Cell {
  BDM_AGENT_HEADER(Person, Cell, 1);

 public:
  Person() {}
  explicit Person(const Double3& position) : Base(position) {}
  virtual ~Person() {}

  /// This attribute stores the current state of the person.
  int state_ = State::kSusceptible;
};
\end{lstlisting}
\caption{New agent class used in the epidemiology use case.}
\label{lst:agent}
\end{lstfloat}

\begin{lstfloat}
\begin{lstlisting}
struct Recovery : public Behavior {
  BDM_BEHAVIOR_HEADER(Recovery, Behavior, 1);

  Recovery() {}

  void Run(Agent* a) override {
    auto* person = bdm_static_cast<Person*>(a);
    if (person->state_ == kInfected) {
      auto* sim = Simulation::GetActive();
      auto* random = sim->GetRandom();
      auto* sparam = sim->GetParam()->Get<SimParam>();
      if (random->Uniform(0, 1) <= sparam->recovery_probability) {
        person->state_ = State::kRecovered;
      }
    }
  }
};
\end{lstlisting}
\caption{New behavior used in the epidemiology use case.}
\label{lst:behavior}
\end{lstfloat}

\section{Performance Analysis}
\label{sec:performance-analysis}

We compare \bdm{}'s performance with two established serial ABS platforms 
  (Cortex3D \citep{zublerdouglas2009framework} and NetLogo \citep{netlogo}),
  analyze \bdm{}'s scalability, and quantify the impact of GPU acceleration.
Display updates are turned off on all platforms for these evaluations.
Cortex3D has the highest similarity in terms of the underlying biological model
  out of all the related works presented in Section~\ref{ch:related-work}.
More specifically, \bdm{} and Cortex3D use the same method to determine mechanical forces
 between agents and the same model to grow neural morphologies. 
This makes Cortex3D the best candidate with which to compare \bdm{}
  and ensure a fair comparison.

We quantify the performance of \bdm{} with four simulations: cell
  growth and division, soma clustering, pyramidal cell growth, and the epidemiology use case.
Table~\ref{tab:systems} details the experimental setup that we used.
\begin{sidewaystable}
\begin{adjustbox}{max height=0.9\textheight}
\centering
 \caption{
   {\bf Experimental setup.}
   Main parameters of the systems that we used to run the benchmarks
    of this chapter. \nameref{si:code} contains more details.
     }
   \begin{tabular}{@{}llp{10cm}l@{}}
     \toprule
     \textbf{System}
     & \textbf{Main memory}
     & \textbf{CPU / GPU}
     & \textbf{OS} \\

     \midrule
     A
     & 504 GB
     & \multirow{2}{10cm}{\begin{tabular}[c]{@{}p{10cm}@{}}Server with four Intel(R) Xeon(R) E7-8890 v3 CPUs @ 2.50GHz with a total of 72 physical
     cores, two threads per core and four NUMA nodes.\end{tabular}}
     & \multirow{2}{2cm}{\begin{tabular}[c]{@{}l@{}}\vspace{3mm}CentOS 7.9.2009\end{tabular}} \\

     \cmidrule{1-2}
     B
     & 1008 GB
     &
     & \\

     \cmidrule{1-4}
     C
     & 191 GB
     & \begin{tabular}[c]{@{}p{10cm}@{}}Server with two Intel(R) Xeon(R) Gold 6130 CPUs @ 2.10GHz with 16 physical cores, two threads per core, and two NUMA nodes.\\ One NVidia Tesla V100 SXM2 GPU with 32 GB memory.\end{tabular}
     & CentOS 7.7.1908 \\

     \midrule
     D
     & 16 GB
     & \begin{tabular}[c]{@{}p{10cm}@{}}Dell Latitude 7480 Laptop from 2017. \\ One Intel(R) Core(TM) i7-7600U CPU @ 2.80GHz with two physical cores and two threads per core. \\ One Intel HD Graphics 620 GPU with 64 MB eDRAM.  \end{tabular}
     & \begin{tabular}[c]{@{}l@{}}\vspace{-3mm}\\Ubuntu 20.04.1 LTS\end{tabular}\\

     \bottomrule
   \end{tabular}
\label{tab:systems}
\end{adjustbox}
\end{sidewaystable}
 We compare the runtime of the first three simulations with Cortex3D and the epidemiology use case with NetLogo 3D.
These simulations have different properties and are, therefore, well suited to
   evaluate \bdm{}'s simulation engine under a broad set of conditions.

\subsection{Benchmark Simulations}
\paragraph{Cell Growth and Division Benchmark.}
The starting condition of this simulation was a 3D grid of cells.
These cells were programmed to grow to a specific diameter and divide afterward.
This simulation had high cell density and slow-moving cells.
This simulation covered mechanical interaction between spherical cells, biological
  behavior, and cell division.

\begin{figure}[h!]
\includegraphics[width=\linewidth]{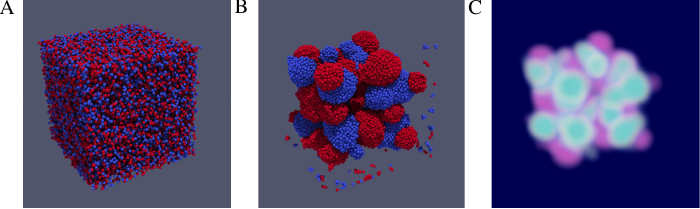}
\caption{{\bf Soma clustering simulation.}
    This simulation contains two types of cells and two extracellular substances.
    Each cell secretes a substance and moves into the direction of the substance
      gradient.
    Cells are distributed randomly in the beginning (A) and form clusters during
      the simulation.
    (B) Cell clusters at the end of the simulation.
    (C) Substance concentrations at the end of the simulation.
    A video is available at \nameref{si:soma-clustering}.
  }
  \label{fig:soma-clustering}
\end{figure}

\begin{figure}
\centering
  \begin{tabular}{c@{\hspace{1mm}}c@{\hspace{1mm}}}

      \includegraphics[width=0.3\linewidth]{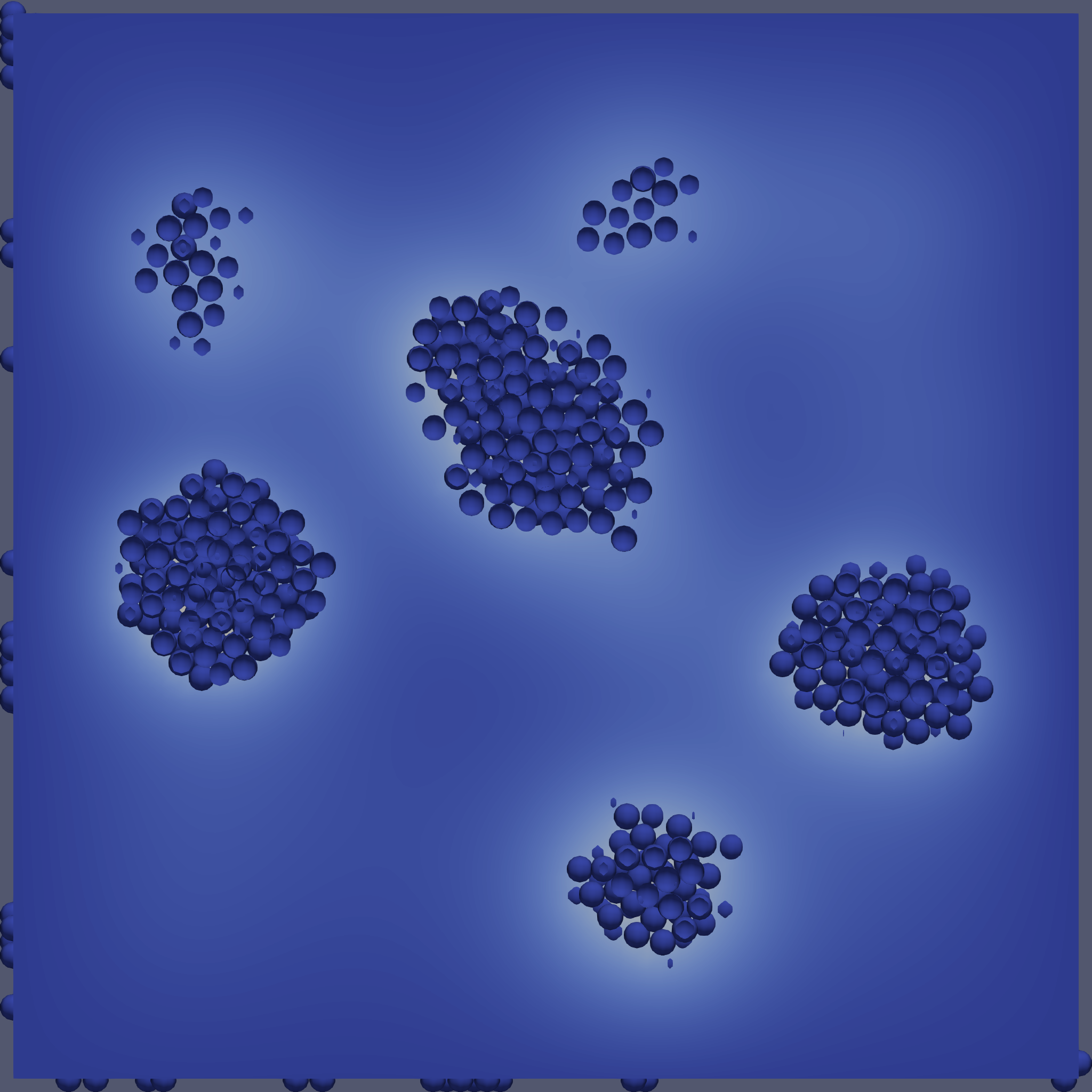} &
      \includegraphics[width=0.3\linewidth]{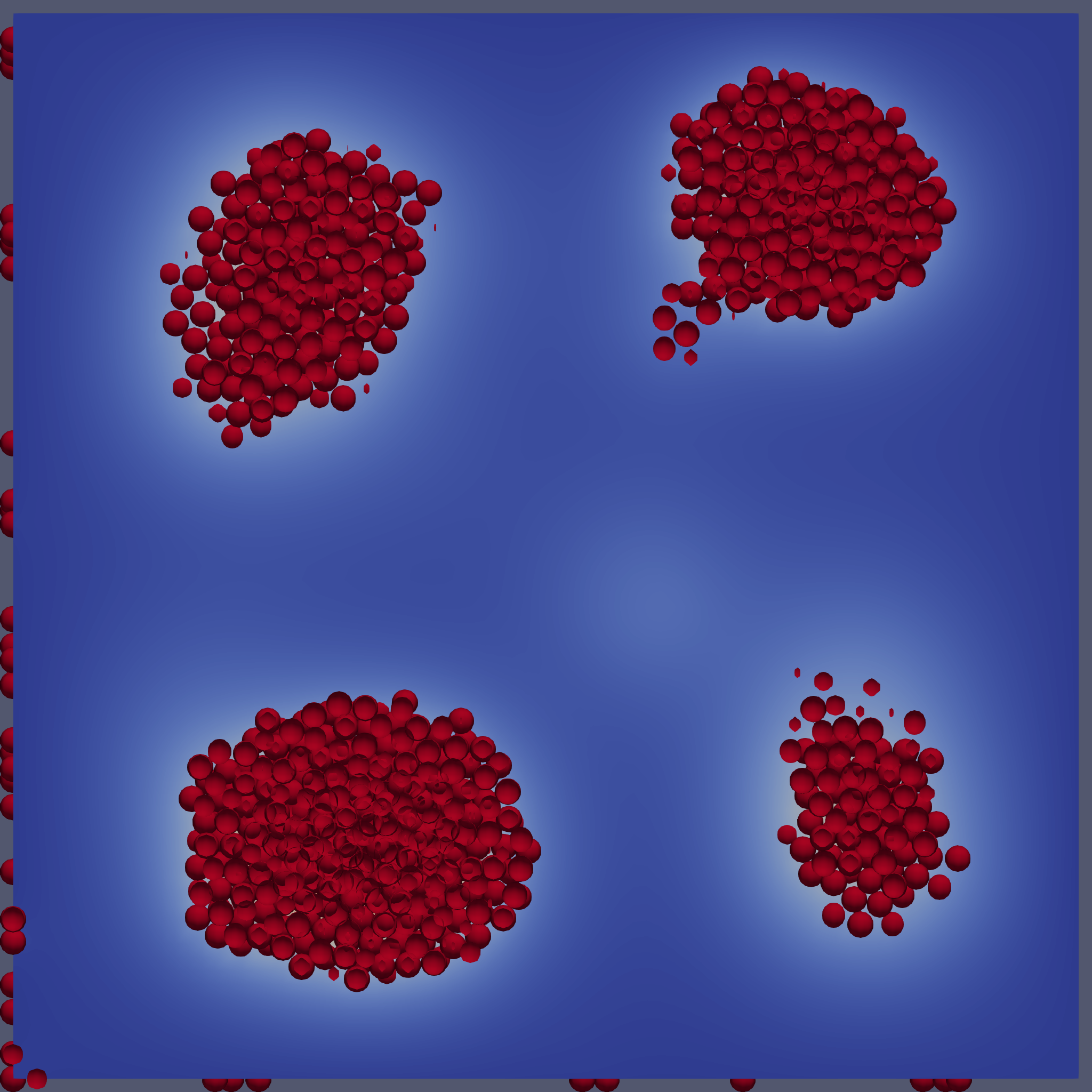} \\
\includegraphics[width=0.3\linewidth]{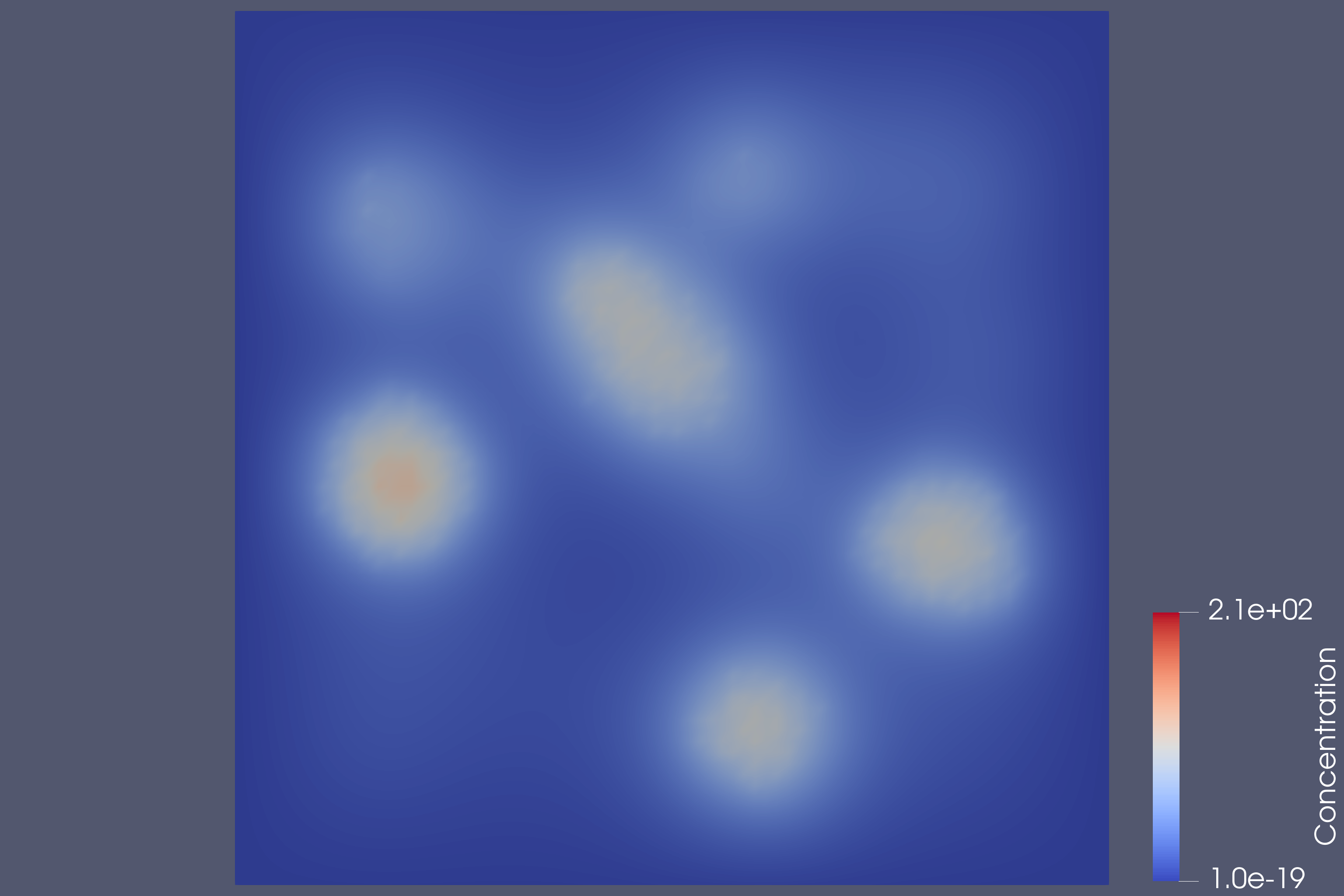} &
      \includegraphics[width=0.3\linewidth]{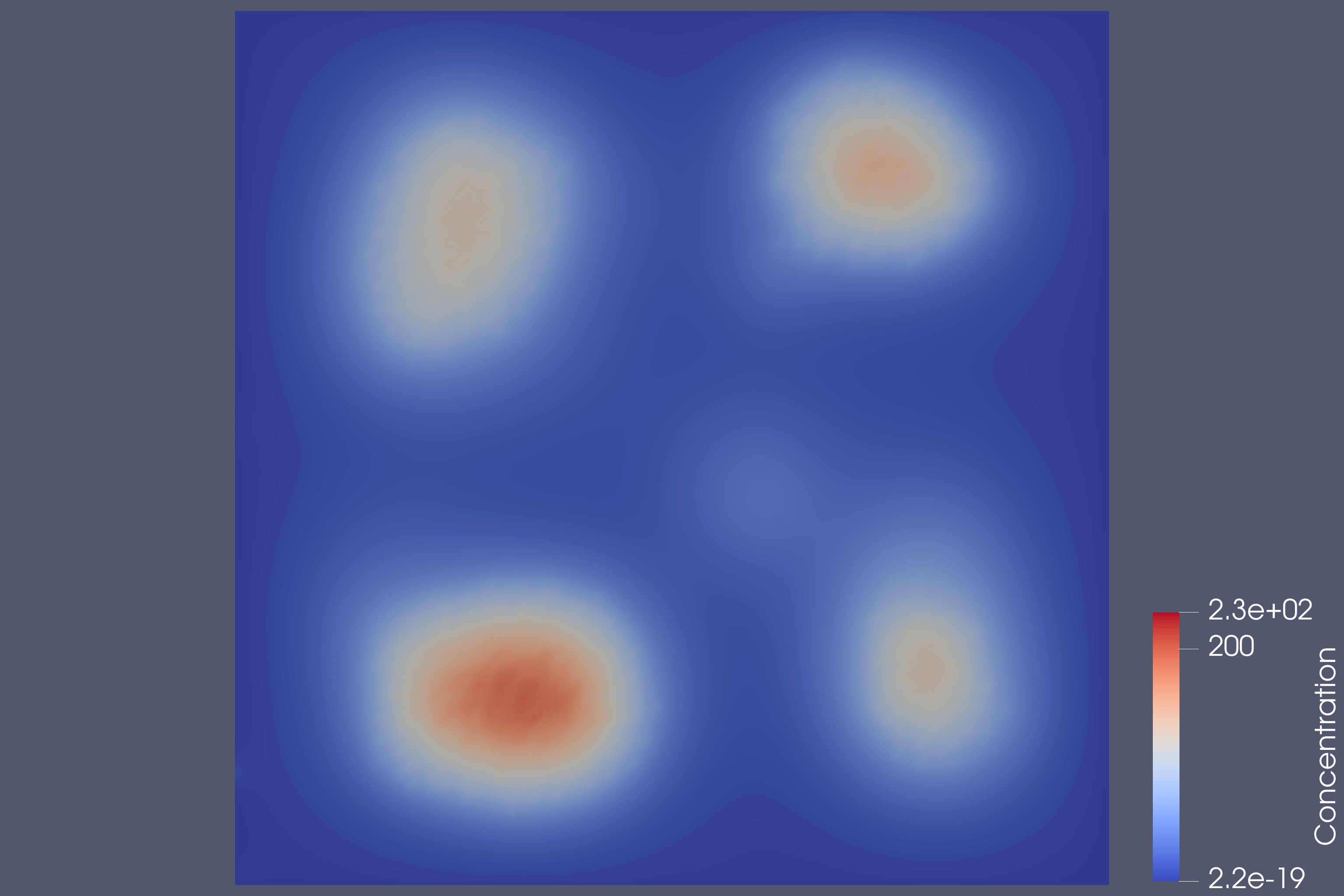} 
\end{tabular}
  \caption{{\bf Soma clustering cross section.}
    Cell positions coincide with regions of high substance concentration. 
    The first row shows substance concentrations and cells, while the second row shows substance concentrations only.
    Columns show cell type with the corresponding substance.
  }
  \label{fig:sc-cross-section}
\end{figure}

\paragraph{Soma Clustering Benchmark.}
\label{sec:platform:soma-clustering}
The goal of this model was to cluster two types of cells that are initially
  randomly distributed.
These cells are represented in red and blue in Figure~\ref{fig:soma-clustering}A
  and B.
Each cell type secreted  a specific
  extracellular substance which attracted homotypic cells.
Substances diffused through the extracellular matrix following
  Eq~\ref{eq:centraldiff}.
We modeled cell processes with two behaviors, ran in sequence: substance
  secretion (Algorithm~\ref{algo:secretion_module}) and
  chemotaxis (Algorithm~\ref{algo:chemotaxis_module}).
We set the parameter \texttt{secretion\_quantity} to $1$ and \texttt{gradient\_weight}
  to $0.75$.
During the simulation, cell clusters formed depending on their type.
The final simulation state after 6000 time steps is shown in
  Figure~\ref{fig:soma-clustering}B.
Clusters were associated with non-homogeneous extracellular substance
  distributions, as shown in Figure~\ref{fig:soma-clustering}C and Figure~\ref{fig:sc-cross-section}.
Besides being used as a benchmark, this example demonstrates the applicability of 
  \bdm{} for modeling biological
  systems, including the dynamics of chemicals such as oxygen or growth factors.
The simulation consisted of 68 lines of C++ code.
Table~\ref{tab:performance-overview} shows the performance on different systems.
There are three main differences comparing this simulation with the previous
  cell growth and division simulation.
First, this simulation covered extracellular diffusion.
Second, cells moved more rapidly.
Third, the number of cells remained constant during the simulation.

\begin{algorithm}
  \caption{Soma clustering substance secretion.}
  \label{algo:secretion_module}

\SetKwInOut{Input}{input}
  \SetKwData{MyCell}{cell}
  \SetKwData{CellType}{cell\_type}
  \SetKwData{Position}{pos}
  \SetKwData{ChemicalSubstanceA}{substance\_A}
  \SetKwData{ChemicalSubstanceB}{substance\_B}
  \SetKwData{DiffusionGrid}{diffusion\_grid}
  \SetKwData{SecretionQuantity}{secretion\_quantity}
  \SetKwFunction{GetCellType}{GetCellType}
  \SetKwFunction{GetPosition}{GetPosition}
  \SetKwFunction{GetDiffusionGrid}{GetDiffusionGrid}
  \SetKwFunction{IncreaseConcentrationBy}{IncreaseConcentrationBy}

  \Input{\MyCell, \DiffusionGrid, \SecretionQuantity}
  \Position $\leftarrow$ \MyCell.\GetPosition{}\;
  \DiffusionGrid.\IncreaseConcentrationBy{\Position, \SecretionQuantity}\;

\end{algorithm}

\begin{algorithm}
  \caption{Soma clustering chemotaxis.}
  \label{algo:chemotaxis_module}

  \SetKwInOut{Input}{input}
  \SetKwData{MyCell}{cell}
  \SetKwData{CellType}{cell\_type}
  \SetKwData{Position}{pos}
  \SetKwData{ChemicalSubstanceA}{substance\_A}
  \SetKwData{ChemicalSubstanceB}{substance\_B}
  \SetKwData{DiffusionGrid}{diffusion\_grid}
  \SetKwData{DiffusionGradient}{grad}
  \SetKwData{GfGradientWeight}{gradient\_weight}
  \SetKwFunction{GetCellType}{GetCellType}
  \SetKwFunction{GetPosition}{GetPosition}
  \SetKwFunction{UpdatePosition}{UpdatePosition}
  \SetKwFunction{GetDiffusionGrid}{GetDiffusionGrid}
  \SetKwFunction{GetGradient}{GetNormalizedGradient}

  \Input{\MyCell, \DiffusionGrid, \GfGradientWeight}

  \Position $\leftarrow$ \MyCell.\GetPosition{}\;
  \DiffusionGradient $\leftarrow$ \DiffusionGrid.\GetGradient{\Position}\;
  \MyCell.\UpdatePosition{\DiffusionGradient $\times \GfGradientWeight$}\;

\end{algorithm}

\newpage

\paragraph{Pyramidal Cell Growth Benchmark.}
We used the pyramidal cell model from the neuroscience use case as a
  building block (see Figure~\ref{fig:pyramidal-cell}).
The simulation started with a 2D grid of initial neurons on the z-plane and started
  growing them.
This simulation has three distinctive features.
First, activity was limited to a neurite growth front, while the rest of the
  simulation remained static.
This introduced a load imbalance for parallel execution.
Second, the neurite implementation modified neighboring agents.
Hence, synchronization was required between multiple threads to ensure
  correctness.
Third, the simulation had only static substances, i.e., substance concentrations
  and gradients did not change over time.

\subsection{Statistical Method}
We perform five measurements for each presented data point in
  Figure~\ref{fig:performance} and Table~\ref{tab:performance-overview}.
We summarize runtimes using the arithmetic mean and rates such as speedup using
  the harmonic mean.

\subsection{Reproducibility}
\label{sec:reproducibility}

We use the latest \bdm{} version \texttt{v1.01-55-gd05111e3} for all use cases and benchmarks in this chapter.
To help other researchers replicate our findings, we provide the following 
  supplementary information for utmost transparency.
First, we publish all source code and data in Supplementary File~SF3 (\nameref{si:code}).
The archive contains six shell scripts that execute all
  simulations, and generate all plots, visualizations and videos shown in this chapter.
Second, we provide a ready-to-use self-contained Docker image to simplify the process of executing our simulations and benchmarks and to guarantee long-term reproducibility (\nameref{si:docker-image}).
Third, we add a step-by-step instruction in \nameref{si:reproduce}.

\subsection{Evaluation}

\begin{figure}[!b]
\includegraphics[width=\linewidth]{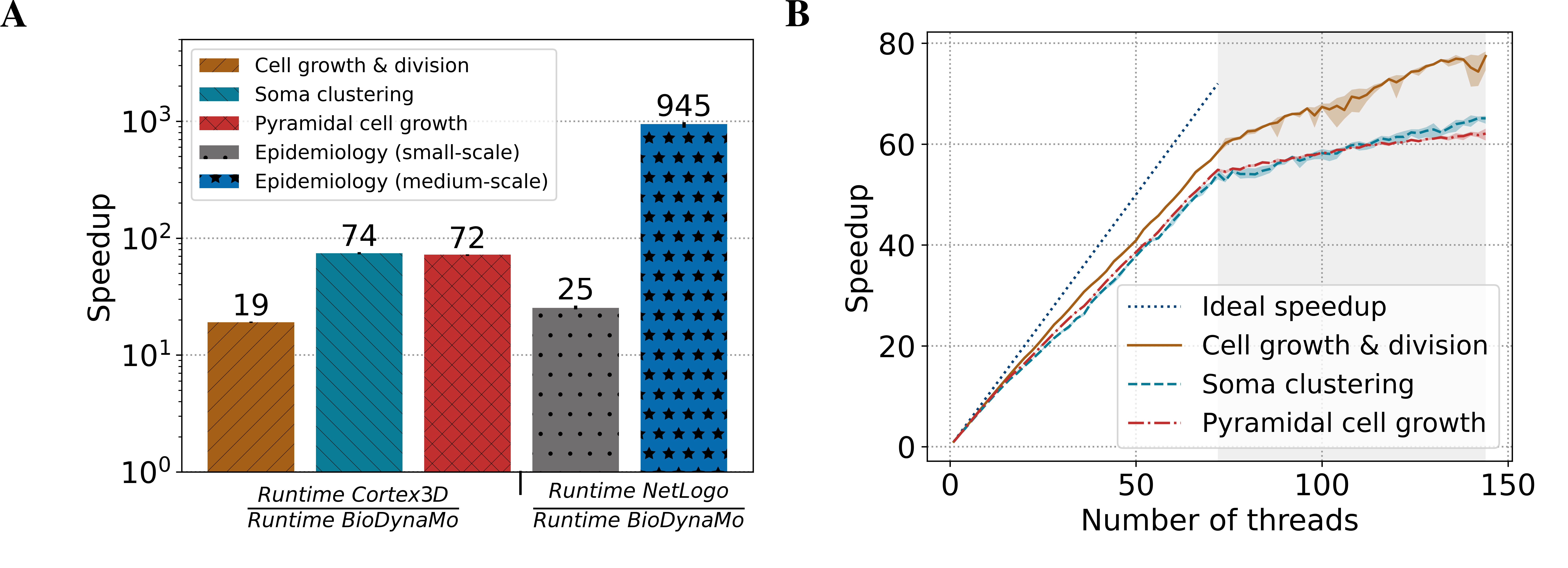}

  \caption{{\bf \bdm{} performance analysis.}
    (A) Speedup of \bdm{} compared to the serial simulation platforms Cortex3D and NetLogo.
    Simulations use one CPU core except for the ``epidemiology (medium-scale)'' benchmark, for which 
    all 72 physical cores were available.
    The comparison with NetLogo uses the same simulation with different numbers of agents.
    Cortex3D supports only small-scale simulations.
    (B) Strong scaling behavior of \bdm{} on a server with 72
      physical cores, two threads per core, and four NUMA domains.
    The grey area highlights hyper-threads.
  }
  \label{fig:performance}
\end{figure}

First, to demonstrate the performance improvements against established ABS platforms, we compared \bdm{} with Cortex3D and NetLogo.
Figure~\ref{fig:performance}A shows the speedup of \bdm{} for four simulations.
We define speedup as the runtime of the compared ABS platform over the runtime of \bdm{}.
We observed a significant speedup between 19 and 74$\times$ for Cortex3D and 25$\times$ for NetLogo.
The speedup was larger when the simulation was more dynamic or more complex.
Note that we set the number of threads available to \bdm{} to one since
  Cortex3D and NetLogo are not parallelized.
In the ``epidemiology (medium-scale)'' benchmark we increased the number of available
  physical CPU cores to 72 and observe a three order of magnitude speedup of 945$\times$.
This result clearly shows the impact of parallelization on the overall performance.
Although NetLogo is not parallelized, it benefits from parallel garbage collection.
We could not perform a medium-scale analysis with Cortex3D, because it only supports simulations 
  with a small number of agents.

Second, to evaluate the scalability of \bdm{}, we measured the simulation
  time with an increasing number of threads.
We increased the number of agents used in the comparison with Cortex3D and
  reduced the number of simulation timesteps to 10.
Figure~\ref{fig:performance}B shows the strong scaling analysis.
We define the term ``strong scaling'' as the property of a 
  simulation platform to reduce the runtime of a simulation with a fixed size $x$ with an
  increasing number of CPU cores $c$: $speedup(c, x) = \frac{time(1, x)}{time(c, x)}$ \citep{hill_what_1990}.

All simulation parameters were kept constant, and the number of threads was
  increased from one to the number of logical cores provided by the benchmark
  server.
The maximum speedup ranged between 62$\times$ and 77$\times$, which corresponds to a parallel
  efficiency of 0.86 and 1.07.
Performance improved even after all physical cores were utilized and hyper-threads were
  used.
Hyper-threads are highlighted in gray in Figure~\ref{fig:performance}B.
We want to emphasize that even the pyramidal cell growth benchmark scaled well,
  despite the challenges of synchronization and load imbalance.

Third, we evaluated the impact of calculating the mechanical forces on the
  GPU using the cell growth and  division, and soma clustering simulations.
We excluded the pyramidal cell growth simulation because the current GPU kernel
  does not yet support cylinder geometry.
The benchmarks were executed on \systemC, comparing an NVidia Tesla V100 GPU
  with 32 CPU cores (64 threads).
We observed a speedup of 1.01$\times$ for cell growth and division, and
  4.16$\times$ for soma clustering.
The speedup correlated with the number of collisions in the simulation.
The computational intensity is directly linked with the number of collisions
  between agents.

In summary, in the scalability test, we observed a minimum speedup of 62$\times$.
Furthermore, we measured a minimum speedup of 19$\times$ comparing \bdm{}
  with Cortex3D both using a single thread.
Based on these two observations, we conclude that on \systemA{} \bdm{} is
  more than three orders of magnitude faster than Cortex3D.
In the comparison with NetLogo we observed a 945$\times$ speedup directly.

Based on these speedups, we executed the neuroscience, oncology, and epidemiology use cases with 
  one billion agents. 
Using all 72 physical CPUs on \systemB, we measured a runtime of 1 hour 26 minutes, 
  6 hours 22 minutes, and 2 hours, respectively. 
One billion agents, however, are not the limit. 
The maximum depends on the available memory and accepted execution duration. 
To be consistent across all use cases and keep our pipeline's total execution 
  time better manageable, we decided to run these benchmarks with one billion agents.
Table~\ref{tab:performance-overview} shows that available memory would permit an 
  epidemiological and neuroscience simulation with two billion agents. 
With enough memory, \bdm{} is capable of supporting hundreds of billions of agents.

\begin{table}[p]
\begin{adjustbox}{max height=0.9\textheight}
\begin{adjustbox}{angle=90}
\centering
  \caption{{\bf Performance data.}
  The values in column ``Agents" and ``Diffusion volumes" are taken from the end of the simulation.
  Runtime measures the wall-clock time to simulate the number of iterations.
  It excludes the time for simulation setup and visualization.
  The entries in column ``System'' correspond to Table~\ref{tab:systems}.
  }
\setlength{\tabcolsep}{6pt}
  \begin{tabular}{@{}lrrrrrrr@{}}
    \toprule
    \textbf{Simulation} &
    \textbf{Agents} &
    \textbf{Diffusion} &
    \textbf{Iterations} &
    \textbf{System} &
    \textbf{Physical} &
    \textbf{Runtime} &
    \textbf{Memory} \\

    &
    &
    \textbf{volumes} &
    &
    \textbf{(Table~\ref{tab:systems})} &
    \textbf{CPUs} &
    &
    \\
    \midrule

    Neuroscience use case\\

    \hspace{3mm} Single (Figure~\ref{fig:pyramidal-cell}A) &
    1 494&
    250 &
    500 &
    A &
    1 &
    0.16 s &
    382 MB \\

    &
    &
    &
    &
    D &
    1 &
    0.12 s &
    479 MB \\

    \hspace{3mm} Large-scale (Figure~\ref{fig:pyramidal-cell}C) &
    9 036 986 &
    65 536 &
    500 &
    A &
    72 &
    35 s&
    6.47 GB\\

    &
    &
    &
    &
    D &
    2 &
    11 min 28 s &
    5.37 GB \\

    \hspace{3mm} Very-large-scale &
    1 018 644 154&
    5 606 442&
    500 &
    B &
    72 &
    1 h 24 min&
    438 GB\\

    \hline

    Oncology use case (Figure~\ref{fig:tumor-spheroids}) \\
    \hspace{3mm} 2000 initial cells &
    4 177 &
    0 &
    312 &
    A &
    1 &
    1.05 s &
    382 MB\\

    &
    &
    &
    &
    D &
    1 &
    0.832 s &
    480 MB \\

    \hspace{3mm} 4000 initial cells &
    5 341 &
    0 &
    312 &
    A &
    1 &
    1.76 s&
    382 MB\\

    &
    &
    &
    &
    D &
    1 &
    1.34 s &
    480 MB \\

    \hspace{3mm} 8000 initial cells &
    7 861 &
    0 &
    288 &
    A &
    1 &
    3.27 s&
    384 MB \\

    &
    &
    &
    &
    D &
    1 &
    2.60 s &
    482 MB \\

    \hspace{3mm} Large-scale&
    1 000 3925&
    0 &
    288 &
    A &
    72 &
    1 min 42 s&
    7.42 GB \\

    &
    &
    &
    &
    D &
    2 &
    43 min 56 s &
    5.84 GB \\

    \hspace{3mm} Very-large-scale&
    986 054 868&
    0 &
    288 &
    B &
    72 &
    6 h 21 min&
    604 GB \\

    \hline

    Epidemiology use case (Figure~\ref{fig:epidemiology}C) \\
    \hspace{3mm} Measles &
    2 010 &
    0 &
    1000 &
    A &
    1 &
    0.53 s &
    381 MB\\

    &
    &
    &
    &
    D &
    1 &
    0.42 s &
    479 MB \\

    \hspace{3mm} Seasonal Influenza&
    20 200 &
    0 &
    2500 &
    A &
    1 &
    16.41 s&
    383 MB\\

    &
    &
    &
    &
    D &
    1 &
    16.40 s &
    479 GB \\

    \hspace{3mm} Medium-scale (measles)&
    100 500 &
    0 &
    1000 &
    A &
    72 &
    1.36 s&
    1 GB \\

    \hspace{3mm} Large-scale (measles)&
    10 050 000&
    0 &
    1000 &
    A &
    72 &
    59.19 s&
    5.87 GB \\

    &
    &
    &
    &
    D &
    2 &
    19 min 18 s &
    5.41 GB \\

    \hspace{3mm} Very-large-scale (measles)&
    1 005 000 000 &
    0 &
    1000 &
    B &
    72 &
    2 h 0 min&
    495 GB \\

    \hline

    Soma clustering (Figure~\ref{fig:soma-clustering}) &
    32 000&
    1 240 000&
    6 000 &
    A &
    72 &
    12.91 s &
    1.02 GB \\

    &
    &
    &
    &
    D &
    2 &
    2 min 7 s &
    522 MB\\

    \bottomrule
  \end{tabular}
\label{tab:performance-overview}
\end{adjustbox}
\end{adjustbox}
\end{table}

\section{Discussion}

This chapter presented \bdm{}, a novel open-source platform for agent-based
  simulations.
Its modular software architecture allows researchers to implement models of distinctly different fields, of which neuroscience, oncology, and epidemiology were demonstrated in this chapter.
Although the implemented models follow a simplistic set of rules, the results that emerge from the simulations are prominent and highlight \bdm{}'s capabilities.
We do not claim that these models are novel, but we rather want to emphasize that \bdm{} enables scientists to (i) develop models in various computational biology fields in a modular fashion, (ii) obtain results rapidly with the parallelized execution engine, (iii) scale up the model to billions of agents on a single server, and (iv) produce results that are in agreement with validated experimental data.
Although \bdm{} is modular, we currently offer a limited number of ready-to-use simulation primitives.
We are currently expanding our library of agents and behaviors to facilitate model development beyond the current capacity.

Ongoing work uses \bdm{} to gain insights into retinal development \cite{demontigny_2023}, cryopreservation \cite{jennings_cryo, jennings_cryo2}, 
multiscale (organ-to-cell) cancer modelling \cite{demontigny_2021, duswald_2024}, radiation-induced tissue damage \cite{cogno_mechanistic_2024, cogno_thesis, cogno_3d_2022, cogno_agent-based_2022}, and more \cite{duswald2024calibrationstochasticagentbasedneuron, hesam_2024, demetriades_interrogating_2022, gazeli_interrogating_2022}.
Further efforts focus on accelerating drug development by replacing in vitro experiments with in silico simulations using \bdm{}.

Our performance analysis showed improvements of up to three orders of magnitude over
  state-of-the-art baseline simulation software, allowing us to scale up simulations to an unprecedented number of agents.
To the best of our knowledge, \bdm{} is the first scalable ABS platform of 
  neural development that scales to more than one billion agents.
The same principles used to model axons and dendrites in the neuroscience use case could 
  also be applied to simulate blood and lymphatic vessels.

We envision \bdm{} to become a valuable tool in computational biology, fostering faster and easier simulation of complex and large-scale systems,
  interdisciplinary collaboration, and scientific reproducibility.

\section{Availability and Future Directions}
\label{sec:future}
\bdm{} is an open-source project under the Apache 2.0 license and can be found on Github (\href{https://github.com/BioDynaMo/biodynamo}{https://github.com/BioDynaMo/biodynamo}).
The documentation is split into three parts: API reference, user guide, and
  developer guide.
Furthermore, a Slack channel is available for requesting assistance or guidance
  from the \bdm{} development team.

\bdm{} officially supports the following operating systems: Ubuntu (18.04, 20.04), CentOS 7, and macOS (10.15, 11.1).
We test \bdm{} on these systems and provide prebuilt binaries for third
  party dependencies: ROOT and ParaView.

All of the results presented in the chapter can be reproduced following the
  instructions in the Supplementary Information.

\par
\begin{figure}[h!]
\includegraphics[width=\textwidth]{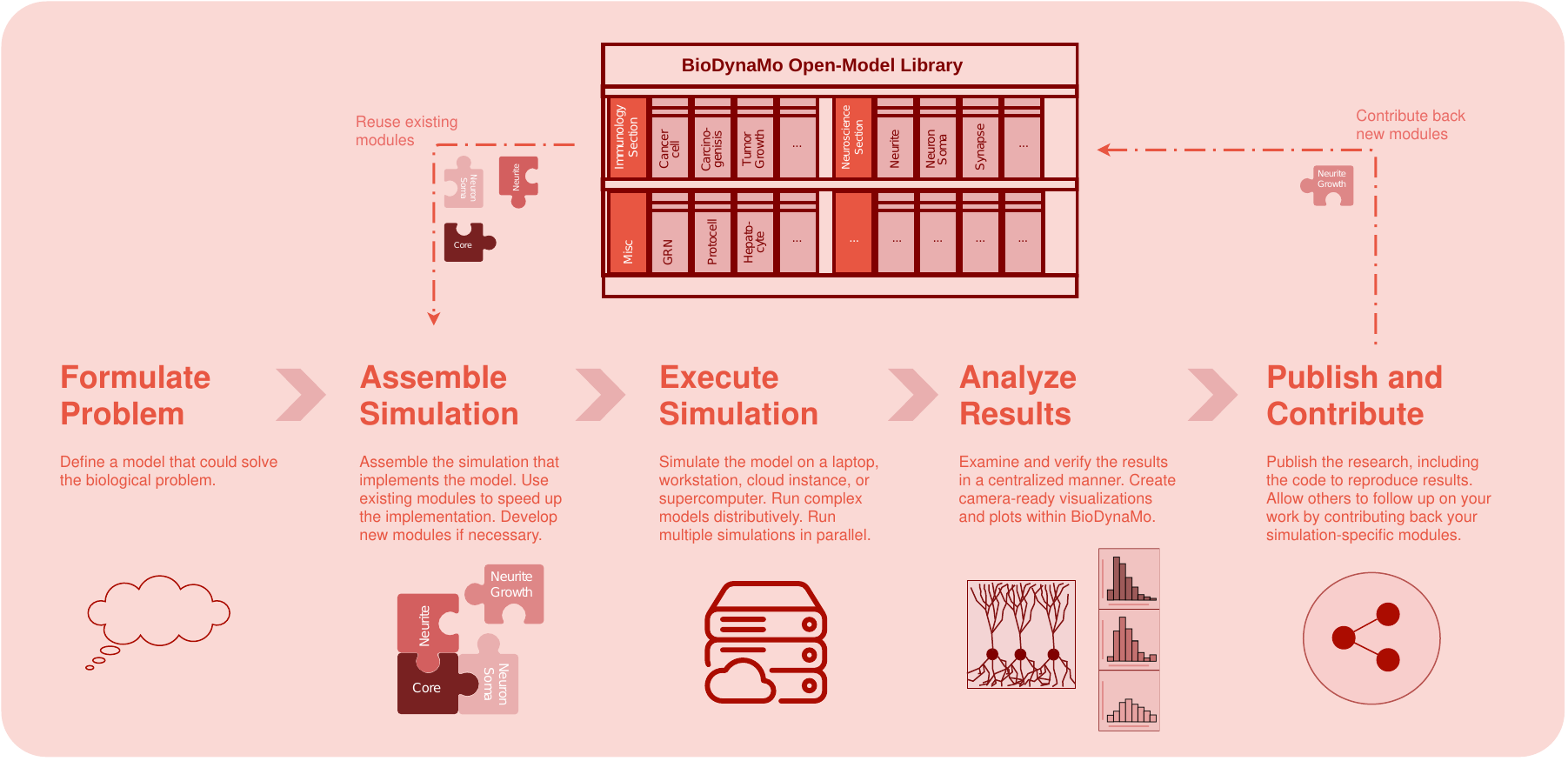}
\caption{{\bf \bdm{} platform.}
      Vision of \bdm{}, a platform to accelerate in silico experiments.
    }
    \label{fig:bdm-platform}
\end{figure}

By designing \bdm{} in a modular and extensible way, we laid the foundation to
  create new functionalities easily.
We encourage the life science community to contribute their developments back to
  the open-source codebase of \bdm{}.
Over time, the accumulation of all these contributions will form the \bdm{}
  open-model library, as shown in Figure~\ref{fig:bdm-platform}.
This library will help scientists accelerate their research by providing the
  required building blocks (agents, biological
  behavior, etc.) for their simulation.
Currently, we collect these contributions in our Github repository (\href{https://github.com/BioDynaMo/biodynamo}{https://github.com/BioDynaMo/biodynamo}).

\newpage
\section{Supporting Information}
 All supporting information is available at: \url{https://doi.org/10.5281/zenodo.3862368}

\paragraph*{SF2-reproduce-results.md}
\label{si:reproduce}
{\bf Instructions on how to reproduce all results presented in this chapter.}

\paragraph*{SF3-code.tar.gz}
\label{si:code}
{\bf Codebase to reproduce all results presented in this chapter.}
This file contains all code necessary to reproduce performance results,
plots, visualizations, and videos shown in this chapter.
Furthermore, it contains more details about the hardware and software
configuration of the different systems described in Table~\ref{tab:systems}.

\paragraph*{SF4-bdm-publication-image.tar.gz}
\label{si:docker-image}
{\bf Docker image to reproduce all results presented in this chapter.}
We provide a Docker image to simplify the process of executing our simulations and benchmarks
  and to guarantee long-term reproducibility.
The only requirement that users must install is a recent version of the Docker engine.
All other prerequisites are already provided in the ready-to-use, self-contained Docker image.
This approach does not rely on content hosted somewhere on the internet that might become
  unavailable in the future. 

\paragraph*{SF5-raw-results.tar.gz}
\label{si:raw-result}
{\bf Raw results.}
This archive contains all raw results from the simulations and benchmarks shown in this chapter.

\paragraph*{SV1-single-pyramidal-cell.mp4}
\label{si:pc-single}
{\bf Single pyramidal cell growth simulation, as shown in Figure~\ref{fig:pyramidal-cell}A.} 

\paragraph*{SV2-large-scale-neuronal-development.mp4}
\label{si:pc-cover}
{\bf Large-scale pyramidal cell growth simulation, as shown in Figure~\ref{fig:pyramidal-cell}C.} 

\paragraph*{SV3-tumor-spheroid.mp4}
\label{si:tumor-spheroids}
{\bf Tumor spheroid growth simulation, as shown in Figure~\ref{fig:tumor-spheroids}B.} 

\paragraph*{SV4-soma-clustering.mp4.}
\label{si:soma-clustering}
{\bf Soma clustering simulation, as shown in Figure~\ref{fig:soma-clustering}.}

\clearpage{}

\cleardoublepage \clearpage{}\chapter{\bdm{}'s Simulation Engine}
\label{ch:engine}

\section{Introduction}

Agent-based modeling (ABM) allows to simulate complex dynamics in a wide range
  of research fields.
ABM has been used to answer research questions in biology
  \cite{metzcar_review_2019, zubler_simulating_2013, hunter_taxonomy_2017},
  sociology \cite{epstein_growing_1996}, economics
  \cite{tesfatsion_chapter_2006}, technology \cite{niazi_agent-based_2009},
  business \cite{rand_agent-based_2011}, and more fields
  \cite{macal_introductory_2014}.
\emph{Agents} are individual entities that, among others, can represent
subcellular
structures to simulate the growth of a neuron, a cell to investigate cancer
development, or a person to simulate the spread of infectious diseases
\cite{breitwieser-bdm}.
The actions of an agent are defined through instances of class \emph{behavior}.
To stay with the examples from before, possible behaviors are neurite
  bifurcation, uncontrolled cell division, or infection.

Agent-based models are developed in an iterative way, during which an initial model 
  is increasingly refined until it matches with observed data 
  \cite{thorne_combining_2007, roberts_mathematical_2016}. 
Model parameters that cannot be derived from the literature are determined
  through optimization. 
An optimization algorithm generates a parameter set, executes the model, and 
  evaluates the error with respect to observed data until the error converges to
  a local or global minimum.
This loop might also contain an uncertainty analysis to evaluate the robustness of 
  a solution \cite{marino_methodology_2008}.
Consequently, the model must be simulated many times. 

The simulation engine's performance limits the scale of the model and determines how often the 
  model can be simulated.
Thus, performance is a key issue for simulating models on extreme scales that might one 
  day be able to simulate all 86 billion neurons in the brain \cite{azevedo_equal_2009}. 
It is also crucial for smaller-scale simulations to explore vast parameter space, 
  analyze parameter uncertainty, repeat the simulation often enough to reach statistical 
  significance, and develop models rapidly.

To achieve these goals, we present a novel simulation engine called \bdm{},
which is optimized for high performance and scalability.
During its development, we identify the following three main performance
  challenges for agent-based simulations.

\textbf{Challenge 1:}
To fully utilize systems with high processor core counts, the parallel part of
  the simulation engine has to be maximized (see Amdahl's law
  \cite{amdahl_validity_1967}).
Although it is easy to parallelize the loop over all agents
  (Algorithm~\ref{algo:main}), our benchmarks revealed two operations whose level
  of parallelization has a significant performance impact.
First, building the environment index, which is used to determine the neighbors
  of an agent.
The literature describes various radial-neighbor search algorithms with
  different design trade offs between build and search performance.
Second, combining thread-local results at the end of each iteration.
In general, attention must also be paid to seemingly minor things, such as
  resizing a large vector, which by default is initialized by a single thread.

\textbf{Challenge 2:}
ABMs are predominantly memory-bound due to two reasons.
First, the behavior of agents often has low arithmetic intensity.
Second, ABM can be very dynamic.
During a simulation, agents move through space, change their behavior, and are
  created and destroyed.
Consequently, the neighborhood of an agent changes continuously, leading to an
  irregular memory access pattern and poor cache utilization.
This results in large data movement between the main memory and the processor
  cores.

\textbf{Challenge 3:}
Under certain conditions, the expensive calculation of mechanical forces
  between agents is redundant (Section~\ref{sec:static-agents}).
These forces are for example used in tissue models to determine the
  displacement of agents.
The challenge is identifying those agents for which the pairwise force
  calculation can be safely omitted.

\bdm{} addresses these challenges with the following new optimizations.
To maximize the parallelization (Challenge~1), we develop an optimized uniform
  grid to search for agent neighbors and fully parallelize the addition and
  removal of agents.
We address the data movement bottleneck (Challenge~2) in software by (i)
  optimizing the iteration over all agents on systems with non-uniform memory
  architecture, (ii) sorting agents and their neighbors to improve the cache hit
  rate and minimize access to remote DRAM, and (iii) introducing a pool memory
  allocator.
To avoid redundant mechanical force calculations (Challenge~3), we add a
  mechanism to detect agents for which we can guarantee that the resulting force
  will not move the agent.

These mechanisms make \bdm{} \result{nearly an order of magnitude} more efficient than Biocellion 
  and three orders of magnitude faster than Cortex3D and NetLogo.
The performance improvements account for a median speedup of
  \result{159$\times$} compared to \bdm{}'s standard implementation with all
  optimizations turned off.
As a result, \bdm{} is able to simulate \result{1.72 billion} agents on one server. 
The main contributions of this chapter are as follows.

\begin{itemize}
  \item We present a novel high-performance agent-based simulation engine.
The engine can be used in many domains due to its modular software design and
          features a specialization for neuroscience, capable of simulating the
          development of neurons.

\item We present six optimizations to maximize performance (Section~\ref{sec:opt:maximize-parallelization}--\ref{sec:static-agents}).
These insights are transferable and can be used to improve the performance of
          other agent-based simulators.

  \item We present an in-depth evaluation of \bdm{}'s performance using five
        different simulations (Section~\ref{sec:se:evaluation}).
This comprehensive analysis provides insights for users of \bdm{} into which
          parameters yield the best performance and hints for developers of future
          agent-based simulation tools.

\end{itemize}

\section{\bdm{}'s Simulation Engine}
\label{sec:pl:design-overview}

This section gives an overview of \bdm{} and its components.
\bdm{} is written in C++, uses OpenMP \cite{openmp} for shared-memory
  parallelism, and is available under the Apache 2.0 open-source license.

Breitwieser et al. \cite{breitwieser-bdm} describe the user-facing features of the \bdm{} platform and detail its modular software design and ease-of-use by means of three use cases in the domains of neuroscience, epidemiology, and oncology. 

\bdm{} is a hybrid framework able to utilize multi-core CPUs and GPUs. 
This dissertation focuses on the CPU version, which has two major advantages.
First, the CPU version can simulate many more agents than a GPU version. 
The reason is that GPUs have typically significantly smaller memory than CPUs. 
For example, our benchmark hardware has 12$\times$ more memory than the current flagship GPU from NVidia, the A100 \cite{nvidia2020a100}. 
Second, the CPU version improves the usability and flexibility for our broad user community, 
who often only have a Matlab \cite{matlab} or R \cite{r} coding background.
In \bdm{}, users create simulations by writing C++ code.
A GPU-only version would require users to write CUDA code to define new agents, behaviors, and other user-defined components.
Therefore, \bdm{} only offloads computations to the GPU, transparently to the user \cite{hesam_gpu}.

The main objects in agent-based simulations are agents, behaviors, and
  operations.
Agents (e.g., a cancer cell) have attributes that are updated through behaviors
  and operations.
Behaviors (e.g., uncontrolled cell division) are functions that can be assigned
  and removed from an agent and give users fine-grained control over the actions
  of an agent.
In contrast, \emph{Agent operations} are executed for each agent.
For example, to calculate the mechanical forces between agents and execute all
  individual behaviors of an agent.
The second type of operation, called \emph{standalone operation} is executed
  once per iteration to perform a specific task (e.g., visualization).
A characteristic property of agent-based simulation is local interaction.
\bdm{} provides a common interface for different neighbor search algorithms
  called \emph{environment}.
Besides the uniform grid detailed in Section~\ref{sec:grid}, \bdm{} features a
  kd-tree based on nanoflann \cite{nanoflann} and octree based on the publication
  of Behley et al.
\cite{unibn}.

The agent-based simulation algorithm (Algorithm~\ref{algo:main}) comprises two
  steps.
First, users have to define the starting condition of the model (L1) in which
  agents, behaviors, operations, and any other resource are created.
Second, the simulation engine executes this model for a number of iterations
  (L2--19).
The engine executes all agent operations for each agent (L7--12) and all
  standalone operations (L12--14).
Standalone operations can be further separated into operations that must be
  executed at the beginning of the iteration (e.g., to update the environment
  index [L3-5]) or the end (e.g., visualization [L16-18]).
There are two barriers synchronizing threads (L6 and L15).

\makeatletter
\newcommand{\removelatexerror}{\let\@latex@error\@gobble}
\makeatother

\newcommand{\myalgorithm}{\begingroup
\removelatexerror \begin{algorithm}[H]
\caption{Simulation algorithm}
  \label{algo:main}

  \SetKwFor{ParallelFor}{parallel for}{do}{end}
  \SetKwFunction{Wait}{wait}
  \SetKwFunction{ModelInitialization}{ModelInitialization}
  \SetKwFunction{PostProcessing}{PostProcessing}
  
  \ModelInitialization{}

  \For{$i\in iterations$}{
    \For{$op\in pre\_standalone\_operations$}{
      $op()$\;
    }

    \Wait{}

    \ParallelFor{$a\in agents$}{
      \For{$op\in agent\_operations$}{
        $op(a)$\;
      }
    }
    \For{$op\in standalone\_operations$}{
      $op()$\;
    }

    \Wait{}

    \For{$op\in post\_standalone\_operations$}{
      $op()$\;
    }
  }

\end{algorithm}
\endgroup}

\begin{figure}[!h]
  \centering
  \begin{minipage}{.75\linewidth}
        \myalgorithm
  \end{minipage}
\end{figure}

\subsection{Alternative Execution Modes}
\label{sec:alternative-execution-modes}

\bdm{} offers different execution modes that influence how simulations are
  executed to cater to different user needs.
These execution modes can also be combined.

\paragraph{Row-Wise vs. Column-Wise Execution Mode.}
For each iteration, the execution engine has to execute a set of operations for
  each agent (Algorithm~\ref{algo:main} L7--12).
We can look at this as a matrix with one column per agent and one row per
  operation.
Each element in the matrix---corresponding to an agent-operation pair---must be
  executed.
Execution can happen in column-wise order, in which all operations are executed
  for one agent before proceeding to the following agent, or in a row-wise order
  in which one operation is executed for each agent before proceeding to the next
  operation.
Algorithm~\ref{algo:main} shows the default column-wise order, which can be
  transformed to the row-wise order by exchanging lines seven and eight.

\paragraph{\texttt{RandomizedRm}.}
The \texttt{ResourceManager}, an essential class in the simulation engine,
  stores raw agent pointers and offers functions to add, remove, get, and iterate
  over agents.
The \texttt{RandomizedRm} class is a decorator for \texttt{ResourceManager} and
  possible subclasses and randomizes the iteration order over all agents after
  each iteration.
This feature is important to avoid an execution order bias for specific models
  \cite{railsback_agent-based_2006}.

\paragraph{Different Execution Contexts.}
The execution context is a facade for the environment interface and the
  resource manager.
There is one instance of this object per thread.
This class aims to encapsulate discretization choices: When should changes to
  an agent be visible to other agents?
When should new or removed agents be visible?
Currently, \bdm{} provides the in-place and copy execution context.
If the user chooses the in-place execution context, changes to an agent are
  immediately visible to its neighbors.
Consequently, neighbor attributes might already be updated to the current
  iteration if the neighbor has been processed before.
Thus the neighbor attributes do not contain the values of the previous
  iteration anymore.
In contrast, the copy execution context updates agent copies and commits the
  changes at the end of the iteration after all agents have been updated.
For both execution context types, agent removals and additions take effect at
  the beginning of the next iteration.
Users can provide additional execution contexts in which different
  discretization choices are made.

\section{Maximize Parallelization}
\label{sec:opt:maximize-parallelization}

\subsection{Grid-Based Neighbor Search}
\label{sec:grid}

Determining the neighbors of an agent is a pre-condition for all agent
  interactions.
For example, the infection behavior in an epidemiological model requires
  information if any of the immediate neighbors is infected.
In this context, it is essential to find neighbors fast and efficiently and
  minimize the build time of the required index.
Building an index in every iteration has a high cost, as shown in the
  evaluation section.
We exploit the fact that the interaction radius is known at the beginning of
  the iteration.
For this fixed-radius search problem, a grid-based solution is a good choice
  because the box of an agent can be determined in constant time using the
  agent's position \cite{unibn}.
This is confirmed by our evaluation in Section~\ref{sec:eval:environment}.
The build stage in which all agents are assigned to a box can be easily
  parallelized.
In the search stage, the grid determines all neighbors by iterating over all
  agents in the same box and the surrounding boxes.
In 3D space, we consider the 3x3x3 cube of boxes surrounding and including the query box. 
All agents inside a box are stored in an array-based linked list.
The box only needs to store the start index and the number of elements it
  contains.
To avoid zeroing all boxes at the beginning of the build stage, we add a
  timestamp attribute to each box, updated whenever an agent is added.
Consequently, we can determine that a box is empty if the simulation and box
  timestamp is different.
Therefore, we can build the grid in $O(\#agents)$ time instead of $O(\#agents +
    \#boxes)$, which is relevant for large simulation spaces that are not fully
    populated.

The array-based linked list uses the same agent indices as in the
  \texttt{ResourceManager}.
The \texttt{ResourceManager}, an essential class in the simulation engine,
  stores raw agent pointers and offers functions to add, remove, get, and iterate
  over agents.
Thus, it also benefits from the memory layout optimization presented in
  Section~\ref{sec:load-balancing}.
This optimization reduces the distance in memory of agents that are close in space. 
Consequently, linked list elements will be closer to each other, improving the 
  cache hit rate of traversing the linked list during the search stage of the grid.
The described grid implementation can be found in the class \texttt{UniformGridEnvironment}.
 
\subsection{Adding and Removing Agents in Parallel}
\label{sec:parallel-remove}

To maximize the theoretically achievable speedup described in Amdahl's law
  \cite{amdahl_validity_1967}, we maximize the parallel part of the simulation by
  parallelizing the addition and removal of agents.
By default, \bdm{} stores a thread-local
  copy of additions and removals and commits them to the \texttt{Resource\-Manager}
  at the end of each iteration.

\begin{figure}[tb]
  \centering
\includegraphics[width=0.7\linewidth]{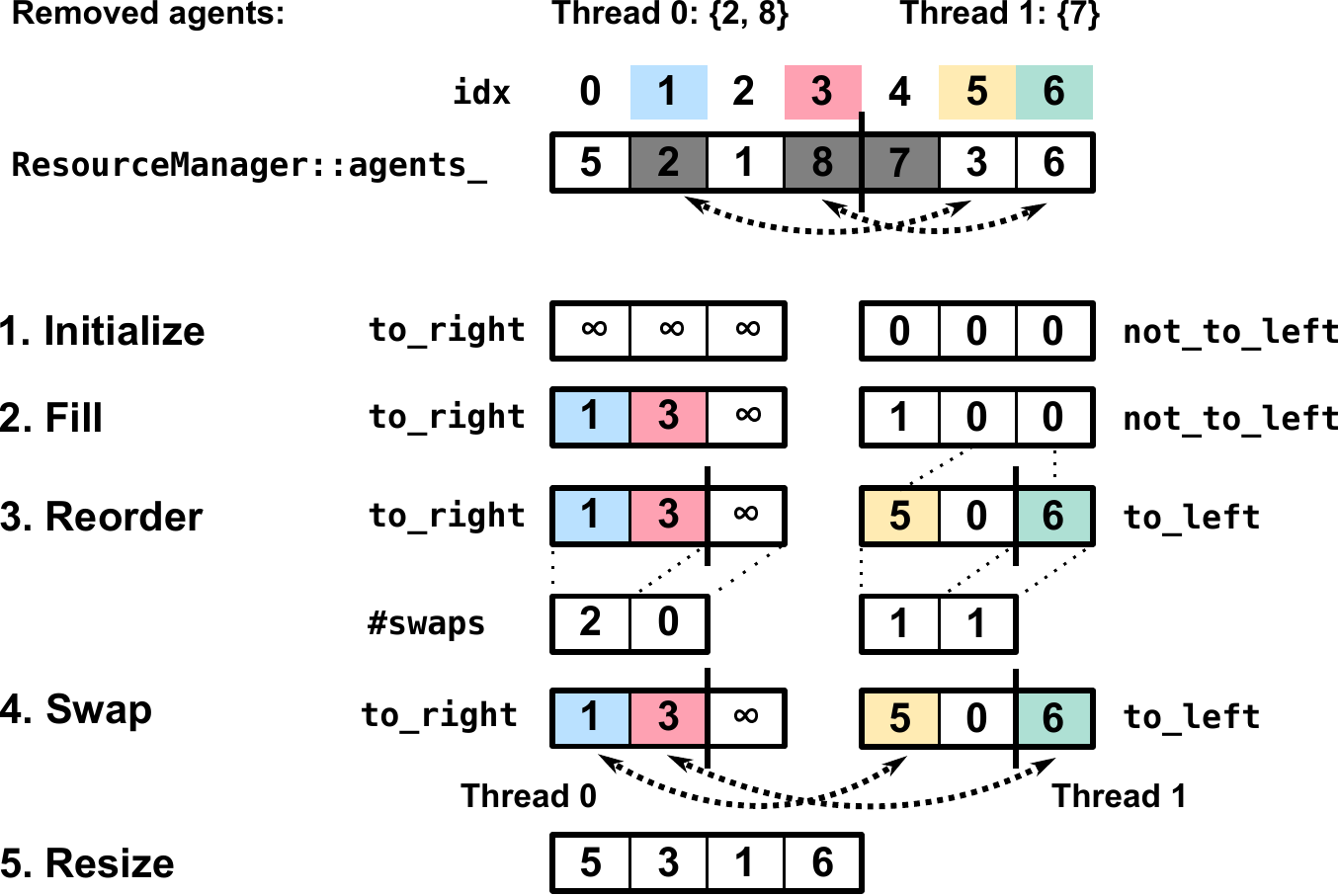}
\caption{Parallel agent removal mechanism}
  \label{fig:parallel-remove}
\end{figure}

Additions are trivial; the engine determines the total number of additions,
  grows the data structures in the \texttt{Resource\-Manager}, and adds the agent
  pointers in parallel.
In contrast, the parallelization of removals is a more elaborate process
  because we disallow empty vector elements in the \texttt{ResourceManager}.
If the simulation engine has to remove an agent stored in the middle of the
  vector, it must swap it with the last element before shrinking it.
The following algorithm aims at performing the necessary swaps and updates in
  dependent data structures in parallel.
Figure~\ref{fig:parallel-remove} illustrates the parallelized algorithm
  simplified for a single NUMA domain.
 This example assumes a simulation with seven agents represented with identifier
  1--7 and two threads, which remove three agents from the simulation.
These agents are highlighted with a grey background.
The other colors serve as a visual aid to track the agents that must be
  swapped.

The algorithm comprises five main steps.
First, the algorithm determines the total number of removed agents, calculates
  the new size of the vector ($old\_size - removed\_agents$), and initializes two
  auxiliary arrays.
The size of the auxiliary arrays equals the number of removed agents.
The new vector size is indicated by the vertical line between indexes three and
  four in Figure~\ref{fig:parallel-remove}.

Second, each thread iterates over its vector of removed agents and fills the
  auxiliary arrays.
If the agent is stored to the left of the new size index, it must be moved to
  the right.
Therefore the algorithm inserts the element index into the array $to\_right$.
If the agent is stored to the right of the new size, we insert a one into array
  $not\_to\_left$ at the index: $idx - new\_size$.
The maximum index used to access elements in the auxiliary array is smaller than 
  the number of removed agents and is independent of the number of remaining agents.

Third, we partition the two auxiliary arrays into blocks corresponding to the
  total number of threads.
A thread iterates over its auxiliary block, moves entries to the beginning if
  they indicate a swap, and stores a counter in the \texttt{\#swaps} array.
For the $to\_right$ array, the algorithm skips or overwrites elements with the
  value $UINT\_MAX$.
In this step, the semantic of the $not\_to\_left$ array changes to $to\_left$.
Thus, the algorithm looks for all zeros in the array, replaces them with the
  value $array\_index + new\_size$, and moves them to the beginning of the block.

In the fourth step, we can finally perform the swaps.
The algorithm calculates the prefix sum of the two $\#swaps$ arrays and
  partitions the swaps among all threads.
Each thread can determine the indices based on the prefix sum of $\#swaps$.

In the last step, the algorithm completes the removal by shrinking the vector
  to $new\_size$.

This algorithm requires $O(removed\_agents)$ time and space and parallelizes
  steps 1--4.
 
\subsection{Visualization}
\label{sec:visualization-design}

Visualization plays an essential role in gaining insights into the dynamics of
  a simulation during its iterative development process.
ParaView \cite{paraview} is a feature-rich and powerful visualization package
  that uses MPI to parallelize work on multiple processes, but its shared-memory
  parallelism is severely limited.
In contrast, \bdm{} only uses one process and is parallelized with OpenMP.
Therefore, the integration of ParaView and \bdm{} performs poorly.
To resolve this performance bottleneck, we present a solution where \bdm{}
  exports visualization files using shared-memory parallelism first, which
  ParaView renders in a second step using multiple processes.

To that extent, we present three optimizations to minimize the visualization
  time.
First, we parallelize ParaView's (p)vtu and (p)vti data writer using OpenMP.
Although ParaView provides a parallelized implementation using MPI, this
  solution is not amenable for shared-memory applications, as mentioned before.
Second, we introduce on-demand export of agent attributes based on just-in-time
  (JIT) compilation with cling \cite{vasilev_cling_2012}.
Instead of exporting all agent attributes, we divide them into obligatory and
  optional ones.
Obligatory attributes are defined in the agent implementation, are essential to
  render its geometry (e.g., position and diameter of a sphere), and are
  therefore exported by default.
The user must explicitly request all other agent attributes by setting a
  parameter in a configuration file, command-line argument, or directly in the
  code.
Third, we introduce an interface that gives third-party software like ParaView
  rapid access to agent attributes.

\begin{figure}[h]
   \centering
   \includegraphics[width=0.8\linewidth]{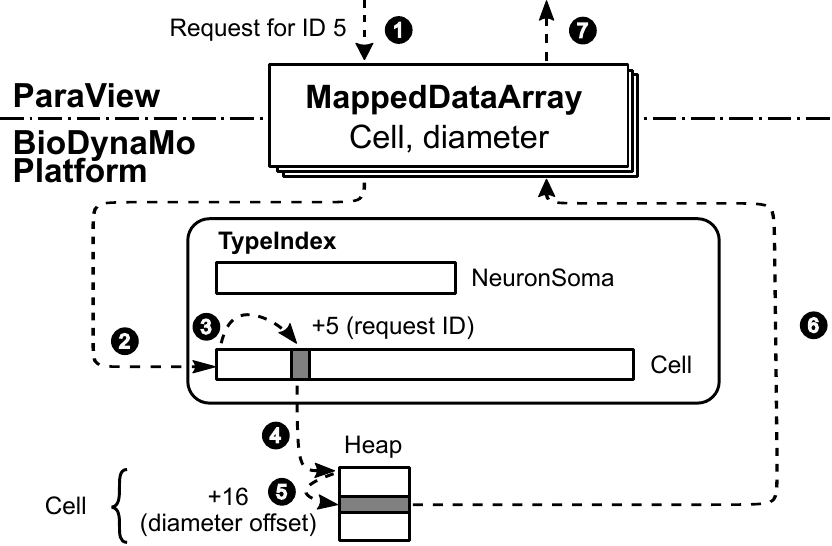}
   \caption{Interface between \bdm{} and ParaView}
\label{fig:visualization-design}
 \end{figure}

Figure~\ref{fig:visualization-design} illustrates how ParaView's request for
  the diameter of cell with ID 5 is handled.
\bdm{} instantiates a \texttt{MappedDataArray} for each selected attribute of
  each agent type, which receives data requests \circled{1} from ParaView's file
  writer.
This request is forwarded to the type index, which holds a vector of agent
  pointers that belong to a specific class \circled{2}.
ParaView's request ID indexes the vector \circled{3}.
The vector element points to the beginning of the agent object on the heap
  \circled{4}.
We add the attribute offset to the base class pointer \circled{5} and return
  the data to the \texttt{MappedDataArray} \circled{6}.
The attribute offset is determined during the build process in which we use
  ROOT \cite{brun_root_1997} to collect reflection information for all relevant
  classes.
The \texttt{MappedDataArray} supports three execution modes: zero-copy, cache,
  and copy.
In the zero-copy mode, the data element is returned to ParaView directly
  \circled{7}.
In the cache and copy mode, each \texttt{MappedDataArray} instance instantiates
  a data vector.
In cache mode, this data vector is filled on-demand in step \circled{6}.
All consecutive requests to the same ID will be serviced from the internal data
  vector, bypassing steps \circled{2}--\circled{6}.
In the copy mode, the internal data vector is completely filled before any
  requests can be made.
Similar to the cache hit scenario, all requests will be served directly from
  the internal data vector.
The described mechanism could also be used to integrate other third-party
  software packages into \bdm{}.

\section{Optimize Memory Layout and Data Access Pattern}
\label{sec:opt:memory-layout}

\subsection{NUMA-Aware Iteration}
\label{sec:numa-iteration}

\bdm{} supports systems with multiple NUMA domains.
We add a mechanism to match threads with agents from the same NUMA domain to
  minimize the traffic to remote DRAM because OpenMP does not provide this
  functionality.

\begin{figure}[tb]
  \centering
\includegraphics[width=0.7\linewidth]{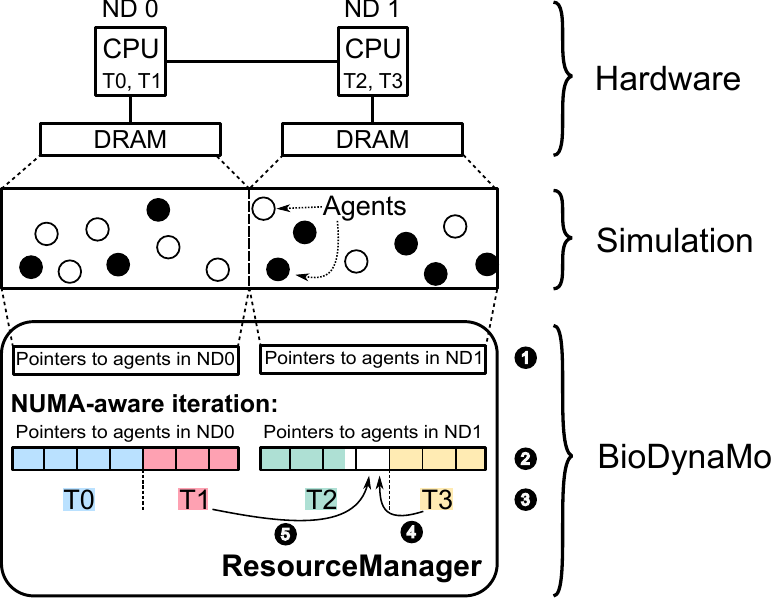}

\caption{NUMA-aware iteration}
\label{fig:numa-iteration}
\end{figure}

Figure~\ref{fig:numa-iteration} shows a server with two NUMA domains
  (\texttt{ND0}, \texttt{ND1}) corresponding to two CPUs with two threads each
  (T0 \& T1, T2 \& T3).
 The CPUs have a local DRAM with shorter memory access latency than the remote
  DRAM.
The agents in the simulation are balanced between the two NUMA domains (see
  Section~\ref{sec:load-balancing}).
The \texttt{ResourceManager} maintains a vector of agent pointers for each NUMA
  domain \circled{1}.
To iterate over agents in a NUMA-aware way, \bdm{} first partitions these
  vectors into blocks of agent pointers of the same \mbox{size \circled{2}.}
Second, these blocks are partitioned among the threads from the matching NUMA
  domain \circled{3}.
Threads process the assigned blocks in parallel.
Figure~\ref{fig:numa-iteration} shows processed blocks with a background color
  of the corresponding thread.

We implement a two-level work-stealing mechanism to avoid imbalanced execution
  times across threads.
First, a thread can steal a block from a different thread from the \emph{same}
  NUMA domain (e.g., \circled{4}).
Second, if the thread's NUMA domain has already finished all work, the thread
  can steal work from a different NUMA domain (e.g., \circled{5}).

\subsection{Agent Sorting and Balancing}
\label{sec:load-balancing}

To accelerate the memory-bound simulations, we must increase the cache hit
  ratio and load balance the agents among NUMA domains to minimize remote DRAM
  accesses.
In Section~\ref{sec:numa-iteration} and Figure~\ref{fig:numa-iteration}, we assume this is already the case.
This section presents an efficient algorithm to achieve this goal by sorting the agents' memory
  locations and preserving the neighborhood relations in 3D.
Preserving the neighborhood relation and reducing the dimensionality is the
  main characteristic of space-filling curves (e.g., Morton order
  \cite{morton1966computer} or Hilbert curve \cite{hilbert_ueber_1891}).

We compared the performance of the Morton order with the Hilbert curve using an oncological 
  simulation \cite{breitwieser-bdm} and observed a negligible performance improvement of 0.54\% from using the Hilbert curve. 
Higher costs to decode the Hilbert curve offset small gains for the agent operations. 
Therefore, we use the Morton order because it results in simpler code.

\begin{figure}[tbp]
  \centering
\includegraphics[width=\linewidth]{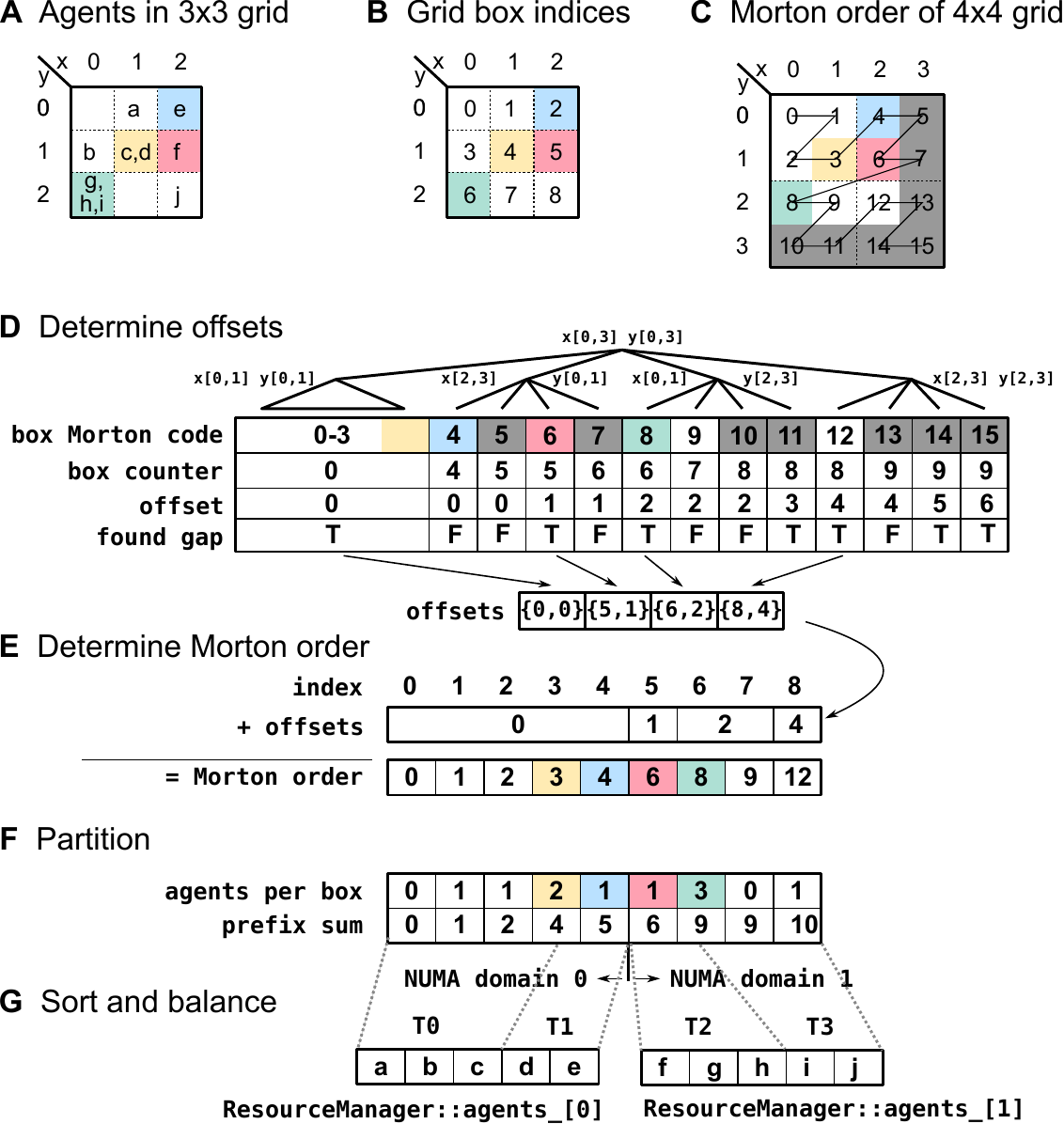}
\caption{Agent sorting and balancing mechanism}
\label{fig:load-balancing-design}
\end{figure}

Figure~\ref{fig:load-balancing-design} and the following description present
  the algorithm in 2D space, but the same principles apply in 3D.
In \bdm{}, the neighborhood information is stored in the implementation of the
  environment interface.
 Since the uniform grid environment performs best, as shown in the evaluation
  section (Section~\ref{sec:eval:environment}), we utilize its characteristics to
  achieve fast sorting and balancing.
We assume the following scenario.
Agents are stored in a 3$\times$3 uniform grid (A).
The simulation runs on a system with two NUMA domains and two threads per
  domain, resulting in four threads.
The grid boxes are stored in a flattened array.
Figure~\ref{fig:load-balancing-design}B shows the box indices for x and y
  coordinates.
We apply a Morton order space-filling curve
  \cite{morton1966computer} on the uniform grid (C).
Our goal is to sort the agents in the nine boxes in increasing Morton order.
The Morton order is only contiguous for quadratic simulation spaces where the
  length is a power of two.
Therefore, C shows a 4$\times$4 grid.
For the 3$\times$3 simulation space, there are gaps between Morton code four
  and six, six and eight, and nine and twelve.

The algorithm comprises three main steps.
First, the algorithm determines the sequence of boxes in Morton order (D, E).
Second, the algorithm partitions the boxes into segments that balance agents
  among NUMA domains and threads (F).
Third, the algorithm stores the agents in their new position in the resource
  manager (G).
In Figure~\ref{fig:load-balancing-design}, we selected four boxes and colored
  them in red, green, blue, and yellow to quickly find the corresponding entries
  for contained agents, box index, and Morton code.
The boxes outside the simulation space have a grey background.

In the first main step (D, E), we determine all gaps between grid boxes in
  simulation space (D) to avoid a costly sorting operation or iteration over all
  $N\times N$ boxes, where $N$ is the next higher power of two of $max(x, y)$.
We exploit the fact that the Morton order corresponds to a depth-first
  traversal of a quadtree, in which each box is a leaf in the tree.
Leaves whose boxes are outside the simulation space are considered empty.
Similarly, an inner node is empty if all of its corresponding leaves are
  outside the simulation space.
If all the corresponding boxes of an inner node are inside the simulation space
  (i.e., the inner node has a perfect subtree), we say the inner node is
  complete.
The quadtree is only an abstraction and does not need to be constructed.
It is only necessary to store the current traversal path, which requires
  $O(log(\#boxes))$ space.

The mechanism in D uses three auxiliary variables: box counter,  offsets, and
  \texttt{found\_gap}, which are initialized to zero, zero, and true.
The matrix in Figure~\ref{fig:load-balancing-design}D shows the three variables
  before the update of the current traversal step.
The algorithm traverses the tree depth-first and continues to the next deeper
  tree level only if the current node is neither complete nor empty.
In this case, the variables are not changed.
If a complete inner node or leave inside the simulation space is found and the
  \texttt{found\_gap} variable is true, the algorithm adds an entry with the
  current box counter and offset values in the offsets array and clears
  \texttt{found\_gap}.
Afterward, the box counter variable is incremented by the number of leaves in
  its subtree or one if it was a leave, irrespectively of the former value of
  \texttt{found\_gap}.
Empty nodes or leaves are handled similarly.
The offset variable is incremented by the number of empty leaves in the subtree
  of an empty node or one if it is an empty leave.
Additionally, the \texttt{found\_gap} variable is set to true.
The algorithm keeps track of the x and y intervals to calculate (in constant
  time) the number of leaves in a subtree and determine if an inner node is
  entirely, partially, or not inside the simulation space.

With the already sorted offsets array, the Morton order can be determined in
  linear time by iterating over all indices and adding the corresponding offset
  (E).

In the second main step (F), the algorithm iterates over all boxes in Morton
  order and fills an auxiliary array with the number of agents in each box.
Afterward, the algorithm calculates the prefix sum of the auxiliary array in a
  parallel work-efficient manner \cite{ladner_parallel_1980} and partitions the
  total number of agents in the simulation such that each NUMA domain receives a
  share corresponding to its number of threads.
Inside a NUMA domain, the agents are further partitioned such that each thread
  in this domain receives an equal share.

In the third main step (G), the threads copy the agents and store the pointer
  in the new position in the resource manager.
The simulation engine can immediately free obsolete agents' memory or delete
  all old copies after the step is finished.
The latter requires more memory but might improve performance due to a more
  optimal memory layout in conjunction with the \bdm{} memory allocator
  (Section~\ref{sec:mem-allocator}).

The presented algorithm runs in $O(\#agents + \#boxes)$ time and space and
  parallelizes steps E--G.
 
\subsection{\bdm{}
Memory Allocator} \label{sec:mem-allocator}

  To improve the
  performance of the simulation engine, we present a custom dynamic memory
  allocator which improves the memory layout of the most frequently allocated
  objects in a simulation: agents and behaviors.
Our solution builds upon pool allocators due to their constant time allocation
  performance.
Pool allocators divide a memory block into equal-sized elements and store
  pointers to free elements in a linked list.

We create multiple instances of these allocators because they can only return
  memory elements of one size.
As a result, agents and behaviors with distinct sizes are separated and stored
  in a columnar way.
We separate the pool allocator into multiple NUMA domains (class
  \texttt{NumaPoolAllocator}) to fully control where memory is allocated.
The \texttt{Numa\-Pool\-Allocator} has a central free-list and thread-private
  ones to minimize synchronization between threads.
List nodes, which correspond to free memory locations, can be migrated between
  thread private and the central list, which is essential to avoid memory leaks.
Migrations are triggered if a thread-private list exceeds a specific memory
  threshold.
Lists minimize these migrations, and thus thread synchronization, by
  maintaining additional skip lists.
These skip lists support additions and removals of a large number of elements
  in constant time.

Memory is allocated in large blocks with exponentially increasing sizes
  controlled by the parameter \texttt{mem\_\-mgr\_\-growth\_\-rate}.
The initialization of these memory blocks, which includes list node generation,
  is performed on-demand in smaller segments to minimize the required worst-case
  allocation time.

\begin{figure}[tb]
  \centering
\includegraphics[width=0.7\linewidth]{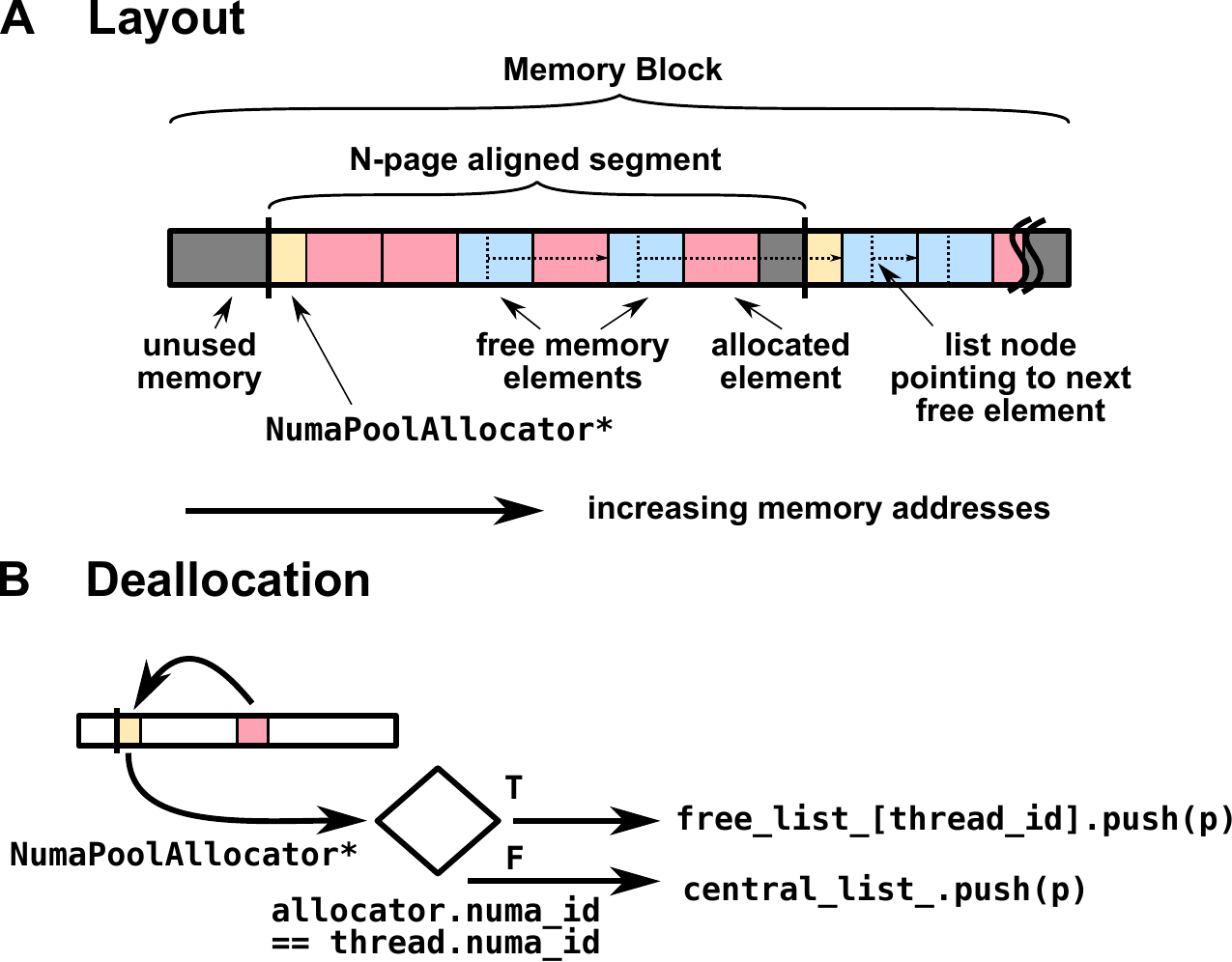}

\caption{\bdm{}'s memory allocator}
\label{fig:mem-mgr}
\end{figure}

Every allocated memory block is divided into N-page aligned segments
  (Figure~\ref{fig:mem-mgr}A), where N can be set with parameter
  \texttt{mem\_mgr\_aligned\_pages\_shift}.
 The linked list nodes are stored inside free memory elements and do not require
  extra space.
At the beginning of each N-aligned segment, we write the pointer to the
  corresponding \texttt{NumaPoolAllocator} instance.
Therefore, allocated memory elements can obtain this pointer in constant time,
  based on their memory address. This solution enables constant time deallocations (see
  Figure~\ref{fig:mem-mgr}b) but wastes memory in three ways.
First, memory blocks are allocated using \texttt{numa\_alloc\_onnode} (libnuma
  \cite{libnuma}).
This function does not return N-page aligned pointers and causes unusable
  regions at the beginning and the end, which sum up to $N * page\_size$ bytes.
Second, there might not be enough space to place a whole element at the end of
  an N-page aligned segment.
Elements must not cross N-page aligned borders because it would overwrite the
  necessary metadata.
Third, the metadata requires the size of a pointer, which is eight bytes on
  64-bit hardware.
The amount of wasted memory is bounded by the following equation: $N *
	  page\_size + element\_size + metadata\_size$.
Despite this memory overhead, our performance evaluation shows that the \bdm{}
  pool allocator uses \result{on average less memory} than ptmalloc2 and jemalloc
  \cite{evans2011scalable}.
Another side effect of this design choice is that the allocation size is
  limited by $N * page\_size - metadata\_size$.

\section{Omit Collision Force Calculation}
\label{sec:static-agents}

The most time-consuming operation in the tissue models presented in
  Section~\ref{sec:se:benchmarks} is the calculation of the displacement of agents
  based on all mechanical forces.
For this purpose, the simulation engine has to calculate pairwise collision
  forces between agents and their neighbors implemented in the class
  \texttt{InteractionForce}.
By default, \bdm{} uses the force calculation method detailed in the Cortex3D
  paper \cite{zublerdouglas2009framework}.
We observe that simulations can contain a significant amount of regions where
  agents do not move.
Neural development simulations (see Section~\ref{sec:se:benchmarks}), for example,
  might only have an active growth front, while the remaining part of the neuron
  is unchanged.

Therefore, we present a mechanism to detect agents for which it is safe to skip
  the expensive force calculation.
We call these agents static.

The following four conditions must be fulfilled in the last iteration: (i) the
  agent and none of its neighbors moved (ii) the agent's and neighbors'
  attributes did not change in a way that could increase the pairwise force
  (e.g., larger diameter), or the resulting displacement, (iii) new agents were
  not added within the interaction radius of the agent, and (iv) there is maximum
  one neighbor force which is non-zero.

The detection mechanism is closely tied to the \texttt{Interaction\-Force}
  implementation (see \cite{breitwieser-bdm}), as condition two implies, and
  might have to be adjusted if a different force implementation is used.

Condition four is needed because we want to allow agents to shrink and to be
  removed from the simulation without setting the agents in this region to not
  static.
Consequently, we have to ensure that two or more neighbor forces did not cancel
  each other out in the previous iteration.

The simulation engine monitors if any of the conditions are violated for each
  agent and sets the affected agents to not static.
In this process, a distinction has to be made whether the changed attribute
  affects only the current agent or also its neighbors.
If, for example, a static agent moves, the agent and all its neighbors will be
  affected, while a change to the agent's force threshold, which must be exceeded
  to move the agent, only affects itself.

\section{Evaluation}
\label{sec:se:evaluation}

\subsection{Benchmark Simulations}
\label{sec:se:benchmarks}

We use five simulations to evaluate the performance of the simulation engine:
  cell proliferation, cell clustering, and use cases in the domains of
  epidemiology, neuroscience, and oncology.
These simulations use double-precision floating point variables and are
  described in detail in \cite{breitwieser-bdm}.
Table~\ref{tab:sim-characteristics} shows that these simulations cover a broad
  spectrum of performance-related simulation characteristics and contain
  information about the number of agents, diffusion volumes, and iterations
  executed.
We set the number of agents between two and 12.6 million to keep the total
  execution time of all benchmarks manageable.
This is necessary due to the slow execution of the various baselines.
In addition, Section~\ref{sec:runtime-complexity} shows a benchmark in which
  each simulation is executed with one billion agents.
Also the comparison with Biocellion in Section~\ref{sec:eval:comparison-with-biocellion} 
  contains a benchmark with 1.72 billion cells.

 \vspace{3em}
\begin{table}[!h]
  \centering
\small
  \caption{Performance-relevant simulation characteristics.}
\begin{tabular}{@{}rrrrrr@{}}
    \toprule
    \textbf{Characteristic} & 
    \rotatebox{90}{\textbf{Cell proliferation}} & 
    \rotatebox{90}{\textbf{Cell clustering}} & 
    \rotatebox{90}{\textbf{Epidemiology}} & 
    \rotatebox{90}{\textbf{Neuroscience}} & 
    \rotatebox{90}{\textbf{Oncology}} \\
    \midrule
    {Create new agents during simulation} &
    \xmark  &
      &
      &
    \xmark  &
    \xmark  \\

    {Delete agents during simulation} &
      &
      &
      &
      &
    \xmark  \\

    {Agents modify neighbors} &
      &
      &
      &
    \xmark  &
      \\

    {Load imbalance} &
      &
      &
    \xmark &
    \xmark &
      \\

    {Agents move randomly} &
      &
      &
    \xmark  &
      &
    \xmark  \\

    {Simulation uses diffusion} &
      &
    \xmark  &
      &
    \xmark  &
      \\

    {Simulation has static regions} & 
      &
      &
      &
    \xmark  &
      \\

    {Number of iterations} &
     500 &
     1000 &
     1000 &
     500 &
     288 \\
    
    {Number of agents (in millions)} &
     12.6 &
      2 &
     10 &
      9 &
     10 \\
    
    {Number of diffusion volumes} &
     0 &
     54m &
     0 &
     65k &
     0 \\

    \bottomrule
  \end{tabular}
\label{tab:sim-characteristics}
\end{table}

\subsection{Experimental Setup and Reproducibility}

All tests were executed in a Docker container with an Ubuntu 20.04 based image.
Table~\ref{tab:se:systems} gives an overview of the main parameters of the three
  servers we used to evaluate the performance of \bdm{}.
If it is not explicitly mentioned, assume that System~A was used to execute a
  benchmark.

We provide all code, the self-contained docker image, more detailed information
  on the hardware and software setup, and instructions to execute the benchmarks
  in the supplementary materials (\url{https://doi.org/10.5281/zenodo.6463816}).

\begin{table}[h!]
\centering
\caption{
	Benchmark hardware
}
\begin{tabular}{@{}llp{9cm}l@{}}
\toprule
\textbf{System}
& \textbf{Memory}
& \textbf{CPU}
& \textbf{OS} \\

\midrule
A
& \vspace{0.5mm}504 GB\vspace{0.5mm}
& \multirow{2}{9cm}{\begin{tabular}[c]{@{}p{9cm}@{}}
		                 Four Intel(R) Xeon(R) E7-8890 v3 CPUs @ 2.50GHz with a total of 72 physical
		                   cores, two threads per core and four NUMA domains.
	\end{tabular}}
& \multirow{2}{1cm}{\begin{tabular}[c]{@{}l@{}}\vspace{1mm}CentOS\\7.9.2009\end{tabular}} \\

\cmidrule{1-2}
B
& \vspace{0.5mm}1008 GB\vspace{0.5mm}
&
& \\

\cmidrule{1-4}
C
& 62 GB
& \begin{tabular}[c]{@{}p{9cm}@{}}
	                 Two Intel(R) Xeon(R) E5-2683 v3 CPUs @ 2.00GHz with a total of 28 physical
	                   cores, two threads per core and two NUMA domains.
\end{tabular}
& \begin{tabular}[c]{@{}l@{}}\vspace{1mm}
CentOS\\Stream 8\end{tabular} \\

  \bottomrule \end{tabular}
  \label{tab:se:systems} \end{table}

  \subsection{General
	  Performance Metrics}
  \label{ch:dse:eval:gpm}

  We characterize the agent-based
  simulation workload by breaking down the operation's execution time, and
  exploring microarchitecture inefficiencies.

\begin{figure}[t]
  \centering
\begin{subfigure}{.49\linewidth}
\centering
    \includegraphics[width=\textwidth]{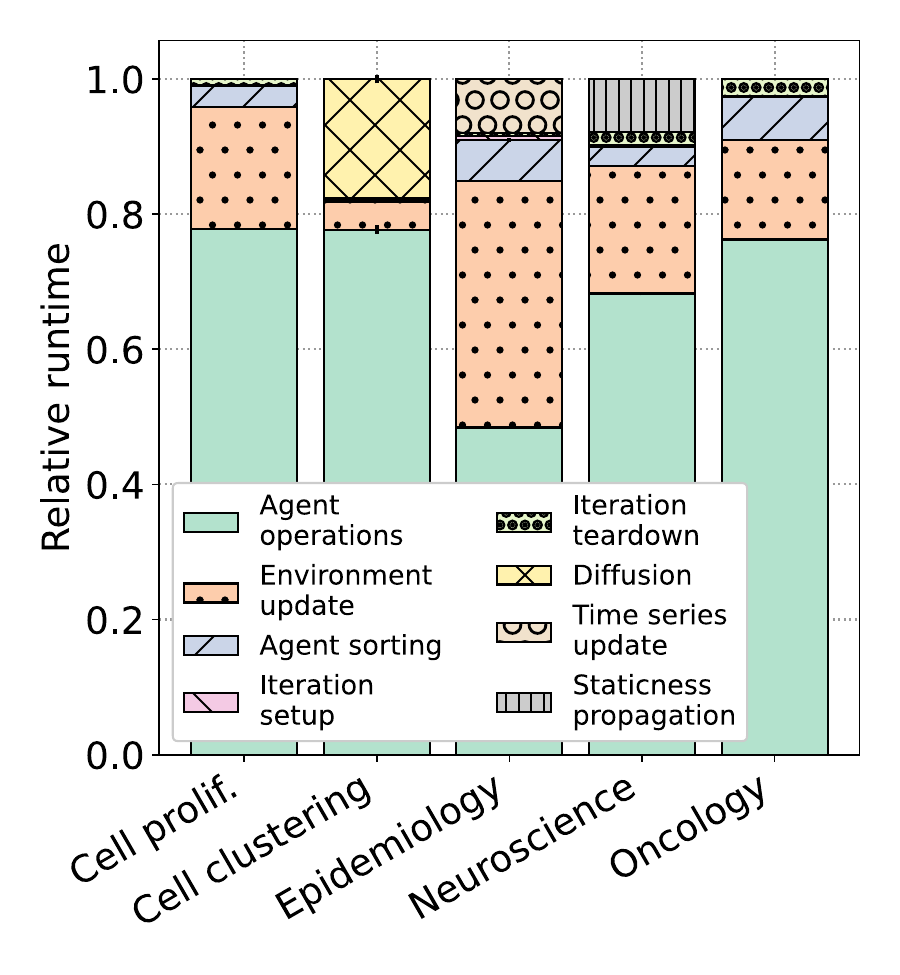}
\end{subfigure}
  \begin{subfigure}{.49\linewidth}
    \centering
    \includegraphics[width=\textwidth]{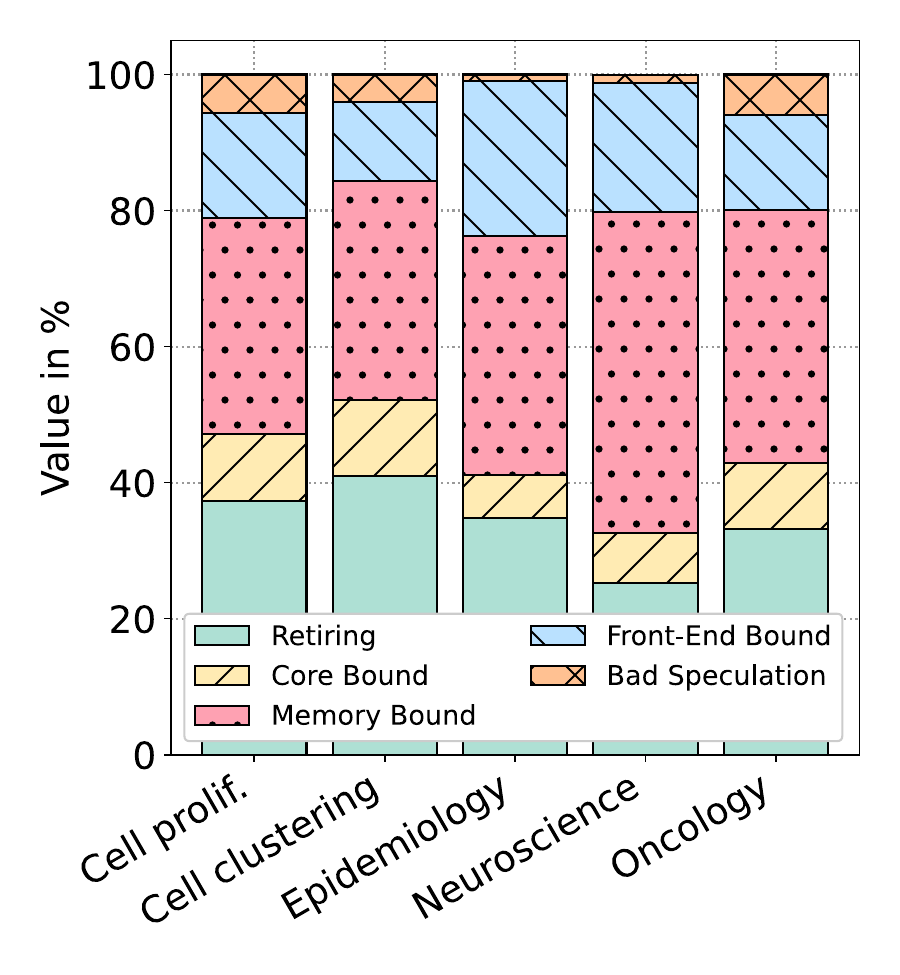}
\end{subfigure}
\caption{Operation runtime breakdown (left) and microarchitecture analysis
    (right)}
  \label{fig:operation-breakdown-uarch-analysis}
\end{figure}

The following benchmarks were performed with all optimizations enabled.
Figure~\ref{fig:operation-breakdown-uarch-analysis} shows a breakdown of all
  operations in our benchmark simulations.
 The majority of the runtime is spent in agent operations (median:
  \result{76.3\%}) which subsumes, among others, the execution of behaviors,
  calculation of mechanical forces, discretization, and detection of static
  regions.
Rebuilding the uniform grid environment at every time step is the second
  biggest runtime contributor, \result{4.09--36.5\%} (median: \result{18.0\%}).
The epidemiology use case considers a wider environment that manifests itself
  in an increased update time.
The average cost of agent sorting in its optimal setting (see
  Figure~\ref{fig:load-balancing}) is \result{0.180\%--6.33\%}.
Since adding and removing agents is parallelized, iterations' setup and tear
  down consume only \result{2.66\% (max)} of the execution time.

In the microarchitecture analysis, we observe that the benchmark simulations
  are primarily memory-bound.
We lose between \result{31.8 and 47.2\%} of processor pipeline slots because
  the operands are not available.

\subsection{Runtime and Space Complexity}
\label{sec:runtime-complexity}

\begin{figure}[tb]
	\centering
\begin{subfigure}{.49\linewidth}
    \centering
    \includegraphics[width=\textwidth]{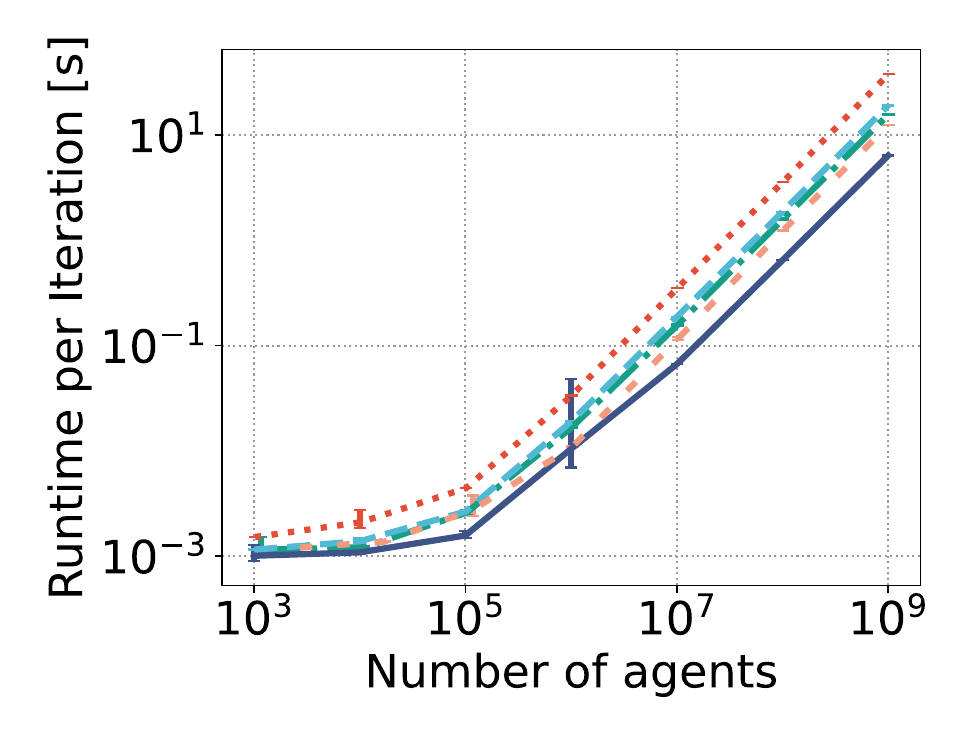}
\end{subfigure}
  \begin{subfigure}{.49\linewidth}
    \centering
    \includegraphics[width=\textwidth]{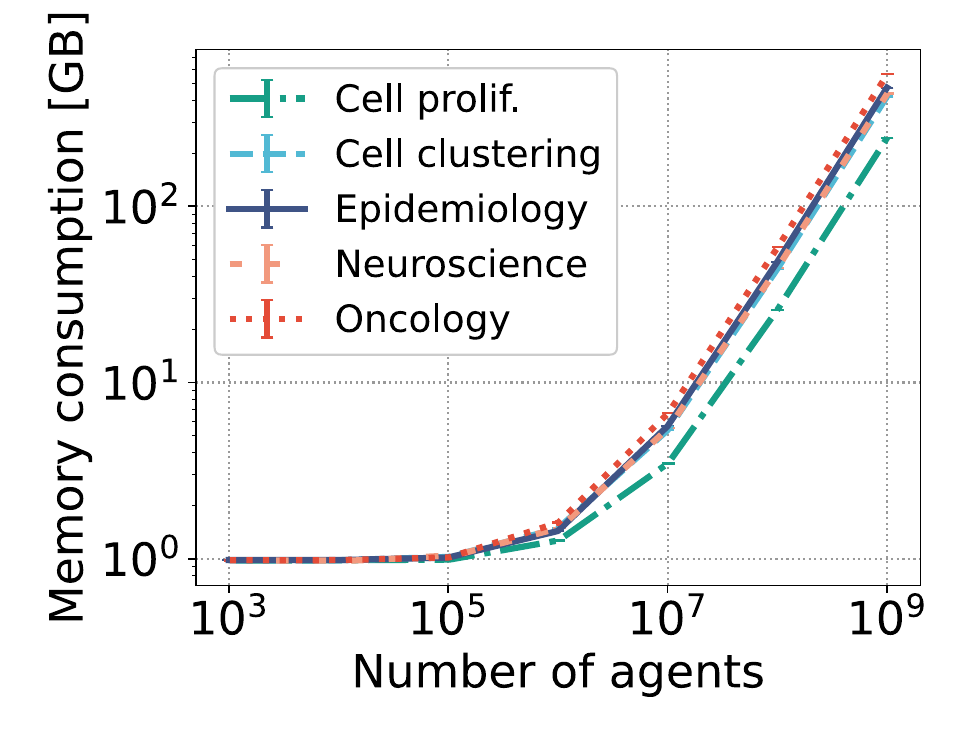}
\end{subfigure}
\caption{Average runtime per iteration and memory consumption analysis as
		the number of agents varies from $10^3$ to $10^9$}.
	\label{fig:runtime-complexity}
\end{figure}

We analyze the runtime and memory consumption of \bdm{} on System~B by
  increasing the number of agents from $10^3$ to $10^9$ for each simulation
  (Figure~\ref{fig:runtime-complexity}).
With one thousand agents, the execution time for one iteration is \result{on
	  average 1.21 ms} and increases only slightly \result{until $10^5$ agents
	  (2.80ms)}.
From there on, runtime increases \emph{linearly} to one billion agents in which
  one iteration takes between \result{6.41 and 38.1 seconds} to execute.
A similar trend can be observed for the memory consumption of \bdm{} (using
  double-precision floating point values), which remains below \result{$1.60$ GB
	  until $10^6$} agents and increases \emph{linearly} to a maximum between
  \result{$245$ and $564$} GB.

The number of agents that \bdm{} can simulate is \emph{not} fundamentally
  limited to one billion.
The maximum depends only on the available memory of the underlying hardware and
  the tolerable execution time.

\subsection{Comparison with Biocellion}
\label{sec:eval:comparison-with-biocellion}

\begin{figure}[tb]
  \centering
  \begin{subfigure}{.25\linewidth}
    \includegraphics[width=\linewidth]{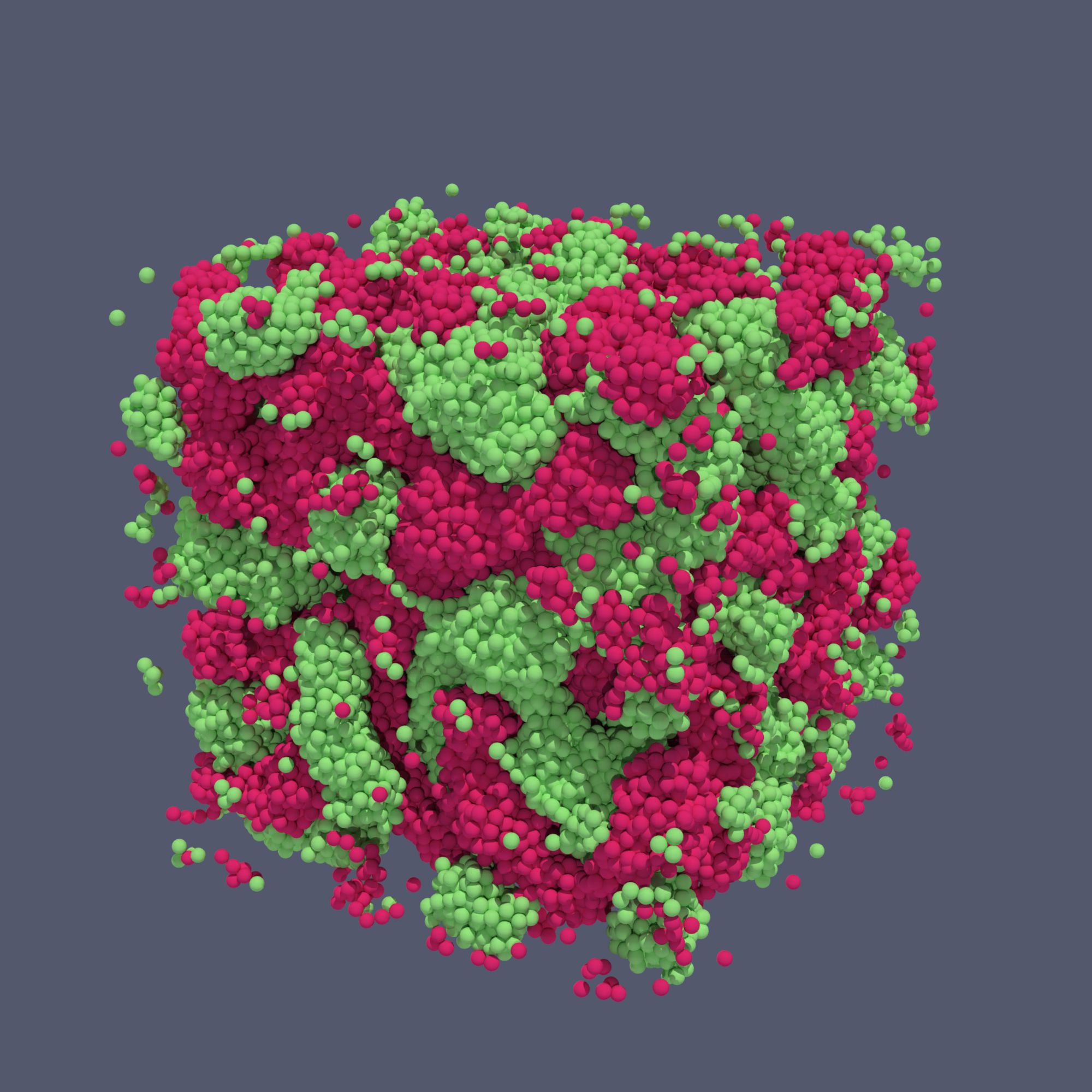}
    \caption{}
    \label{fig:biocellion-rendering}
  \end{subfigure}
  \begin{subfigure}{.72\linewidth}
    \includegraphics[width=0.325\linewidth]{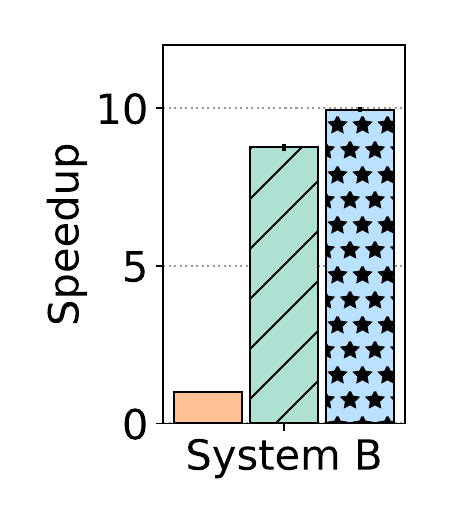}
    \includegraphics[width=0.655\linewidth]{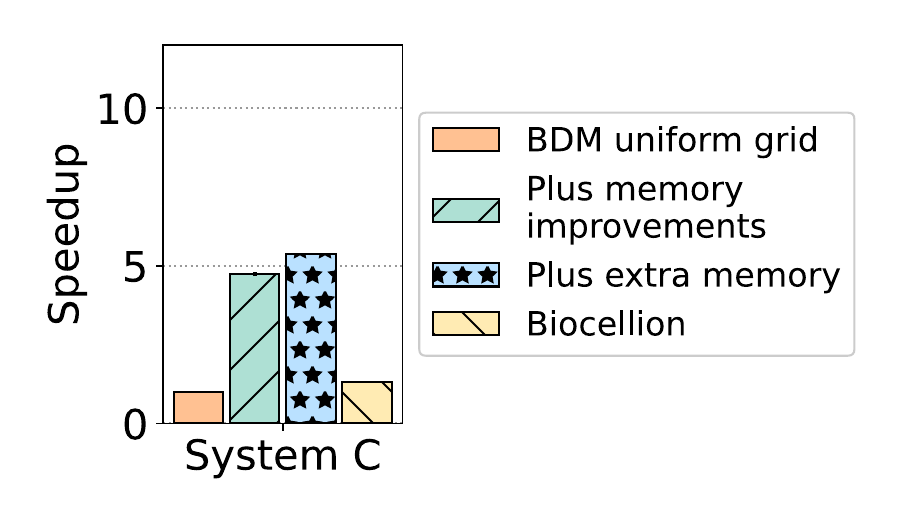}
    \caption{}
    \label{fig:biocellion-comparison}
  \end{subfigure}
    \caption{
      (a) Final simulation state after executing the Biocellion cell sorting model on \bdm{}.
      (b) Performance evaluation of the \bdm{} optimizations with a model with 28.6 million cells 
      on System~B (left) and System~C limited to 16 physical CPU cores (right).
      The Biocellion paper \cite{biocellion} provides only a performance measurement for the latter benchmark.
    }
  \label{fig:biocellion}
\end{figure}

We compare \bdm{} with Biocellion \cite{biocellion}, an agent-based framework
  for tissue models optimized for performance.
We implement the cell sorting simulation presented in the Biocellion paper
  (Section~3.1) in \bdm{} and use identical model parameters.
The visualization of the \bdm{} simulation with 50k cells
  (Figure~\ref{fig:biocellion-rendering}) demonstrates a good agreement with the
  Biocellion results in Figure~3a in \cite{biocellion}.

Since we do not have access to the Biocellion code, because it is proprietary
  software, we compare \bdm{} to the performance results provided in
  \cite{biocellion}.
First, we replicate the benchmark with 26.8 million agents using 16 CPU cores.
For Biocellion, Khang et al.
\cite{biocellion} used a system with two Intel Xeon E5-2670 CPUs with 2.6 GHz.
We execute \bdm{} on System~C with a comparable CPU and limit the number of CPU
  cores to 16 to ensure a fair comparison.
We observe that \bdm{} is \result{4.14$\times$} faster than Biocellion.
\bdm{} executes one iteration in 1.80s (averaged over 500 iterations), while
  Biocellion requires 7.48s.

Second, we consider the Biocellion benchmark in which Kang et al.
executed 1.72 billion cells on a cluster with 4096 CPU cores
(128 nodes with two AMD Opteron 6271 Interlago 2.1 GHz CPUs per node).
We execute the \bdm{} simulation with the same number of cells on a
  \emph{single} node (System~B).
Although \bdm{} requires \result{26.3s} per iteration, which is
  \result{5.90$\times$} slower than Biocellion, \bdm{} uses 56.9$\times$ fewer
  CPU cores.
Therefore, we conclude that the performance per CPU core of \bdm{} is
  \result{9.64$\times$} more efficient than Biocellion.
We repeat the experiment with 281.4 million cells to verify the last
  observation.
Biocellion requires 4.37s per iteration (extracted from Figure~3b in
  \cite{biocellion}) using 21 nodes with a total of 672 CPU cores.
The \bdm{} simulation on System~B with 72 CPU cores runs in almost identical
  \result{4.24s} per iteration.
This result confirms our observation that \bdm{} is an order of magnitude more
  efficient than Biocellion.

We evaluate the impact of our optimizations to provide insights into the
  question of why \bdm{} processes \result{4.14$\times$} more agents per CPU core
  in the first benchmark and \result{9.64$\times$} in the second.
Therefore, we execute the relevant optimizations with 26.8 million cells on
  System~C limited to 16 CPU cores and System~B with 72 CPU cores.
Figure~\ref{fig:biocellion-comparison} shows that the difference can largely be
  explained by the memory optimizations having a more significant impact on
  machines with higher CPU core count.

\subsection{Comparison with Cortex3D and NetLogo}
\label{sec:eval:comparison-with-cx3d-netlogo}

\begin{figure}[t]
  \centering
    \includegraphics[width=0.8\linewidth]{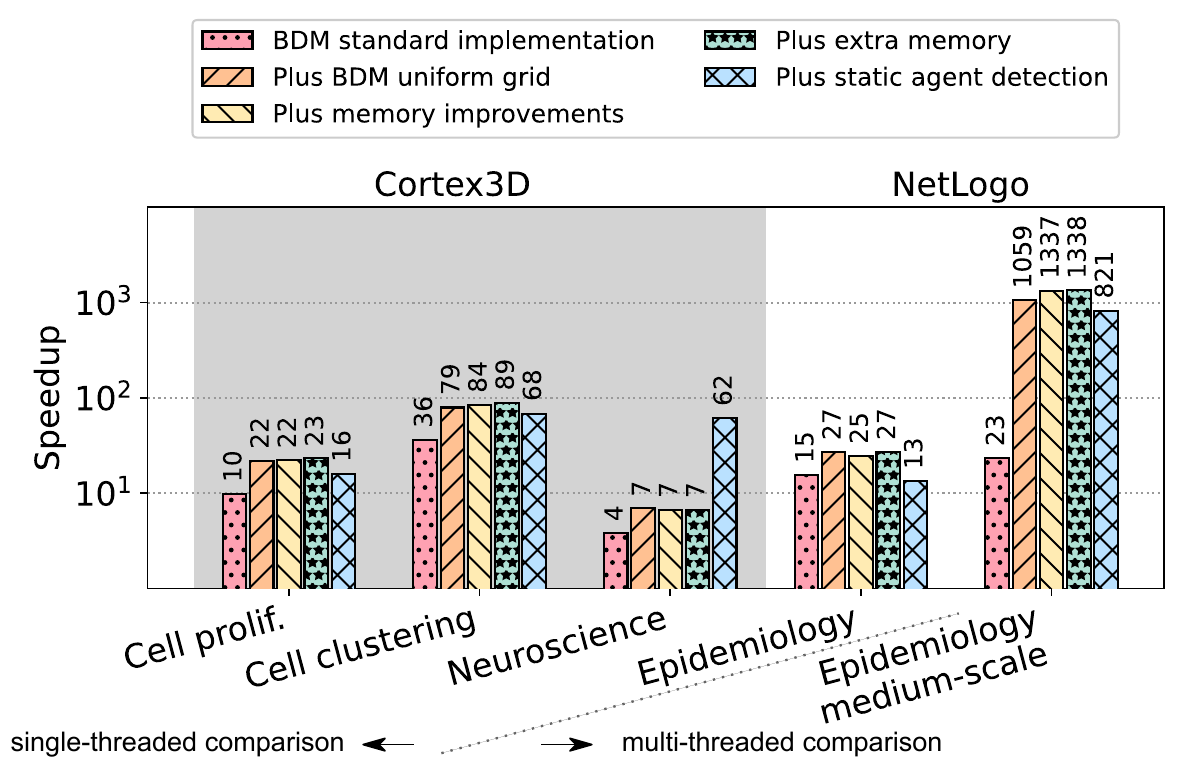}
\caption{Performance comparison with Cortex3D and NetLogo after the optimizations are progressively switched on.
}
  \label{fig:comparison-with-others}
\end{figure}

We also compare with capable single-thread tools to evaluate the parallel
  overhead of the \bdm{} implementation \cite{scalability_cost}.
We choose Cortex3D \cite{zublerdouglas2009framework} due to its similarity with
  the neuroscience features of \bdm{} and select NetLogo \cite{netlogo} as a
  representative for an easy-to-use general-purpose tool.
We extend the experiments from \cite{breitwieser-bdm} by analyzing the impact
  of the presented performance improvements and comparing the memory consumption.
This benchmark uses different simulation parameters for agents, diffusion
  volumes, and iterations, than shown in Table~\ref{tab:sim-characteristics}.
The first four benchmarks in Figure~\ref{fig:comparison-with-others} are
  small-scale benchmarks using between 2k and 30k agents and 0--128k diffusion
  volumes.
These benchmarks run for 100--1000 iterations and only use one thread because
  Cortex3D and NetLogo are not parallelized.
The ``epidemiology (medium-scale)'' benchmark contains 100k agents and uses 144
  threads.
NetLogo only benefits from parallel garbage collection in this scenario.
In the ``\bdm{} standard implementation'', all optimizations are turned off,
  and the kd-tree environment is used.

We make the following observations.
For the small-scale simulations using one thread, \bdm{} achieves a speedup of
  up to \result{78.8$\times$} while using \result{2.49$\times$} less memory.
We observe three orders of magnitude speedup and two orders of magnitude
  reduction in memory consumption for the medium-scale benchmark in which all
  threads were used.

The median speedup of the \bdm{} standard implementation is
  \result{15.5$\times$}.
The optimized uniform grid of \bdm{} boosts performance in all benchmarks
  (median: \result{2.18$\times$}) but has the most significant impact if
  parallelization is used (45.5$\times$).
Memory layout optimizations improve the runtime of medium-scale simulations by
  \result{26.2\%}, but not for small-scale ones.
The memory layout optimizations comprise the NUMA-aware iteration
  (Section~\ref{sec:numa-iteration}), agent sorting and balancing
  (Section~\ref{sec:load-balancing}), and memory allocator
  (Section~\ref{sec:mem-allocator}).
Due to the interdependency between these individual optimizations, we subsumed
  them into one category.
Similarly, extra memory usage during the agent sorting and balancing stage
  (Section~\ref{sec:load-balancing}) has only a slight performance impact (median
  speedup: \result{4.82\%}).
However, the static region optimization dramatically improves the performance
  in the neuroscience use case (speedup \result{9.22$\times$}).
Although the mechanism's overhead reduces the speedup for simulations without
  static regions, this is not problematic.
The modeler usually knows this characteristic a priori and only enables the
  mechanism if static regions are expected (see parameter
  \texttt{detect\_static\_agents}).

\subsection{Optimization Overview}
\label{sec:se:eval:optimization-overview}

\begin{figure}[t]
  \centering
\begin{subfigure}{0.75\linewidth}
    \centering
    \includegraphics[width=\linewidth]{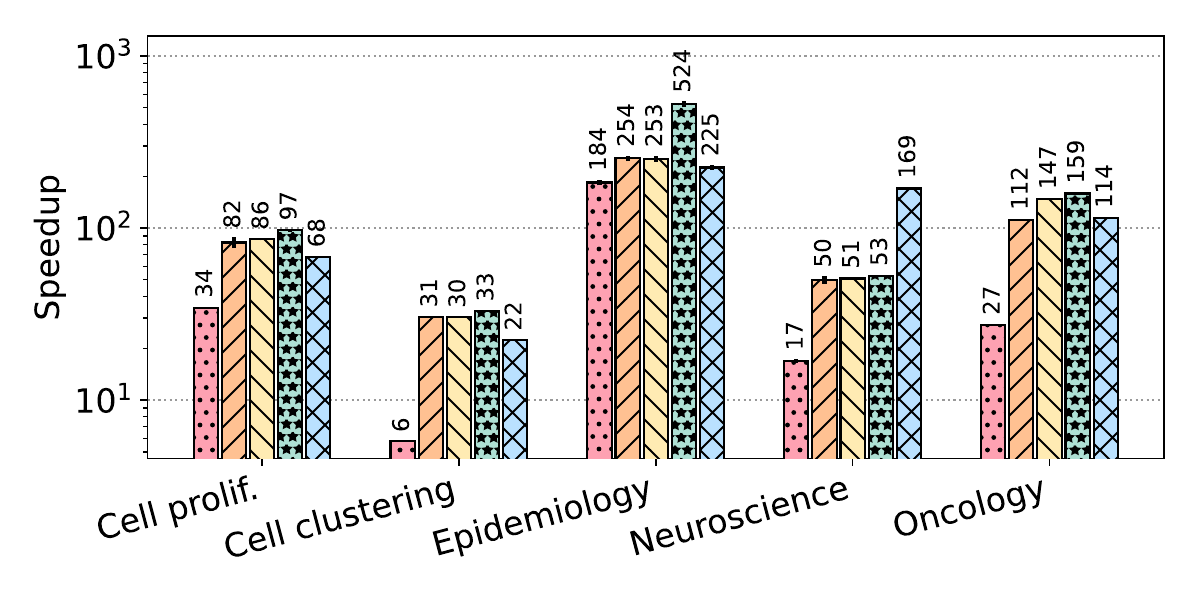}
\end{subfigure}
  \begin{subfigure}{0.75\linewidth}
    \centering
    \includegraphics[width=\linewidth]{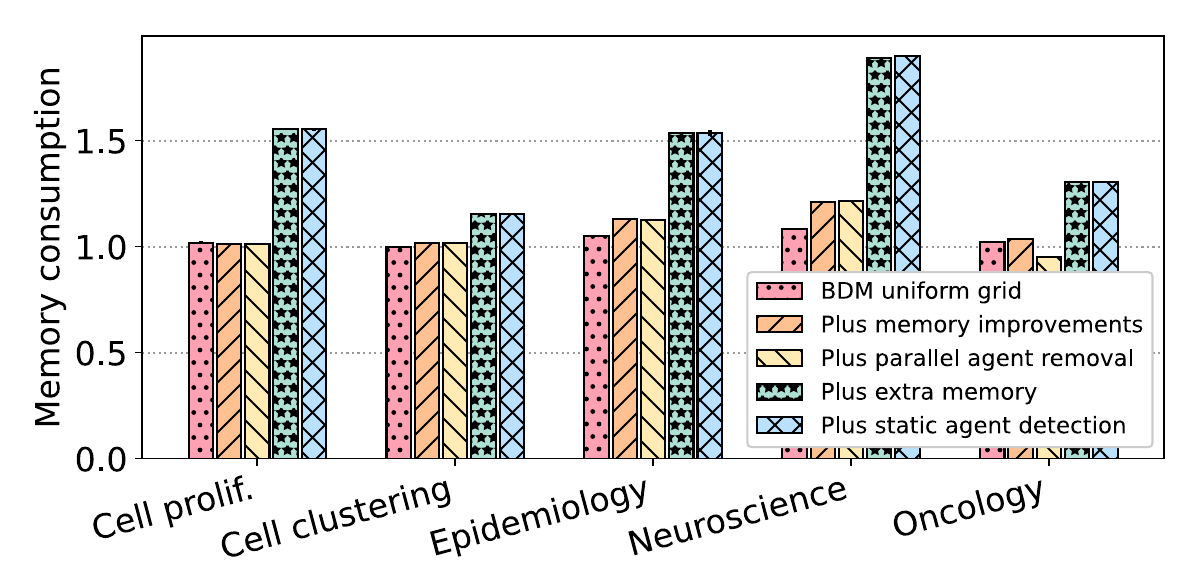}
\end{subfigure}
\caption{Speedup (top) and memory consumption (bottom)
	compared with the \bdm{} standard implementation after the
	optimizations are progressively switched on.
	The legend is shared between the plots.
}
\label{fig:optimization-overview}
\end{figure}

We assess the performance of the presented optimizations using larger-scale
  simulations (Table~\ref{tab:sim-characteristics}) by enabling optimizations
  step-by-step (Figure~\ref{fig:optimization-overview}).
The baseline in this comparison is the \bdm{} standard implementation
  introduced in Section~\ref{sec:eval:comparison-with-cx3d-netlogo}.

We make the following observations.
The \bdm{} optimizations improve overall performance between
  \result{33.1$\times$} and \result{524$\times$} (median: \result{159$\times$}).
These benchmarks confirm the speedup of \bdm{}'s optimized uniform grid that we
  observed in comparison with Cortex3D and NetLogo.
For these larger-scale simulations, the magnitude of the speedup increases up
  to \result{184$\times$} with a median of \result{27.4$\times$}.
A similar observation can be made for the static region detection mechanism,
  albeit with reduced magnitude (speedup: \result{3.22$\times$}).
The main difference between the comparison with Cortex3D and NetLogo and this
  benchmark is the impact of the memory layout optimizations of agents and
  behaviors and the usage of extra memory during agent sorting.
The maximum speedup is up to \result{5.30$\times$} (median:
  \result{2.96$\times$}) and up to \result{2.07$\times$} (median:
  \result{1.09$\times$}), respectively.
Only the Biocellion benchmark in Figure~\ref{fig:biocellion-comparison} shows a
  bigger impact.

The simulation time of the oncology use case, the only benchmark that removes
  agents from the simulation is reduced by \result{31.7\%} using the ``parallel
  removal'' optimization described in Section~\ref{sec:parallel-remove}.
The optimizations increase the median memory consumption by a mere
  \result{1.77\%}, which increases to \result{55.6\%} by enabling the use of
  extra memory during agent sorting.

\subsection{Scalability}
\label{ch:se:eval:scalability}

We evaluate the scalability of \bdm{} using the complete simulations lasting
  between 288 and 1000 iterations and perform a strong and weak scaling analysis with
  different optimizations enabled.
The strong scaling analysis is performed with ten iterations.

\begin{figure}[tbp]
	\centering
\begin{subfigure}{.3\linewidth}
    \centering
    \includegraphics[width=0.95\textwidth]{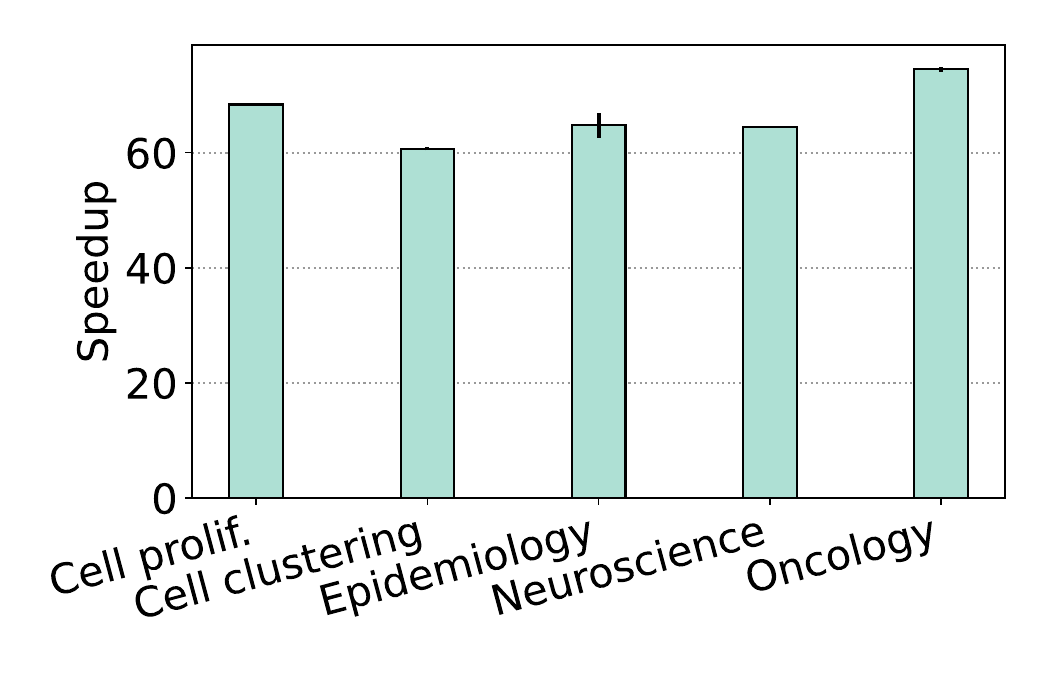}
    \caption{Whole simulation scalability}
    \label{fig:full-scalability}
  \end{subfigure}
  \begin{subfigure}{.3\linewidth}
    \centering
    \vspace{4mm}
    \includegraphics[width=0.9\textwidth]{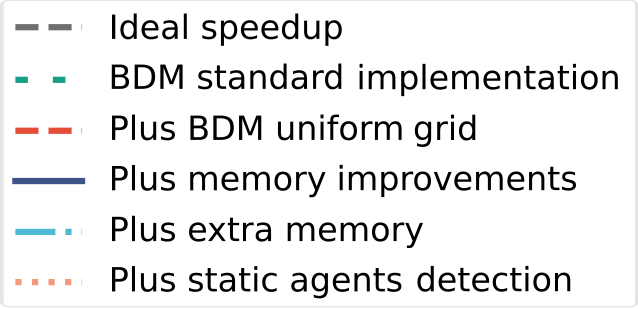}
    \vspace{4mm}
    \caption{Legend for (c)--(g)}
  \end{subfigure}
  \begin{subfigure}{\linewidth}
    \centering
    \vspace{2mm}
    \includegraphics[width=0.3\textwidth,trim=6mm 8mm 8mm 8mm, clip]{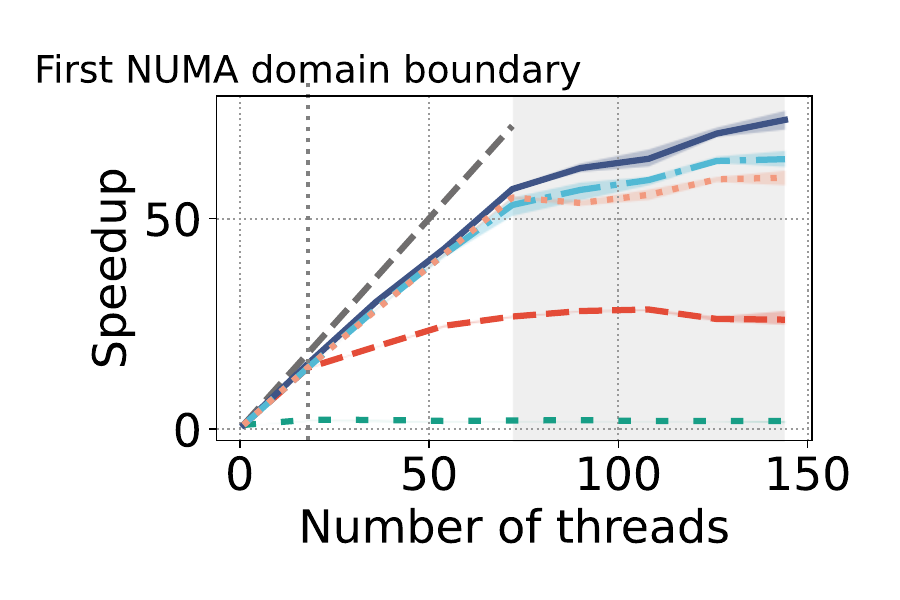}
    \includegraphics[width=0.3\textwidth,trim=6mm 8mm 8mm 8mm, clip]{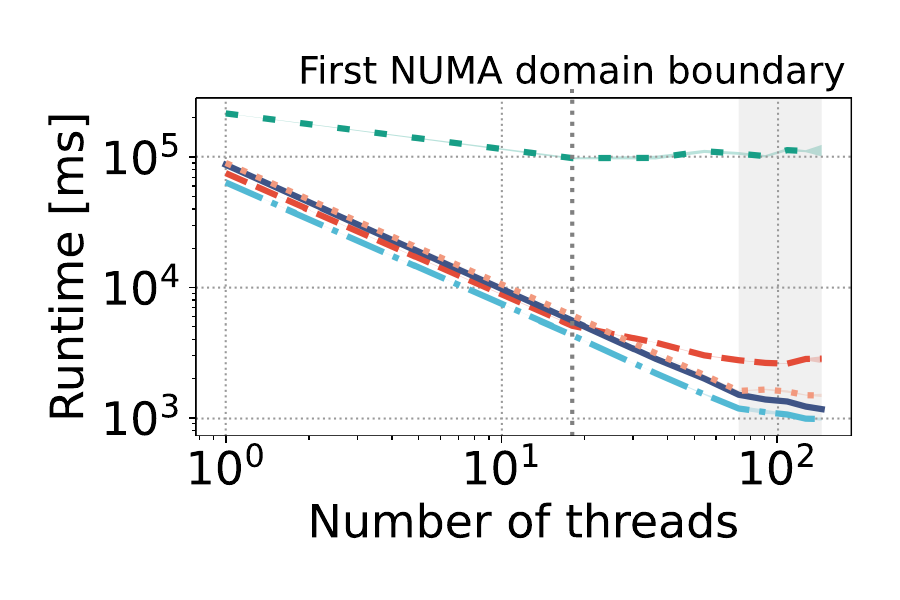}
    \caption{Cell proliferation}
    \label{fig:scalability-cgd}
  \end{subfigure}
  \begin{subfigure}{\linewidth}
    \centering
    \includegraphics[width=0.3\textwidth,trim=6mm 8mm 8mm 8mm, clip]{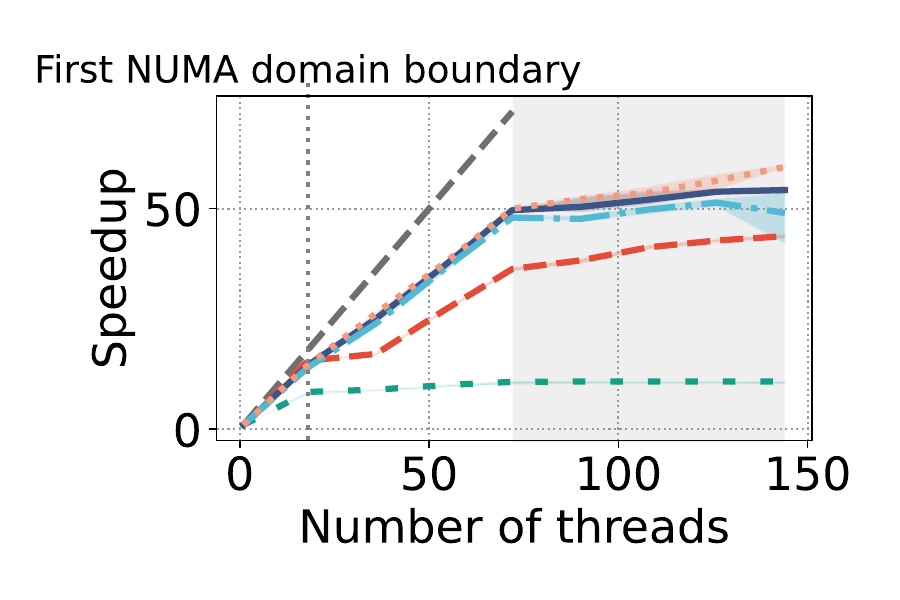}
    \includegraphics[width=0.3\textwidth,trim=6mm 8mm 8mm 8mm, clip]{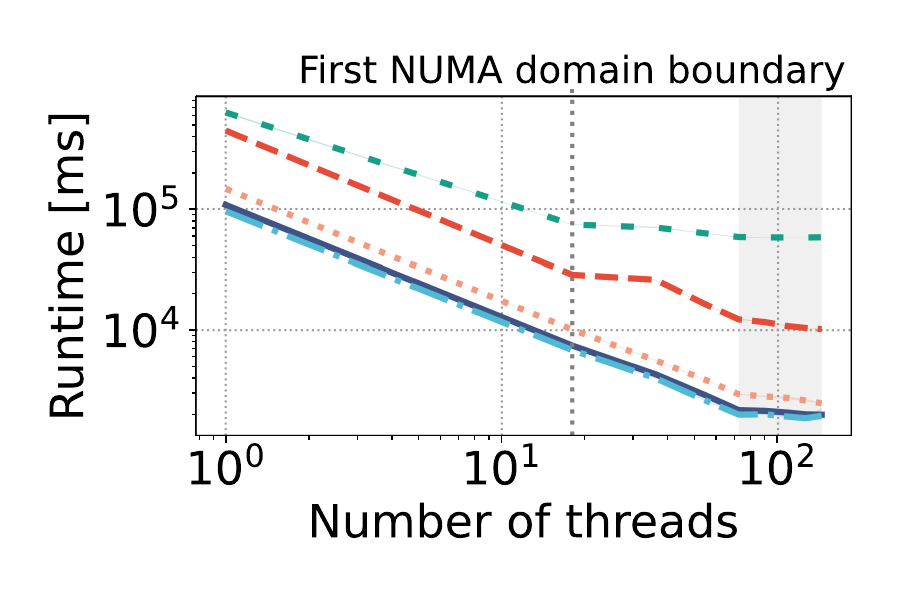}
    \caption{Cell clustering}
  \end{subfigure}
  \begin{subfigure}{\linewidth}
    \centering
    \includegraphics[width=0.3\textwidth,trim=6mm 8mm 8mm 8mm, clip]{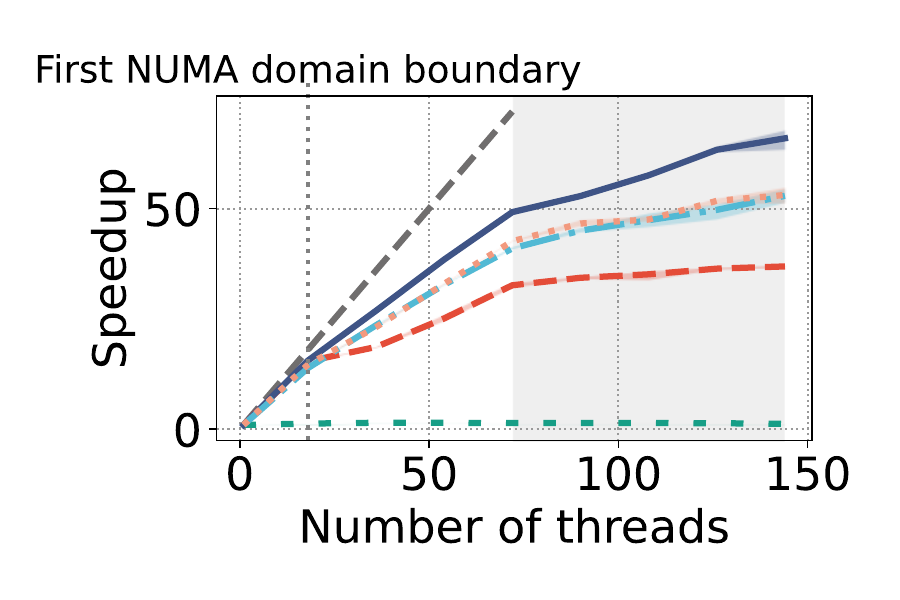}
    \includegraphics[width=0.3\textwidth,trim=6mm 8mm 8mm 8mm, clip]{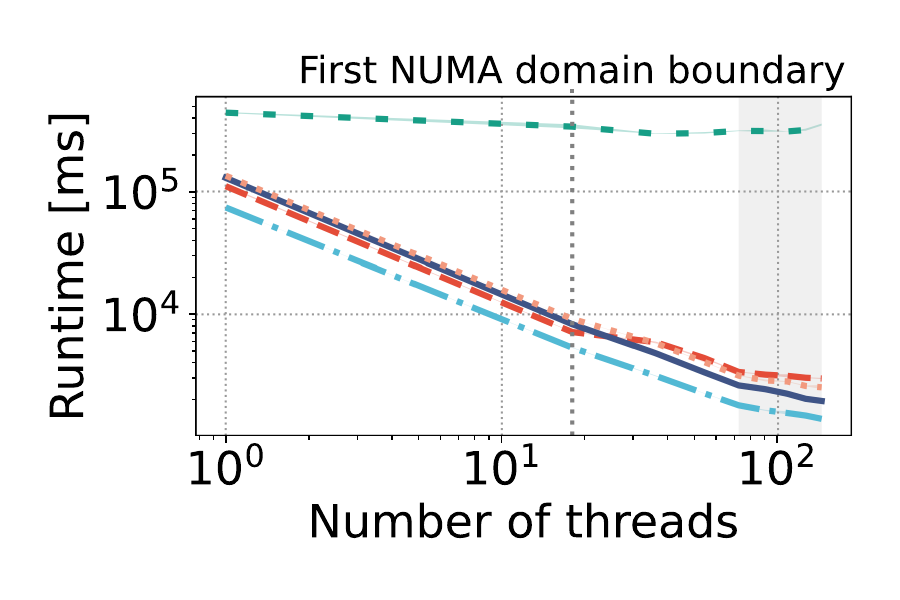}
    \caption{Epidemiology}
  \end{subfigure}
  \begin{subfigure}{\linewidth}
    \centering
    \includegraphics[width=0.3\textwidth,trim=6mm 8mm 8mm 8mm, clip]{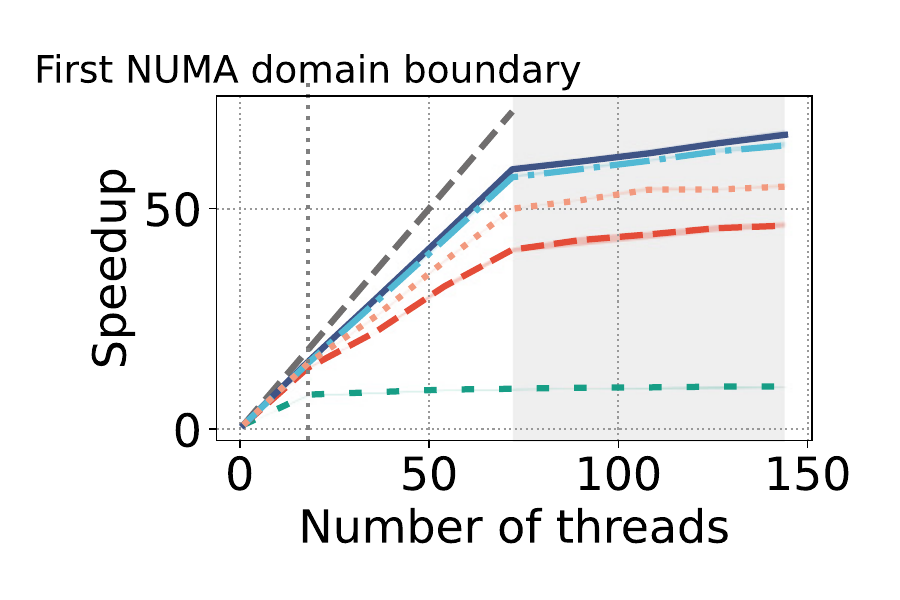}
    \includegraphics[width=0.3\textwidth,trim=6mm 8mm 8mm 8mm, clip]{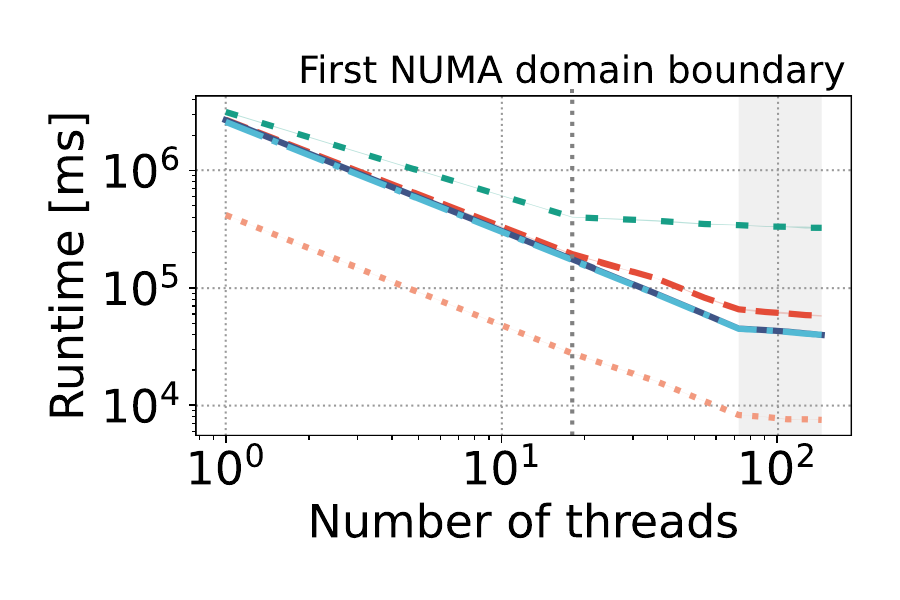}
    \caption{Neuroscience}
  \end{subfigure}
  \begin{subfigure}{\linewidth}
    \centering
    \includegraphics[width=0.3\textwidth,trim=6mm 8mm 8mm 8mm, clip]{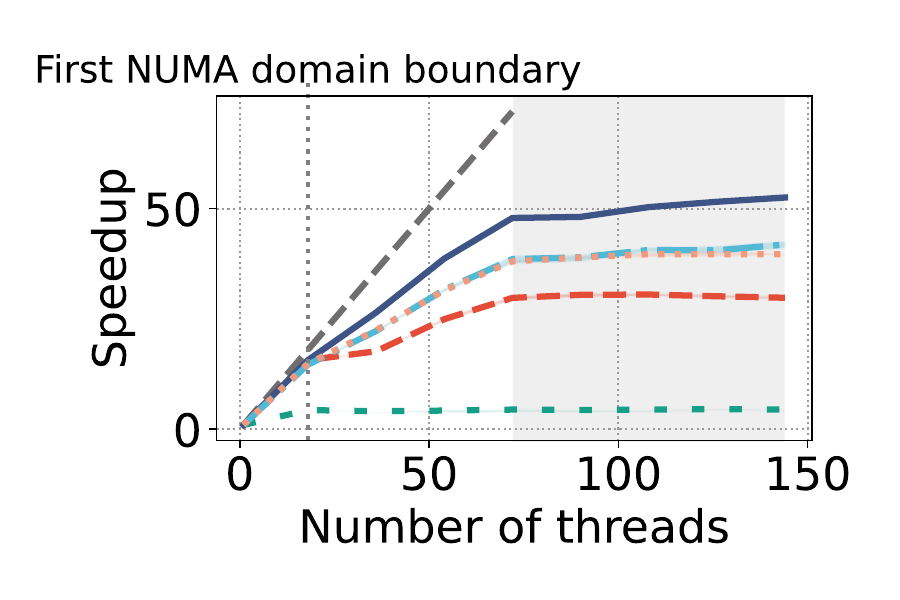}
    \includegraphics[width=0.3\textwidth,trim=6mm 8mm 8mm 8mm, clip]{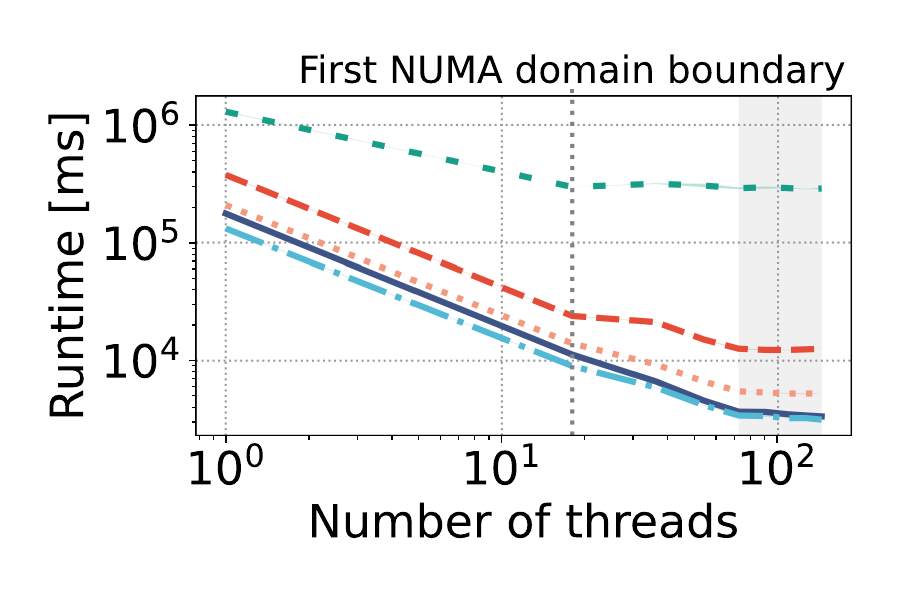}
    \caption{Oncology}
    \label{fig:scalability-ts}
  \end{subfigure}
\caption{
		(a) Simulation scalability using the whole simulation.
		(b--g) Detailed strong scaling analysis using only ten time steps.
		The left column shows the speedup with respect to a single-thread execution,
		  while the right column presents the total runtime.
}
\label{fig:scalability}
\end{figure}

Figure~\ref{fig:full-scalability} illustrates the excellent scalability of
  \bdm{} for complete simulations (i.e., executing all iterations).
The speedup using 72 physical cores with hyperthreading enabled is between
  \result{60.7$\times$ and 74.0$\times$ (median 64.7$\times$)} compared to serial
  execution.
Section~\ref{sec:eval:comparison-with-cx3d-netlogo} shows that \bdm{} with one
  CPU core is more than \result{23$\times$} faster than Cortex3D.
If we combine this result with the scalability analysis, which shows that
  \bdm{} with 72 CPU cores is more than \result{60$\times$} faster than one CPU
  core, we can conclude that \bdm{} is up to three orders of magnitude faster
  than Cortex3D.

Figures~\ref{fig:scalability-cgd}--\ref{fig:scalability-ts} show the \emph{strong}
  scaling analysis for each benchmark simulation with ten iterations after
  progressively switching on the presented optimizations.
The left column shows the speedup with respect to a single-thread execution,
  and the right column presents the average runtime in milliseconds to highlight
  the absolute differences between various optimizations and the reduction in
  runtime with increasing threads.
We make the following observations.
The \bdm{} standard implementation scales poorly due to the serial build of the
  kd-tree environment, which is improved considerably by using \bdm{}'s optimized
  uniform grid (Section~\ref{sec:grid}).
The presented memory optimizations (Section~\ref{sec:opt:memory-layout}) fully
  achieve their desired effect and allow \bdm{} to scale across NUMA domains and
  high CPU-core counts.

\begin{figure}[tbp]
	\centering
\begin{subfigure}{.47\linewidth}
    \centering
    \vspace{2mm}
    \includegraphics[width=\textwidth]{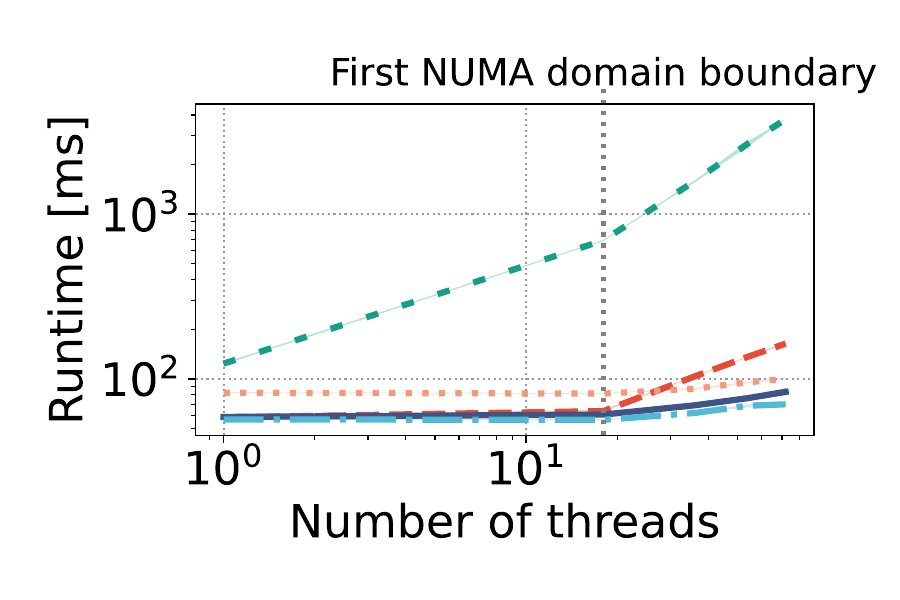}
    \caption{Cell proliferation}
    \label{fig:weak-scalability-cgd}
  \end{subfigure}
  \begin{subfigure}{.47\linewidth}
    \centering
    \includegraphics[width=\textwidth]{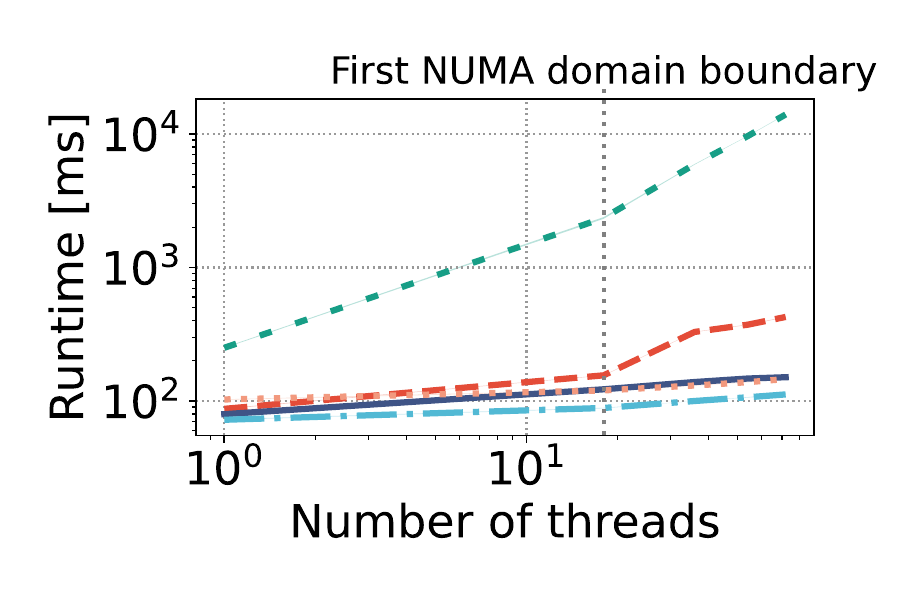}
    \caption{Cell clustering}
  \end{subfigure}
  \begin{subfigure}{.47\linewidth}
    \centering
    \includegraphics[width=\textwidth]{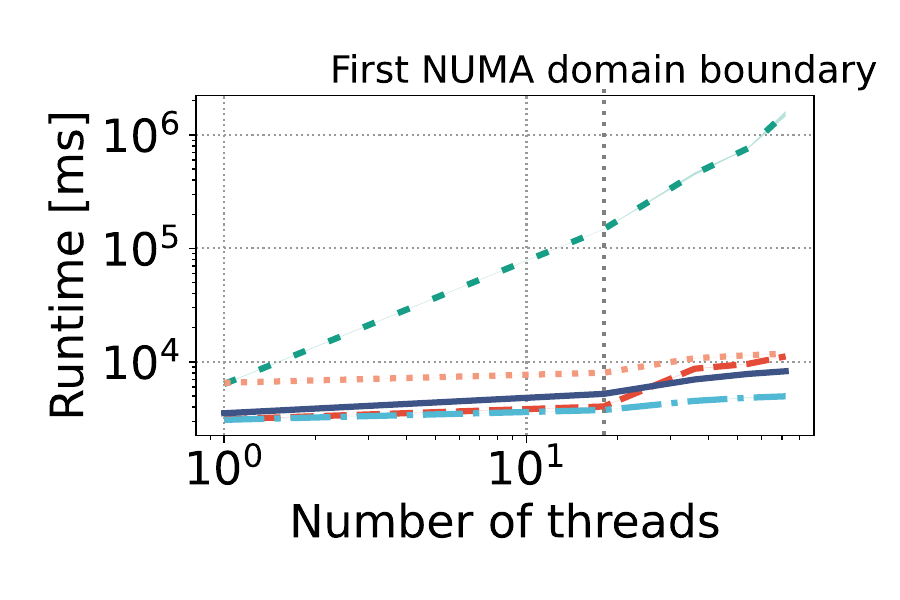}
    \caption{Epidemiology}
  \end{subfigure}
  \begin{subfigure}{.47\linewidth}
    \centering
    \includegraphics[width=\textwidth]{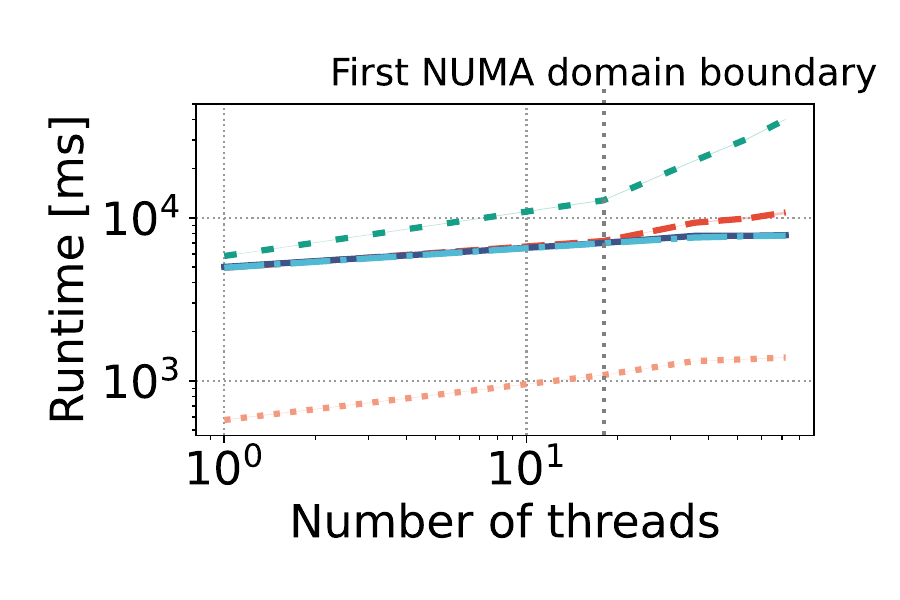}
    \caption{Neuroscience}
  \end{subfigure}
  \begin{subfigure}{.47\linewidth}
    \centering
    \includegraphics[width=\textwidth]{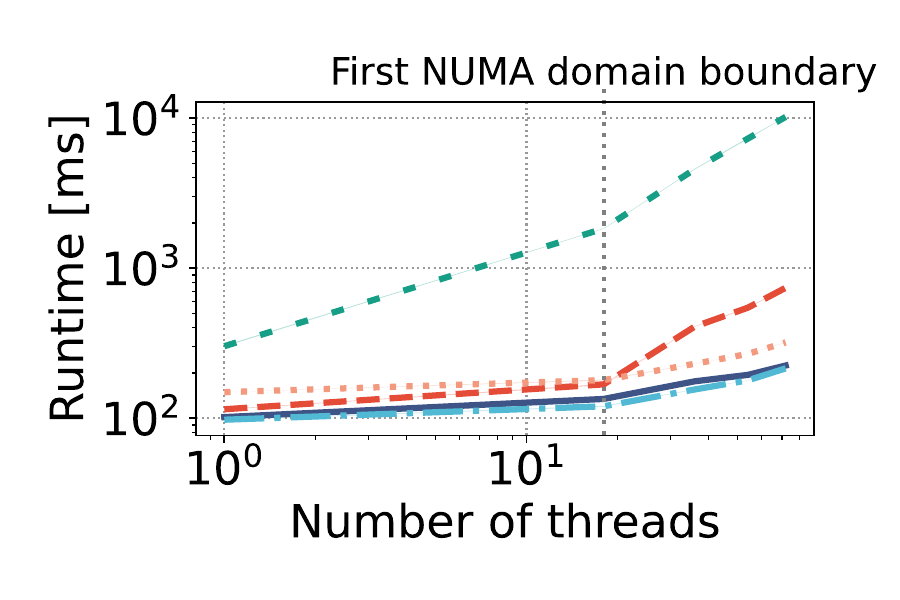}
    \caption{Oncology}
  \end{subfigure}
  \begin{subfigure}{.47\linewidth}
    \centering
    \includegraphics[width=0.6\textwidth]{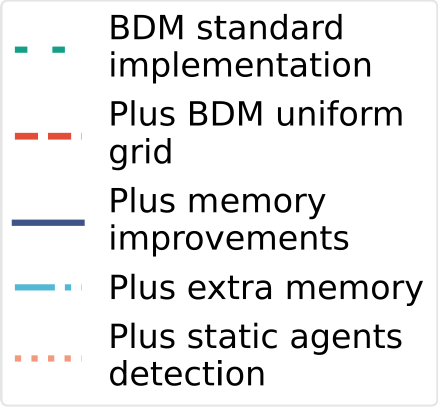}
    \caption{Legend for (a)--(e)}
  \end{subfigure}
\caption{
		(a--e) Weak scaling analysis using the whole simulation.
    (f) Legend shared between the figures (a--e). 
}
\label{fig:weak-scalability}
\end{figure}

We make the same observations for the \emph{weak} scaling analysis (Figure~\ref{fig:weak-scalability}), in which the number of agents is increased proportionally to the number of threads. Under ideal conditions, we expect the runtime to remain constant for each thread count. Also for this benchmark, we switch on the optimizations one after the other to highlight their impact on the runtime and see significant improvements over the baseline.

\subsection{Neighbor Search Algorithm Comparison}
\label{sec:eval:environment}

\begin{figure}
	\centering
\begin{subfigure}{\linewidth}
    \centering
    \includegraphics[width=0.49\textwidth]{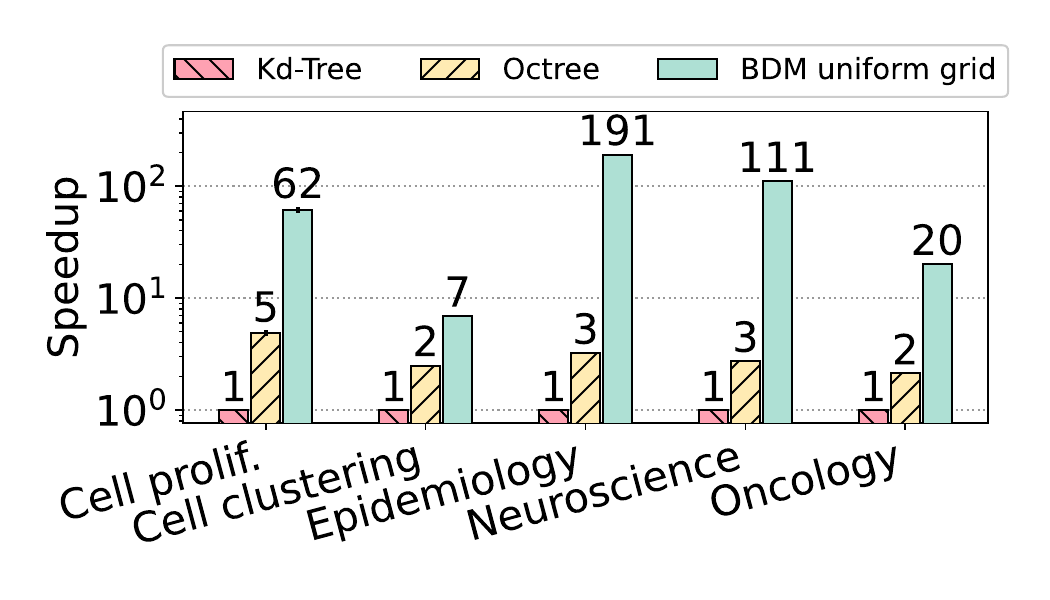}
    \includegraphics[width=0.49\textwidth]{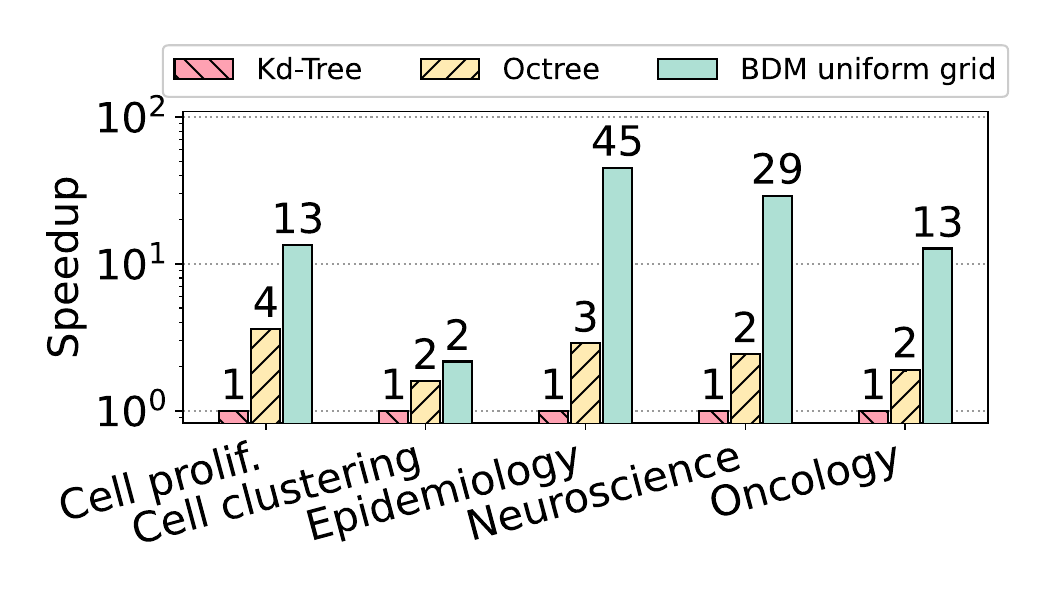}
     \caption{Whole simulation}
   \end{subfigure}
  \begin{subfigure}{\linewidth}
    \centering
    \includegraphics[width=0.49\textwidth]{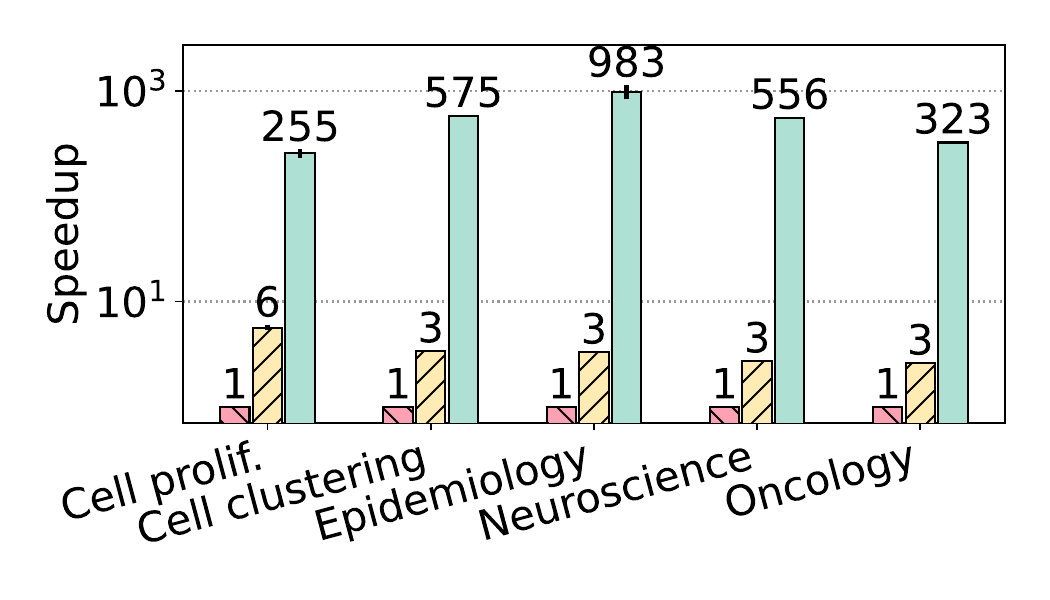}
    \includegraphics[width=0.49\textwidth]{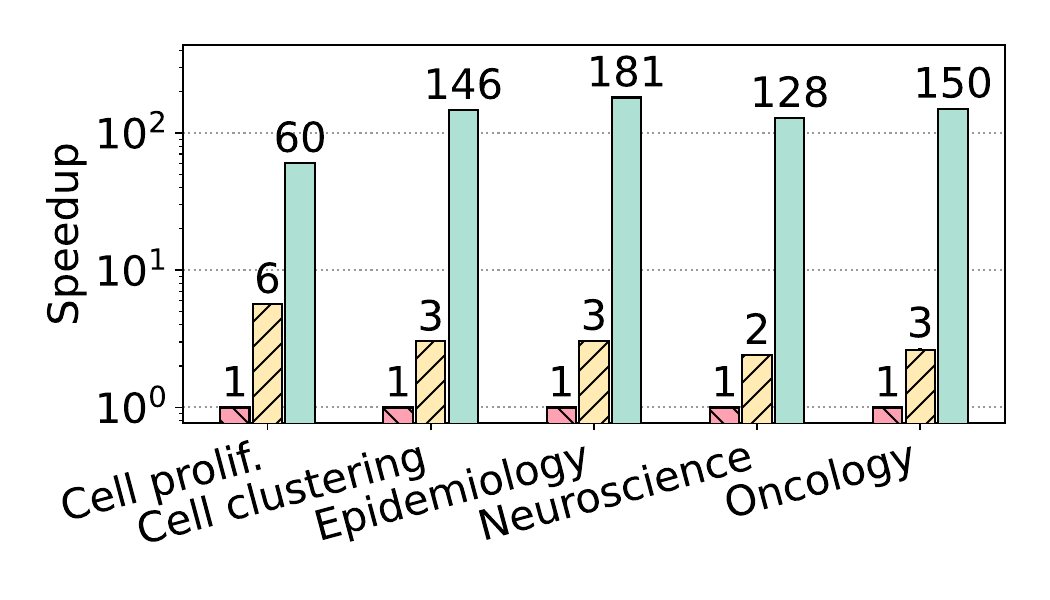}
     \caption{Build time}
   \end{subfigure}
  \begin{subfigure}{\linewidth}
    \centering
    \includegraphics[width=0.49\textwidth]{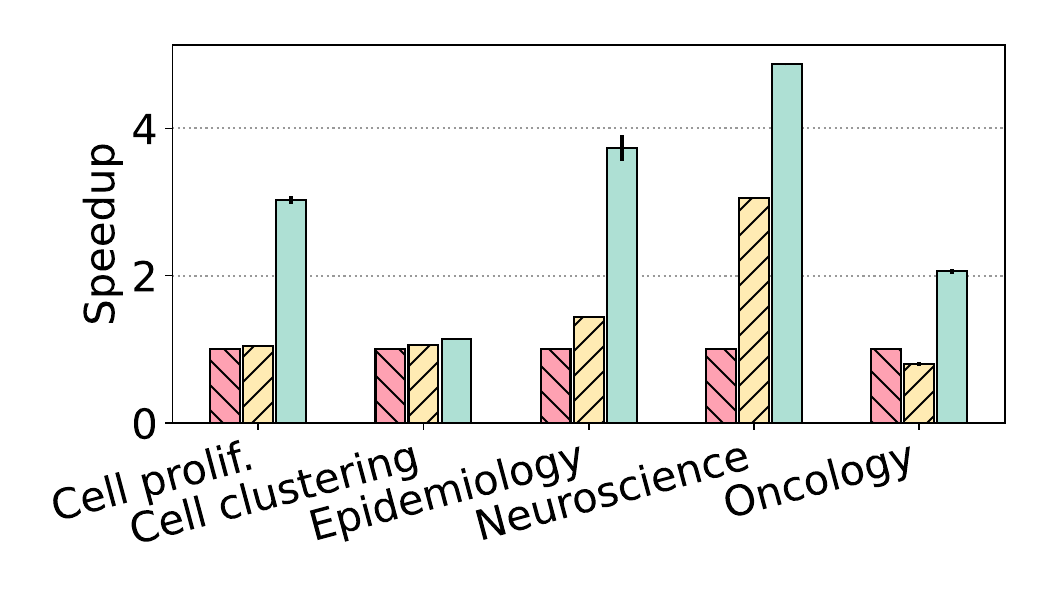}
    \includegraphics[width=0.49\textwidth]{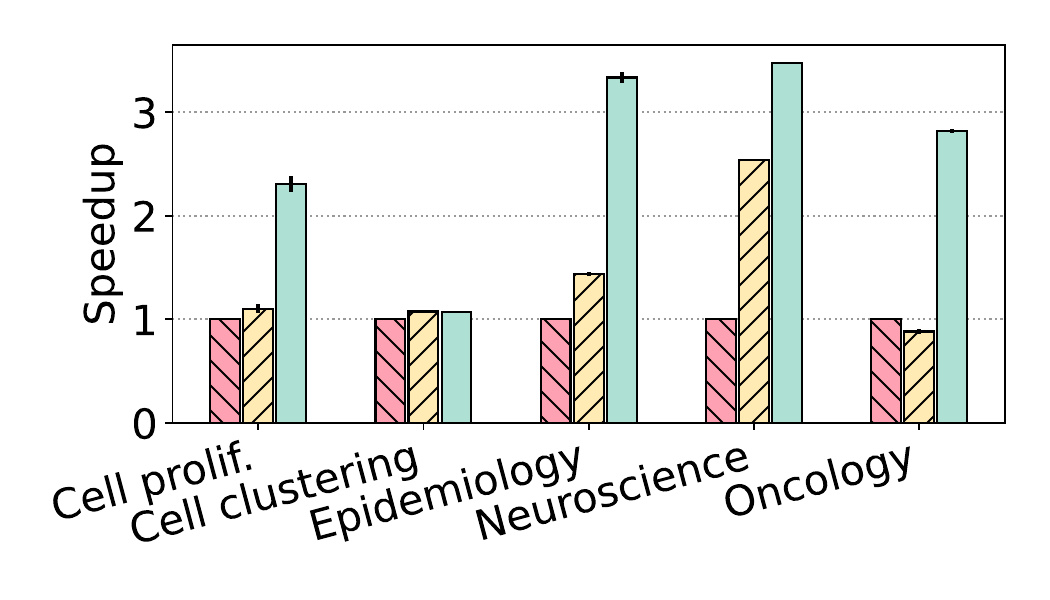}
     \caption{Search time (indirect)}
   \end{subfigure}
  \begin{subfigure}{\linewidth}
    \centering
    \includegraphics[width=0.49\textwidth]{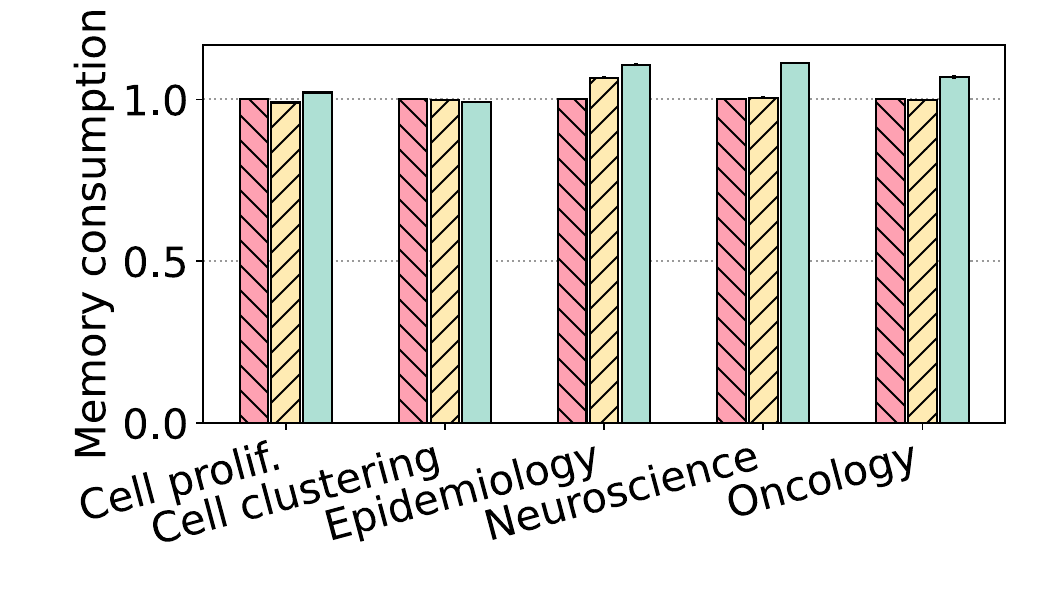}
    \includegraphics[width=0.49\textwidth]{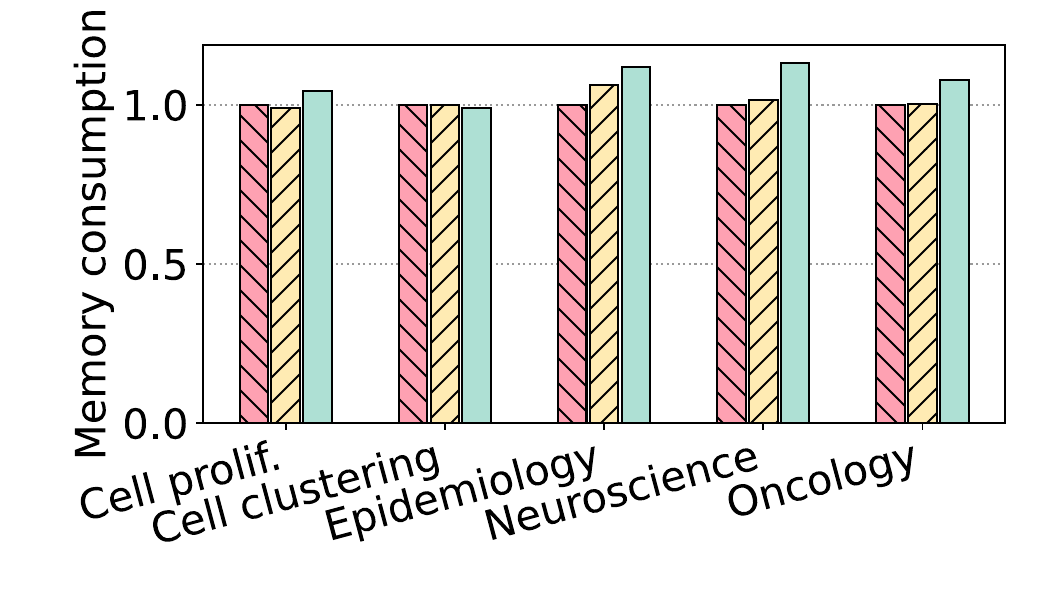}
    \caption{Memory consumption}
   \end{subfigure}
\caption{Neighbor search algorithm comparison
		(left column: four NUMA domains and 144 threads, right column: one NUMA
		domain
		and 18 threads).
		The legend is shared between the plots.
	}
\label{fig:environment}
\end{figure}

Figure~\ref{fig:environment} compares three different neighbor search
  algorithms: \bdm{}'s uniform grid, UniBN's octree \cite{unibn}, and the kd-tree
  from nanoflann \cite{nanoflann}.
To ensure a fair comparison, we turned off agent sorting for all algorithms
  because it is currently only implemented for the uniform grid.
We validate our choice for the octree bucket size and nanoflann depth parameter
  and observe that the used parameters are within \result{4.20\%} of the optimum
  runtime.
The left column in Figure~\ref{fig:environment} shows the result for four NUMA
  domains and 144 threads, while the right column shows results for one NUMA
  domain and 18 threads.
We analyzed four properties of these radial neighbor search methods: runtime
  impact on the whole simulation, build and search time of the index, and memory
  consumption (Figure~\ref{fig:environment}).
We measure the search time indirectly by comparing the agent operation
  runtimes.
This operation contains the initial neighbor searches and thus provides
  information on how fast searches are executed.

The \bdm{} uniform grid implementation shows its benefits not only in the pure
  build time comparison but also in the full simulation analysis.
Although a significant build time difference in comparison to the kd-tree and
  octree is expected (because the build process is serial), the magnitude between
  \result{255 and 983$\times$} on four NUMA domains is surprising.
The uniform grid outperforms the other algorithms also during the search stage
  for all simulations.

Simulations using \bdm{}'s uniform grid implementation are up to
  \result{191$\times$} faster than the kd-tree implementation while consuming
  only \result{11\%} more memory in the worst case.

\subsection{NUMA-Aware Iteration}

We evaluate the individual performance impact of NUMA-aware iteration
(Section~\ref{sec:numa-iteration}).
In the other benchmarks, this optimization was included in the 
  ``memory layout optimization'' group.
We compare the simulation runtime with all optimizations enabled, 
  to executions in which ``NUMA-aware iteration'' is turned off.
This benchmark shows that this mechanism reduces the runtime between
  \result{1.07$\times$} and \result{1.38$\times$} (median:
  \result{1.30$\times$}).

\subsection{Agent Sorting and Balancing}

\begin{figure}[h!]
  \centering
\begin{subfigure}{.49\linewidth}
    \includegraphics[width=\textwidth]{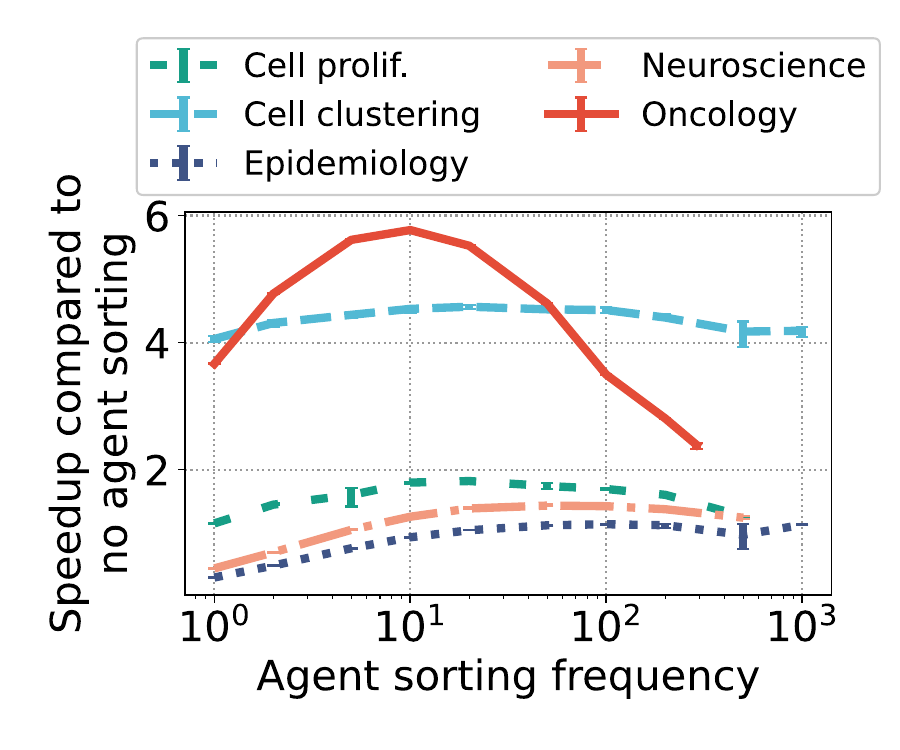}
\end{subfigure}
  \begin{subfigure}{.49\linewidth}
  \includegraphics[width=\textwidth]{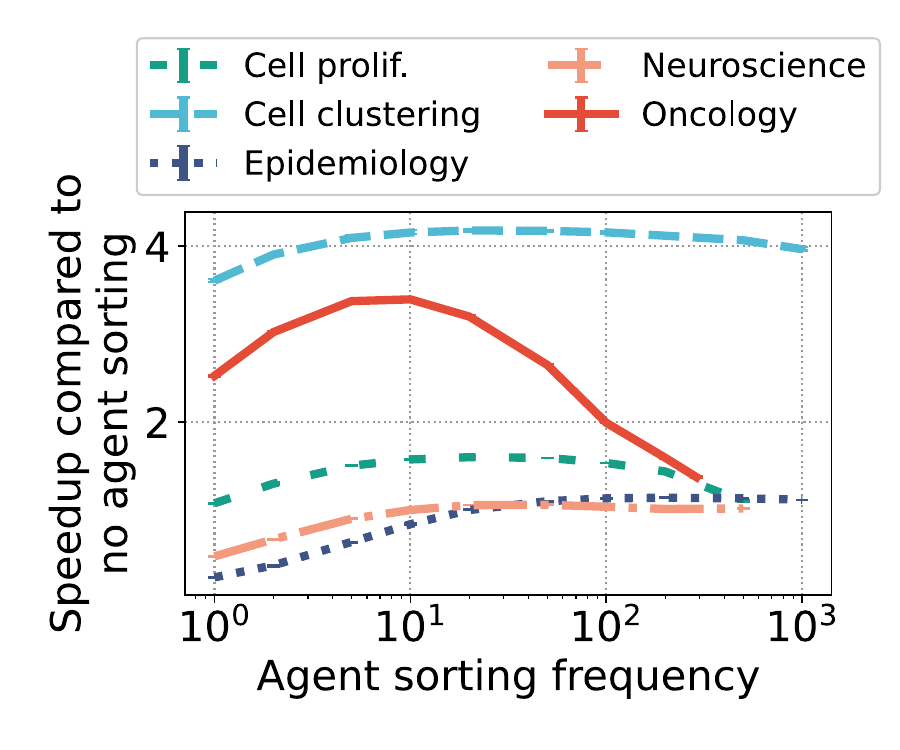}
\end{subfigure}
\caption{Agent sorting and balancing speedup for different execution
    frequencies (left: four NUMA domains and 144 threads, right: one NUMA
    domain
    and 18 threads).}
\label{fig:load-balancing}
\end{figure}

This section evaluates the impact of agent sorting and balancing
  (Section~\ref{sec:load-balancing}) on the simulation runtime for one and four
  NUMA domains.
To this extent, we perform a parameter study with varying agent sorting
  frequencies for each simulation.
Figure~\ref{fig:load-balancing} shows the speedup for four NUMA domains (left)
  and one NUMA domain (right).
The baselines in both cases are simulations without agent sorting.
An agent sorting frequency of one means that the operation is executed in every
  iteration; similarly, a frequency of ten would mean that the operation is
  executed every ten iterations.

Load balancing of agents among NUMA domains greatly impacts performance even on
  systems without NUMA architecture.
This stems from the fact that the agent sorting operation also aligns agents
  that are close in space also in memory.

The oncology and cell clustering simulations benefit most of this performance
  improvement (peak speedup of \result{5.77 and 4.56$\times$} for four NUMA
  domains).
Both simulations are initialized with a random distribution of agents.
Although the epidemiology simulation is also initialized randomly, its agents
  also move randomly with large distances between iterations.
This behavior reduces the alignment improvements significantly (peak speedup
  \result{1.14$\times$} for four NUMA domains).
The cell proliferation simulation is initialized with a 3D grid of cells, which
  improves the alignment compared to the worst-case random initialization.
Therefore, the maximum obtained speedup is reduced to \result{1.82$\times$}
  (four NUMA domains).
Suppose we change the initialization of the cell proliferation simulation to
  random, the maximum speedup increases to \result{4.68$\times$}.
This optimization performs below average for the neuroscience simulation.
This simulation only has an active growth front, while the remaining part
  remains static.
The static agent detection mechanism exploits this fact and avoids calculating
  mechanical forces for the static regions.
Therefore, the number of neighbor accesses is significantly reduced, and thus
  the benefits of aligned agents.
If static region detection is disabled, agent sorting and balancing improve the
  runtime by \result{3.80$\times$ at a frequency of 20}.

\subsection{\bdm{}
Memory Allocator} 

  To evaluate the performance of the \bdm{} memory allocator, we compare it with
  glibc's version of ptmalloc2 \cite{gloger2006ptmalloc} and jemalloc
  \cite{evans2011scalable} using our five benchmark simulations.
A comparison with tcmalloc \cite{tcmalloc} was impossible due to deadlock issues that we discovered during benchmarking.
Only the epidemiology use case uses additional memory during agent sorting and
  balancing.
Since the \bdm{} memory allocator only covers agents and behaviors, we need to
  use another allocator for the remaining objects.

\begin{figure}[h!]
  \centering
  \begin{subfigure}{0.49\linewidth}
    \includegraphics[width=\textwidth]{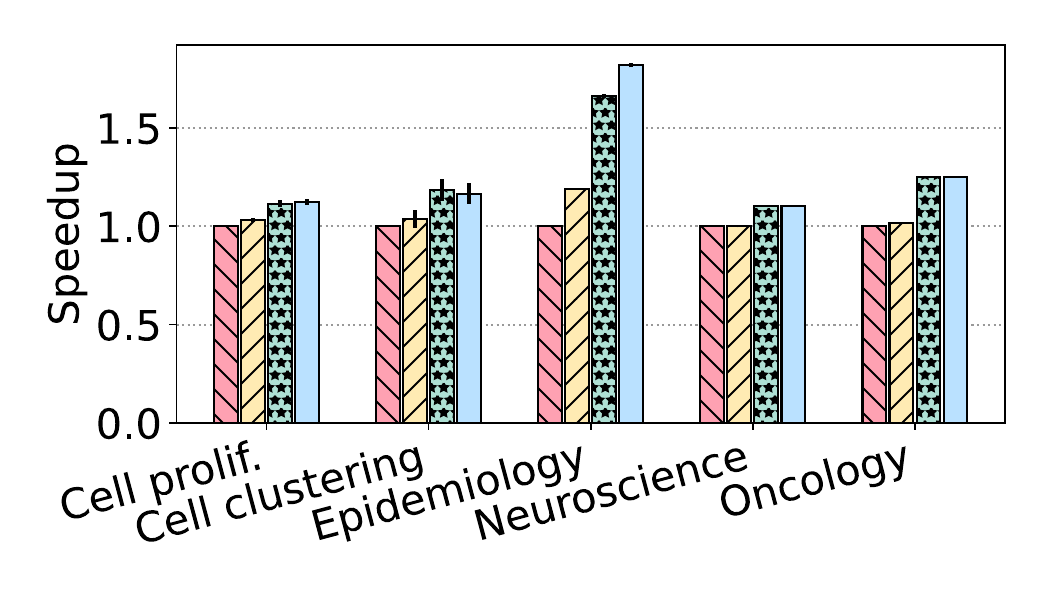}
\end{subfigure}
  \begin{subfigure}{0.49\linewidth}
    \includegraphics[width=\textwidth]{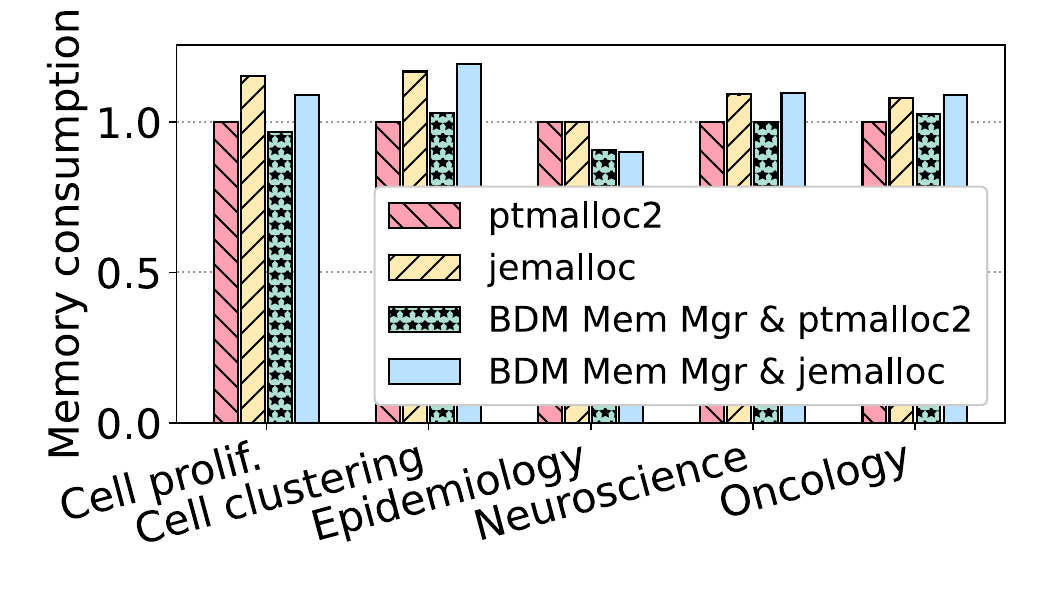}
\end{subfigure}
\caption{Memory allocator comparison (left: speedup, right: memory consumption). The legend is shared between the plots.}
\label{fig:mem-mgr-pc}
\end{figure}

This requirement results in four tested configurations per simulation, as
  illustrated in Figure~\ref{fig:mem-mgr-pc}.
The \bdm{} memory allocator improves the overall simulation runtime up to
  \result{1.52$\times$} over ptmalloc2 (\result{median: 1.19$\times$}) and up to \result{1.40$\times$} over jemalloc
  (\result{median 1.15$\times$}).
The allocator consumes \result{1.41\%} less memory than ptmalloc2 and \result{2.43\%} less memory than
  jemalloc on average.

\subsection{Visualization}

This section evaluates the performance of the visualization integration which
  comprises the export of visualization files and rendering.
We compare the performance of the file export on four dimensions.
We distinguish between different data access methods (zero-copy, copy, and
  cache), storage location (SSD and in-memory file system [DRAM]), number of
  exported agent attributes (all attributes vs.
required ones),
and investigate the impact of file compression.
The distinction between data access methods is specific to the file export.
Therefore, the rendering evaluation has only three comparison dimensions.

\begin{figure}[h!]
 \centering
 \begin{subfigure}{\linewidth}
   \centering
   \includegraphics[width=0.56\textwidth]{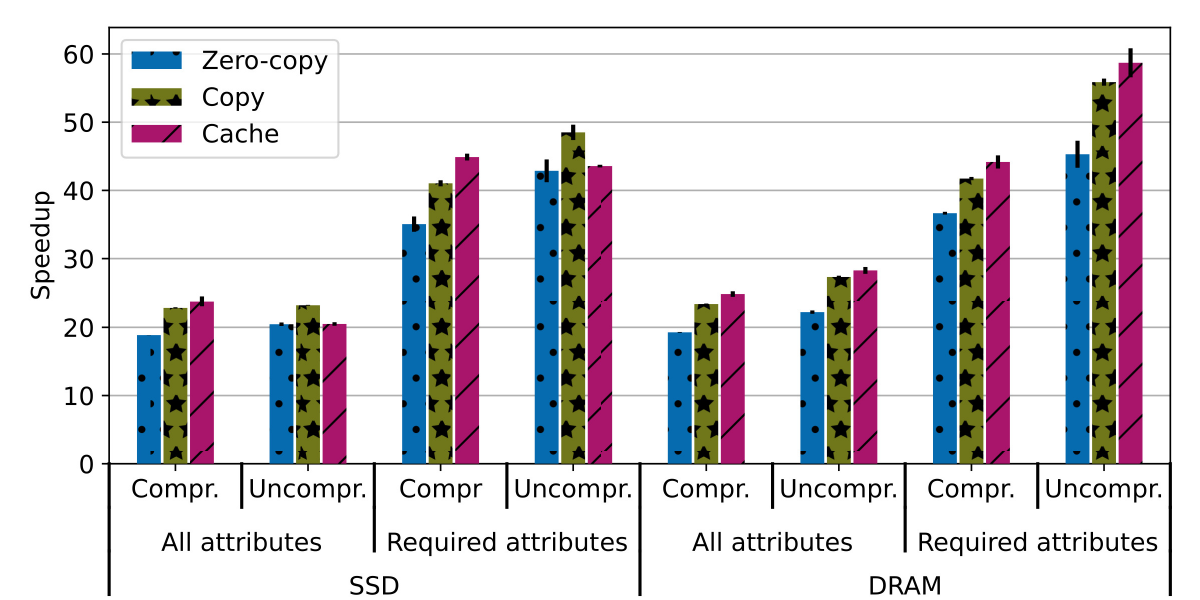}
   \includegraphics[width=0.34\textwidth]{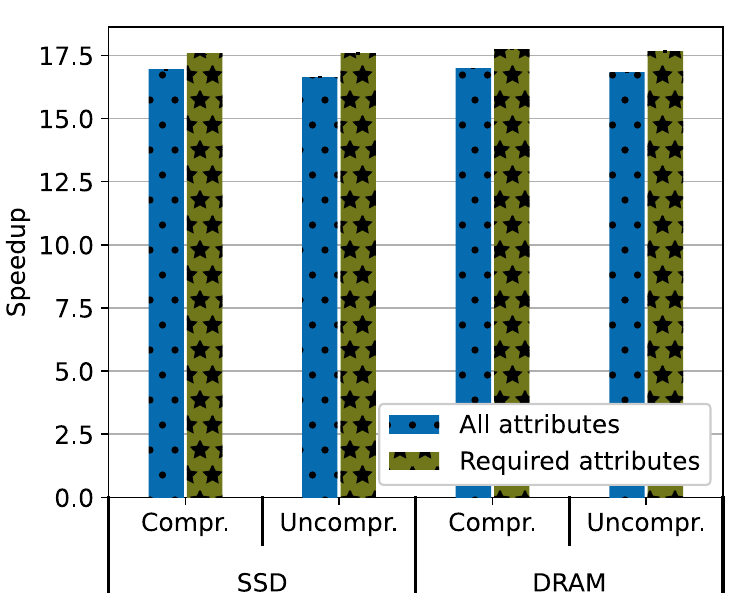}
   \caption{Cell clustering}
 \end{subfigure}
 \begin{subfigure}{\linewidth}
   \centering
   \includegraphics[width=0.56\textwidth]{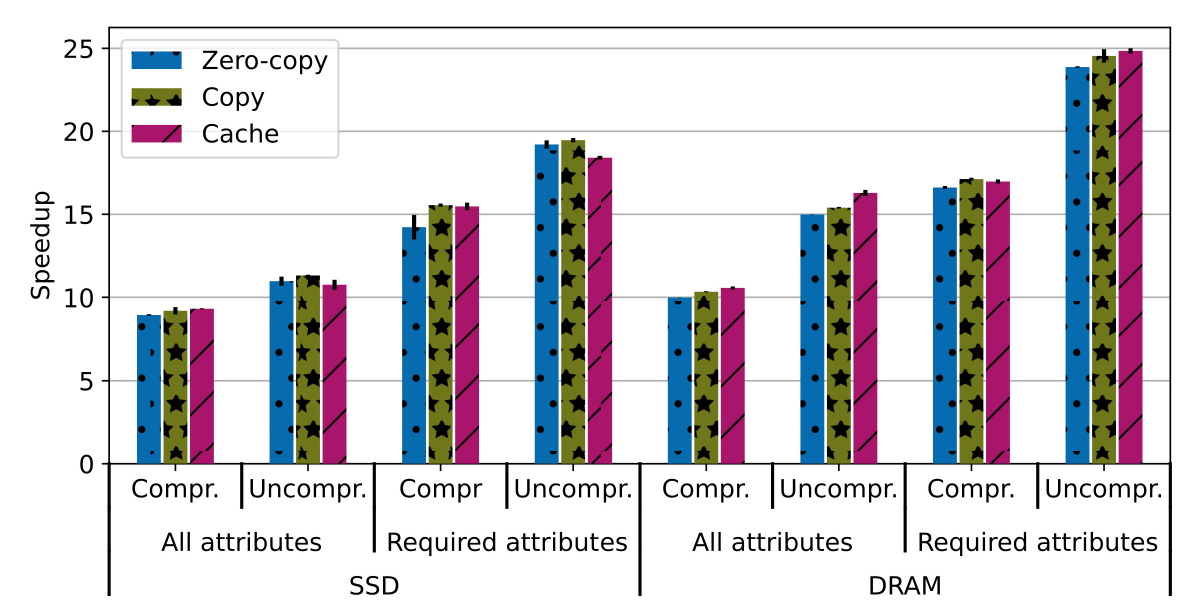}
   \includegraphics[width=0.34\textwidth]{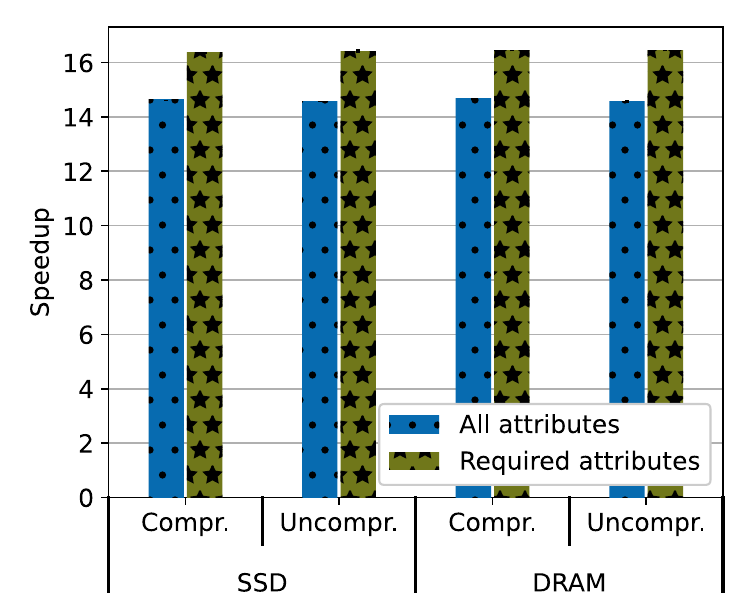}
   \caption{Neuroscience use case}
 \end{subfigure}
 \caption{Visualization performance analysis. Left column: Speedups of the visualization operation using different parameters. Right column: Speedup of the rendering stage.}
 \label{fig:visualization}
\end{figure}

Figure~\ref{fig:visualization} shows the speedups for the visualization
  operation for the cell clustering and neuroscience use case.
The baseline does not use the parallel file exporter and compression, and
  exports all agent attributes.
The left column shows the file export and the right column the rendering
  results.

We make the following observations.
The parallel file writers speedup the export between \result{9$\times$ and
	  19$\times$}.
Although compression reduces the file size between \result{1.62$\times$ }and
  \result{3.7$\times$} the costs outweigh the benefits on our test system.
This result might change for slower hard drives.
Exporting required agent attributes on-demand reduces the file size by
  \result{2.47$\times$ and 2.67$\times$}.
The caching data access mode performs best for the in-memory file system and
  the cell clustering benchmark, while the differences for the neuroscience use
  case are minimal.
The differences depend on the memory access pattern and computational intensity
  of the workload.
Thus, bigger differences might be observed if this mechanism is used to
  integrate other third-party applications.
As expected writing files to the in-memory file system is faster due to the
  higher bandwidth compared to the SSD.

In the rendering stage, no significant difference between compressed and
  uncompressed files, and DRAM and SSD could be observed.
Also the difference between files that contain all agent attributes and files
  that contain the required ones can be observed.
We attribute this effect to the columnar storage of the file format.

\subsection{Alternative Execution Modes}
\label{sec:eval:alternative-execution-modes}

\begin{figure}[h!]
  \centering
  \begin{subfigure}{0.49\linewidth}
    \includegraphics[width=\textwidth]{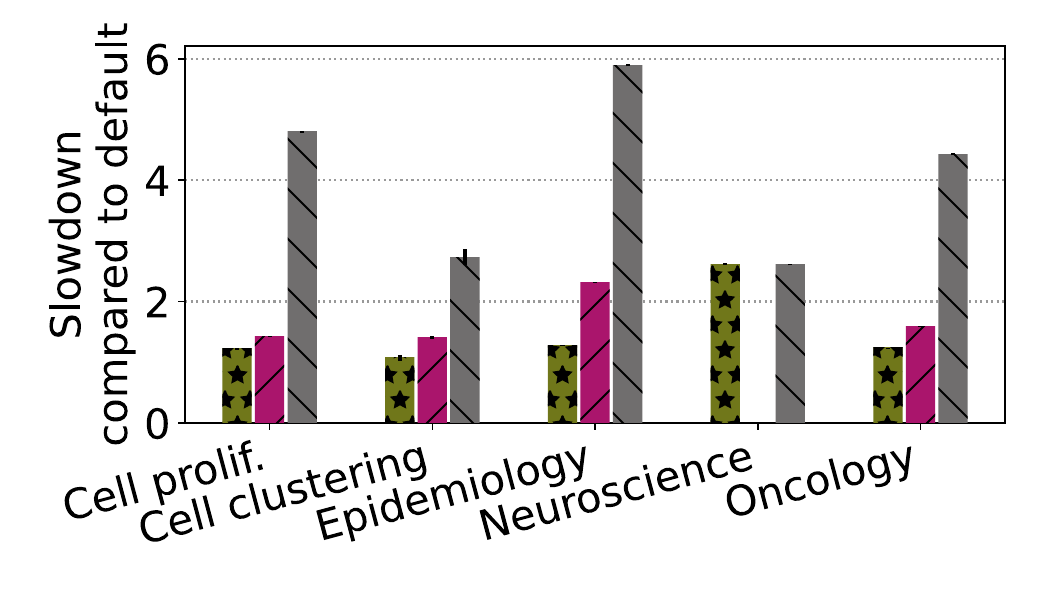}
\end{subfigure}
  \begin{subfigure}{0.49\linewidth}
    \includegraphics[width=\textwidth]{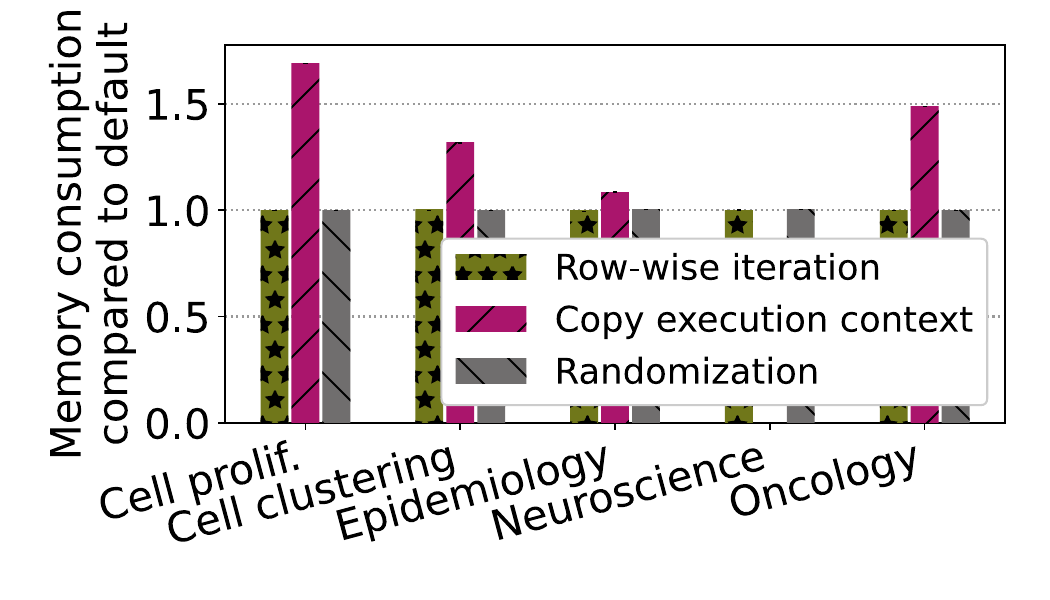}
\end{subfigure}
  \caption{Comparison of alternative execution modes compared to \bdm{}'s default settings (left: slowdown, right: memory consumption). The legend is shared between the plots.
  }
  \label{fig:alternative-exec-modes}
\end{figure}

This section evaluates the slowdown and memory consumption of different
  execution modes presented in Section~\ref{sec:alternative-execution-modes}
  compared to \bdm{}'s default settings.
These results are shown in Figure~\ref{fig:alternative-exec-modes}.
The copy execution context does not support the modification of neighbors yet.
Therefore, this comparison is missing for the neuroscience use case.

As expected, row-wise iteration and randomization have the same memory
  consumption as \bdm{} in its default setting.
Up to \result{1.67$\times$} more memory is needed for the copy execution
  context to store the temporary agents.
Row-wise execution is up to \result{2.67$\times$} slower (median:
  \result{1.27$\times$}) due to poor temporal cache utilization and increased
  synchronization overhead.
Locks must be acquired and released for each operation instead of once.
The copy execution context slows down execution by \result{1.49$\times$}
  (median) because every agent is duplicated before it is processed.
Randomization counteracts the memory layout improvements and hence degrades
  median performance by \result{3.97$\times$}.

\section{Conclusion and Future Work}

This chapter presents a novel agent-based simulation engine optimized for high
  performance and scalability.
\bdm{} enables not only larger-scale simulations, but also helps researchers of
  small scale studies with accelerated parameter space exploration, and faster
  iterative development.

We identify general agent-based performance challenges and provide six
  solutions to maximize parallelization, reduce memory access latency and data
  transfers, and avoid unnecessary work.
These solutions are transferable and can be used to accelerate other
  agent-based simulation tools.

We present a comprehensive performance analysis to provide insights into the
  agent-based workload and to give our users a better understanding of \bdm{}'s
  capabilities.
We find that on our system, the presented optimizations improve performance up
  to \result{524$\times$ (median 159$\times$)} and allow \bdm{} to scale to 72
  physical processor cores with a parallel efficiency of \result{91.7\%}.
A comparison with state-of-the-art tools shows that \bdm{} is up to three
  orders of magnitude faster.
These performance characteristics enable simulations with billions of agents,
  as demonstrated in our analysis.

Our performance optimizations, which are effective on machines with one or more NUMA domains,  are an important stepping stone towards a distributed simulation engine with a 
hybrid MPI/OpenMP design. 
Ongoing work focuses on realizing this distributed simulation engine capable of dividing
  the computation among multiple nodes to push the boundaries of agent-based
  simulation even further.

\clearpage{}

\cleardoublepage \clearpage{}\chapter{\ta{}: A Distributed Simulation Engine for Simulating Half a Trillion Agents}
\label{ch:dse}

\section{Introduction}

	Agent-based modeling is a bottom-up simulation method to study complex systems.
	Early agent-based models, dating back as far as 1971, studied segregation in
	  cities \cite{schelling_segration_1971}, flock of birds
	  \cite{reynolds_flocks_1987}, and complex social phenomena
	  \cite{epstein1996growing}.
	These models follow the same three-step structure, although the studied systems differ
	  widely.
	First, the researcher has to define what the abstract term agent should
	  represent.
	In the flock model from Craig Reynolds, an agent represents a bird.
	The bird has a current position, and a velocity vector.
	Second, the researcher has to define the interactions between agents.
	For example, birds avoid collision, align their heading with neighbors, and try
	  to stay close to each other.
	Third, the researcher has to define the starting condition of the simulation.
	How should the birds be distributed in space in the beginning?
	What are their values for the initial velocity, and parameter values for the
	  three behaviors?
	Once all these decisions are made, the model is given to the simulation engine,
	  which executes it for a number of iterations, or until a specific condition is
	  met.

Agent-based models exhibit two characteristics: local interaction and emergent
  behavior.
First, agents only interact with their local environment and do \emph{not} have
  knowledge about agents that are ``far'' away.
This characteristic makes a distributed execution feasible via spatial
  partitioning of the agents, which results in data exchanges across partitions
  at the bordering regions.
Second, agent-based models show emergent behavior or in other words they
  demonstrate that: ``the whole is more than the some of its parts''.
To come back to the flock of birds example, the swarm dynamics were not
  programmed into the model, but arose solely through the interactions of the
  birds following the three behaviors.

Since these early success stories, agent-based models have been used in
  numerous domains including biology \cite{\abmBiology}, medicine
  \cite{\abmMedicine},
  epidemiology \cite{\abmEpidemiology}, finance
  \cite{\abmFinanceAndEconomics}, policy making \cite{farmer2009} and many more
  \cite{macal_introductory_2014,\abmOther}.
These systems often comprise a very large number of agents.
Especially biomedical systems (the human cortex consists of approximately 86
  billion neurons \cite{azevedo_equal_2009}) or country-scale epidemiological
  simulations \cite{hesam_2024}.

Enabling simulations on these scales requires an efficient, high-performance,
  and scalable simulation platform.
\ifthesis
Chapter~\ref{ch:platform} and \ref{ch:engine}, have
  demonstrated that our simulation platform \bdm{} outperforms state-of-the-art
  simulators and is capable of simulating 1.7 billion agents on a single server.
\fi
However, \bdm{} only leverages shared-memory parallelism with OpenMP
  \cite{openmp} and is therefore limited by the compute resources of \emph{one}
  server.
More precisely, researchers that use \bdm{} are facing three critical restrictions:
\begin{itemize}
  \item Limited simulation size and complexity. As demonstrated in \ThesisPaper{Chapter~\ref{ch:engine}}{\cite{breitwieser_biodynamo_2023}} 
    a server with 1TB of main memory---significantly larger than typical memory capacities found in today's data centers---cannot support more than two billion agents.
  \item It is not feasible to integrate with third party software that is parallelized predominantely with MPI, because the resulting performance would be prohibitively slow. 
    Examples include the lattice Boltzmann solver OpenLB \cite{krauseOpenLBOpenSource2021} and the visualization tool ParaView \cite{paraview}, particularly the in-situ visualization mode \cite{paraview-insitu} that eliminates the need to export files to disk.
\item Limited hardware flexibility.
    Connecting less powerful hardware together may be more cost-effective than relying on a high-end server for large-scale simulations that surpass the capabilities of typical commodity servers.
\end{itemize}

Our goals are two fold.

\paragraph{Goal 1:}
We address the afore listed performance restrictions by presenting \ta{} a novel
  distributed simulation engine that is able to execute a \emph{single}
  simulation on \emph{multiple} servers by dividing the simulation space.
Distributed execution requires the exchange of agent information between
  servers, to obtain the local environment for an agent update, or migrate agents
  to a new server.
These exchanges comprise a serialization stage (packing
  agents into a contiguous buffer) and a transfer stage, which incur high performance and energy overheads.
To alleviate such overheads, we address both stages
  with the following key improvements.
First, we introduce a serialization mechanism for \ta{}, tailored to the needs of agent-based simulations.
Our serialization mechanism allows
  using the objects directly from the received buffer.
Second, we extend the serialization mechanism with delta encoding and
  compression to exploit the iterative nature of agent-based simulations to
  minimize the required data transfers.

\paragraph{Goal 2:}
Second, analogous to the OpenMP parallelization \ThesisPaper{presented in Chapter~\ref{ch:platform}}{by Breitwieser \etal{}
  \cite{breitwieser-bdm}}, we incorporate the distribution into the
  platform, such that the additional parallelism is mostly hidden from the user.
Thus, we support seamless model execution starting from laptops,
  workstations, and high-end servers \cite{breitwieser_biodynamo_2023} up to
  supercomputers and clouds without the need to modify the simulation code, 
  for a wide range of simulations.

The main contributions of this \ThesisPaper{chapter}{paper} are:

\begin{itemize}
  \item We present a novel distributed simulation engine called \ta{} capable of simulating \ESMaxAgentsExact{} agents and scaling
to \ESNumCPUs{} CPU cores.
  \item We show that it is possible to make scale-out agent-based simulations practical by alleviating their data movement overheads 
        across nodes via tailored serialization and delta encoding based compression.
	\item We present a novel serialization method for the agent-based use case, which significantly reduces the time spent on packing and unpacking agents during agent migrations and aura exchanges.
	      We observe a median speedup of \result{110}$\times$ for serialization and
	        \result{37}$\times$ for deserialization.
	\item We extend our serialization method to support delta encoding to reduce the amount of data that has to be transferred between servers, which reduces the message size by up to	\result{3.5}$\times$.

\end{itemize}

\section{\ta{}'s Distributed Simulation Engine}
\label{sec:design-overview}

This section provides an overview of the required steps in distributed
  execution (Section~\ref{sec:design:distribution}), details the \ta{} IO
  serialization mechanism (Section~\ref{sec:design:serialization}), presents our
  delta encoding scheme to reduce the amount of data that needs to be transferred
  (Section~\ref{sec:design:delta-encoding}), explains implementation details
  (Section~\ref{sec:design:implementation}), and further improvements over \bdm{}
  (Section~\ref{sec:design:smpbdm}).

\subsection{Distribution Overview}
\label{sec:design:distribution}

\begin{figure}[bth]
	\includegraphics[width=0.7\linewidth]{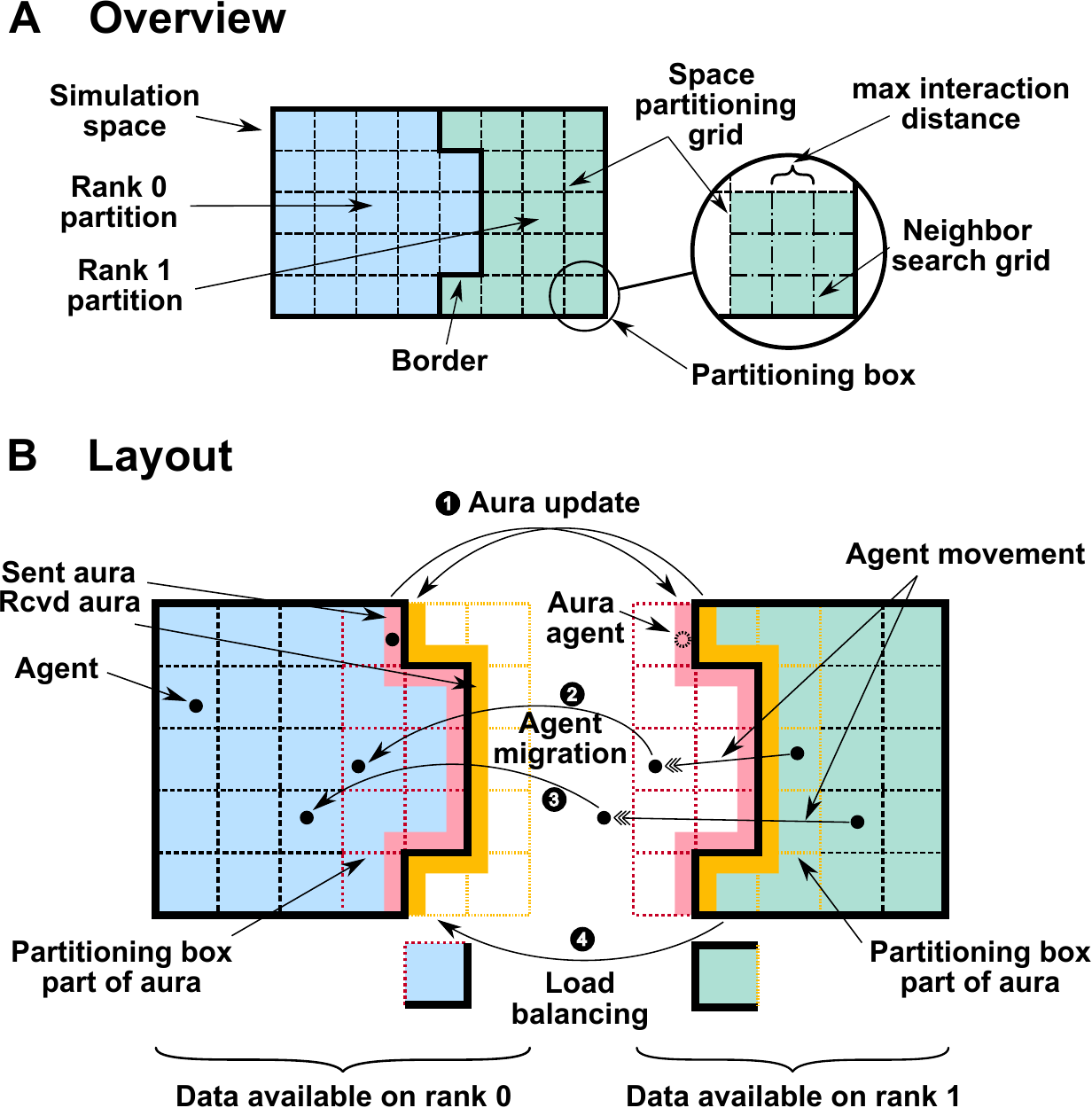}
	\caption{Distributed execution overview}
	\label{fig:design:overview}
\end{figure}

Figure~\ref{fig:design:overview} shows an overview of the steps involved in the
  distributed execution of an agent-based simulation.
The figure is simplified for a 2D simulation, but also applies for simulations
  in 3D.
Figure~\ref{fig:design:overview}A shows a simulation that is executed on two processes 
  (ranks in MPI terminology \cite{mpi}) that do \emph{not}
  share their memory space.
We apply a partitioning grid (consisting of partitioning boxes) on the simulation space to divide 
  the space into mutually exclusive volumes corresponding to the number of
  ranks.
It follows that each rank is authoritative of the assigned volume (or area
  in 2D) and the agents inside it.
Figure~\ref{fig:design:overview}B shows these regions in blue and green together with 
  the interactions between the two ranks at the bordering regions.
The specifics of these interactions, which happen in each simulation iteration are outlined below.

\paragraph{Aura Update.}
In this section, we focus on how agents close to the bordering region between two 
  ranks interact with their local environment, which, for our purposes, consists 
  only of other agents. 
In simulations conducted in Euclidean space, the local environment is defined by a 
  radius centered around each agent. 
The modeler sets the radius value at the beginning of the simulation. 

For agents located near the boundary between ranks, part of their local environment 
  may reside on a different rank. 
Therefore, agents situated near these boundaries must have their data transmitted to neighboring 
  ranks to enable the update to the next simulation iteration. 
Figure~\ref{fig:design:overview}B illustrates the bordering region sent from rank~0 to 
  rank~1 in red and yellow for the opposite direction. 
These areas are commonly referred to in the literature as aura, halo, or ghost regions. 
Consequently, these exchanges of data are known as \emph{aura updates}.

It is important to note that the partitioning grid's boxes can be larger than the maximum 
  interaction distance of the agents. 
The zoomed-out detail in Figure~\ref{fig:design:overview}A shows that, in this specific example, 
  the partitioning boxes are three times larger than the maximum interaction distance. 
Consequently, regions outside this interaction distance do \emph{not} need to be transferred. 
Therefore, the aura regions shown in red and yellow in Figure~\ref{fig:design:overview}B are 
  narrower than the partitioning box.

\paragraph{Agent Migration.}
Agents that change their position might also move out of the locally-owned
  simulation space and must therefore be moved to the rank which is authoritative for 
  the agent's new position.
For this scenario we have to distinguish two cases.
First, the current rank can determine the destination rank itself \circled{2}
  because the new position lies inside a partitioning box that is locally
  available.
Second, if the agent lies outside all locally-available partitioning boxes, a
  collective lookup stage is necessary to determine the rank which is
  authoritative for the position \circled{3}.

\paragraph{Load Balancing.}
The current space partitioning might be adjusted during the simulation
  \circled{4} to avoid load imbalances between MPI ranks and a therefore
  suboptimal resource utilization.

\subsection{Serialization}
\label{sec:design:serialization}

For all required steps presented in Section~\ref{sec:design:distribution}
  (agent migration, aura updates, and load balancing), agents are moved or copied
  to other ranks.
Serialization is necessary to pack agents and their attributes to a contiguous
  buffer which can then be sent with MPI.

We initially utilized ROOT I/O, because it is already used in \bdm{}
  \cite{breitwieser-bdm} for backing up and restoring whole simulations.
ROOT serves as the main data analysis framework for high-energy physics
  \cite{brun_root_1997}.
ROOT's serialization (called ROOT I/O) is used at CERN to store more than one
  exabyte scientific data from the large hadron collider experiments
  \cite{root-files}.
According to Blomer \cite{blomer_quantitative_2018}, ROOT I/O outperforms other
  serialization frameworks like Protobuf \cite{protobuf}, HDF5 \cite{hdf5},
  Parquet \cite{parquet}, and Avro \cite{avro}.

However, we observed that agent serialization was a significant performance
  bottleneck and made the following four observations.

First, ROOT I/O keeps track of already seen pointers during serialization.
Thus, ROOT can skip over repeated occurrences of the same pointer and ensure
  that upon deserialization all pointers point to the same instance.
\ta{} does not need this feature, because \bdm{} does not allow multiple agents having pointers to the same object.
In the distributed setting, this would lead to further challenges, because the
  pointed object could be required on more than one rank and updates to this
  object would need to be kept in sync.
We enable pointers to other agents with an indirection implemented in the smart-pointer class \texttt{AgentPointer}.
An \texttt{AgentPointer} stores the unique agent identifier of the pointed agent and obtains the raw pointer from a map stored in the \texttt{ResourceManager}.
Therefore, the serialization of \texttt{AgentPointers} reduces to the serialization of the unique agent identifiers.
The current \ta{} version only supports \texttt{const AgentPointers} to avoid merging changes from multiple ranks.

Second, we observe that deserialization takes a significant amount of time. 
Therefore, we analyze the design of Google's FlatBuffer serialization library which provides 
  ``access to serialized data without parsing/unpacking'' \cite{flatbuffers}.
FlatBuffer serialization library even supports mutability, but is limited to changing the value of
  an attribute.
For example, adding or removing an element from a vector is not supported.

Third, the current goal is to execute \ta{} on supercomputers and clouds on
  machines with the same endianness.
Thus, if the condition holds, there should \emph{not} be any CPU cycles spent on
  endian conversions.

Fourth, ROOT I/O ensures that data that was captured and stored decades ago,
  can still be read and analyzed today.
Schema evolution is therefore needed to deal with changes that occurred since
  the data was written.
In \ta{}, the transferred data exists only during the execution of the
  simulation, sometimes only for one iteration.
In this time frame, the schema, i.e., the agent classes with their attributes do
  not change.
Therefore, \ta{} skips schema evolution to reduce the performance and energy overheads.

Based on these four observations, we design \ta{} to avoid pointer deduplication, 
  deserialization, endianness conversions, and schema evolution to alleviate the 
  performance and energy overheads of agent serialization.

\subsubsection{\ta{}
IO} \label{sec:design:bdmio}

  Following one of the design
  principles of C++ ``What you don't use you don't pay for''
  \cite{stroustrup1994design}, we create a serialization method that
  take these four observations into account and thus avoids spending compute
  cycles on unnecessary steps.

\begin{figure}[bth]
	\includegraphics[width=.6\linewidth]{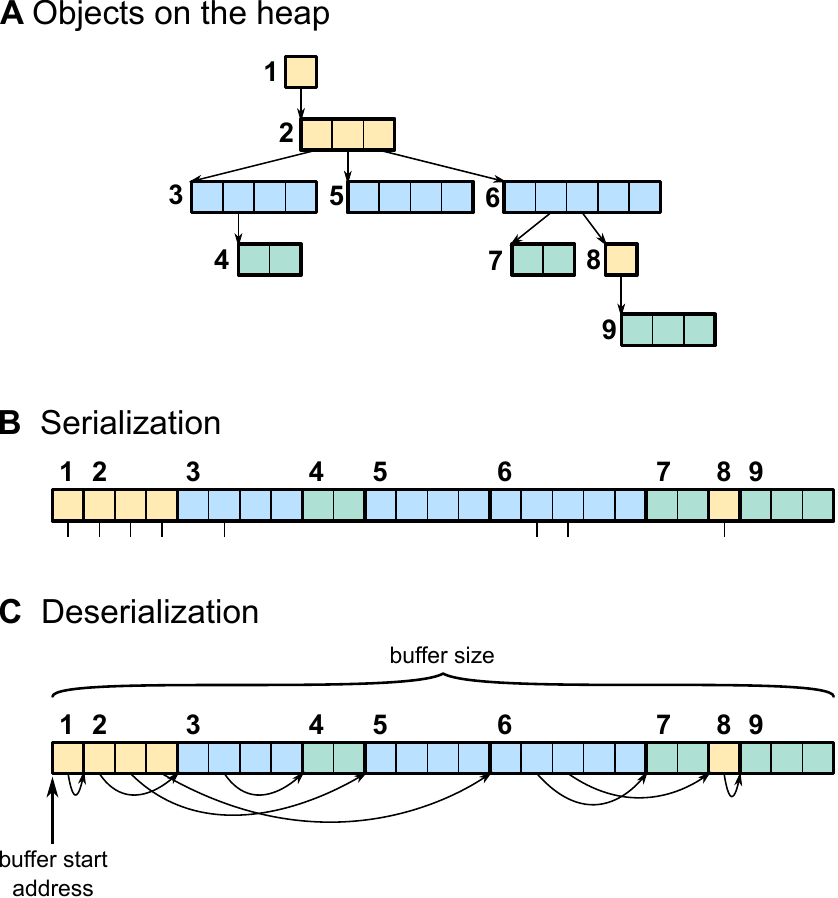}
	\caption{\ta{} serialization mechanism}
	\label{fig:design:serialization}
\end{figure}

Figure~\ref{fig:design:serialization}A shows an abstract representation of the 
  objects in memory.
Starting from a root object (e.g., a container of agent pointers), a tree
  structure can be observed, because multiple pointers to the same memory block are disallowed.
The nodes of the tree are formed by contiguous memory blocks that were
  allocated on the heap.
The given example could be interpreted as having a pointer to a container with three different agents, that
  have zero, one, and two attached behaviors.

\paragraph{Serialization.}
\label{sec:design:bdmio-serialization}
This structure can be serialized by an in-order tree traversal (B).
Each tree node (i.e., memory block) is copied to the serialized buffer.
Fields inside each memory block that point to other memory blocks are labeled
  as such, but point to the invalid address $Ox1$.
Furthermore, the virtual table pointer of polymorphic classes is replaced by a
  unique identifier of the most derived class.
This step is necessary because we cannot guarantee that the virtual table
  pointers will be the same on all ranks.

\paragraph{Deserialization.}
From the communication subsystem, the deserialization implementation receives a
  buffer with a starting address, length, and type of the root object.
Deserialization consists out of four steps that can be performed in a single
  traversal of the buffer.
First, we traverse the tree from the beginning of the buffer and restore the
  virtual table pointer for polymorphic objects.
The offsets of the memory blocks are determined by their type and retrieved from
  the compiler.
Second, if a pointer is encountered, we set it to the next memory block in the
  buffer.
Third, as we traverse, we count the number of memory blocks, which will be
  required for memory deallocation (see next paragraph).
Fourth, reinterpret the buffer's starting address as a pointer to the root
  object and return it to the caller.
No other memory reads or writes are taking place.
Furthermore, there are no calls to allocate memory besides the single
  contiguous receive buffer that holds the data.

\paragraph{Mutability.}
By returning the root object pointer to the higher-level code, we create the
  illusion that all contained memory blocks have been allocated separately on the
  heap.
Therefore, higher-level code is not aware that these objects were deserialized
  using \ta{} IO and can change them in any way.
This includes setting value of attributes, but extends to, for example, adding
  elements to containers, even if there is not sufficient space in the buffer.
In this scenario, the vector implementation notices the capacity is reached,
  allocates a new memory block on the heap (separately from the buffer) and
  deallocates the obsolete memory block inside the deserialized buffer.

\paragraph{Deallocation.}
As we have seen in the vector example above, higher-level code will at some
  point try to deallocate memory blocks by calling delete.
These delete calls, however, would crash the memory allocator, due to the
  missing corresponding new call.
To maintain the illusion, we intercept all calls to delete and filter those
  that fall into the memory range of the deserialized buffer.
If the number of expected delete calls (determined during deserialization)
  matches the intercepted delete calls, we deallocate the whole buffer and remove
  the filter rule.

The disadvantage that memory is leaked if not all memory blocks are freed in a
  deserialized buffer can be solved for \ta{}.
First, the aura region is completely rebuilt in each iteration, which means
  that the previous aura information is completely destroyed.
Second, for agent migrations and load balancing data, we rely on the periodic
  agent sorting mechanism in \bdm{}.
Agent sorting changes the memory location of agents to improve the cache hit
  rate.
During this process all agents are copied to a new location and the old ones
  are deallocated.
Figure~\ref{fig:eval:serialization} shows that the memory consumption does not 
  increase by using the \ta{} serialization mechanism.

\subsection{Data Transfer Minimization}
\label{sec:design:delta-encoding}

\begin{figure}[tbh]
	\includegraphics[width=.32\linewidth]{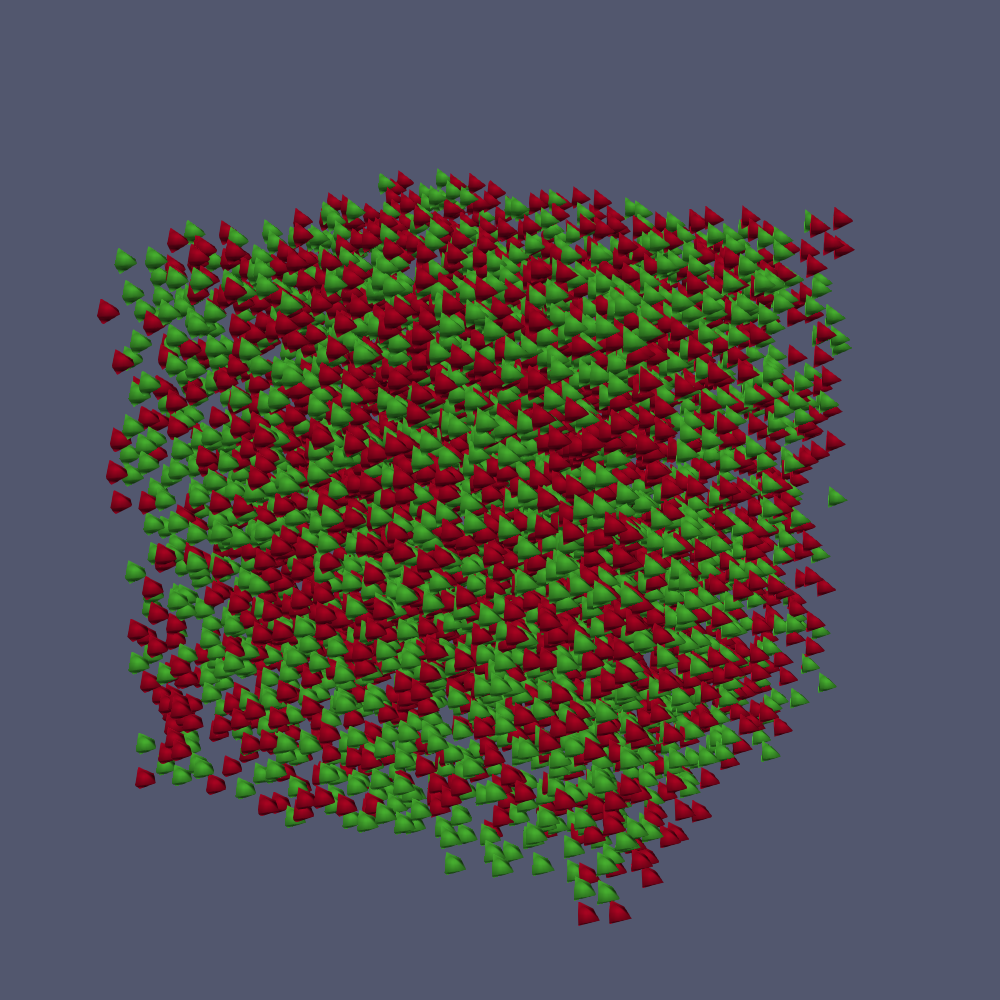}
	\includegraphics[width=.32\linewidth]{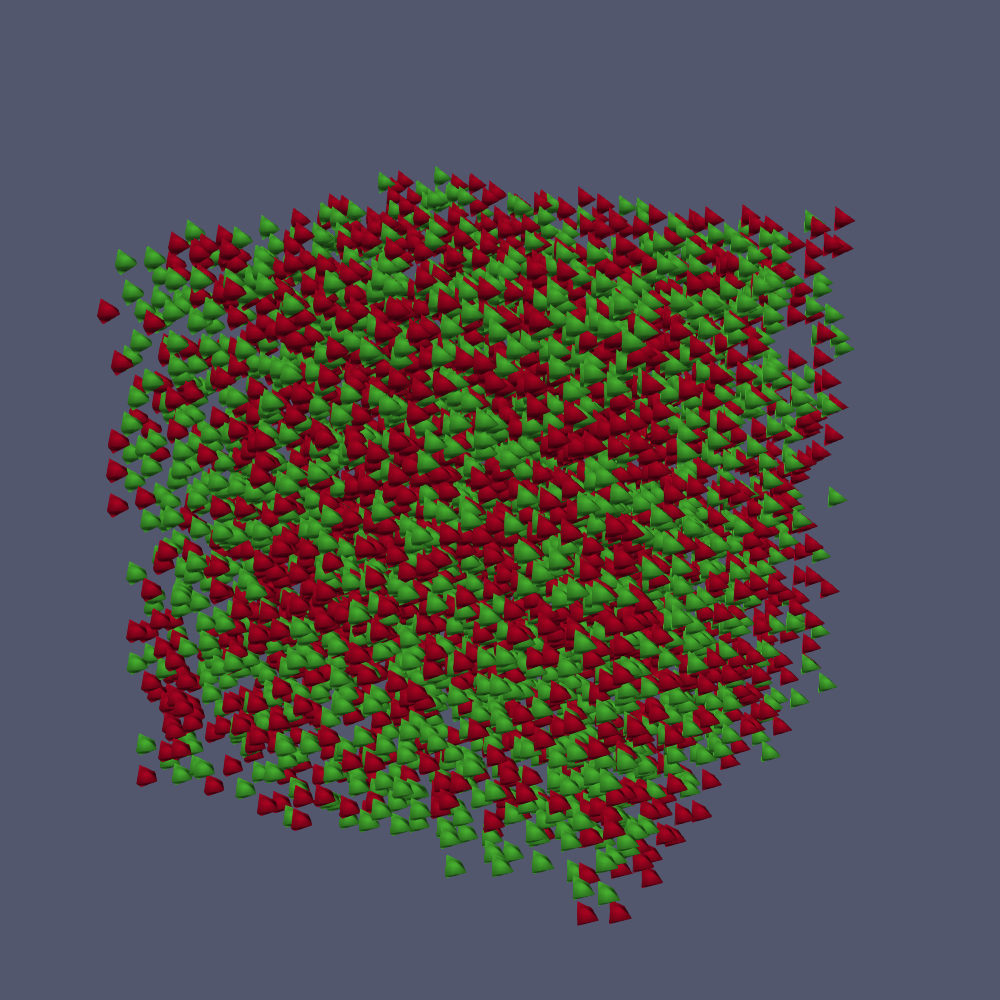}
	\includegraphics[width=.32\linewidth]{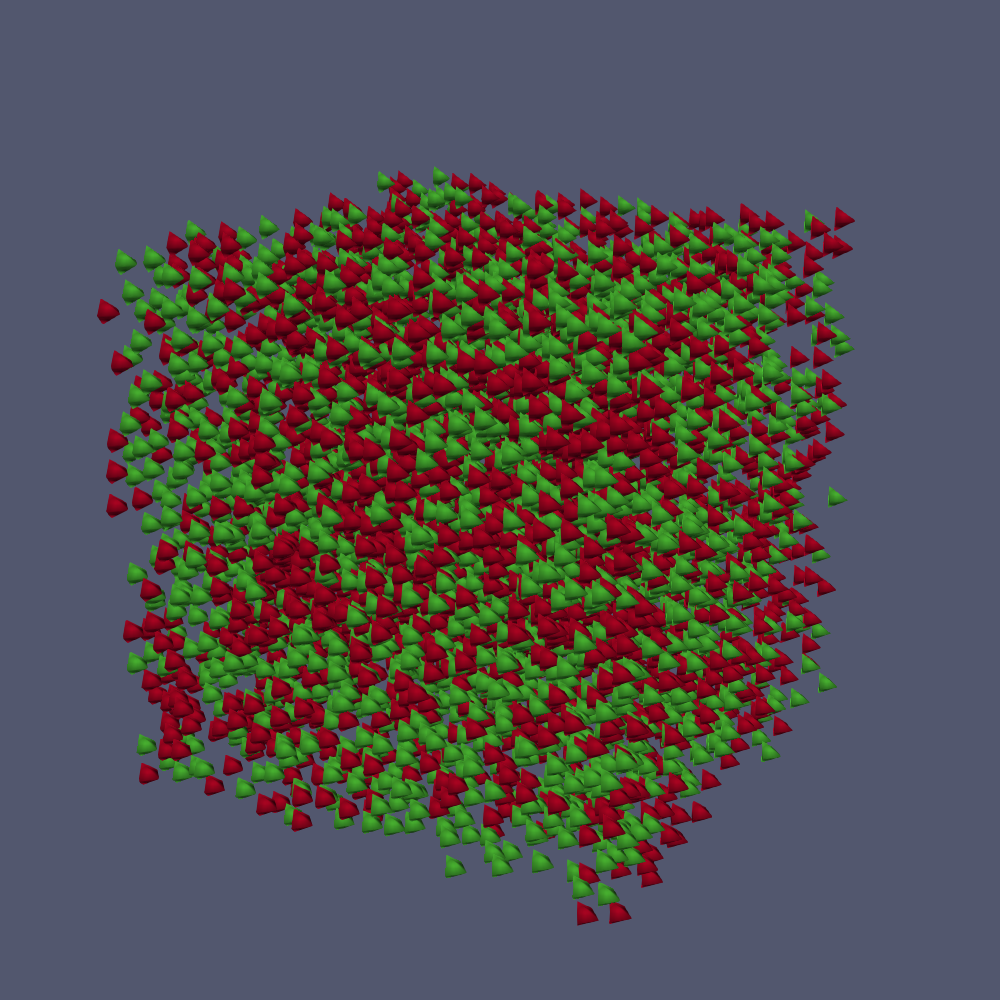}
	\caption{Cell clustering visualization for the first three iterations}
	\label{fig:design:successive-iterations}
\end{figure}

\begin{figure}[tbh]
	\includegraphics[width=0.8\linewidth]{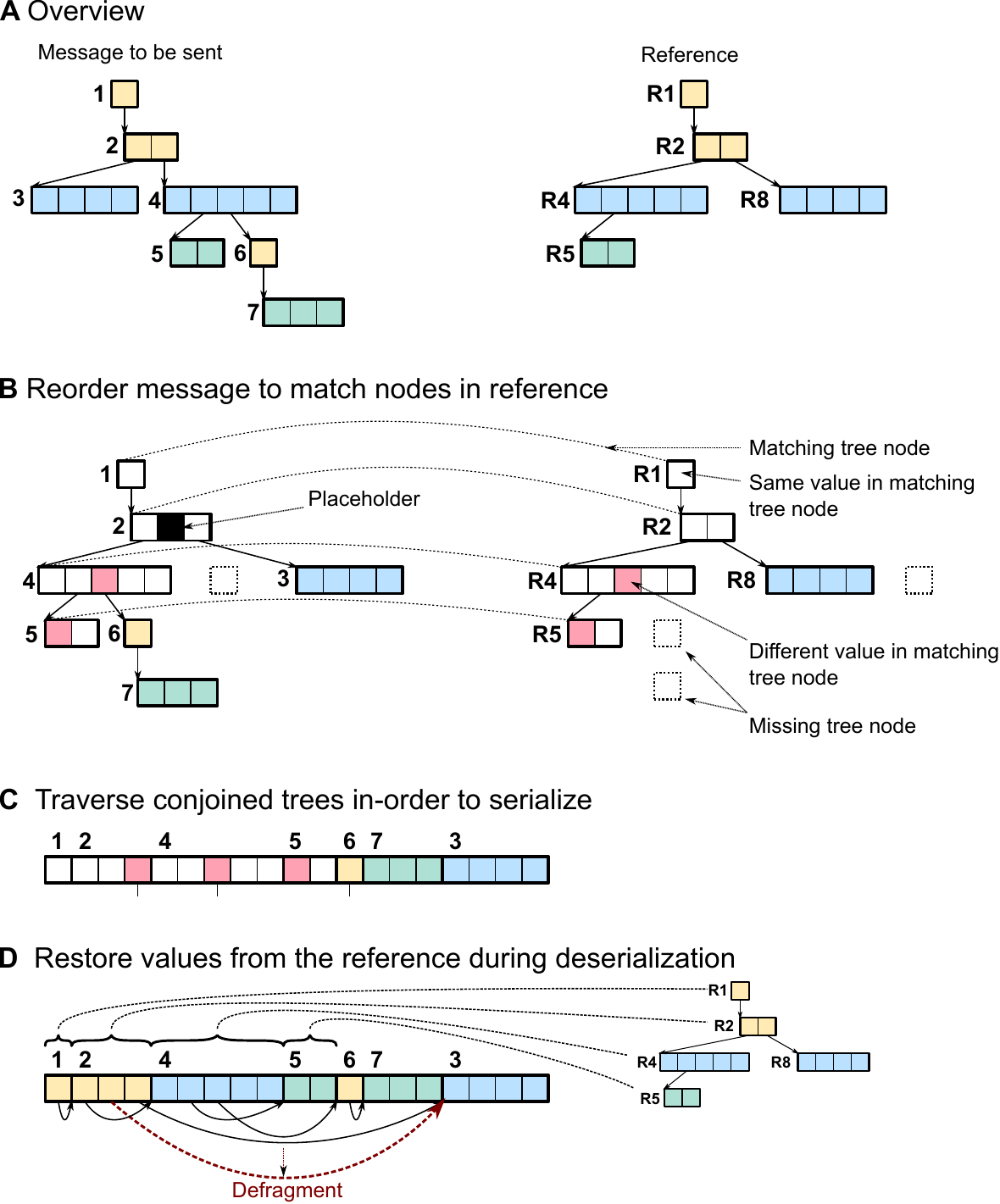}
	\caption{\ta{} delta compression}
	\label{fig:design:delta-compression}
\end{figure}

Agent-based modeling is an iterative method.
Figure~\ref{fig:design:successive-iterations} shows three successive iterations
  of the cell clustering simulation.
We can observe that the cell position change only gradually between iterations
  while the cell type and diameter do not change at all.
We leverage this observation to reduce the amount of data that must be
  transferred between MPI ranks to update the aura region by using delta
  encoding.
The implementation of delta encoding is built on top of the \ta{}
  serialization mechanism (see Figure~\ref{fig:design:delta-compression}).

Each sender and receiver pair stores the same reference.
The sender calculates the difference between the message and the reference,
  compresses it and sends it to the receiver.
On the receiving side, the process is reversed to restore the original message.
The receiver uncompresses the buffer and restores the original message by
  inverting the difference operation using the received buffer and the stored
  reference as operands.
At regular intervals, sender and receiver update their reference.

More precisely, the sender reorders the message at the agent pointer level (B).
Agents which exist both in the message and the reference are moved to the same
  position that the agent has in the reference.
Agents that exist in the reference and not in the message are indicated by an
  empty placeholder, a value that cannot occur at the same tree depth in the
  message.
In the \ta{} case, this is simply a null pointer.
Lastly, agents that exist in the message and not in the reference are appended
  at the very end.
Reordering agents does not affect the correctness of a \ta{} simulation.

After the matching stage (B), the \ta{} serialization mechanism (C) traverses
  the two conjoined trees, in the same way as described in
  Section~\ref{sec:design:bdmio-serialization}.
The only difference is that if a match is found in the reference, the
  difference between the two is written to the result buffer.
Since the message is reordered at the sender, we do \emph{not} need to send additional
  information about the agent order.

During deserialization (D), we use the data stored in the reference to restore
  the original message.
Lastly, we defragment the message by removing any placeholders that were
  inserted during the matching stage (B).
Note that defragmentation will not restore the original agent ordering as shown
  in (A).
The defragmented message is then passed to higher-level code.

By using this mechanism, \ta{} reduces the data movement bottleneck and increases the efficiency 
  of distributed execution.

\subsection{Implementation Details}
\label{sec:design:implementation}

\subsubsection{Partitioning Grid}
One important data structure in the distributed simulation engine is the
  partitioning grid, which is responsible for the domain decomposition of the
  simulation space as shown in Figure~\ref{fig:design:overview}.
We use Sierra Toolkit (STK) \cite{sierra-toolkit}, because it is (i)
  an established tool maintained as part of Trilinos \cite{trilinos-website},
  (ii) well integrated with the load balancing framework Zoltan2 \cite{zoltan2},
  and (iii) has the ability to export the grid in exodus format \cite{exodus},
  which our main visualization framework ParaView \cite{paraview} can read.
However, since STK is a generic mesh library, capable of much more than
  processing rectilinear grids, these functionality comes with a performance
  penalty in terms of compute and required memory.
Furthermore, although STK uses Kokkos \cite{kokkos} for shared memory
  parallelism, we observe that if \ta{} is executed in MPI hybrid mode
  (MPI/OpenMP), most threads are idle during balancing and mesh modification
  calls into STK.
As an alternative, we also considered the parallel grid library
  \cite{honkonen_parallel_2013}, but experienced difficulties generating very
  large grid sizes.

To reduce the memory and compute footprint of STK's partitioning grid, we
  introduce a parameter to make the partitioning box length a multiple of the
  neighbor search grid (see Figure~\ref{fig:design:overview}).
The higher this factor is chosen the less memory and compute resources are
  needed for the space partitioning, but at the cost of increased granularity at
  which load balancing decisions can be made.

\subsubsection{Serialization}

The \ta{} IO mechanism (Section~\ref{sec:design:bdmio}) requires that
  (de)serialization methods (also known as \ta{} IO functions) are available for
  all types in the message.
\ta{} provides the necessary implementation for its internal classes, the
  standard template library (STL) classes, and important groups of types (e.g.,
  polymorphic types, pointers, and plain arrays).
Due to the design decision to opt for a very lightweight deserialization
  method, the implementation of the (de)serialization methods depends on its
  internal attributes (i.e., the concrete implementation), rather than the public
  interface. 
This design decision might increase the maintenance effort.
However, the \ta{} IO methods follow a regular structure and can be generated
  automatically in most of the cases.
This is especially true for all user-defined classes in this \ThesisPaper{chapter}{paper}.
Although the (de)serialization methods are currently written by hand, they can
  easily be integrated into \bdm{}'s existing code generation stages.
\bdm{} uses code generation during compile time, and during runtime using the
  C++ just-in-time compiler cling \cite{vasilev_cling_2012}.

Custom \ta{} IO functions are necessary if classes contain pointer attributes,
  whose memory is not owned.
An example could be an array-based stack implementation with an attribute that
  points to the current top of the stack.
To this end, \ta{} allows for custom implementations of the IO functions that
  will replace the automatically generated ones.

To intercept delete calls, we overwrite all global C++ delete operators and
  insert a call to the delete filter, which returns if the delete should be
  filtered out or executed.
This design choice causes some challenges in combination with performance or
  correctness tools (e.g., valgrind) that use \texttt{LD\_PRELOAD} to inject
  their own implementation of new and delete and prevent the \ta{} ones from
  being executed.
For valgrind, we solved this issue by patching its code.

\subsubsection{Communication}

For the majority of data transfers, we use non-blocking point to point
  communication (\texttt{MPI\_Isend}, \texttt{MPI\_Irecv}, and
  \texttt{MPI\_Probe}).
This choice allows for overlapping communication with computation to hide
  latency.
For regular communication patterns that occur between neighbors, we issue
  speculative receive requests right after the previous transfer finished, to
  avoid delay through late receivers.
If the neighbors of a rank change due to load balancing, obsolete speculative
  receives are cancelled.

Furthermore, we transmit large messages in smaller batches to reduce the memory
  needed for transmission buffers, compression, and serialization.

\subsubsection{Distributed Initialization}

Although the distributed simulation engine has the capability to migrate agents
  to any other MPI rank (Section~\ref{sec:design:distribution}), we try to avoid
  a costly mass migration of agents during the initialization stage, by trying to
  create agents on the authoritative rank.
This is straightforward for regular geometric shapes (e.g., agents created on a
  surface defined by a function), but we also address agent populations that are
  created using a uniform random number distribution within a specific space.
Each rank determines the fraction of the given target space with its
  authoritative volume, and adjusts the number of agents for this space and the
  bounds accordingly, if the number of agents for each node is sufficiently high.

\subsubsection{Load Balancing}

Load balancing has two goals.
First, partition the simulation space in a way that simulating one iteration
  takes the same amount of time on all ranks.
Second, the partitioning should minimize the distributed overheads, e.g., the
  number of aura agents that must be exchanged.

\ta{} provides two classes of load balancing methods to achieve these goals:
  global and diffusive.

The global balancing method is based on STK and Zoltan2.
We provide their functionality for \ta{} users and choose the recursive
  coordinate bisection (RCB) algorithm as default.
We apply a weight field on the partitioning grid and set the weight of each
  partitioning box based on the number of agents contained and scale it by the
  runtime of the last iteration.
Zoltan2 now partitions the space in a way that the sum of all owned partition
  boxes is distributed uniformly between all ranks.
This approach might lead to a new partitioning that differs substantially from the
  previous one, causing mass migrations of agents to their new authoritative
  rank.

Therefore, we also implement a diffusive approach in which neighboring ranks
  exchange partition boxes.
Ranks whose runtime exceed the local average, send boxes to neighbors that were
  faster than the local average.

\subsection{Improvements and Modifications to \bdm{}}
\label{sec:design:smpbdm}

This section describes the necessary modifications and improvements to the
  existing OpenMP parallelized \bdm{} version to enable distributed execution.

\paragraph{Parallelization Modes.}

By building upon and extending the shared-memory capabilities, we provide two
  distributed execution modes: MPI hybrid (MPI/OpenMP), and MPI only.
For the MPI hybrid mode, we launch one MPI rank for each NUMA domain on a
  compute node, while for MPI only we launch one rank for each CPU core.
On today's modern hardware with constantly increasing CPU core counts, the MPI
  hybrid mode can reduce the number of ranks currently by almost two orders of
  magnitude.
We therefore expect the MPI hybrid mode to be more efficient.
The MPI only mode, on the other hand, provides benefits when interfacing with
  third party applications that are only parallelized using MPI.
In the MPI hybrid mode, these applications, would leave all but one thread per
  rank idle.

Switching between parallelization modes does not require recompilation of
  \ta{}.

\paragraph{Unique Agent Identifiers.}
\label{sec:design:agent-identifiers}

\bdm{} uses unique agent identifiers to address agents, since the actual memory
  location might change due to agent sorting.
Agent sorting improves the performance by reordering agents in a way that
  agents that are close in 3D space are also close in memory (see
  \cite{breitwieser_biodynamo_2023}).
The identifier is comprised of two fields: $\langle index, reused\_counter
	  \rangle$ and has the following invariants: At any point in time, there is only
  one (active) agent in the simulation with the same index.
If an agent is removed from the simulation, this index will be reused, but to
  satisfy uniqueness the \texttt{reuse\_counter} is incremented.
This design allows the construction of a vector-based unordered map, where the
  first part of the identifier (the index) is used to index the vector.
This map allows for lock-less additions and removals to distinct map elements
.

However, this solution does not work without changes for distributed executions
  with agent migrations and aura updates, because the invariant that the $index$
  of the identifiers are almost contiguous does not hold anymore.
This would waste a considerable amount of memory in the vector-based unordered
  map.

Therefore, we rename the existing identifier to ``local identifier'' and
  introduce also a global identifier $\langle rank, counter \rangle$.
Rank is set to the rank where the agent was created, and ``counter'' is a
  strictly increasing number.
The global identifier of an agent is constant, but the agent might have various
  different local identifiers during the whole simulation.

We implement this change minimally invasive.
The translation between local and global identifier happens only during
  serialization if the agent is transferred to another rank, or written to disk
  as part of a backup or checkpoint.
Global identifiers are only generated on demand.
If there are no backups or checkpoints and the agent stays on the same rank
  throughout the simulation, the agent will only have a local identifier.

\paragraph{Incremental Updates to the Neighbor Search Grid (NSG).}
\ifthesis
Chapter~\ref{ch:engine} showed that our
  optimized uniform grid implementation performed best for the benchmarked
  simulations.
\fi
Updates to the NSG required a complete rebuild, which was adequate so far.
The distributed simulation engine, however, relies on the NSG not only to
  search for neighbors of an agent, but also to accurately determine the agents
  in a specific sub volume for agent migrations, aura updates, and load
  balancing.
Rebuilding the complete NSG after each of these steps would be prohibitively
  expensive.
Therefore, we adapted the implementation to allow for incremental changes,
  i.e., the addition, removal, and position update of single agents.

\paragraph{Modularity Improvements.}
Alongside high-performance, modularity is another key design aspect of \ta{}.
We have therefore made additional efforts in this direction during the
  development of \ta{} and made three main changes.
First, we introduce the \texttt{VisualizationProvider} interface to facilitate
  rendering of additional information besides agents and scalar/vector fields.
We use an implementation of this interface to render the partitioning grid as
  can be seen in Figures~\ref{fig:eval:validation}.
Second, we add the \texttt{SimulationSpace} interface to explicitly gather
  information about whole and local simulation space in one place.
Third, we refactor the code and introduce the
  \texttt{SpaceBoundaryCondition} interface and refactor the implementations
  for ``open'', ``closed'', and ``toroidal''.

\section{Evaluation}
\label{sec:evaluation}

\subsection{Benchmark Simulations}
\label{sec:benchmarks}

We use four simulations from \cite{breitwieser_biodynamo_2023, breitwieser-bdm}
  to evaluate the performance of the simulation engine: cell clustering, cell
  proliferation, and use cases in the domains of epidemiology, and oncology.
We did not include the neuroscience use case, since it requires the ability to
  modify neighbors, which is currently not implemented yet and chose only one of
  the two cell clustering implementations.
Table~1 in \cite{breitwieser_biodynamo_2023} shows that these simulations cover
  a broad spectrum of performance-related simulation characteristics.

\subsection{Experimental Setup and Reproducibility}
All tests were executed in a Singularity container with an Ubuntu 22.04 based
  image.
We used two systems to evaluate the distributed simulation engine.
First, we used the Dutch national supercomputer Snellius (genoa partition)
  where each node has two AMD Genoa 9654 processors with 96 physical CPU cores
  each, 384 GB total memory per node, and Infiniband interconnect (200 Gbps
  within a rack, 100 Gbps outside the rack) \cite{snellius_hardware}.
Second, we use a two node system (System B) where each node has four Intel Xeon
  E7-8890 v3 CPUs processors each with 72 physical CPU core, 504 and 1024 GB main
  memory, and Gigabit Ethernet interconnect.
We are therefore able to evaluate the performance with a commodity and high-end
  interconnect.

We use the term ``simulation runtime'' for the wall-clock time needed to
  simulate a number of iterations, which excludes the initialization step and
  tear down.

We will provide all code, the self-contained Singularity image, more detailed
  information on the hardware and software setup, and instructions to execute the
  benchmarks upon publication on Zenodo.

\subsection{Correctness}

\begin{figure}[bth]
  \includegraphics[width=.48\linewidth]{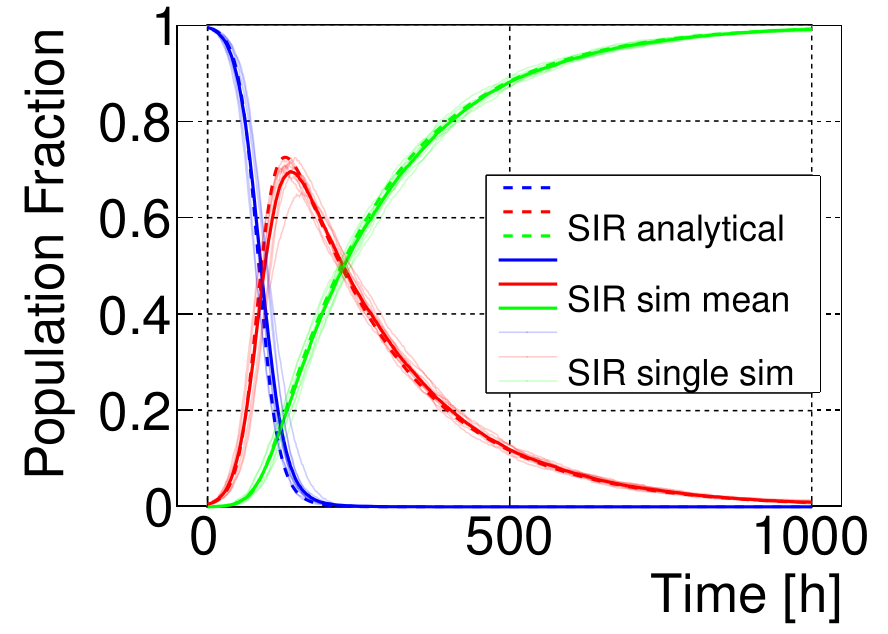}
  \includegraphics[width=.48\linewidth]{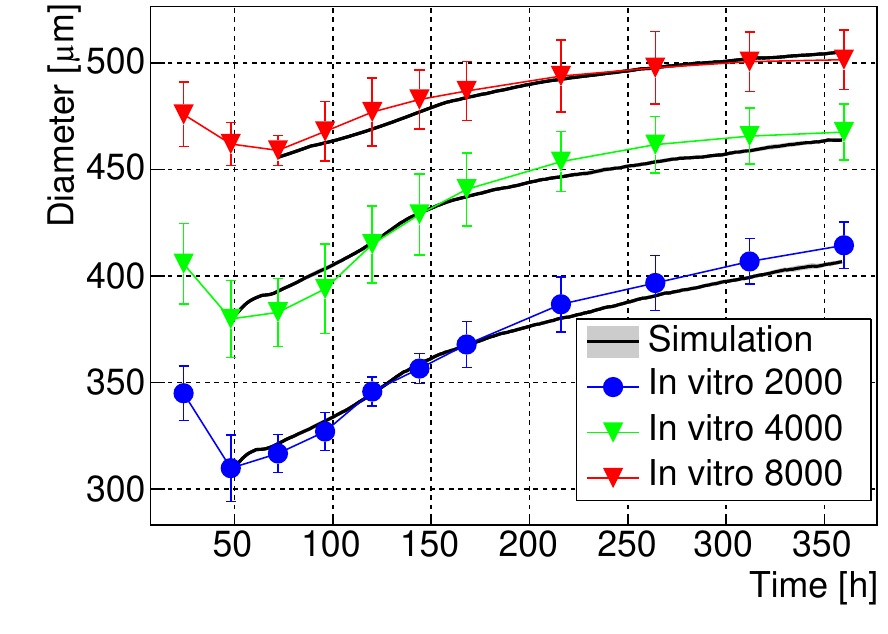}
  \includegraphics[width=.4\linewidth]{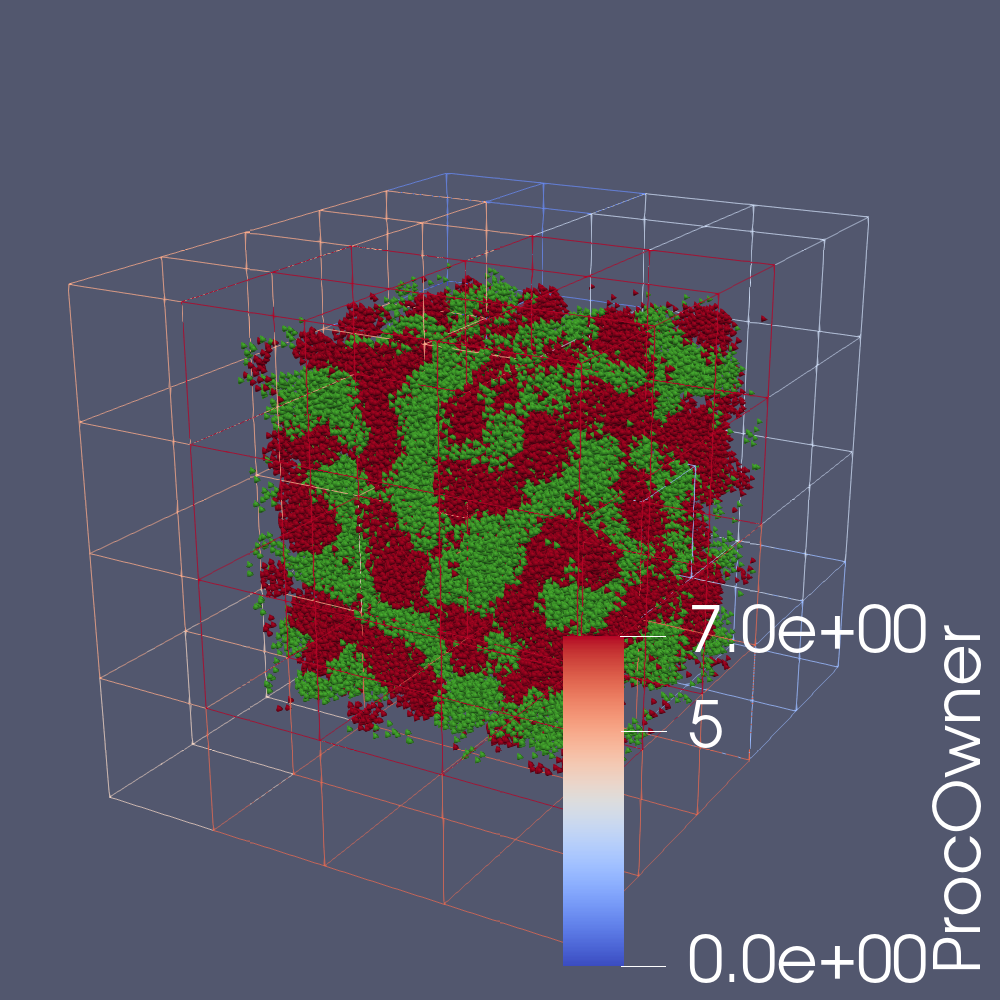}
  \caption{Result verification of \ta{}}
	\label{fig:eval:validation}
\end{figure}

To ensure the correctness of the distributed simulation engine, we added 180
  automated tests, and replicate the results from \cite{breitwieser-bdm,
	  breitwieser_biodynamo_2023}.
Figure~\ref{fig:eval:validation} shows the quantitative comparison of the
  simulation compared to analytical (epidemiology use case) or experimental data
  (oncology use case), and a qualitative comparison for the cell sorting
  simulation.
We can observe that \ta{} produces the same results as \bdm{}.

\subsection{Seamless Transition From a Laptop to a Supercomputer}
\label{sec:eval:laptop-to-sc}

User-friendliness is a key design aspect, alongside our focus on performance. Regarding distributed computing, the model definitions for the four benchmark simulations are completely transparent to the user. Only the evaluations for the epidemiology and oncology use cases require additional code for distributed execution.

The changes in the epidemiology simulation, are limited to \emph{two} lines of
  code.
To create the graph shown in Figure~\ref{fig:eval:validation}, the engine has
  to count the number of agents for the three groups (susceptible, infected, and
  recovered).
For distributed executions the rank local results have to be summed up to
  obtain the correct result.
To do that, we provide the function \texttt{SumOverAllRanks}, which hides the
  MPI related function call.
Furthermore, to save the generated plot to disk, we have to make sure that only
  one rank writes to the same file location.
This goal can be achieved by telling the other ranks to exit the function
  before the save function is executed.
We can achieve this with the preprocessor macro
  \texttt{IF\_NOT\_RANK0\_RETURN}.

Also the tumor spheroid simulation requires extra code to generate the result
  plot (Figure~\ref{fig:eval:validation}).
To accurately measure the tumor diameter in the simulation, we determine the
  volume of the convex hull, from which we can calculate the diameter by assuming
  a spherical shape.
\ifthesis
Chapter~\ref{ch:engine} uses libqhull \cite{libqhull}, a
  library which is not distributed.
\fi
Consequently, the simulation contains a couple lines of extra code to transmit
  agent positions to the master rank to perform the diameter calculation.
For simulations with a larger number of agents, we use a more approximate
  method, by determining the enclosing bounding box.
The approximate method is provided by \ta{} and is the same
  whether executed distributed or not.

To summarize, for many simulations our users do not need to know about
  distributed computing to scale out a simulation to tens of thousands of CPU
  cores, as demonstrated in this chapter.
Researchers can start the development on a laptop and seamlessly transition to
  more powerful hardware as the model size and complexity grows.

\subsection{Comparison with \bdm{}}

\begin{figure}[h]
	\includegraphics[width=.49\linewidth]{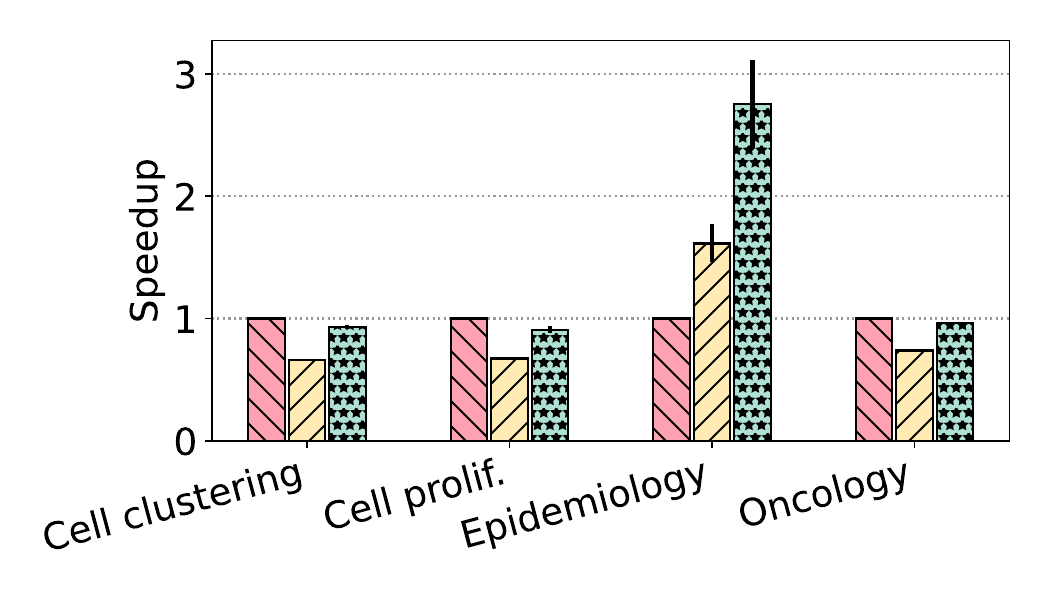}
	\includegraphics[width=.49\linewidth]{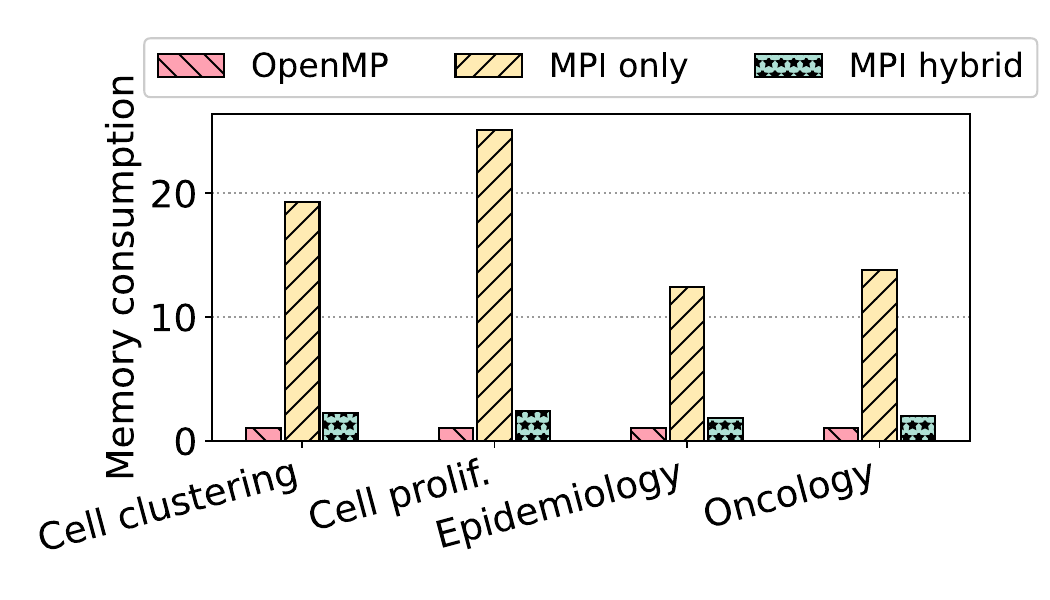}
  \caption{Speedup (left) and normalized memory consumption (right) of \ta{} in MPI only and MPI hybrid configuration with respect to \bdm{} (OpenMP).}
	\label{fig:eval:mpi-vs-openmp}
\end{figure}

We compare the performance of \ta{} in MPI hybrid and MPI only
  mode with \bdm{}.
The benchmarks were executed on one node of System B with $10^7$ agents and for
  all iterations of the simulations.
Figure~\ref{fig:eval:mpi-vs-openmp} shows that MPI hybrid mode performs close
  to the OpenMP version (slowdown between 4--9\%) despite the extra steps
  required in distributed execution (Figure~\ref{fig:design:overview}).
The performance drops significantly for MPI only (slowdown between 26--34\%),
  which uses 18$\times$ more MPI ranks and also does not use hyperthreading.

For the epidemiology simulation, both distributed modes outperform (speedup of
  MPI hybrid: 2.8$\times$) \bdm{}.
\ifthesis
Chapter~\ref{ch:engine} shows that the performance of the
  epidemiology simulation is sensitive to the memory layout.
\fi
Although, \bdm{} is NUMA-aware, not all data structures (e.g., the neighbor
  search grid) are separated for each NUMA domain.
We attribute the observed performance increase to reduced cross-CPU traffic of
  the distributed engine, which outweighs the overheads of aura updates and agent
  migrations.

The memory consumption increases approximately by $2\times$ for the MPI hybrid
  mode due to the additional data structures.
A large part of the memory required by the MPI only mode can be attributed to
  the use of cling \cite{vasilev_cling_2012}.
In the current implementation, each rank has its own instance of cling, which
  requires several hundred MB of memory on its own.

\subsection{Improved Interoperability}
\label{sec:de:eval:interoperability}

This section demonstrates how the distributed simulation engine improves
  interoperability with third party software by enabling the interaction from the
  performance aspect.
A well-suited example to demonstrate this improvement is ParaView, the main
  visualization provider of \bdm{}.
ParaView offers two visualization modes: export mode in which the simulation
  state is exported to file during the simulation and visualized afterwards, and
  the situ mode in which ParaView accesses the agent attributes directly in
  memory and generates the visualization while the simulation is running.

\begin{figure}[h!]
	\includegraphics[width=.4\linewidth]{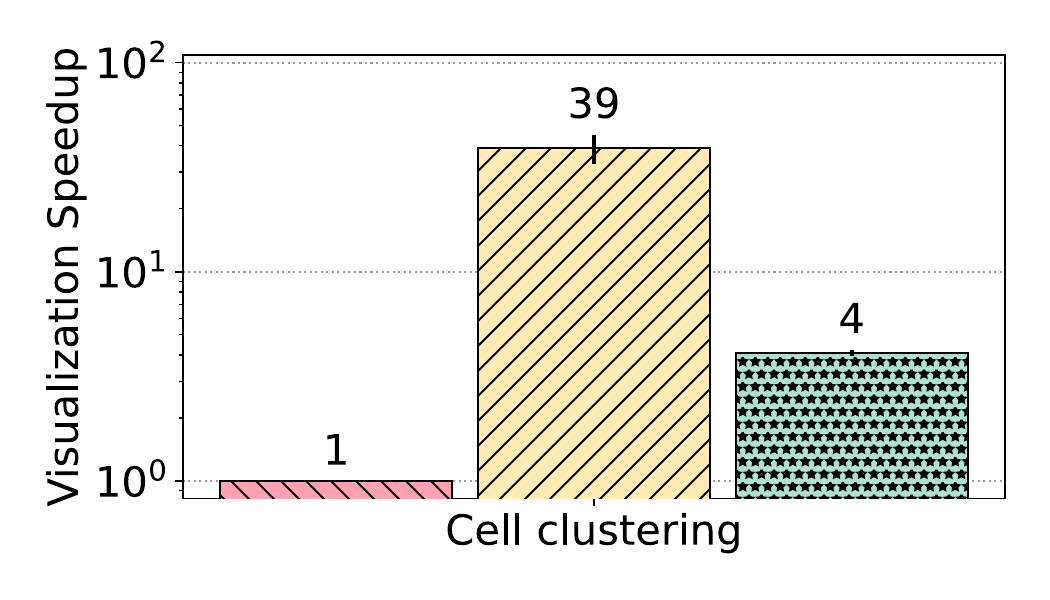}
	\includegraphics[width=.4\linewidth]{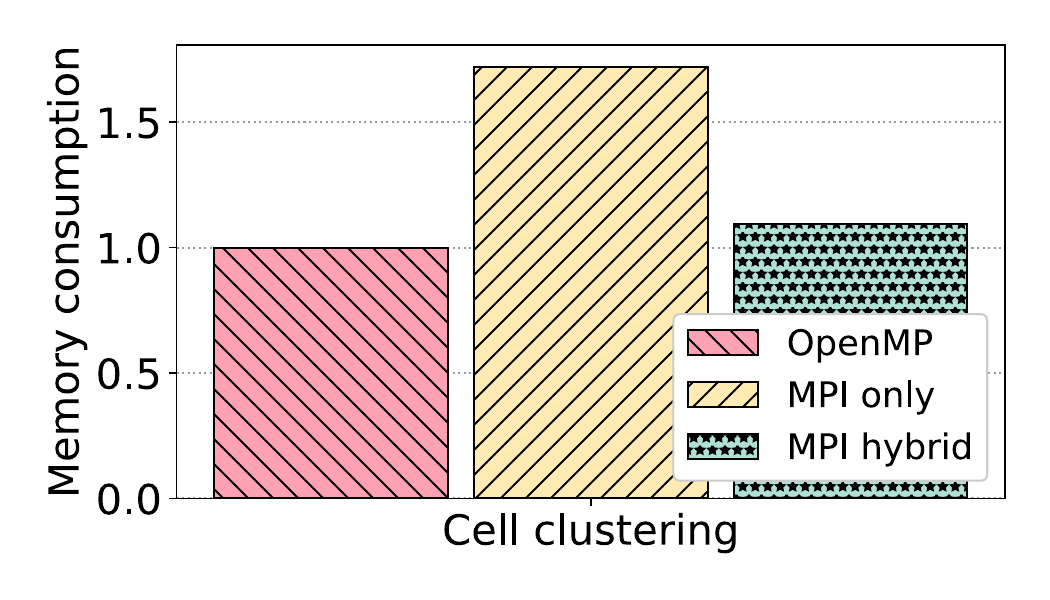}
	\caption{Performance comparison of in situ visualization with ParaView on System~B comparing \ta{} in MPI only and MPI hybrid configuration with \bdm{} (OpenMP).}
	\label{fig:eval:insitu-performance}
\end{figure}

Although ParaView uses distributed and shared-memory parallelism, the
  scalability of using threads alone is very limited (see
  Figure~\ref{fig:eval:insitu-performance}).
Therefore, \bdm{} used mainly the export mode.
\ta{}, however, is now able to leverage the unused potential of in situ
  visualization.

On one System B node, we execute the cell clustering simulation with 10 million agents
  for 10 iterations (Figure~\ref{fig:eval:insitu-performance}).
Each iteration renders one image.
We evaluate three configurations: \bdm{} using OpenMP (i.e., one rank and 144 threads), and two \ta{} configurations: MPI
  only (i.e., 72 ranks with one thread each) and MPI hybrid (i.e., 4 ranks and 36
  threads each).

We can clearly see that ParaView's in situ mode scales mainly with the number
  of ranks.
\ta{}'s MPI only configuration visualizes 39$\times$ faster than \bdm{} although
  it is only using half the number of threads.
The memory consumption is dominated by ParaView and therefore shows a less
  pronounced difference between MPI hybrid and MPI only mode.

\subsection{Scalability}
\label{sec:de:eval:scalability}

\begin{figure}[h!]
	\includegraphics[width=.49\linewidth]{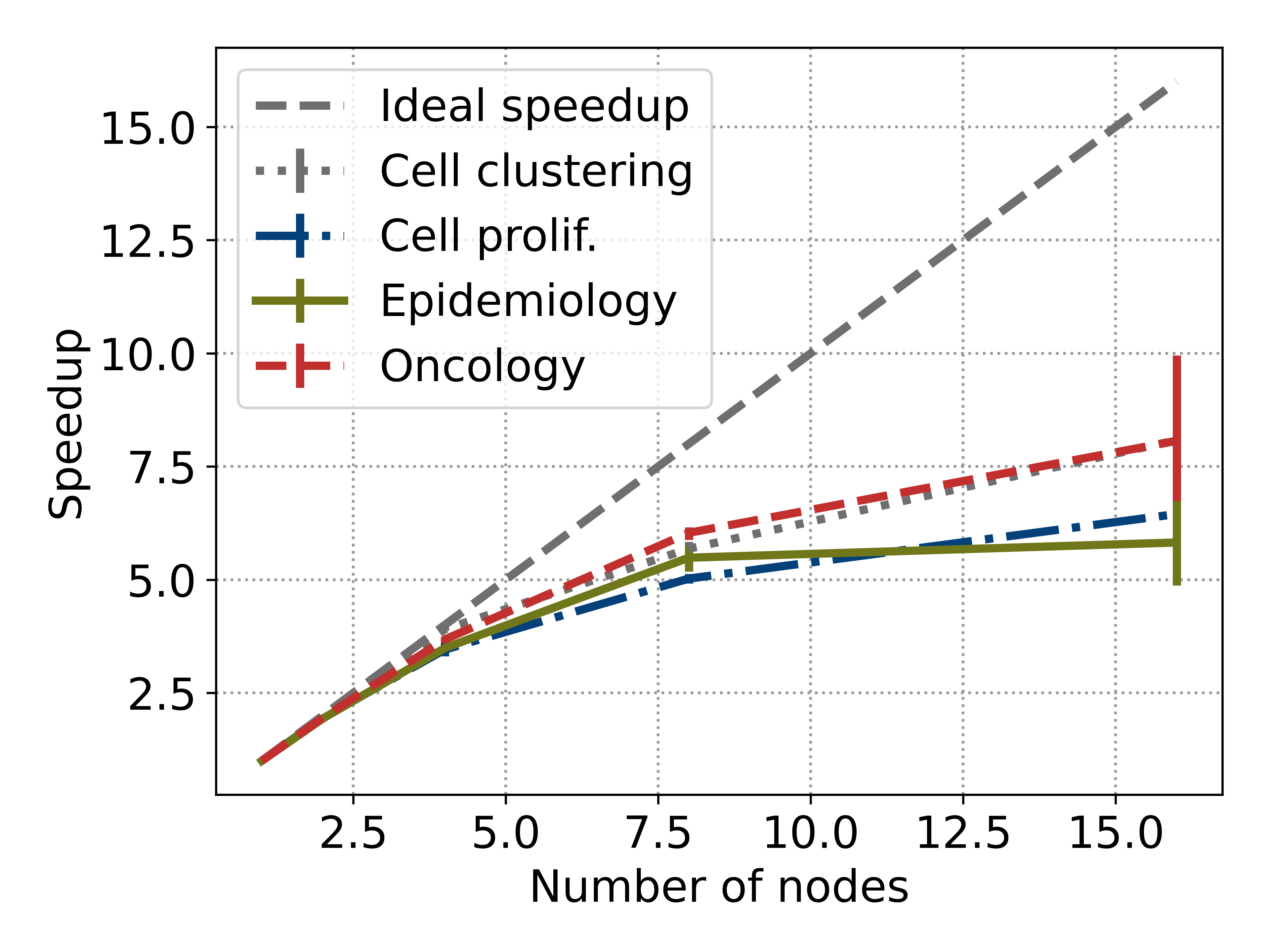}
	\includegraphics[width=.49\linewidth]{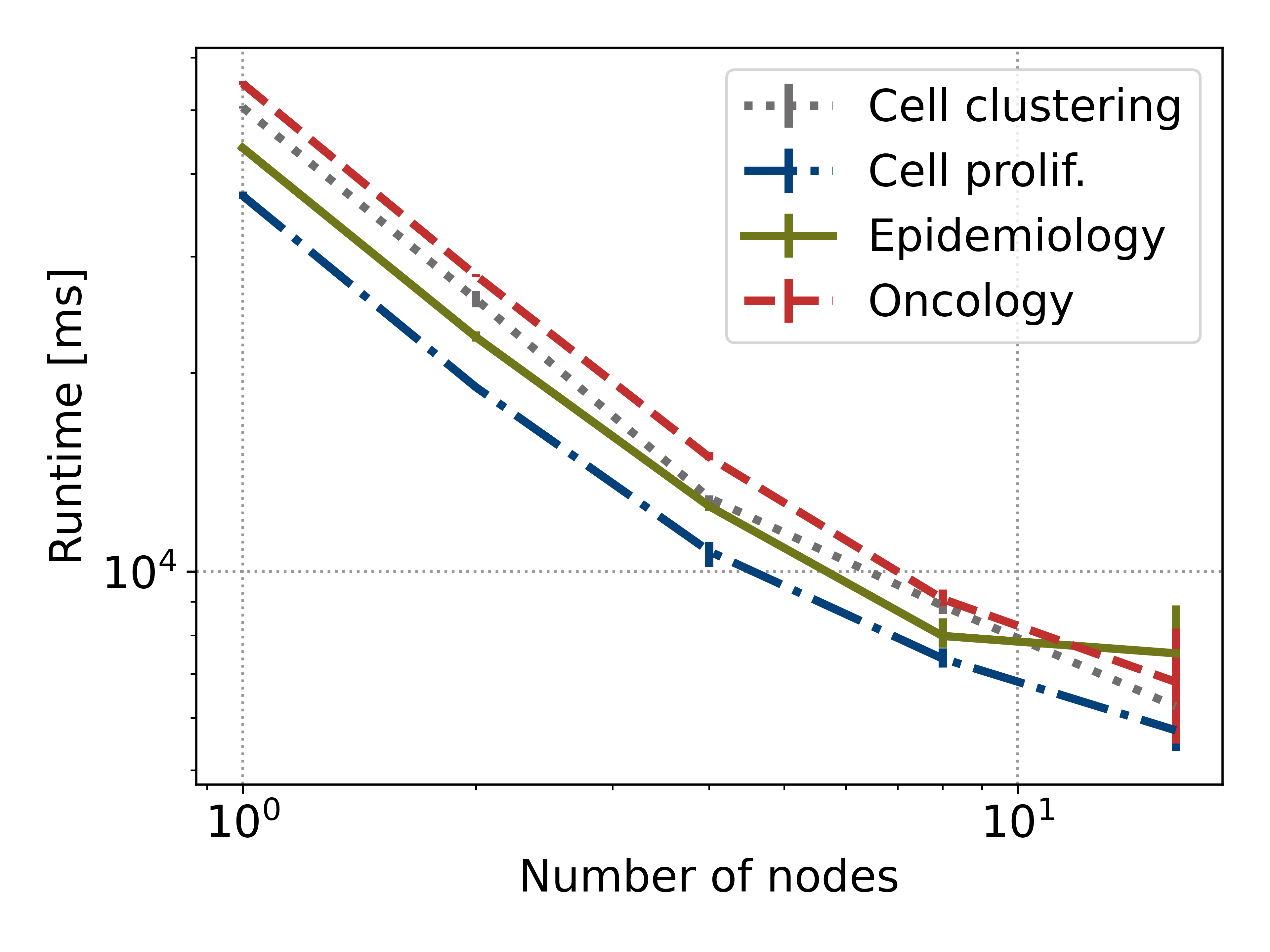}
	\caption{Strong scaling analysis: speedup with respect to an execution on a single node (left), absolute runtime (right).}
	\label{fig:eval:strong-scaling}
\end{figure}

We analyze the scalability of the distributed engine under a strong and weak
  scaling benchmark.
First, for the strong scaling analysis, we examine how much we can reduce the
  runtime of a simulation with a fixed problem size by adding more compute nodes.
We chose the simulation size such that it fills one server and run it for 10
  iterations with node counts ranging from one to 16 (3072 CPU cores) increased
  by powers of two.
Figure~\ref{fig:eval:strong-scaling} shows good scaling until 8 nodes (1536 CPU
  cores), which slows down due to load imbalances and the associated wait times
  for the slowest rank.

\begin{figure}[h]
	\centering
	\includegraphics[width=.49\linewidth]{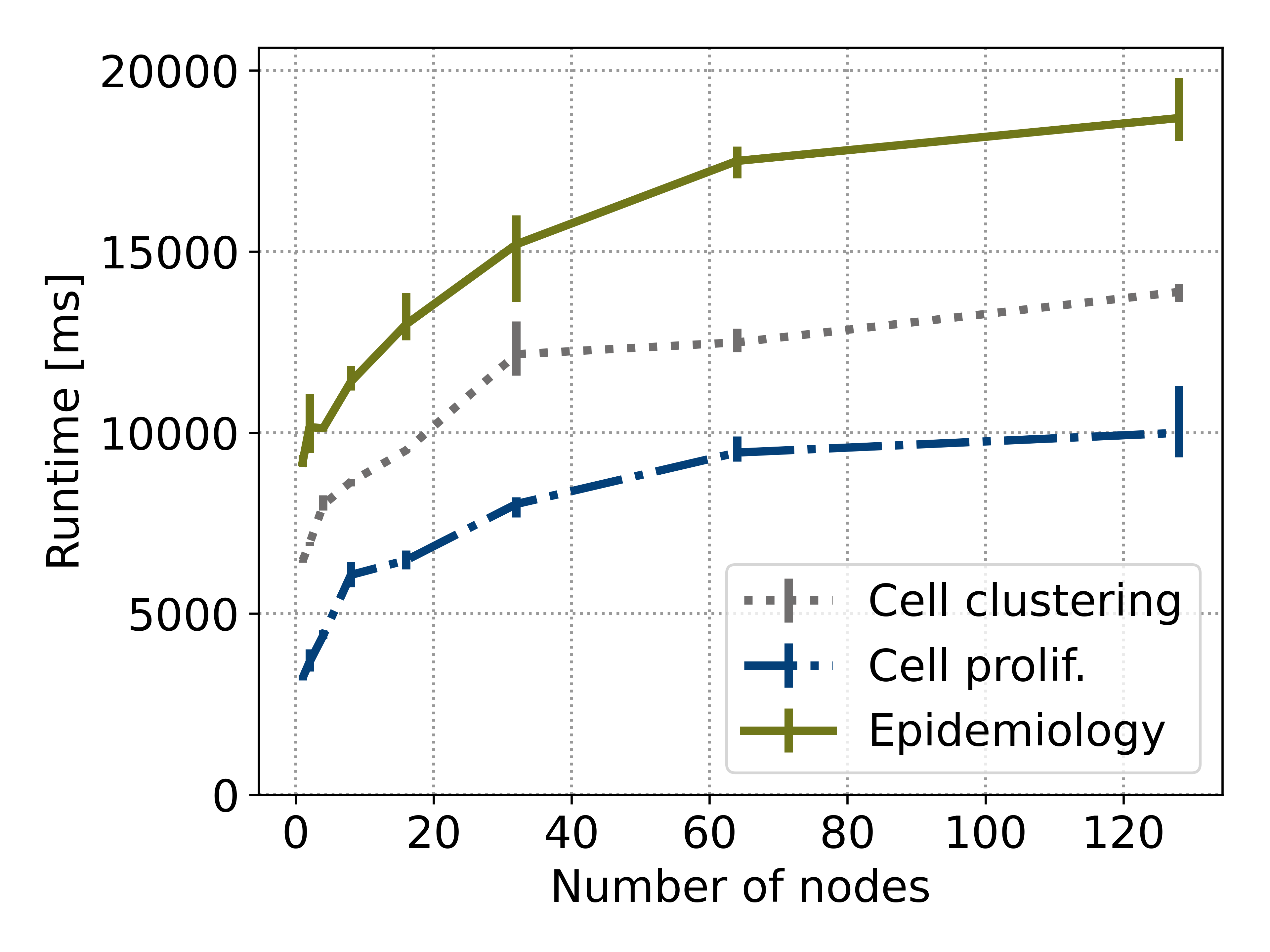}
	\caption{Weak scaling analysis}
	\label{fig:eval:weak-scaling}
\end{figure}

Second, we investigate the performance of the distributed engine by simulating
  larger models with proportionally increased compute resources.
We use $10^8$ agents per node and increase the node count from one to 128,
  which corresponds to 24'576 CPU cores.
Figure~\ref{fig:eval:weak-scaling} shows that after an initial increase in
  runtime, a plateau is reached.

\subsection{Comparison with Biocellion}

To compare the performance of the distributed simulation engine with Biocellion
  \cite{biocellion}, we replicate the benchmark from
  \cite{breitwieser_biodynamo_2023}.
We execute the cell clustering simulation with 1.72 billion cells on System B.
In contrast to \cite{breitwieser_biodynamo_2023}, we use two nodes due to the
  additional required memory.
We measure a runtime of \result{15.8s} averaged over all iterations using 144
  physical CPU cores.
This results in \result{$7.56e5$} $\frac{agent\_updates}{s \times CPU\_core}$.
As \cite{breitwieser_biodynamo_2023}, we use the result from the Biocellion
  paper, because the software is not available under an open source license.
Kang \etal{} report 4.46s per iteration on 4096 CPU cores (AMD Opteron 6271
  Interlago), resulting in $9.42e4$ $\frac{agent\_updates}{s \times CPU\_core}$.
We therefore conclude that \ta{} is \result{8}$\times$
  more efficient than Biocellion.

\subsection{Extreme-Scale Simulation}
\label{sec:de:eval:extreme-scale}

To demonstrate that \ta{} can substantially
  increase the state-of-the-art in terms of how many agents can be simulated, we perform two experiments.

First, we execute the cell clustering simulation with 102.4 billion cells (or agents) for
  10 iterations.
The simulation was executed on 24'576 CPU cores on 128 Snellius nodes using 40~TB 
  of memory and taking on average \SI{7.08}{\second} per iteration.

Second, we increase the number of agents even further to \emph{\ESMaxAgentsExact{}} and use \ESNumNodes{} nodes with \ESNumCPUs{} CPU cores.
To fit this amount of agents into the available main memory, we reduce the engine's memory consumption by disabling all optimizations that require extra memory, use 
  single-precision floating point numbers, reduce the agent's size by changing the base class, and reduce the memory consumption of the neighbor search grid.
These adjustments reduce the memory consumption to \ESMemoryConsumptionInTB{}~TB, but also increase the average runtime per iteration to \SI{\ESRuntimePerSecond}{\second}.

\subsection{Serialization}
\label{sec:de:eval:serialization}

\begin{figure}
	\centering
\begin{subfigure}{\linewidth}
    \centering
    \includegraphics[width=0.49\textwidth]{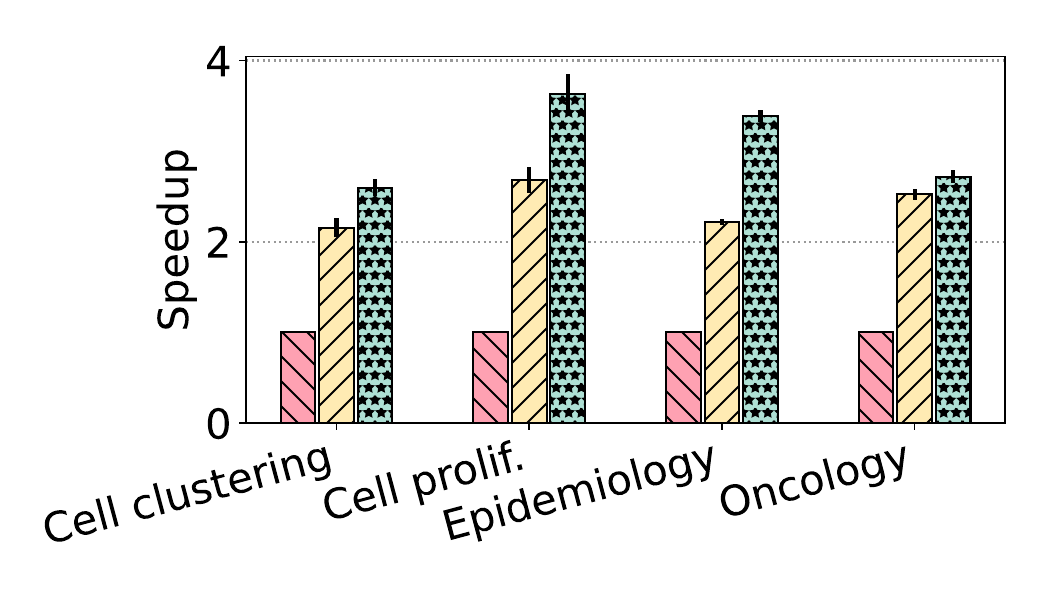}
    \includegraphics[width=0.49\textwidth]{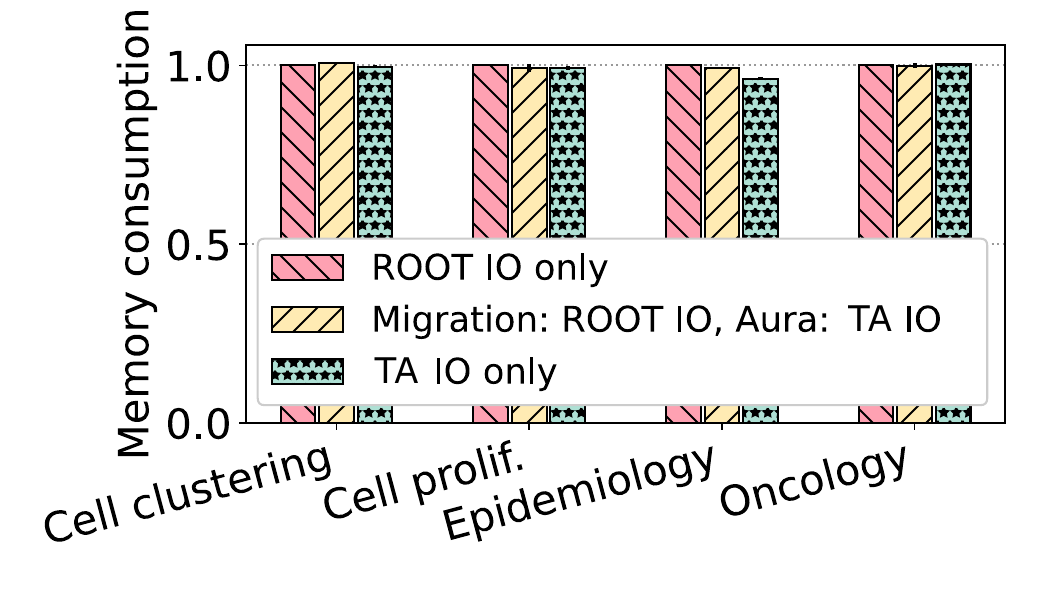}
    \caption{Simulation runtime (left), memory consumption normalized (right)}
   \end{subfigure}
  \begin{subfigure}{\linewidth}
    \centering
    \includegraphics[width=0.49\textwidth]{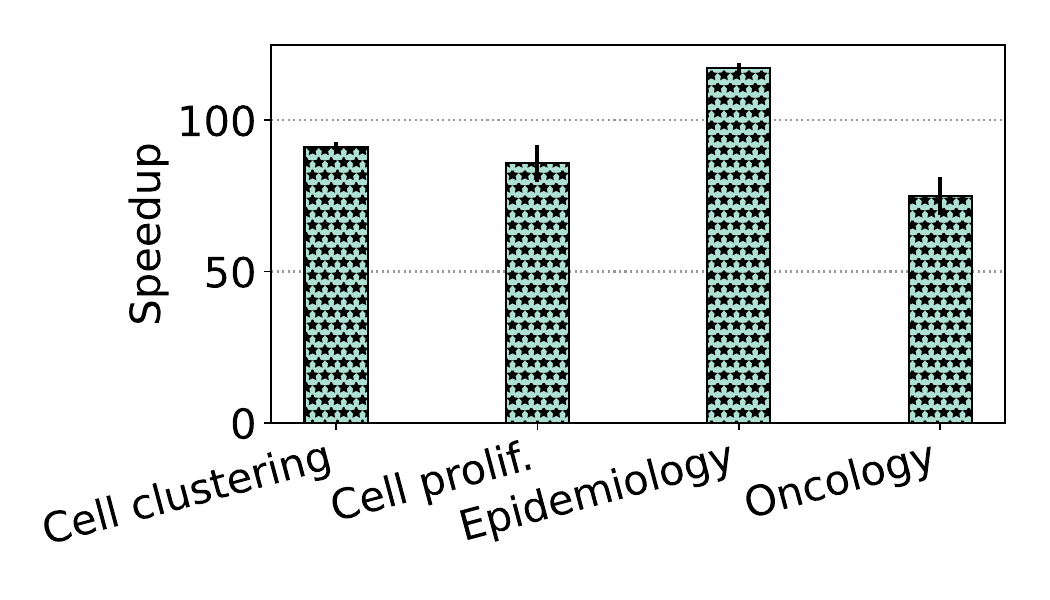}
    \includegraphics[width=0.49\textwidth]{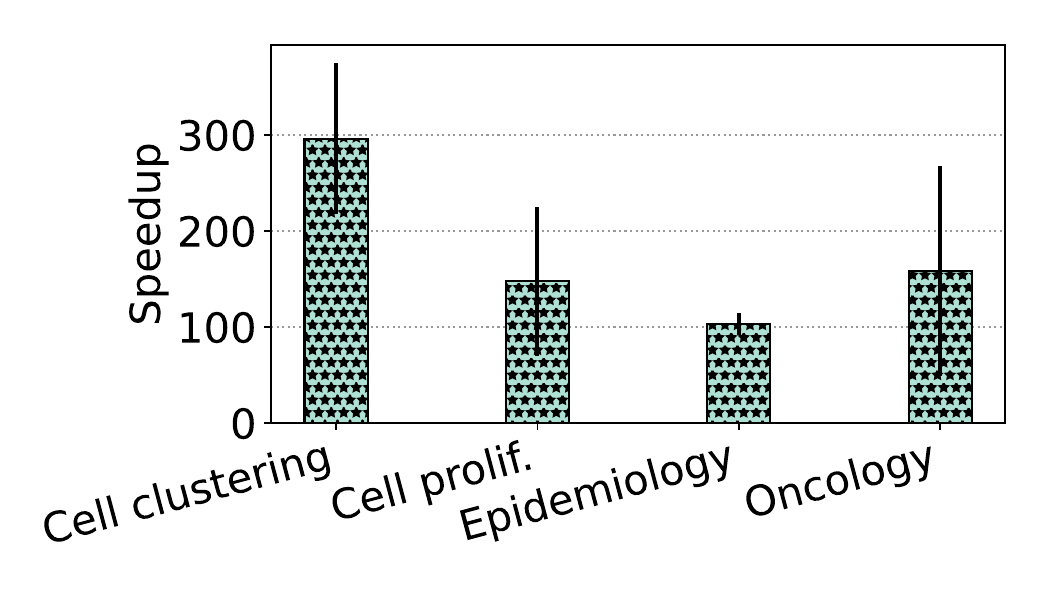}
    \caption{Serialization: aura (left), migrations (right)}
   \end{subfigure}
  \begin{subfigure}{\linewidth}
    \centering
    \includegraphics[width=0.49\textwidth]{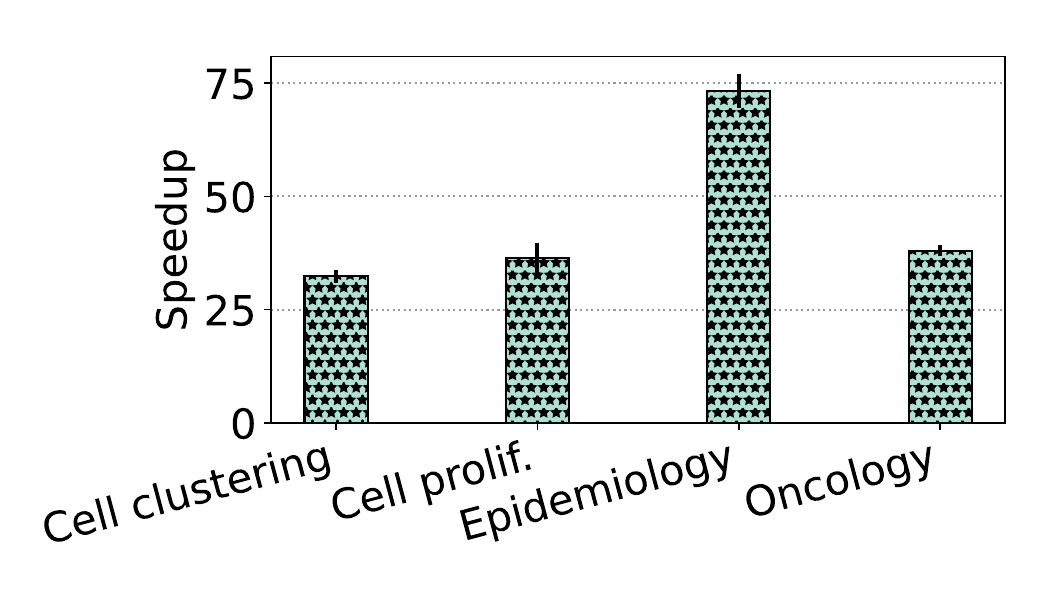}
    \includegraphics[width=0.49\textwidth]{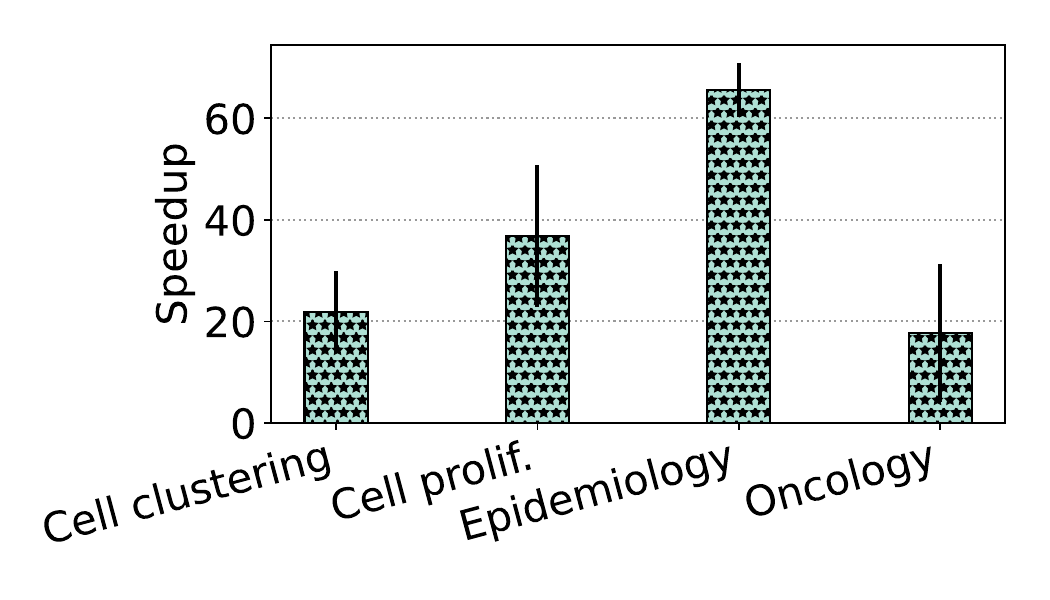}
    \caption{Deserialization: aura (left), migrations (right)}
   \end{subfigure}
  \begin{subfigure}{\linewidth}
    \centering
    \includegraphics[width=0.8\textwidth]{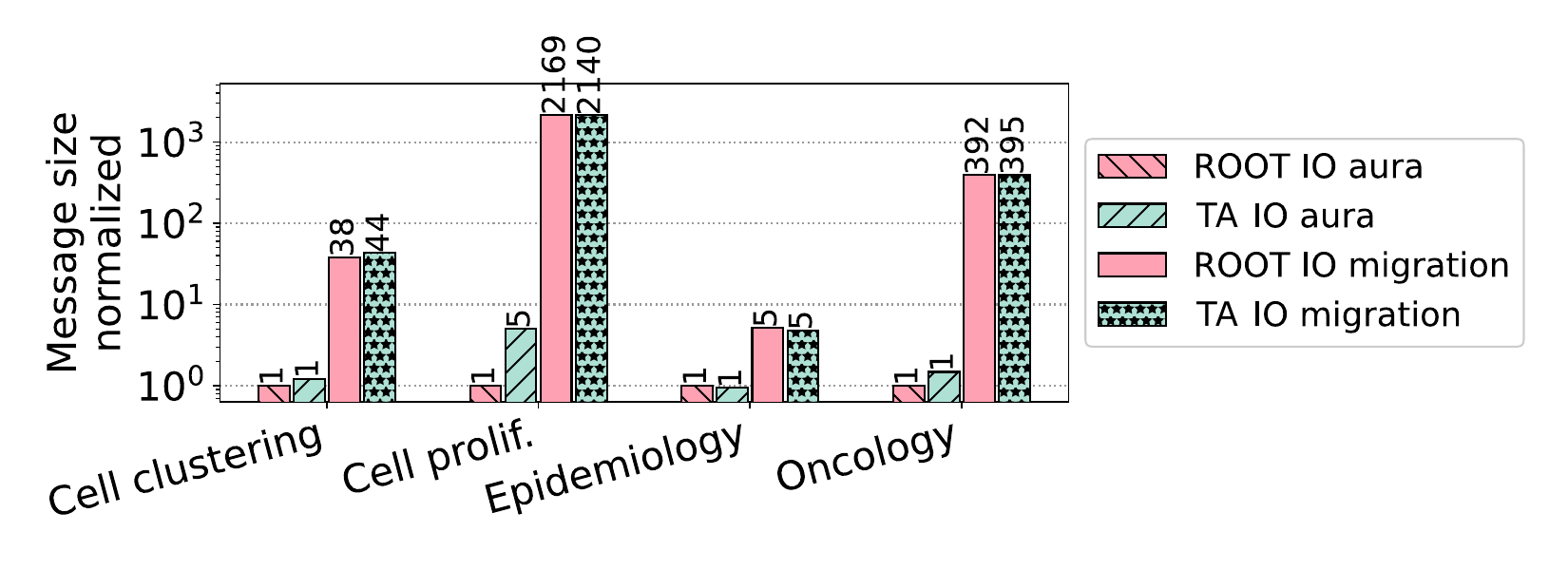}
    \caption{Message size normalized}
    \label{fig:eval:serialization:message-size}
   \end{subfigure}
\caption{Comparison of the \ta{} serialization mechanism (TA IO) and ROOT IO on Snellius.}
	\label{fig:eval:serialization}
\end{figure}

This section compares the tailor-made \ta{} serialization mechanism for the
  agent-based use case presented in Section~\ref{sec:design:serialization} with
  the baseline ROOT IO (see Figure~\ref{fig:eval:serialization}).
We execute the benchmark simulations on four Snellius nodes in MPI hybrid mode
  with $10^8$ agents for 10 iterations.
Figure~\ref{fig:eval:serialization} shows that simulation runtime was reduced
  by up to 3.6$\times$, while keeping the memory consumption constant.
\ta{} IO serializes the agents up to \result{296$\times$} faster (median
  \result{110$\times$}) and also significantly improves the deserialization
  performance: maximum observed speedup of \result{73$\times$} (median
  \result{37$\times$}).
Figure~\ref{fig:eval:serialization:message-size} shows that the resulting
  message sizes are equivalent.
The only outlier in cell proliferation is due to the small message size and
  does not impact the performance negatively.

\subsection{Data Transfer Minimization}
\label{sec:de:eval:delta-encoding}

\begin{figure}[tbh!]
	\centering
\begin{subfigure}{\linewidth}
    \centering
    \includegraphics[width=0.49\textwidth]{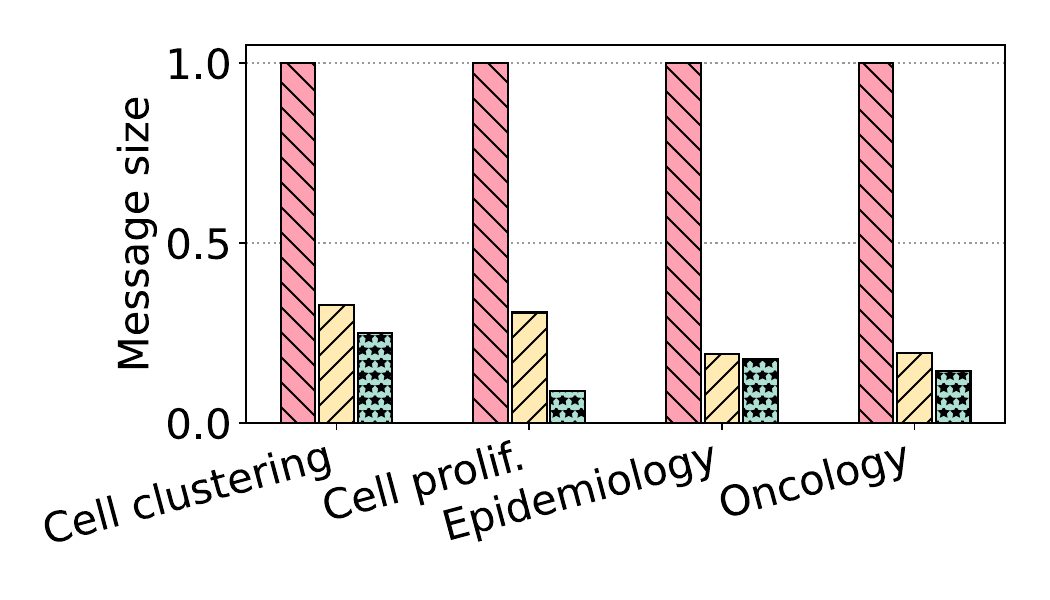}
    \includegraphics[width=0.49\textwidth]{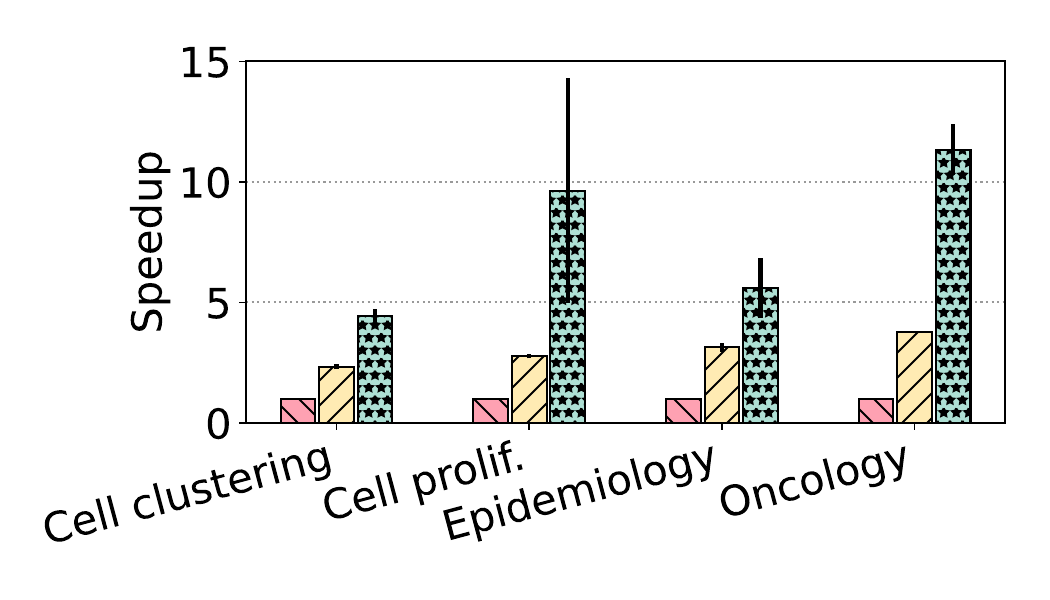}
    \caption{Normalized message size (left) and distribution operation speedup (right)}
   \end{subfigure}
  \begin{subfigure}{\linewidth}
    \centering
    \includegraphics[width=0.49\textwidth]{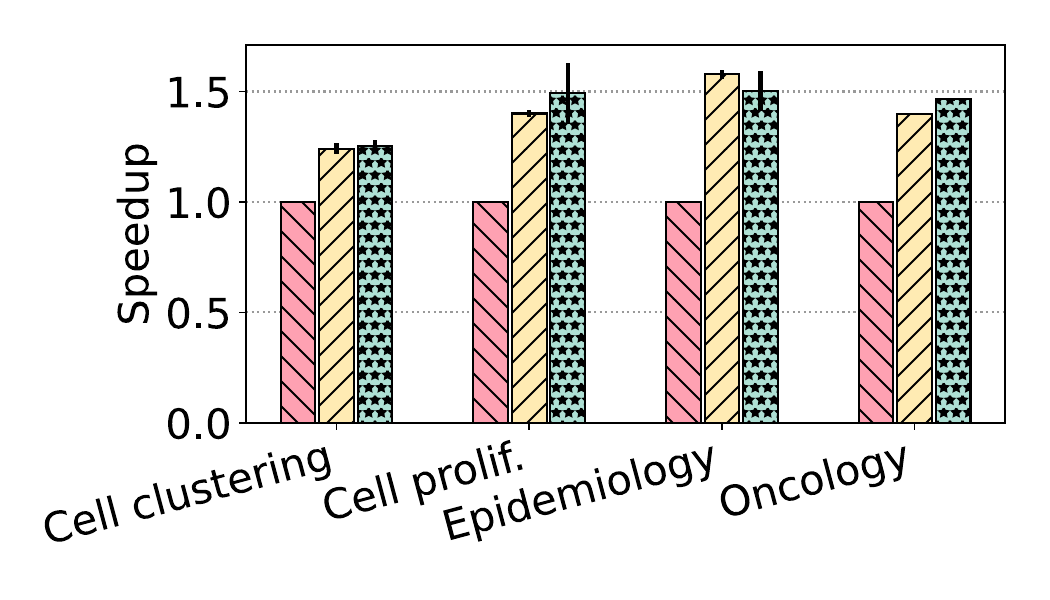}
    \includegraphics[width=0.49\textwidth]{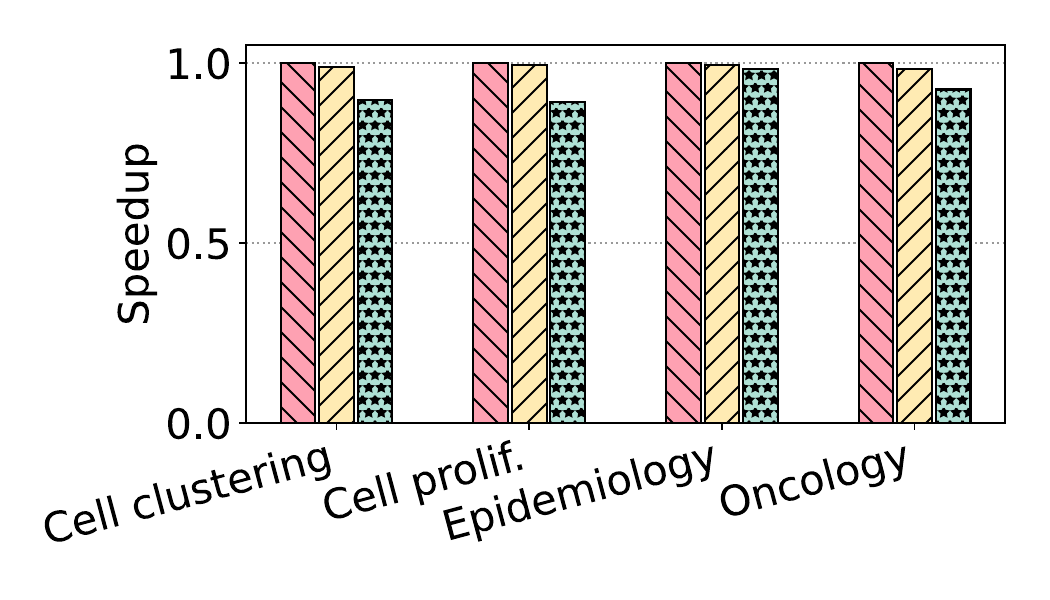}
    \caption{Simulation runtime speedup (left) and agent operations speedup (right)}
     \label{fig:eval:serialization-compression:ops}
   \end{subfigure}
  \begin{subfigure}{\linewidth}
    \centering
    \includegraphics[width=0.49\textwidth]{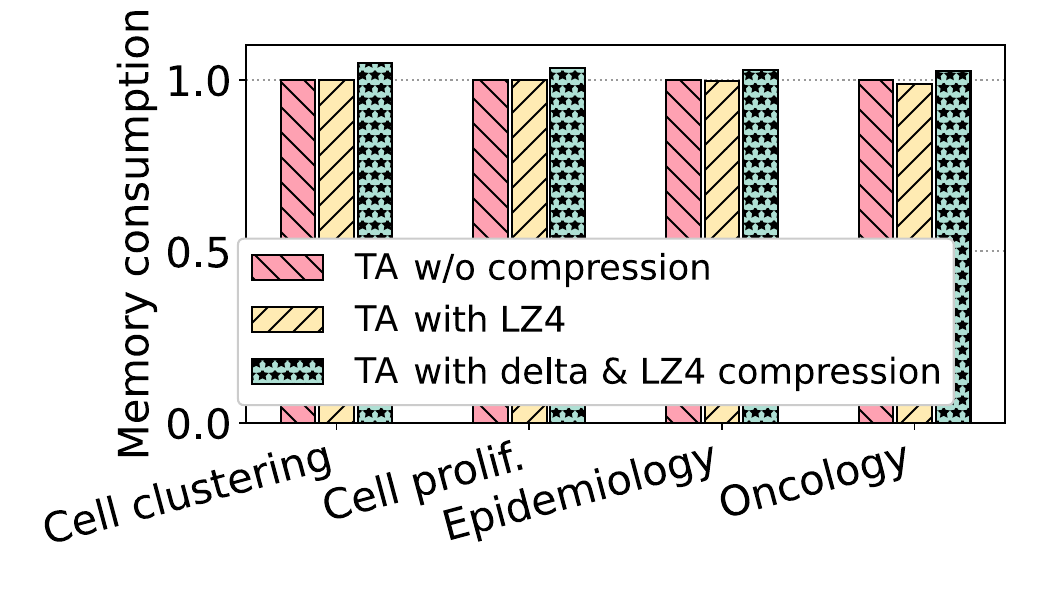}
    \caption{Normalized memory consumption}
   \end{subfigure}
\caption{Comparison of the \ta{}
		IO serialization mechanism with and without compression enabled.
	}
	\label{fig:eval:serialization-compression}
\end{figure}

To evaluate the performance improvements of LZ4 compression \cite{lz4} and
  delta encoding, we execute all benchmark simulations with $10^8$ agents on two
  System B and four Snellius nodes for 10 iterations.
Figure~\ref{fig:eval:serialization-compression} shows the comparison on Snellius with
  \ta{} IO as baseline.
The message size is reduced between \result{3.0--5.2}$\times$ by LZ4
  compression and by another \result{1.1--3.5}$\times$ for adding the delta
  encoding scheme described in Section~\ref{sec:design:delta-encoding}.
This improvement speeds up the distribution operation, which subsumes aura
  updates and agent migrations, up to \result{11}$\times$.
However, the significant speedups of delta encoding do not translate to the
  whole simulation.
The reason is shown in Figure~\ref{fig:eval:serialization-compression:ops}
  (right).
Delta encoding reduces the performance of agent operations (i.e., the main
  functionality of the model), caused by agent reordering.
The memory consumption is increased slightly for delta encoding enabled by the
  data structures that holds the reference (median: \result{3\%}).

On Snellius, simulation runtime is reduced in three out of four simulations by
  using LZ4 compression (median improvement: \result{1.8\%}).
However, due to Snellius' low latency and high bandwidth interconnect, delta
  encoding does not lead to further runtime reductions, because the overheads
  outweigh the benefits.

\section{Conclusion and Future Work}

This chapter presents \ta{} a distributed simulation engine which addresses the scaling
  limitations of the state-of-the-art agent based simulation platform, \bdm{}.
To do so, our distributed simulation engine 1)~enables extreme-scale
  simulations with \ESMaxAgentsInTrillion{} agents, 2)~reduces time-to-result by adding
  additional compute nodes, 3)~improves interoperability with third-party tools
  from the performance side, and 4)~gives users more hardware flexibility.
Our distributed simulation engine allows researchers to seamlessly scale out their execution environment
  from laptops and workstations to clouds and supercomputers.
We demonstrate (Section~\ref{sec:eval:laptop-to-sc}) that such scale-out does \emph{not} require any model code changes.
These results clearly show the benefits of distributed computing
  capabilities.
Researchers can leverage these performance improvements and gain new insights
  into complex systems by building models on an unprecedented scale that has not
  been possible before.

\clearpage{}

\cleardoublepage \clearpage{}\chapter{Conclusion \& Future Work}
\label{ch:conclusion-future-work}

In summary, this dissertation demonstrates that creating a
  high-performance, scalable, and modular agent-based simulation platform from
  the ground up can lead to a drastic reduction in simulation runtime.
This reduction enables the execution of larger and more complex simulations,
  facilitates faster iterative development, and allows for more extensive
  parameter exploration.
This dissertation also demonstrates that a modular platform can significantly
  enhance adoption across various domains.
These major findings come as a result of the dissertation's three main contributions, which involve the
  design, implementation, analysis, and optimization of a novel extreme-scale
  agent-based simulation platform.

\textbf{Contribution 1:}
We design, implement, test, and validate \bdm{}.
This process involves defining abstraction layers, developing infrastructure,
  and setting up a rigorous testing framework.
We create a modular design with 1) hidden low-level features so users can focus
  on the bigger picture without worrying about implementation details, 2)
  high-level features that provide critical agent-based functionality across
  various domains, and 3) model building blocks containing domain-specific
  functions.
\bdm{}'s features are extensively tested with over 600 automated unit, system,
  and integration tests executed for each change to \bdm{}'s code repository.
We demonstrate \bdm{}'s capabilities with three simple yet representative use
  cases in neuroscience, oncology, and epidemiology.
We validate these models and \bdm{} with experimental data and an analytical
  solution.
We show that \bdm{}'s functionality and flexibility lead to succinct model
  definitions ranging from 128 to 181 lines of C++ code for the three presented
  use cases.

\textbf{Contribution 2:}
We build a highly efficient multithreaded simulation engine for the
  \bdm{} platform.
Based on a detailed performance analysis of \bdm{}, we identify three major
  areas for improvement and provide solutions to improve \bdm{} in these three areas.
1)~To maximize the parallel part of the simulation, we develop an optimized grid for neighbor searches and parallelize the merging of thread-local results and visualization.
2)~Agents often follow computationally inexpensive behaviors, which leads to a memory-bound workload on today's state-of-the-art CPU hardware.
We optimize the memory access pattern and data layout to mitigate this
  performance issue.
We (i) reduce memory accesses to remote DRAM regions (on systems with
  non-uniform memory architecture) that have higher access latency, (ii) increase
  the cache-hit rate by bringing agents and their local environment closer
  together in memory, and (iii) add a faster memory allocator.
3)~Based on the observation that some simulations contain regions where agents do not move, we skip the expensive pair-wise force calculation under specific conditions.
The results show that our improved single-node \bdm{} simulation engine is not
  only highly scalable (with a parallel efficiency of 91.7\%) but also efficient
  in terms of absolute runtime.
A single-thread comparison between the improved \bdm{} and Cortex3D and NetLogo
  shows a 23$\times$ and 27$\times$ speedup with \bdm{}.
By running \bdm{} on all 72 CPU cores on our benchmark hardware, while Cortex3D
  and NetLogo use only one core, we can achieve a speedup of three orders of
  magnitude.
When compared to Biocellion using a simulation with 1.72 billion cells, \bdm{}
  accomplishes the same wall-clock runtime but with 9$\times$ fewer CPU cores.

\textbf{Contribution 3:}
We further extend the capabilities of \bdm{} by introducing \ta{}, a distributed
  simulation engine that executes one simulation on multiple servers.
This capability allows for even larger and more complex simulations, improves
  interoperability with third-party software, and gives users more flexibility
  regarding what type of hardware they want to run their simulations on.
We identify agent serialization as the main bottleneck and develop a tailored
  serialization mechanism that reduces unnecessary overheads like checks for
  object duplication, schema evolution, and unpacking during deserialization.
We extend this mechanism with delta encoding to reduce data transfers between
  processes, which improves performance on systems with commodity interconnect.
We demonstrate that the \ta{} distributed simulation engine can simulate
  extreme-scale simulations with \ESMaxAgentsInTrillion{} agents using
  \ESNumCPUs{} CPU cores.
Our optimizations reduce the time spent on serialization by 110$\times$
  (median) and time spent on deserialization by 37$\times$.
Delta encoding reduces the amount of data transferred between servers between
  1.1--3.5$\times$.
The \ta{} distributed simulation engine also improves interoperability with
  third-party libraries that are predominantly parallelized with MPI.
Thus the parallel speedup of these libraries depends on the number of processes.
Since \bdm{} executes simulations with one process only, these libraries often
  utilize only one CPU core.
\ta{} addresses this issue by providing an execution mode that creates one process per CPU core.
We demonstrate \ta{}'s improved interoperability with our visualization library
  ParaView for the cell clustering benchmark and observe a 39$\times$ visualization speedup of \ta{} over \bdm{}.

In conclusion, \bdm{} and especially \ta{} set new standards for agent-based simulation platforms
  by combining high performance, scalability, modularity and user-friendliness.
\ta{} is the first agent-based simulation platform that supports more than
  \ESMaxAgentsInTrillion{} agents.

\newpage

\section{Future Work}

We identify multiple opportunities for future work that can extend the library
  of model building blocks, make \bdm{}/\ta{} accessible to a wider community, build a vibrant community,
  achieve bit-reproducible simulation results, add support for implicit differential
  equation solvers, extend GPU support, automated performance parameter tuning, and 
  taking advantage of processing in memory systems.

\subsection{Extending the Library of Model Building Blocks and Demos}

The time required to implement a model in \bdm{} significantly depends on 1)
  whether the required model building blocks
  (Section~\ref{sec:model-building-blocks}), e.g., agents, behaviors, the
  environment, and other functionality, are readily available, and 2) if these
  building blocks can be easily understood through documentation and their use in
  well-explained demos and tutorials (see for example
  Appendix~\ref{app:tutorials}).
Our initial approach to motivating users to spend extra time generalizing their
  functionality and contributing it back has had limited success.
Therefore, an organized approach to extending the library of model building blocks is needed to increase
  the adoption of \bdm{}.
We believe there are at least two directions.

First, although \bdm{} provides building blocks for its core domains (e.g.,
  neuroscience, cancer research, and epidemiology), their number needs to be
  extended to further strengthen \bdm{}'s position.
The following additions would be valuable: 1) the ability to simulate the
  electrophysiology of neurons
  \cite{diesmann_nest_2001,jordan_closed-loop_2019,layer_effect_2024,jimenez_romero_emergent_2024,steiner_neural_2024},
  2) support for more agent shapes besides spheres and cylinders, 3) and
  map-based environments for epidemiological models
  \cite{ozik_citycovid_2021,\abmUrbanAreas}.

Secondly, \bdm{} should expand to other areas that may require a large number
  of agents and which would, therefore, greatly benefit from \bdm{}'s performance
  characteristics.
Immunological models would benefit from having cell definitions for
  macrophages, T- and B-lymphocytes and their subtypes, memory cells, antibodies,
  and more \cite{\abmImmuneSystem}.
In finance and economics, users may need a virtual stock exchange component for
  financial market models, agents that can behave irrationally, agents who learn,
  and a social network of agents that influence their behavior
  \cite{\abmFinanceAndEconomics}.

\subsection{Making \bdm{}
Accessible to a Wider Community} \label{ch:conclusion:wider-community}

  We have observed that simulation definitions in \cpp{}
  challenge computational researchers with a more Matlab- or R-based background.
Although we have put substantial effort into making \bdm{} easy to use,
  choosing the \cpp{} programming language is a limiting factor.
\cpp{} is an excellent choice for performance-oriented software but is more challenging to learn than, for example, Python.
Future research focusing on alternative ways to define a model could make the
  \bdm{} platform more appealing to a broader audience, while hopefully keeping
  the same performance and efficiency of the platform.

A promising key idea for this work is to enable users to define their model in higher
  level languages like Python.
ROOT, for example, uses dynamic Python bindings \cite{cppyy} and lets
  physicists define their data analysis in Python code.
Python bindings present a promising solution to assemble a simulation from an
  extensive model library (Figure~\ref {fig:bdm-platform}).
At present, however, most \bdm{} models necessitate user-defined agents,
  behaviors, or operations.
When user-defined classes are written in a higher level language, a key
  challenge is the performance overhead that arises from frequent crossings of
  the language boundaries.
This issue can be observed between Java and \cpp{} in \cite{master-thesis}.
The performance challenge of language interaction is not specific to
  agent-based simulation.
Recent work by Kundu \etal{} \cite{kundu_efficient_2023} addresses this issue
  by making JIT-compiled Python (using Numba \cite{numba}) available to \cpp{}.

Kundu \etal{}'s work could be a starting point for \bdm{} to be even easier to use.
Advanced Python integration would simplify simulation development and allow
  users to access Python's vast library ecosystem, while ideally keeping the
  performance and scalability of \bdm{}.

\subsection{Community Building}

This dissertation presents an agent-based simulation platform with performance capabilities exceeding state-of-the-art competitors by a large margin. Despite these capabilities and successful use cases \cite{\bdmUsage}, a larger developer and user base is needed to unfold the project's innate potential.

Without an active and vibrant development community, any open-source project fears stagnation and decline. Therefore, a larger development community is required to keep the project alive and thriving through timely user support, active maintenance, frequent releases, and further feature development to consolidate \bdm{}'s position in simulation fields like computational biology, while expanding into new areas where \bdm{} could make a difference.

An active development community that follows these guidelines builds the required trust with new users to invest their time and resources to base their models on the \bdm{} platform. On the other hand, a growing number of users provide essential feedback and direction for further improvements in the code base, motivate the developers, and, last but not least, legitimate the investments made into the project. 

However, attracting, onboarding, and retaining talent is a challenging endeavor. There is no shortage of other open-source projects. According to Github, the largest open-source code platform, over 300 million code repositories compete for approximately 120 million registered developers on the platform \cite{github-innovationgraph}.

To differentiate in this crowded marketplace and attract the necessary talent, a project needs to be convincing on several dimensions described in the literature. These works highlight the importance of managing the first impression, providing fast feedback, welcoming community, extensive documentation, transparent processes, and many other sometimes subtle hints that make a difference \cite{qiuSignalsThatPotential2019,choiManagingFirstImpressions2010,riehleEconomicMotivationOpen2007,riehleHowOpenSource2015,smirnovaWhatMakesRight2022,steinmacherLetMeGuidelines2019}.

Riehle also points out that: ``[e]conomically rational developers strive to become committers to high-profile open source projects to further their careers ...'' \cite{riehleEconomicMotivationOpen2007}. Therefore, the project should follow the SAFARI Research Group's motto of ``Thinking Big and Aiming High'' \cite{safari-research-group,safari-newsletter-june2023} to achieve high-impact results, which will also help to increase the project's visibility and capability to attract new highly-motivated and capable members.

\subsection{Achieving Bit-Reproducible Simulation Results}

Currently, only two execution modes (Figure~\ref{fig:parallelism-modes})
  produce deterministic simulation results.
First, serial executions \circled{B}, and second, distributed executions
  \circled{G} without dynamic load balancing where each rank is limited to one
  thread (MPI only mode).
All the other parallel modes are non-deterministic but produce
  \emph{statistically} reproducible results.
Non-determinism impedes debugging and testing and causes unstable gradients
  during parameter optimizations.

The most significant source of non-determinism is different random number
  sequences an agent observes.
\bdm{} has thread-private random number generators.
A different execution order will lead to different random numbers during the
  agent update.
A potential solution can happen in four steps.
First, we can create a pseudo-random number generator (PRNG) for each agent
  that is only used during its update.
Counter-based PRNGs \cite{random123} could be a good choice and address the
  large state size of, e.g., Mersenne Twister \cite{matsumoto_mersenne_1998}.
Second, we can replace the \texttt{InPlaceExecutionContext}, with the
  \texttt{CopyExecutionContext} (Section~\ref{sec:alternative-execution-modes}).
The latter execution context commits all changes at the end of the simulation,
  providing a consistent view of the neighbor's attributes independent of the
  order in which agents are updated.
However, this comes at higher memory size and performance penalties, as shown
  in Section~\ref{sec:eval:alternative-execution-modes}.
Third, simulation initialization must be ensured to be deterministic.
Fourth, further attention must be given to other non-associative reductions and
  transcendental functions \cite{arteaga_designing_2014}.
In summary, addressing the sources of randomness improves debugging and testing
  and leads to stable gradients for parameter optimization.

\subsection{Adding Support for Implicit Differential Equation Solvers}
\label{sec:future-work:solvers}

Many agent-based models (e.g., tissue models) incorporate a continuum-based
  component to simulate processes such as substance diffusion
  \cite{duswald_2024}, heat dissipation \cite{jennings_cryo,jennings_cryo2}, and
  other physical processes.
The interaction between the agent-based and continuum-based parts occurs
  through the agents’ ability to query or modify values at their respective
  positions.
These modifications, known as source and sink terms, are crucial in tissue
  models, where cells can secrete or uptake substances, e.g., through
  endocytosis.

Numerical solvers of differential equations can be categorized into explicit
  and implicit methods.
Explicit methods calculate the update for the next timestep based on the
  current state (e.g., the Euler method); implicit methods must solve a system of
  equations (e.g., the backward Euler method).
Explicit methods can be implemented as a stencil code, a method that updates
  values in a grid-based system based on neighboring grid points.
Implicit methods, which solve a system of equations, are computationally more
  expensive but offer greater stability.
Explicit methods require 1) a stability criterion that restricts parameter
  choices and 2) potentially shorter time steps to maintain accuracy and
  stability.

\bdm{} currently provides a stencil-based explicit Euler solver
  (Section~\ref{ch:platform:diffusion}).
To extend the capabilities of \bdm{} and address a wider range of applications,
  incorporating a finite element framework such as MFEM
  \cite{mfem,andrej_high-performance_2024}, which provides implicit solvers, would
  be advantageous.
The key challenge in doing so is how to efficiently integrate source and sink
  terms in the implicit method's system of equations.
This integration is much simpler for explicit methods, which allow direct value
  modification after identifying the nearest grid point.

Besides better numerical stability, the use of MFEM or similar tools would
  offer a higher level of abstraction, simplifying the process for users to solve
  differential equations that extend beyond the built-in diffusion model.

\subsection{Extending GPU Support}

Although \bdm{} can offload computations to the GPU (see
  Section~\ref{sec:performance-analysis} and \cite{hesam_gpu}), this
  functionality is currently limited to force calculations between spheres
  \cite{hesam_gpu}.
Substance diffusion (Section~\ref{ch:platform:diffusion}) is another operation
  that could see a significant benefit from GPU acceleration.
As explained in Section~\ref{sec:future-work:solvers}, the explicit stencil code might require
  short update intervals to produce accurate results.
This attribute allows multiple GPU kernel executions without intermittent data
  transfers between the host and GPU memory, increasing the expected speedup.

Executing agent behavior on the GPU would be the next promising step in this
  research line, opening up additional performance gains.
To enable a broader user base as described in
  Section~\ref{ch:conclusion:wider-community}, we aim to relieve researchers from
  the burden of writing CUDA or OpenCL code themselves.
The \bdm{} platform could be extended with the necessary GPU implementation for
  functions to query neighbors or generate random numbers.
User-defined classes would need a compiler-based approach.
As discussed in Section~\ref{ch:conclusion:wider-community}, Python's JIT compiler Numba
  \cite{numba} and cling \cite{vasilev_cling_2012} might be a good starting
  point.

\subsection{Automated Performance Parameter Tuning}

The performance of \bdm{} depends on configuration parameters that control parallelism, memory layout, data compression, optimizations, and more. For example, the frequency with which the agent sorting mechanism is executed may significantly affect the simulation performance. Our analysis shows that a single parameter choice can account for a difference as high as 2.4$\times$ (Figure~\ref{fig:load-balancing}). 
Although we defined suitable default values based on our extensive benchmark studies, we expect but cannot guarantee equally good results for future simulations running on \bdm{}. The more a simulation or the underlying hardware differs from those analyzed in this thesis, the higher the likelihood that different parameter values may yield better performance.

Therefore, the modelers are responsible for assessing their simulation's performance and adjusting the parameter values to reduce time-to-result. This situation is suboptimal for the following three reasons. First, modelers are often domain scientists without a background in high-performance computing, making it difficult for them to understand the underlying process a performance parameter is modifying.
Second, finding better performance parameters takes time away from creating the model and generating novel insights.
Third, the interdependence of different parameters makes it difficult to find near optimal parameter values manually.

To mitigate these issues, further research could investigate automated parameter-tuning techniques to achieve good performance while freeing the modeler from the burden of adjusting performance-related parameters themselves. 

The literature contains a large body of work focusing on automated parameter tuning for 
big data processing systems \cite{baoLearningbasedAutomaticParameter2018,linAdaptiveCodeLearning2022,herodotouSurveyAutomaticParameter2020}, 
database systems \cite{vanakenAutomaticDatabaseManagement2017,zhangEndtoEndAutomaticCloud2019,mozaffariSelftuningDatabaseSystems2024,diasAutomaticPerformanceDiagnosis,leeArtBalanceRateupDB2021,zhaoAutomaticDatabaseKnob2023}, 
linear algebra \cite{nugterenCLBlastTunedOpenCL2018,clintwhaleyAutomatedEmpiricalOptimizations2001,sorensenAutotuningLevel12013}, 
HPC applications \cite{wolfAutomaticPerformanceModeling2016,tilletInputawareAutotuningComputebound2017a,menonAutotuningParameterChoices2020,balaprakashAutotuningHighPerformanceComputing2018,balaprakashCanSearchAlgorithms2011},
and more \cite{mebratuAutomaticTuningTensorflows2021,huangSurveyAutomaticParameter2020,mirkovicAutomaticPerformanceTuning2001,panFastEffectiveOrchestration2006,tapusActiveHarmonyAutomated2002}. 
Herodotou \etal{} classify these approaches into the following categories: ``rule-based, cost modeling, simulation-based, experiment-driven, machine learning, and adaptive'' \cite{herodotouSurveyAutomaticParameter2020}.
The challenges that need to be addressed in this line of research include high-dimensional parameter spaces, difficulty in obtaining enough samples \cite{baoLearningbasedAutomaticParameter2018,balaprakashAutotuningHighPerformanceComputing2018} (e.g., due to long-running simulations), and integration into the application workflow \cite{balaprakashAutotuningHighPerformanceComputing2018}.

\subsection{Taking Advantage of Processing in Memory Systems}

Off-chip memory accesses are a fundamental bottleneck of today's computing
  systems \cite{mutlu_processing_2019,mutlu2022modern}, leading to reduced
  performance for memory-intensive workloads such as agent-based modeling.
So far, in the processor-centric view of computing systems, this challenge has
  been addressed by adding multi-level cache hierarchies and prefetchers to bring
  data into on-chip memory and access it from there, thus avoiding costly accesses to the main memory.
Caches and prefetchers exploit temporal and spatial data locality and regular
  data access patterns.
We also addressed the memory bottleneck in this dissertation with several
  optimizations that improve the memory layout and access patterns of agents,
  which lead to significant performance improvements, as demonstrated.
However, the mentioned mitigation mechanisms in the processor and \bdm{}'s
  simulation engine merely treat the symptoms rather than curing the underlying
  problem.
We deliberately made this choice for \bdm{}, because we prioritized developing
  a high-performance platform for \emph{today's} hardware.

However, future research could be more forward-looking by investigating the
  impact of processing-in-memory (PIM) for agent-based modeling.
PIM \cite{mutlu2022modern, mutlu_processing_2019,
	  ghose_processing--memory_2019} is an emergent research field in computer
  architecture that addresses the memory bottleneck at its root by bringing
  processing close to the data
  \cite{kautz1969cellular,stone1970logic,mutlu2022modern,
	  mutlu_processing_2019,ghose_processing--memory_2019}.
PIM reduces not only the memory access latency but also the energy consumption.
The literature contains many works that explore the design space of PIM using
  hardware simulation.
These works can be divided into processing-near-memory (i.e., bringing
  processing elements close to data)
  \cite{oliveira_damov_2021,boroumand_google_2018,boroumand2019conda,ahn_scalable_2015,giannoula2021syncron,singh2019napel,zhu2013accelerating,
	  akin2015data, augusta_jafar_2015, kim_grim-filter_2018,
	  cali2020genasm,fernandez2020natsa,hsieh_accelerating_2016,singh2020nero,
	  ahn_pim-enabled_2015,singh_fpga-based_2021,denzler_casper_2023,oliveira_accelerating_2022,cali_segram_2022,ghiasi_genstore_2022,besta_sisa_2021,ghiasi_megis_2024,pattnaik_scheduling_2016,
	  boroumand_lazypim_2017,boroumand_google_2021,guPIMAllYou2025,hePAPIExploitingDynamic2025}, or processing-using-memory
  (i.e., performing computations directly with the memory circuitry by exploiting
  their specific properties) \cite{seshadri_-dram_2019, seshadri_ambit_2017,
	  seshadri_rowclone_2013,kim_dram_2018,kim_d-range_2019,gao_computedram_2019,chang_low-cost_2016,
	  li_drisa_2017, hajinazar_simdram_2021, olgun_quac-trng_2021,wang_figaro_2020,
	  olgun_pidram_2022,ferreira_pluto_2022,park_flash-cosmos_2022,mao_genpip_2022,fernandez_matsa_2024,yuksel_simultaneous_2024,seshadri_fast_2015,yuksel_simultaneous_2024,besta_sisa_2021,oliveira_accelerating_2022,mutlu_memory-centric_2024,seshadriFastBulkBitwise2015a,liPinatuboProcessinginmemoryArchitecture2016,yuksel_simultaneous_2024,yuksel_functionally-complete_2024,oliveiraProteusAchievingHighPerformance2025,eckertNeuralCacheBitSerial2018,agaComputeCaches2017}.
The promising results of simulation-based research of PIM led to recent
  releases of real hardware, such as UPMEM
  \cite{upmemwebsite2024,gomez2021benchmarking,giannoula2021syncron,gomez-luna_benchmarking_2022,gomez-luna_evaluating_2023,item_transpimlib_2023,diab_framework_2023,rhyner_pim-opt_2024,giannoulaPyGimEfficientGraph2024,hyunPathfindingFuturePIM2024},
  Samsung FIMDRAM \cite{kimAquaboltXLSamsungHBM2PIM2021,kwon25420nm6GB2021,leeHardwareArchitectureSoftware2021}, SK Hynix \cite{lee1ynm125V8Gb2022}.

Future research on using PIM for ABM could focus on accelerating operations
  that are executed on an agent's local neighborhood.
Due to the dynamic nature of many agent-based models, this neighborhood is
  constantly changing, sometimes so rapidly that sorting agents based on a
  space-filling curve does not improve the cache-hit rate.
There are at least two promising research directions: 1) examining the usage of real
  PIM hardware in \bdm{}, dealing with the current limitations of these early
  hardware versions (e.g., UPMEM emulates floating-point operations, uses an
  ``accelerator architecture'' similar to a GPU, which requires explicit data
  transfers and prevents fine-grained collaborative work between PIM and CPU) and
  2) contribute to the design space exploration of future PIM systems (e.g.,
  based on 3D-stacked memory+logic chips or near-data processing chips with
  specialized and sophisticated accelerators) focusing on the agent-based
  workload.

\section{Concluding Remarks}

The primary objective of our research was to design and develop a
  high-performance agent-based simulation platform capable of pushing the
  boundary in terms of performance, scalability, efficiency, and modularity.
We believe our optimized versions of \bdm{} and \ta{} address this primary objective, as
  evidenced by the positive results presented in this dissertation.

There is also significant external validation of the \bdm{} platform and our
  primary objectives.
A significant number of simulations have already been built upon the \bdm{}
  platform \cite{\bdmUsage}
  (see also Appendix~\ref{app:other-works} and~\ref{app:complex-simulations}),
  including the prize-winning radiotherapy model \cite{cogno_mechanistic_2024},
  which was recognized by PhysicsWorld \cite{physicsworld-top10} as one of the top
  10 breakthroughs in physics in 2024.
While the majority of the simulations presented in this dissertation are
  situated in Euclidean (3D) space, \bdm{} is also capable of simulating
  non-Euclidean models, as demonstrated in two epidemiological studies
  \cite{hesam_2024, hiv_malawi}.

Our single-node performance improvements comprise detailed parallelization,
  memory layout improvements, and exploiting simulation-specific properties.
These improvements make \bdm{} up to three orders of magnitude faster than
  state-of-the-art simulators \cite{zublerdouglas2009framework,netlogo,
	  kang_biocellion:_2014} and enable fast iterative development and
  high-throughput calibration studies.
Duswald \etal{} employed \bdm{} to execute 50 million simulations, with each
  simulation running for 2.06 to 2.35 seconds per CPU core
  \cite{duswald2024calibrationstochasticagentbasedneuron}.

Our distributed engine \ta{}, featuring a custom serialization mechanism
  and a delta-encoding scheme to reduce data transfers, allows for simulations at
  an unprecedented scale—supporting up to \ESMaxAgentsInTrillion{} agents across
  \ESNumCPUs{} CPU cores.

We hope that the unique qualities of \bdm{} and \ta{} will not only inspire but also
  motivate researchers to model systems that were previously beyond computational
  reach, unlocking new insights into complex systems in many domains.

\clearpage{}

\appendix
\cleardoublepage \clearpage{}
\cleardoublepage \chapter{Other Works of the Author}
\label{app:other-works}

In my Ph.D.
at ETH Zurich and CERN, I led the design, development, analysis, and optimization of \bdm{} and \ta{}.
This work led to three manuscripts comprising
  Chapters~\ref{ch:platform}--\ref{ch:dse} in this dissertation.
The first main paper detailing the \bdm{} platform,
  its features, modular software design, use cases, and high-level performance metrics was published in the
  Bioinformatics journal \cite{breitwieser-bdm}.
Afterwards, I continued my work on performance characterization of
  agent-based simulations and a series of performance improvements for
  shared-memory parallelism.
These results were published at the ACM Annual Symposium on Principles
  and Practice of Parallel Programming (PPoPP) \cite{breitwieser_biodynamo_2023}.
I also ensured that our publications' benchmarks, visualizations, and videos
  can be easily reproduced with the pipeline provided in supplementary materials
  \cite{breitwieser_supporting_2021, breitwieser_supporting_2022}.
I am happy that the PPoPP reviewers independently reproduced our results and
  awarded us all available reproducibility badges.
In addition, my effort to go the extra mile was recognized with the Best
  Artifact Award at PPoPP'23.
In my third main work, I designed, implemented, analysed, and optimized a distributed simulation engine for 
extreme-scale simulations with \ESMaxAgentsInTrillion{} agents.
This research used computing resources from the Dutch National Supercomputer called
  Snellius and will be submitted to SC'25.
In addition to the scientific publications, I spent much time and effort on maintenance, user support, and improving
  software quality and user-friendliness.

I also contributed to Ahmad Hesam's research on GPU acceleration of \bdm{}
  \cite{hesam_gpu}.
We focused on offloading the calculation of pair-wise mechanical forces between
  agents.
In tissue models, this operation can be the dominant factor determining the
  simulation's overall runtime.

Besides my work on \bdm{} and \ta{}, I was working on several models
  together with researchers from the Technical University of Munich, Delft
  University of Technology, the University of Newcastle upon Tyne, the University
  of Cyprus, the University of Surrey, the University of Geneva, and ScimPulse.
In these work packages, I supported the lead researcher with design- and
  implementation-specific questions of the model and \bdm{}.

\begin{figure}
  \centering
\includegraphics[width=\linewidth]{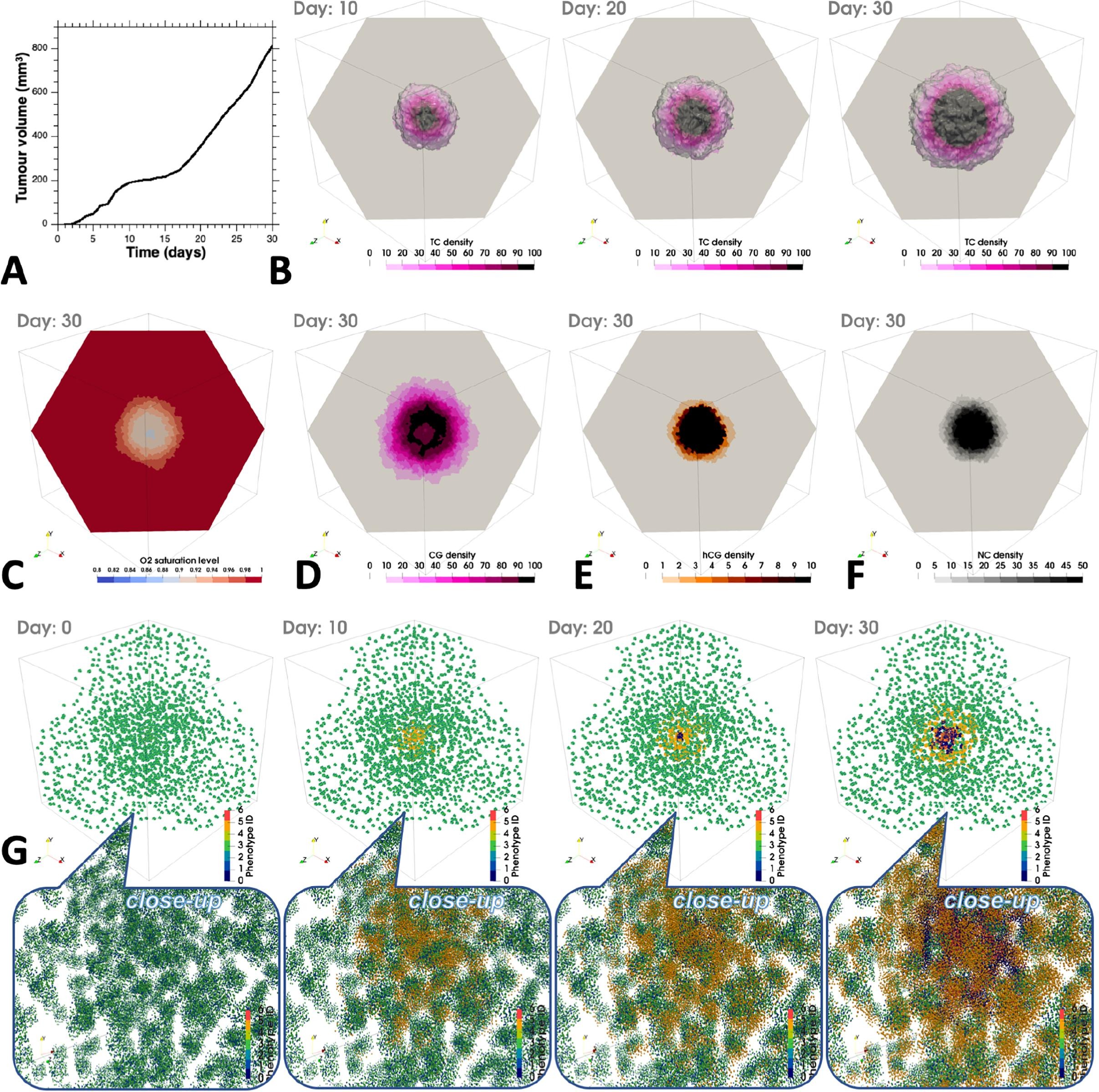}
\caption{
  Glioma growth simulation combining an agent-based and continuum-based model
   from Jean de Montigny \etal{} \cite{demontigny_2021}. 
  Figure taken from
  \cite{demontigny_2021}
  without modification and used under 
  \href{https://creativecommons.org/licenses/by/4.0/}{CC BY 4.0}.}
  \label{fig:app:other-works:glioma}
\end{figure}

I worked with Jean de Montigny on a model that studies gliomas, a specific type
  of brain tumor.
This work aimed to combine the benefits of continuum and agent-based modeling
  to simulate the invasion of tumor cells into the surrounding tissue (Figure~\ref{fig:app:other-works:glioma}).
The continuum-based part simulates the macroscopic properties of the tumor
  using the finite element method, while the agent-based model simulates detailed
  cell-level dynamics.
A key aspect of this work was the connection between these two modeling
  approaches using averaging and downsampling.
The published work shows that the hybrid approach allows for a ``detailed but
  computationally cost-eﬀective'' \cite{demontigny_2021} way of simulating
  gliomas.

I supervised Jack Jennings on developing additional functionality for \bdm{} to
  enable cryogenics simulations \cite{jennings_cryo, jennings_cryo2}.
Freezing and thawing cells with high survival rates is essential for several
  medical applications.
The survival rate depends on many factors during the freezing and thawing
  process, which are time-consuming to test in a wet lab for each cell type.
Simulation can remedy this challenge.
The work adds support for heat transfer and osmotic cell membrane properties.

I cooperated with Tobias Duswald on calibrating pyramidal cells, i.e., neurons
  from the cerebral cortex, using approximate Bayesian computation
  \cite{duswald_2024}.
This work aimed to find parameter distributions that most likely lead to
  specific observables, in this case, 3D neuron geometries.
To this extent, we integrated \bdm{} with the tool SMCABC (Sequential Monte
  Carlo Approximate Bayesian Computation) \cite{del_moral_adaptive_2012},
  focusing on reducing the overhead of simulation startup.
With this issue addressed, we could execute high-throughput calibration studies
  with 50 million individual simulations, which not only output parameters that
  lead to specific geometries but also quantify the uncertainties.

With Ahmad Hesam, I worked on a country-scale COVID simulation demonstrating
  \bdm{}'s performance translates into more accurate models \cite{hesam_2024}.
The work builds upon a model where one agent represents 100 persons.
This simplification was necessary to allow for simulation execution within a
  reasonable time.
With \bdm{}'s superior performance characteristics, we were able to lift this
  restriction and achieve a one-to-one mapping between agents and persons for the
  whole Dutch population of 17 million people.
This high resolution also allows the integration of individual-based data
  (i.e., microdata) from the Dutch National Statistics Bureau, improving the
  model accuracy by 38\% and explaining hospital admission at a subnational
  level.

A similar research direction was taken in collaboration with Tobias Duswald and
  Janne Estill to accelerate an R-based model that simulates the spatial spread
  of HIV in Malawi \cite{estill2020}.
The key issue of this work is the long model execution time of several hours,
  which prevents meaningful parameter optimization or sensitivity analysis.
In this collaboration, we re-implemented the model in \bdm{} \cite{hiv_malawi}.
Our preliminary evaluation shows impressive performance gains in which a \bdm{}
  simulation runs in only minutes.

\cleardoublepage \chapter{Complete List of the Author's Contributions}
\label{app:contribution}

This section contains a list of contributions that were led by the author (Section~\ref{app:contribution:lead}), and those as a co-author (Section~\ref{app:contribution:lead}) in reverse chronological order.

\section{Contributions Led by the Author}
\label{app:contribution:lead}

\ifx \showCODEN    \undefined \def \showCODEN     #1{\unskip}     \fi
\ifx \showDOI      \undefined \def \showDOI       #1{#1}\fi
\ifx \showISBNx    \undefined \def \showISBNx     #1{\unskip}     \fi
\ifx \showISBNxiii \undefined \def \showISBNxiii  #1{\unskip}     \fi
\ifx \showISSN     \undefined \def \showISSN      #1{\unskip}     \fi
\ifx \showLCCN     \undefined \def \showLCCN      #1{\unskip}     \fi
\ifx \shownote     \undefined \def \shownote      #1{#1}          \fi
\ifx \showarticletitle \undefined \def \showarticletitle #1{#1}   \fi
\ifx \showURL      \undefined \def \showURL       {\relax}        \fi
\providecommand\bibfield[2]{#2}
\providecommand\bibinfo[2]{#2}
\providecommand\natexlab[1]{#1}
\providecommand\showeprint[2][]{arXiv:#2}

\begin{itemize}

\item
\bibfield{author}{\bibinfo{person}{\underline{Lukas Breitwieser}}, \bibinfo{person}{Ahmad
  Hesam}, \bibinfo{person}{Abdullah Giray Yaglikci}, \bibinfo{person}{Mohammad Sadrosadati}, 
  \bibinfo{person}{Fons Rademakers},
  {and} \bibinfo{person}{Onur Mutlu}.} \bibinfo{year}{2024}\natexlab{}.
\newblock \showarticletitle{TeraAgent: A Distributed Agent-Based Simulation Engine for Simulating Half a Trillion Agents}. 
\emph{Manuscript to be submitted to:} \bibinfo{booktitle}{\emph{
the International Conference for High Performance
Computing, Networking, Storage, and Analysis}} (St. Louis, MO, USA) \emph{(\bibinfo{series}{SC '25})}.

\item
\bibfield{author}{\bibinfo{person}{\underline{Lukas Breitwieser}}, \bibinfo{person}{Ahmad
  Hesam}, \bibinfo{person}{Fons Rademakers}, \bibinfo{person}{Juan~G\'{o}mez
  Luna}, {and} \bibinfo{person}{Onur Mutlu}.} \bibinfo{year}{2023}\natexlab{}.
\newblock \showarticletitle{High-Performance and Scalable Agent-Based
  Simulation with BioDynaMo}. In \bibinfo{booktitle}{\emph{Proceedings of the
  28th ACM SIGPLAN Annual Symposium on Principles and Practice of Parallel
  Programming}} (Montreal, QC, Canada) \emph{(\bibinfo{series}{PPoPP '23})}.
  \bibinfo{publisher}{Association for Computing Machinery},
  \bibinfo{address}{New York, NY, USA}, \bibinfo{pages}{174–188}.
\newblock
\showISBNx{9798400700156}
\urldef\tempurl \url{https://doi.org/10.1145/3572848.3577480}
\showDOI{\tempurl}
\showeprint[arxiv]{2301.06984}~[cs.DC]\\
Supplementary Materials: \url{https://doi.org/10.5281/zenodo.5121618}

\item
\bibfield{author}{\bibinfo{person}{\underline{Lukas Breitwieser}}, \bibinfo{person}{Ahmad
  Hesam}, \bibinfo{person}{Jean de Montigny}, \bibinfo{person}{Vasileios
  Vavourakis}, \bibinfo{person}{Alexandros Iosif}, \bibinfo{person}{Jack
  Jennings}, \bibinfo{person}{Marcus Kaiser}, \bibinfo{person}{Marco Manca},
  \bibinfo{person}{Alberto Di~Meglio}, \bibinfo{person}{Zaid Al-Ars},
  \bibinfo{person}{Fons Rademakers}, \bibinfo{person}{Onur Mutlu}, {and}
  \bibinfo{person}{Roman Bauer}.} \bibinfo{year}{2021}\natexlab{}.
\newblock \showarticletitle{{BioDynaMo: a modular platform for high-performance
  agent-based simulation}}.
\newblock \bibinfo{journal}{\emph{Bioinformatics}} \bibinfo{volume}{38},
  \bibinfo{number}{2} (\bibinfo{date}{09} \bibinfo{year}{2021}),
  \bibinfo{pages}{453--460}.
\newblock
\showISSN{1367-4803}
\urldef\tempurl \url{https://doi.org/10.1093/bioinformatics/btab649}
\showDOI{\tempurl}\\
Supplementary Materials: \url{https://doi.org/10.5281/zenodo.6463816}

\end{itemize}

\section{Other Contributions}
\label{app:contribution:other}

\begin{itemize}

\item
\bibfield{author}{\bibinfo{person}{Tobias Duswald}, \bibinfo{person}{\underline{Lukas
  Breitwieser}}, \bibinfo{person}{Thomas Thorne}, \bibinfo{person}{Barbara
  Wohlmuth}, {and} \bibinfo{person}{Roman Bauer}.}
  \bibinfo{year}{2024}\natexlab{a}.
\newblock \showarticletitle{Calibration of stochastic, agent-based neuron
  growth models with approximate Bayesian computation}.
\newblock \bibinfo{journal}{\emph{Journal of Mathematical Biology}}
  \bibinfo{volume}{89}, \bibinfo{number}{5} (\bibinfo{date}{08 Oct}
  \bibinfo{year}{2024}), \bibinfo{pages}{50}.
\newblock
\showISSN{1432-1416}
\urldef\tempurl \url{https://doi.org/10.1007/s00285-024-02144-2}
\showDOI{\tempurl}

\item
\bibfield{author}{\bibinfo{person}{Ahmad Hesam}, \bibinfo{person}{Frank~P
  Pijpers}, \bibinfo{person}{\underline{Lukas Breitwieser}}, \bibinfo{person}{Peter
  Hofstee}, {and} \bibinfo{person}{Zaid Al-Ars}.}
  \bibinfo{year}{2024}\natexlab{}.
\newblock \showarticletitle{Country-Wide Agent-Based Epidemiological Modeling
  Using 17 Million Individual-Level Microdata}.
\newblock \bibinfo{journal}{\emph{medRxiv}} (\bibinfo{year}{2024}).
\newblock
\urldef\tempurl \url{https://doi.org/10.1101/2024.05.27.24307982}
\showDOI{\tempurl}

\item
\bibfield{author}{\bibinfo{person}{Jack~L. Jennings}, \bibinfo{person}{Sanja
  Bojic}, \bibinfo{person}{\underline{Lukas Breitweiser}}, \bibinfo{person}{Alex Sharp},
  {and} \bibinfo{person}{Roman Bauer}.} \bibinfo{year}{2022}\natexlab{}.
\newblock \showarticletitle{Computational modelling and optimisation of slow
  cooling profiles for the cryopreservation of cells in suspension}.
\newblock \bibinfo{journal}{\emph{Cryobiology}}  \bibinfo{volume}{109}
  (\bibinfo{date}{Dec.} \bibinfo{year}{2022}), \bibinfo{pages}{21--22}.
\newblock
\showISSN{0011-2240}
\urldef\tempurl \url{https://doi.org/10.1016/j.cryobiol.2022.11.068}
\showDOI{\tempurl}

\item
\bibfield{author}{\bibinfo{person}{Jean {de Montigny}},
  \bibinfo{person}{Alexandros Iosif}, \bibinfo{person}{\underline{Lukas Breitwieser}},
  \bibinfo{person}{Marco Manca}, \bibinfo{person}{Roman Bauer}, {and}
  \bibinfo{person}{Vasileios Vavourakis}.} \bibinfo{year}{2021}\natexlab{b}.
\newblock \showarticletitle{An in silico hybrid continuum-/agent-based
  procedure to modelling cancer development: Interrogating the interplay
  amongst glioma invasion, vascularity and necrosis}.
\newblock \bibinfo{journal}{\emph{Methods}}  \bibinfo{volume}{185}
  (\bibinfo{year}{2021}), \bibinfo{pages}{94--104}.
\newblock
\showISSN{1046-2023}
\urldef\tempurl \url{https://doi.org/10.1016/j.ymeth.2020.01.006}
\showDOI{\tempurl}
\newblock
\shownote{Methods on simulation in biomedicine}.

\item
\bibfield{author}{\bibinfo{person}{Ahmad Hesam}, \bibinfo{person}{\underline{Lukas
  Breitwieser}}, \bibinfo{person}{Fons Rademakers}, {and} \bibinfo{person}{Zaid
  Al-Ars}.} \bibinfo{year}{2021}\natexlab{}.
\newblock \showarticletitle{GPU Acceleration of 3D Agent-Based Biological
  Simulations}. In \bibinfo{booktitle}{\emph{2021 IEEE International Parallel
  and Distributed Processing Symposium Workshops (IPDPSW)}}.
  \bibinfo{publisher}{{IEEE}}, \bibinfo{address}{New York, NY, USA},
  \bibinfo{pages}{210--217}.
\newblock
\urldef\tempurl \url{https://doi.org/10.1109/IPDPSW52791.2021.00040}
\showDOI{\tempurl}

\item
\bibfield{author}{\bibinfo{person}{Jack Jennings}, \bibinfo{person}{Roman
  Bauer}, {and} \bibinfo{person}{\underline{Lukas Breitwieser}}.}
  \bibinfo{year}{2019}\natexlab{}.
\newblock \showarticletitle{Computational Modelling Of Slow Freezing Using The
  Biodynamo Software Package Cryodynamo}.
\newblock \bibinfo{journal}{\emph{Cryobiology}}  \bibinfo{volume}{91}
  (\bibinfo{year}{2019}), \bibinfo{pages}{191--192}.
\newblock
\showISSN{0011-2240}
\urldef\tempurl \url{https://doi.org/10.1016/j.cryobiol.2019.10.175}
\showDOI{\tempurl}

\end{itemize}

\cleardoublepage \chapter{Selected \bdm{} Simulations Without the Author's Involvement}
\label{app:complex-simulations}

\begin{figure}[h]
  \centering
\includegraphics[width=\linewidth]{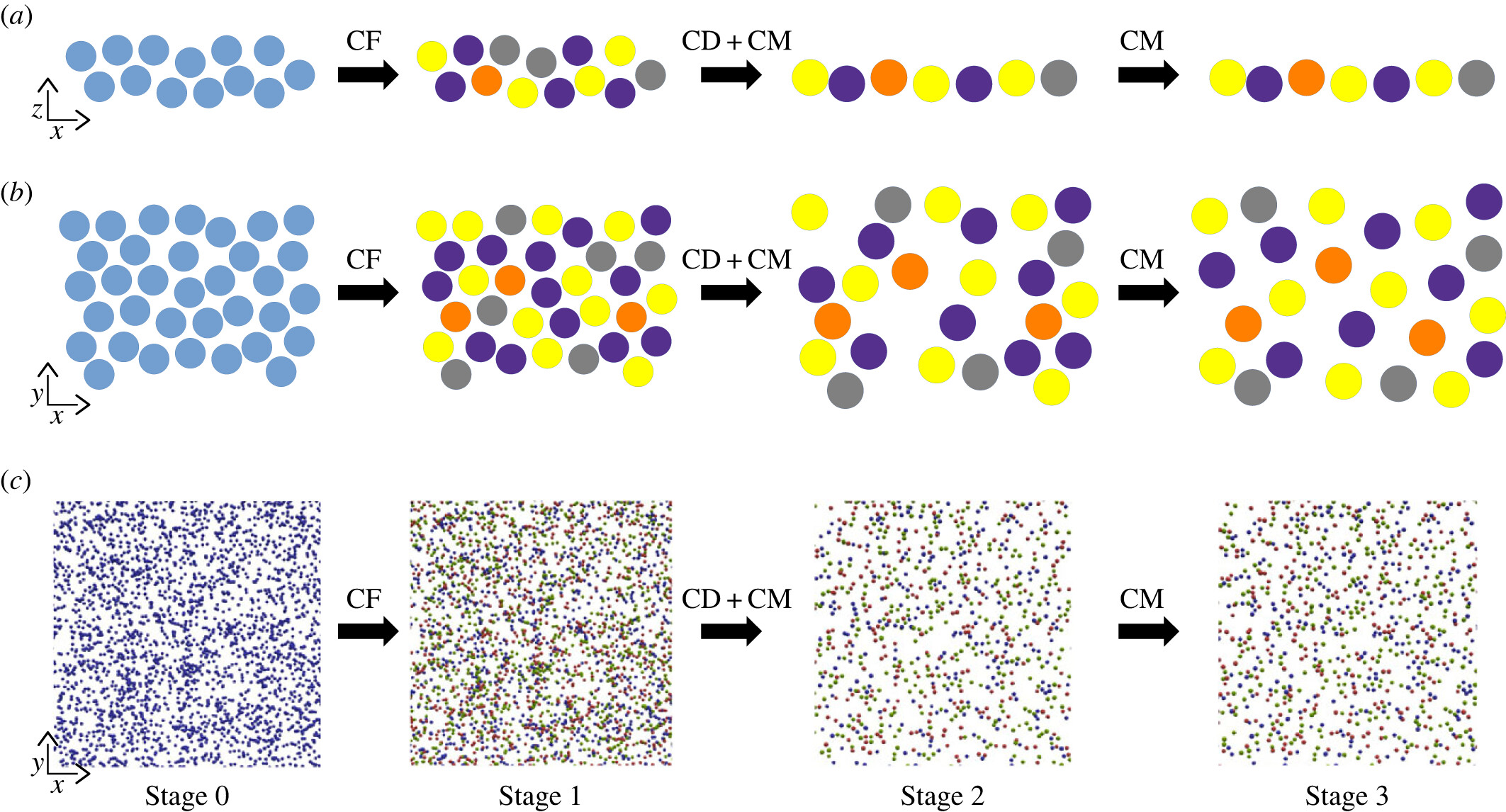}
\caption{
    Retinal mosaics are a specific spatial configuration of cells essential for the eye's function. 
  This model from Jean de Montigny \etal{} simulates different mechanisms that could be responsible for the development of retinal mosaics \cite{demontigny_2023}. 
CF: cell fate, CD: cell death, CM: cell migration.
  Figure taken from 
  \cite{demontigny_2023} 
  without modification and used under 
  \href{https://creativecommons.org/licenses/by-sa/4.0/}{CC BY-SA 4.0}.}
  \label{fig:app:gallery:retinal-mosaics}
\end{figure}

\begin{figure}
  \centering
\includegraphics[width=\linewidth]{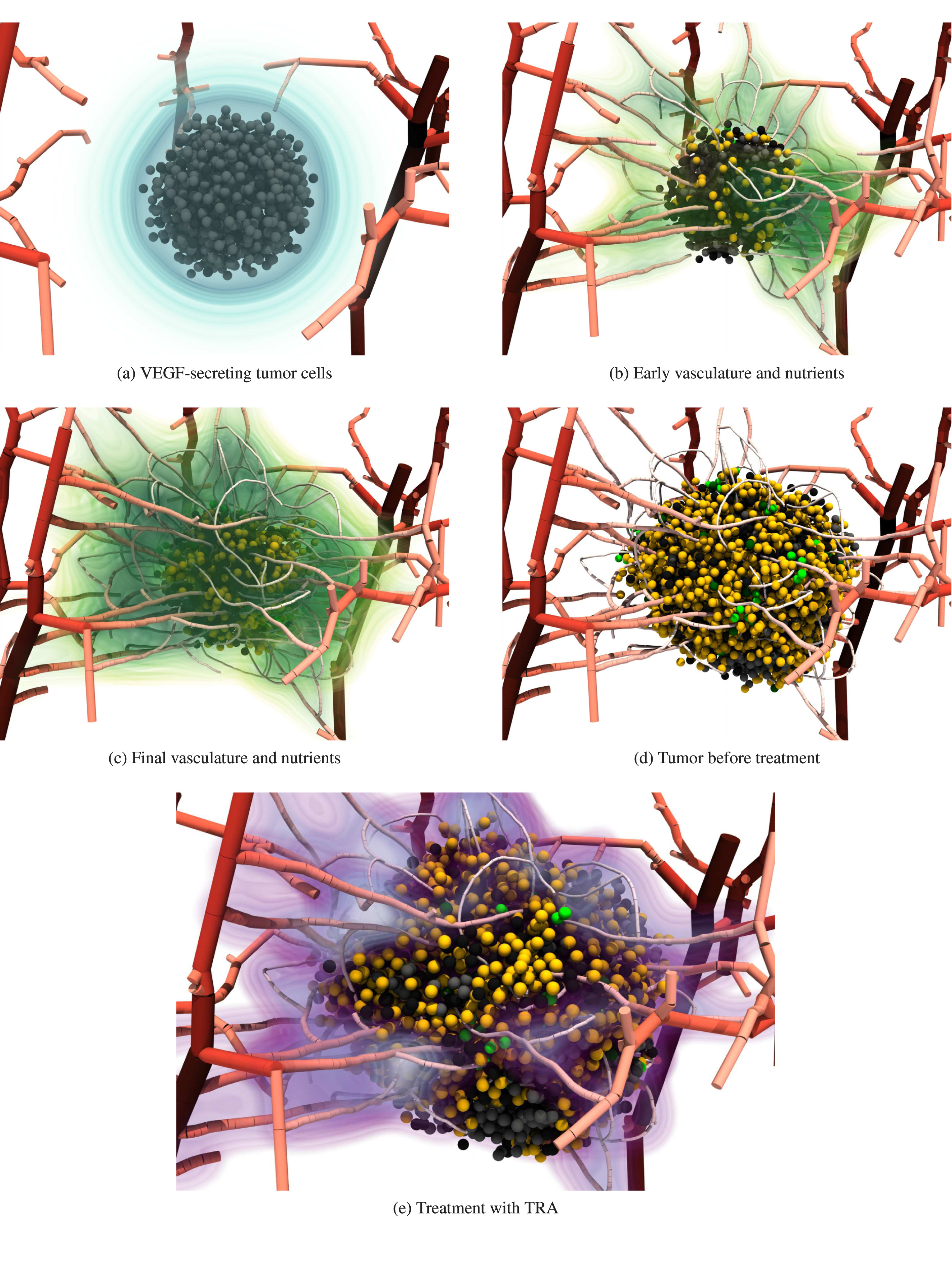}
\caption{Model simulating vascular tumor growth and treatment 
   from Tobias Duswald \etal{} \cite{duswald_2024}. 
  Figure taken from
  \cite{duswald_2024}
  without modification and used under 
  \href{https://creativecommons.org/licenses/by/4.0/}{CC BY 4.0}.}
  \label{fig:app:gallery:bridging-scales}
\end{figure}

\begin{figure}
  \centering
\includegraphics[width=\linewidth]{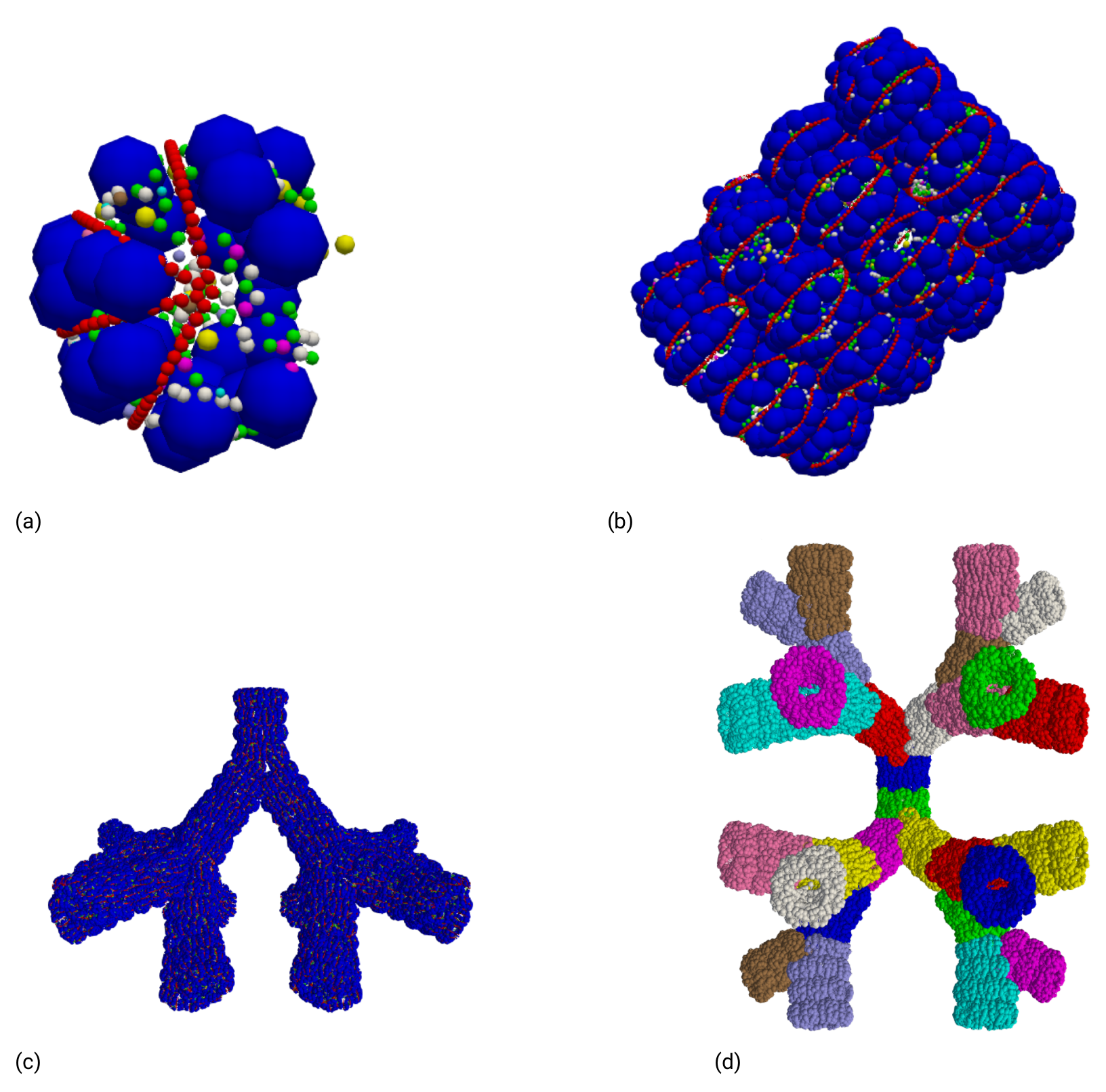}
\caption{
    Lung tissue model from Nicol{\`o} Cogno's dissertation \cite{cogno_thesis} about radiation-induced lung injuries. 
    (a) Alveolus with nine different cell types, (b) alveolar segment, pulmonary acinus from the front (c), and bottom (d) \cite{cogno_thesis}.
    This work was recognized by PhysicsWorld as one of the top 10 breakthroughs in physics in 2024 \cite{physicsworld-top10}.
  Figure taken from
  \cite{cogno_thesis}
  without modification and used under 
  \href{https://creativecommons.org/licenses/by-sa/4.0/}{CC BY-SA 4.0}.}
  \label{fig:app:gallery:lung-tissue}
\end{figure}

\begin{figure}
  \centering
\includegraphics[width=\linewidth]{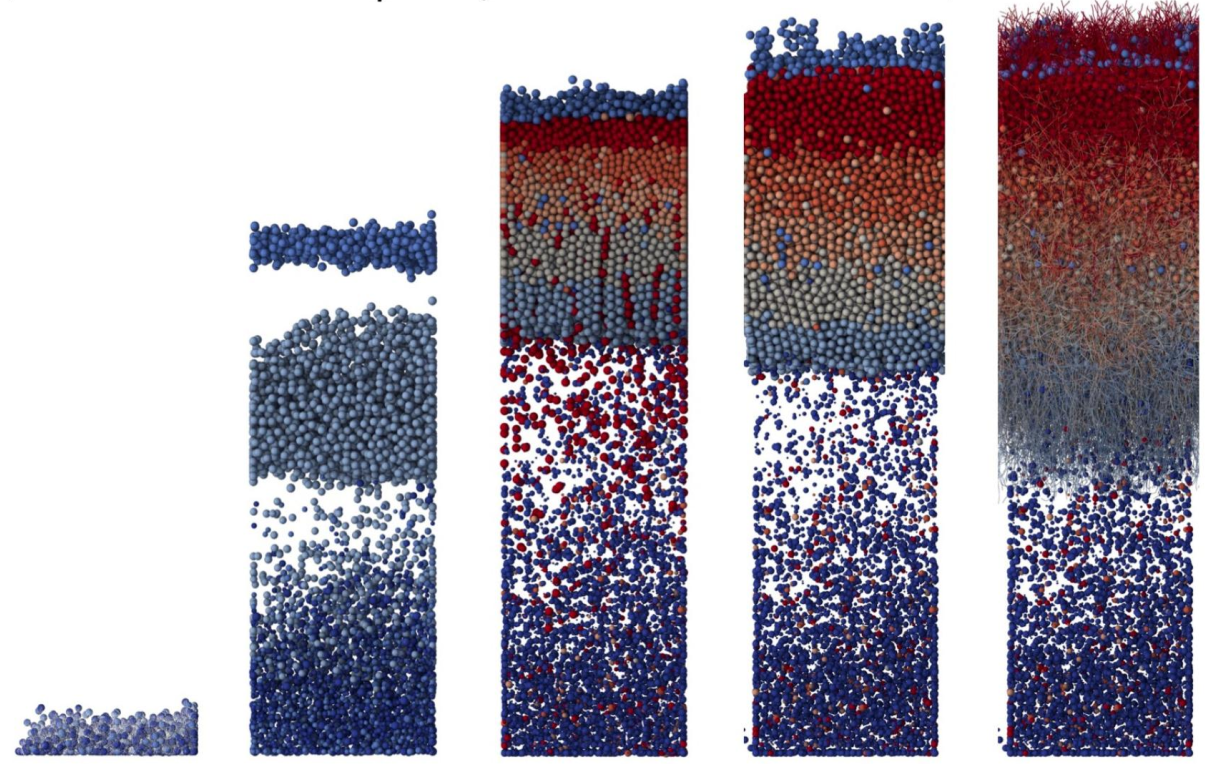}
\caption{Development of the cerebral cortex from Umar Abubacar and Roman Bauer \cite{abubacar_neuronal_growth}.
  Figure taken from 
  \cite{cogno_agent-based_2024} and used under 
  \href{https://creativecommons.org/licenses/by/4.0/}{CC BY 4.0}.
  The figure was extracted from a collage of different agent-based simulation visualizations.
  }
  \label{fig:app:gallery:cortex}
\end{figure}

\begin{figure}
  \centering
\includegraphics[width=\linewidth]{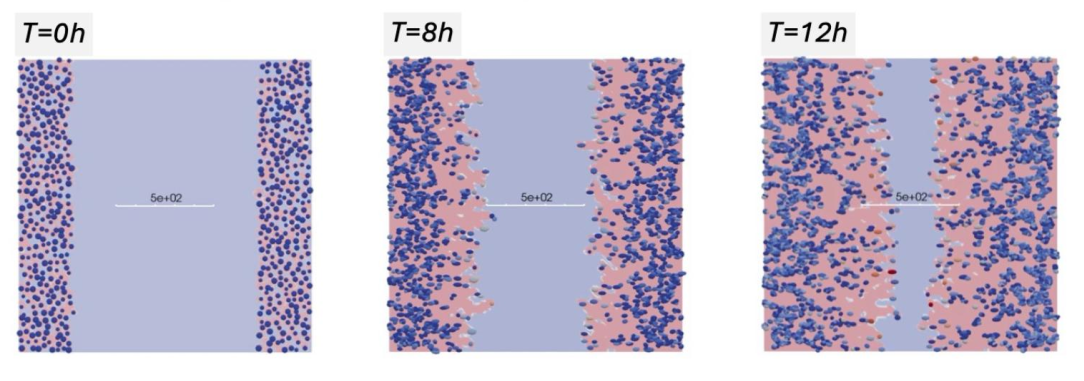}
\caption{Wound healing simulation by Vasileios Vavourakis, which was inspired by the work of \cite{mascheroni_impact_2019}.
  A video of the simulation is available at \cite{wound-healing-video}.
  Figure taken from 
  \cite{cogno_agent-based_2024} and used under 
  \href{https://creativecommons.org/licenses/by/4.0/}{CC BY 4.0}.
  The figure was extracted from a collage of different agent-based simulation visualizations.
  }
  \label{fig:app:gallery:wound-healing}
\end{figure}

\begin{figure}
  \centering
\includegraphics[width=\linewidth]{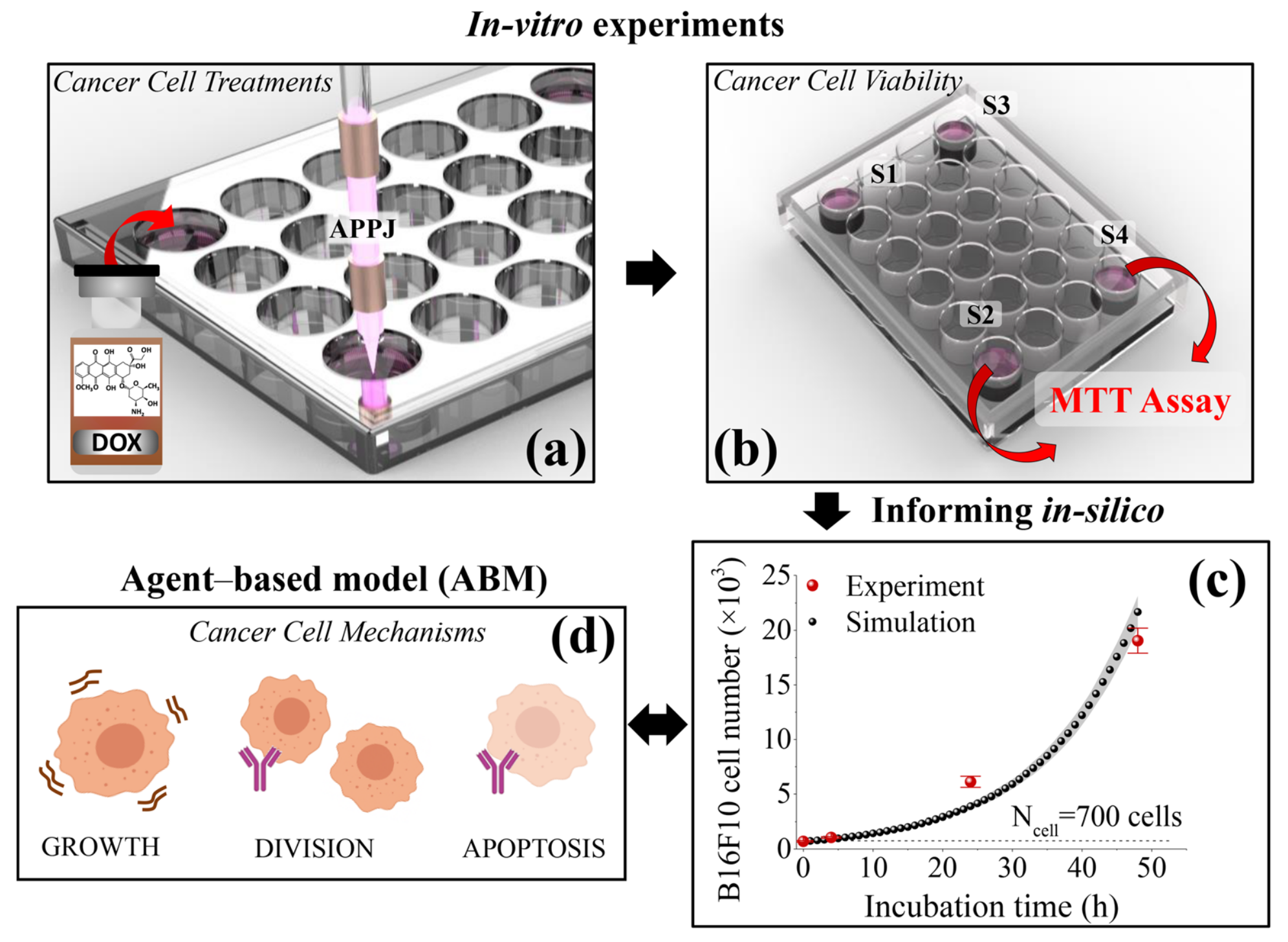}
\caption{
    In silico study of cancer cell treatment with a helium plasma jet (APPJ) and the doxyrobucin drug (DOX) from Kristaq Gazeli \etal{} \cite{gazeli_interrogating_2022}. 
  Figure taken from
  \cite{gazeli_interrogating_2022}
  without modification and used under 
  \href{https://creativecommons.org/licenses/by/4.0/}{CC BY 4.0}.}
  \label{fig:app:gallery:helium-plasma-jet}
\end{figure}

\begin{figure}
  \centering
\includegraphics[width=\linewidth]{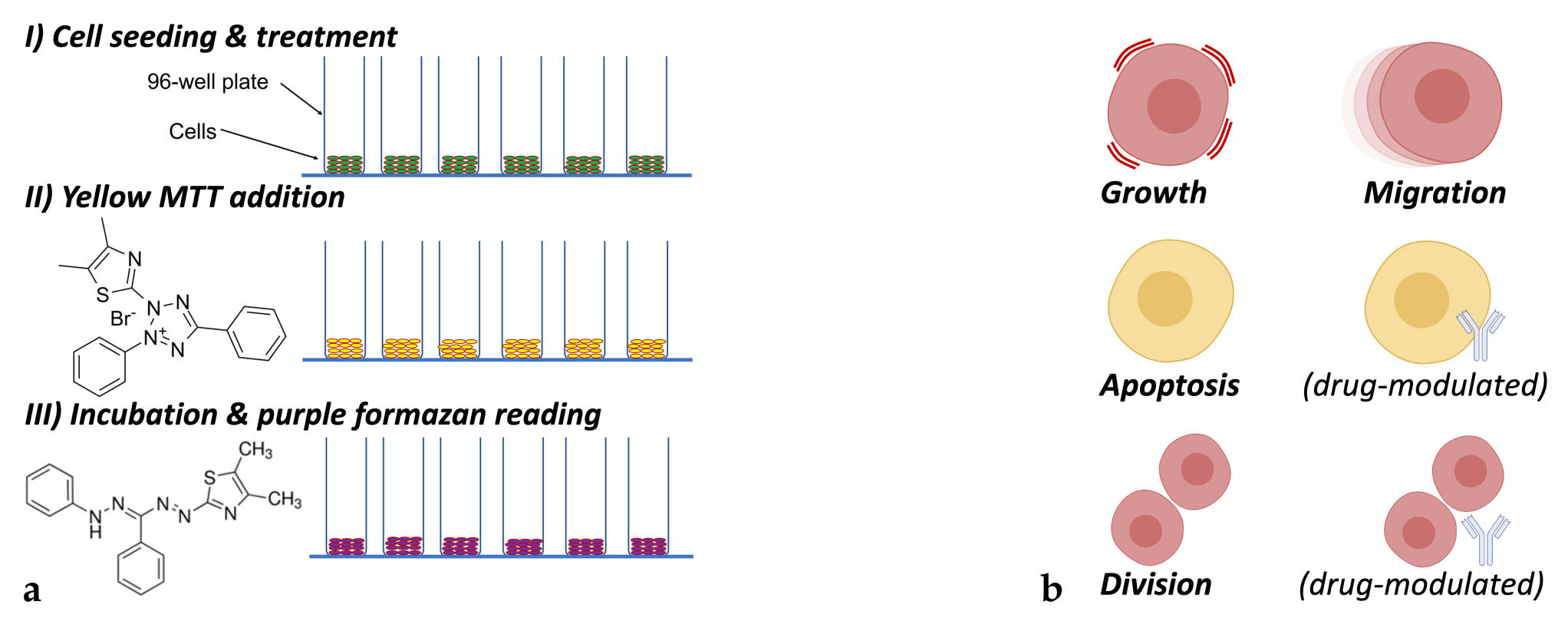}
\caption{
    Overview of the work of Marios Demetriades \etal{} \cite{demetriades_interrogating_2022} to study cancer drug pharmacodynamics by combining in vitro (a) and in silico (b) methods. 
  Figure taken from
  \cite{demetriades_interrogating_2022}
  without modification and used under 
  \href{https://creativecommons.org/licenses/by/4.0/}{CC BY 4.0}.}
  \label{fig:app:gallery:pharmacodynamics}
\end{figure}

This section shows selected simulation visualizations from \bdm{} users, without the author's
  involvement (Figure~\ref{fig:app:gallery:retinal-mosaics}--\ref{fig:app:gallery:pharmacodynamics}).
The simulations demonstrate that the functionality provided in the \bdm{} platform can be used to model
  dynamic systems whose complexity exceeds the presented use cases in this
  dissertation.

\cleardoublepage \chapter{List of Agents, Events, and Operations}
\label{app:table}

\begin{longtable}{@{}lp{11cm}@{}}
\caption{{\bf List of agents, events, and operations that \bdm{} currently provides.}} \\
\toprule
     &
    \textbf{Description} \\
    \midrule
    \textbf{Agents} & \\
    Agent &
    Agent is the base class for all agents in \bdm{}.
    This class has a unique id that remains constant during the whole simulation
      and a collection of behaviors that have been attached to this
      agent.
    Agent contains functions to manage behaviors, and to remove itself
      from the simulation. \\

    SphericalAgent &
    SphericalAgent extends Agent and adds a spherical agent geometry.\\

    Cell &
    Cell extends Agent and represents a generic cell with a spherical shape.
    It includes attributes to describe its geometry, density, and adherence.
    Cell provides member functions to change its volume, move it in space, 
     calculate mechanical forces, and
      divide it into two daughter cells.
    This cell division function creates a new daughter cell and distributes the volume of the
      mother cell according to the volume ratio parameter.
    The position of the daughter cell is determined based on the division axis.\\

    NeuronSoma &
    NeuronSoma extends Cell.
    This class represents the cell body of a neuron and, like Cell, has a
      spherical shape.
    NeuronSoma has a list of neurite elements that extend from the cell body
      together with their attachment points.
    NeuronSoma adds a function to extend a new neurite element from the soma. 
    This function takes two parameters: the diameter of the new neurite
      element, and the orientation of the cylinder in spherical coordinates. 
    \cite{zublerdouglas2009framework} contains more details.\\

    NeuriteElement &
    NeuriteElement extends Agent and has a cylindrical shape with a proximal
      and distal end.
    A dendrite is modeled as a binary tree of neurite elements that are
      internally connected with springs to transmit forces to its proximal
      connection.
    The proximal connection can be a NeuriteSoma or a NeuriteElement.
    This class contains attributes to describe the cylindrical geometry,
      the spring, pointers to its proximal and distal connections, density, and adherence.
    A NeuriteElement can elongate, retract, split, branch, bifurcate, and extend a new
      side neurite.
    The following list describes these functions in more detail.  
    \begin{itemize}
      \item \emph{Split neurite element.}
              This function splits a neurite element into two segments.
              The neurite element whose split neurite element function was called becomes the distal one.
              The new neurite element will be the proximal one.
      \item \emph{Extend side neurite.}
              This function adds a side neurite, if one of the distal connections is empty.
      \item \emph{Bifurcate.}
              This function creates two new neurite elements and assigns them to the distal
                connections.
              This function can only be called for terminal neurite segments because both
                connections must be empty.
      \item \emph{Branch.}
              This function splits the current neurite element into two elements
                and adds a new side branch at the proximal segment.
              It is, therefore, a combination of split neurite element and
                extend side neurite.
    \end{itemize}
    \cite{zublerdouglas2009framework} contains more details.
    \\

    \midrule
    \textbf{Behaviors} & \\
    
    Chemotaxis &
    This behavior moves agents along the diffusion gradient (from low concentration to high). \\

    Secretion &
    This behavior increases the substance concentration at the position of the agent.\\

    GrowthDivision &
    This behavior grows cells to a specific size and divides them if they exceed the threshold.
    \\

    GeneRegulation &
    This behavior calculates protein concentrations which are defined as differential equations.\\

    StatelessBehavior &
    This behavior reduces the amount of code that has to be written for behaviors without attributes.\\

    \midrule
    \textbf{Agent operations} & \\

    BehaviorOp &
    This operation runs all behaviors which are attached to the agent. \\

    BoundSpaceOp & 
    This operation enforces the space boundary condition (see Section~\ref{sec:space-boundary}) which can be open, closed or toroidal.\\

    DiscretizationOp &
    This operation calls the agent's discretization function. NeuriteElement uses this function to split itself if it becomes too long, or merge with another segment if it is too short.\\

    MechanicalForcesOp &
    This operation calls the agent's calculate displacement function and moves the agent accordingly. The calculate displacement function contains the implementation how an agent moves based on all forces that act on it.\\

    \midrule
    \textbf{Standalone operations} & \\

    DiffusionOp &
    This operation calls the update function of all substances in the simulation.\\

    UpdateEnvironmentOp &
    This operation calls the update function of the environment algorithm.\\

    UpdateTimeSeriesOp &
    This operation calls the update function of the TimeSeries object which collects relevant data from the current iteration which can be analysed at the end of the simulation.\\

    VisualizationOp &
    This operation updates the live visualization or generates visualization files for later analysis.\\

    \bottomrule
\label{tab:biological-features}
\end{longtable}
 
\cleardoublepage \chapter{Supplementary Tutorials}
\label{app:tutorials}

\includepdfset{pagecommand=\thispagestyle{plain}}



\section*{Supplementary material}

# Generate random samples from a user-defined distribution

**Author: Lukas Breitwieser**

In this tutorial we demonstrate how to create a random number generator that draws samples from a user-defined distribution.

Let's start by setting up BioDynaMo notebooks

In [1]:

```
%jsroot on
gROOT->LoadMacro("${BDMSYS}/etc/rootlogon.C");
```

INFO: Created simulation object 'simulation' with UniqueName='simulation'.

We have to create a `TCanvas` object to draw results in this notebook.

In [2]:

```
TCanvas c("", "", 400, 300);
c.SetGrid();
```

Let's assume that we want to generate random numbers from a student-t distribution.

Class `Random` (https://biodynamo.org/api/classbdm_1_1Random.html) does not provide a direct function for that.

Therefore, we use the user-defined distribution feature `Random::GetUserDefinedDistRng1D`.

Fortunately, ROOT already provides a function called `tdistribution_pdf` that we can use.

Have a look at the following two links for more math functions: TMath (https://root.cern/doc/master/namespaceTMath.html) and statistical functions (https://root.cern/doc/master/group__StatFunc.html)

In [3]:

```
auto* random = simulation.GetRandom();
auto distribution = [](const double* x, const double* param) {
    return ROOT::Math::tdistribution_pdf(*x, 1.0);
};
auto udd_rng = random->GetUserDefinedDistRng1D(distribution, {}, -5, 10);
```

The returned random number generator has a function to draw the distribution.

```
udd_rng.Draw();
c.Draw();
```

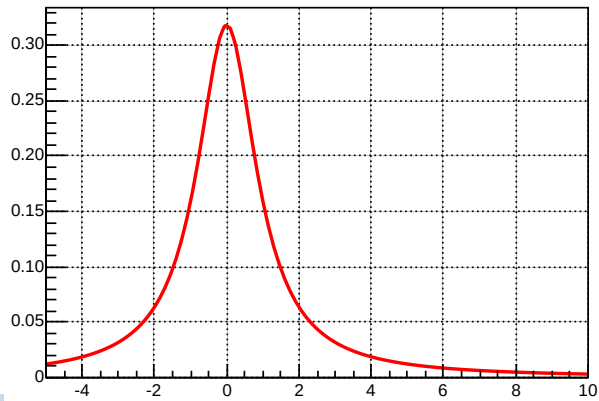

In the next step we want to verify that the created random number generator follows the specified distribution. Therefore, we create a histogram with 100 bins in the range [-5, 10] and fill it with 10000 samples.

In [5]:

```
TH1F h("",""", 100, -5, 10);
for (int i = 0; i < 10000; ++i){
    auto rndm_sample = udd_rng.Sample();
    h.Fill(rndm_sample);
}
```

Let's draw the result:

In [6]:

```
h.SetFillColor(kBlue-10);
h.Draw();
c.Draw();
```

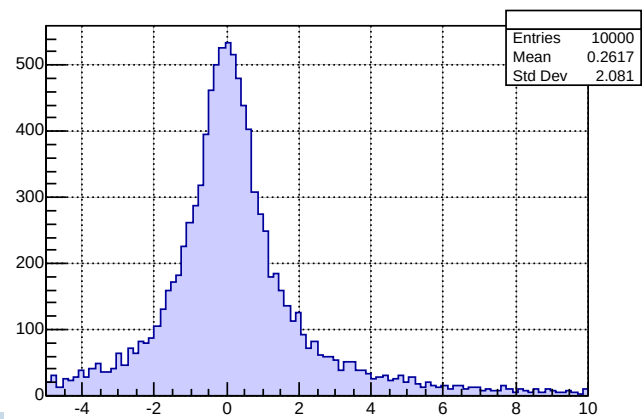

As we can see from the last figure the samples from our random number generator fit our distribution.

# Agent reproduction and mortality

**Author: Lukas Breitwieser**

In this tutorial we want to demonstrate how to add and remove agents from the simulation.

Let's start by setting up BioDynaMo notebooks.

In [1]:

```
%jsroot on
gROOT->LoadMacro("${BDMSYS}/etc/rootlogon.C");
```

INFO: Created simulation object 'simulation' with UniqueName='simulation'.

In [2]:

```
auto* ctxt = simulation.GetExecutionContext();
auto* scheduler = simulation.GetScheduler();
auto* rm = simulation.GetResourceManager();
```

Let's define our initial model: One cell at origin.
We also create an agent pointer for our cell, because raw pointers might be invalidated after a call to
`Scheduler::Simulate`

In [3]:

```
auto* cell = new Cell();
ctxt->AddAgent(cell);
auto cell_aptr = cell->GetAgentPtr<Cell>();
scheduler->FinalizeInitialization();
VisualizeInNotebook();
```

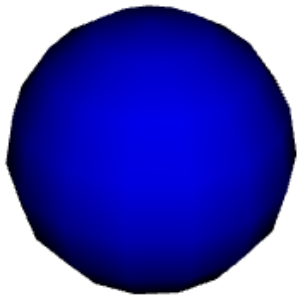

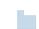

Adding an agent to the simulation is as easy as constructing one and adding it to the execution context.
Our default execution context will add the new agent to the simulation at the end of the iteration.
Therefore, the visualization still shows only one agent.

```
ctxt->AddAgent(new SphericalAgent({3, 0, 0}));
VisualizeInNotebook()
```

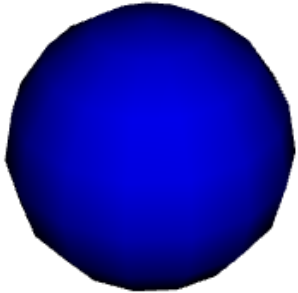

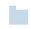

Let's simulate one time step and see what happens.

```
scheduler->Simulate(1);
VisualizeInNotebook()
```

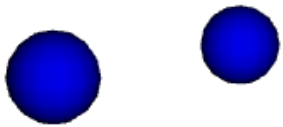

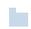

Our new agent has been added to the simulation.

Usually, new agents will be created based on some process, e.g. cell division, neurite extension from soma, neurite branching, etc. The following example shows cell division. We specify the division axis.

```
cell_aptr->Divide({1, 0, 0});
VisualizeInNotebook()
```

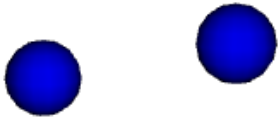

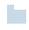

Again, the new cell is not visible yet. We have to finish one iteration.

```
scheduler->Simulate(1);
VisualizeInNotebook()
```

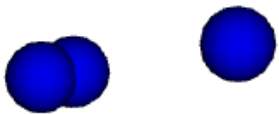

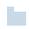

Removing agents from the simulation works similarly.
The default execution context will remove it at the end of the iteration.

```
cell_aptr->RemoveFromSimulation();
rm->GetNumAgents();
VisualizeInNotebook()
```

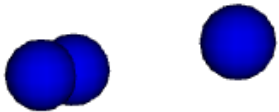

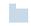

We expect that after the `Simulate` call only 2 agents are shown in the visualization.

In [9]:

```
scheduler->Simulate(1);
VisualizeInNotebook()
```

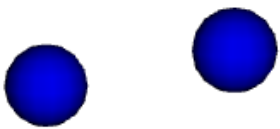

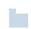

# Agent reproduction with behaviors

**Author: Lukas Breitwieser**

In tutorial `ST3-agent-reproduction-mortality` we have explored how to add and remove agents from the simulation. In this tutorial we want to explore different behavior options to control if a new agent gets a behavior from the original agent, and if a behavior will be removed from the original one.

Let's start by setting up BioDynaMo notebooks.

In [1]:

```
%jsroot on
gROOT->LoadMacro("${BDMSYS}/etc/rootlogon.C");
```

INFO: Created simulation object 'simulation' with UniqueName='simulati
on'.

We define a simple behavior which prints `has print behavior`.

In [2]:

```
StatelessBehavior print_behavior([](Agent* agent) {
  std::cout << " has print behavior" << std::endl;
});
```

We define the following experiment which we will run with different options of the `print_behavior`.
We create a cell, add a copy of the `print_behavior`, and run all behaviors. We expect that the following output is created.

```
    mother:
      has print behavior
```

Afterwards we print a seperator `-------` to indicate cell division, divide the mother cell and run the behaviors of daughter 1 and daughter 2. By definition the original mother cell turns into daugther 1 and the new agent becomes daughter 2.

In [3]:

```
void Experiment() {
    Simulation sim("my-simulation");
    auto* mother = new Cell();
    mother->AddBehavior(print_behavior.NewCopy());
    std::cout << "mother: " << std::endl;
    mother->RunBehaviors();
    std::cout << "---------------------" << std::endl;
    auto* daughter2 = mother->Divide();
    std::cout << "mother = daughter 1: " << std::endl;
    mother->RunBehaviors(); // mother = daughter 1
    std::cout << "daughter 2: " << std::endl;
    daughter2->RunBehaviors();
}
```

Let's run the experiment with default parameters and see what happens.

```
Experiment();
```

```
mother:
  has print behavior
--------------------
mother = daughter 1:
  has print behavior
daughter 2:
```

The `print_behavior` was **not copied** to the daughter 2 cell and was **not removed** from the mother cell.

---

Let's try to copy the behavior from the mother cell to daughter 2.

```
print_behavior.AlwaysCopyToNew();
Experiment();
```

```
mother:
  has print behavior
--------------------
mother = daughter 1:
  has print behavior
daughter 2:
  has print behavior
```

Now the `print_behavior` **was copied** to the daughter 2 cell and was **not removed** from the mother cell.

---

Let's try to remove the behavior from the mother cell.

```
print_behavior.AlwaysCopyToNew();
print_behavior.AlwaysRemoveFromExisting();
Experiment();
```

```
mother:
  has print behavior
--------------------
mother = daughter 1:
daughter 2:
  has print behavior
```

Now the `print_behavior` **was copied** to the daughter 2 cell and **was removed** from the mother cell.

---

Let's reset the values to the default.

```
print_behavior.NeverCopyToNew();
print_behavior.NeverRemoveFromExisting();
Experiment();
```

```
mother:
  has print behavior
--------------------
mother = daughter 1:
  has print behavior
daughter 2:
```

Behaviors provide also more fine-grained distinction. Some agents support multiple [new agent events (https://biodynamo.org/docs/userguide/new_agent_event/)](https://biodynamo.org/docs/userguide/new_agent_event/): neurite branching, neurite bifurcation, side neurite extension, etc. For each event we can specify if the behavior should be copied to the new, or removed from the existing agent.

```
print_behavior.CopyToNewIf({CellDivisionEvent::kUid});
print_behavior.RemoveFromExistingIf({CellDivisionEvent::kUid});
Experiment();
```

```
mother:
  has print behavior
--------------------
mother = daughter 1:
daughter 2:
  has print behavior
```

# Agent reproduction advanced

**Author: Lukas Breitwieser**

In the tutorials so far we used `Cell::Divide` to create new agents. In this demo we want to show how to define your own "process" that creates a new agent. Furthermore, we will explain the purpose of the functions `Agent::Initialize` and `Agent::Update`.

Assume that we want to create a new agent type `Human` which should be able to `GiveBirth`.

Let's start by initializing BioDynaMo notebooks.

In [1]:

```
%jsroot on
gROOT->LoadMacro("${BDMSYS}/etc/rootlogon.C");
```

INFO: Created simulation object 'simulation' with UniqueName='simulation'.

In [2]:

```
auto* ctxt = simulation.GetExecutionContext();
auto* scheduler = simulation.GetScheduler();
```

Let's start by creating the `ChildBirthEvent`. In this example we do not need any attributes.

In [3]:

```
struct ChildBirthEvent : public NewAgentEvent {
  ChildBirthEvent() {}
  virtual ~ChildBirthEvent() {}
  NewAgentEventUid GetUid() const override {
      static NewAgentEventUid kUid =
          NewAgentEventUidGenerator::GetInstance()->GenerateUid();
      return kUid;
  }
};
```

We continue by defining the class `Human` which derives from `SphericalAgent`.

In [4]:

```
class Human : public SphericalAgent {
  BDM_AGENT_HEADER(Human, SphericalAgent, 1);

 public:
  Human() {}
  explicit Human(const Double3& position) : Base(position) {}
  virtual ~Human() {}

  void GiveBirth();
  void Initialize(const NewAgentEvent& event) override;
};
```

The implementation of `GiveBirth` only requires two lines of code.

```
void Human::GiveBirth() {
    ChildBirthEvent event;
    CreateNewAgents(event, {this});
}
```

First, creating an instance of the event.
Second, invoking `CreateNewAgents` function which is defined in class `Agent`.

The first parameter of `CreateNewAgents` takes an event object, and the second a vector of agent prototypes. The size of this vector determines how many new agents will be created. In our case: one. If twins should be born we could change it to `CreateNewAgents(event, {this, this});`.

But why do we have to pass a list of agent pointers to the function?

The answer is simple: we have to tell `CreateNewAgents` which agent type it should create. In our use case we want to create another instance of class `Human`. Therefore, we pass the `this` pointer.

The only part missing is to tell BioDynaMo how to initialize the attributes of the new child. This decision is encapsulated in the `Initialize` function which we override from the base class. Don't forget to also call the implementation of the base class using `Base::Initialize(event)`. Otherwise the intialization of the base class is skipped.

In our example we define that the child should be created next to the mother in 3D space.

```
void Human::Initialize(const NewAgentEvent& event) {
    Base::Initialize(event);
    auto* mother = bdm_static_cast<Human*>(event.existing_agent);
    SetPosition(mother->GetPosition() + Double3{2, 0, 0});
}
```

This concludes all required building blocks. Let's try it out!

```
auto* human = new Human();
ctxt->AddAgent(human);
```

```
human->GiveBirth();
```

```
scheduler->Simulate(1);
VisualizeInNotebook();
```

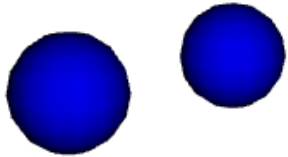

As expected the simulation consists of two "humans".

---

Let's take this one step further. Let's assume that class `Human` was provided in a library that we don't want to modify. However, we want to add two more attributes:

- the number of offsprings
- the mitochondiral dna (Note: the mitochondrial dna is inherited solely from the mother)

Let's create a new class called `MyHuman` which derives from `Human` and which adds these two attributes.

In [10]:

```
using MitochondrialDNA = int;
```

In [11]:

```
class MyHuman : public Human {
  BDM_AGENT_HEADER(MyHuman, Human, 1);

 public:
  MyHuman() {}
  explicit MyHuman(const Double3& position) : Base(position) {}
  virtual ~MyHuman() {}

  void Initialize(const NewAgentEvent& event) override;
  void Update(const NewAgentEvent& event) override;

  int num_offsprings_ = 0;
  MitochondrialDNA mdna_;
};
```

As in the example above, the `Initialize` method is used to set the attributes during new agent events. In this example, we have to set the mitochondrial dna of the child to the value from the mother. The following function definition does exactly that and prints out the value.

```cpp
void MyHuman::Initialize(const NewAgentEvent& event) {
    Base::Initialize(event);
    auto* mother = bdm_static_cast<MyHuman*>(event.existing_agent);
    mdna_ = mother->mdna_;
    std::cout << "Initialize child attributes: mitochondrial dna set to "
              << mdna_ << std::endl;
}
```

The only task left is to update the attributes of the mother. This is done by overriding the `Update` method. Again, do not forget to call the implementation of the base class for correctness. We increment the `num_offsprings_` attribute by the number of newly created agents. Although we could just have incremented the attribute by one, the solution below is generic enough to handle e.g. twin births.

```cpp
void MyHuman::Update(const NewAgentEvent& event) {
    Base::Update(event);
    num_offsprings_ += event.new_agents.size();
    std::cout << "Update mother attributes: num_offsprings incremented to "
              << num_offsprings_ << std::endl;
}
```

Let's create a new `MyHuman`, set its mitochondrial dna to `123` and output the current value of `num_offsprings_`, which we expect to be `0`.

```cpp
auto* my_human = new MyHuman();
my_human->mdna_ = 123;
my_human->num_offsprings_
```

```
(int) 0
```

Now we can call `GiveBirth` again. We expect the output of two lines.

- The first coming from the child informing us about the initialization of its `mdna_` attribute
- and the second from the mother telling us about the update of `num_offsprings_`

```cpp
my_human->GiveBirth();
```

```
Initialize child attributes: mitochondrial dna set to 123
Update mother attributes: num_offsprings incremented to 1
```

To double check, let's output the value of `num_offsprings`, which we expect to be 1

```cpp
my_human->num_offsprings_
```

```
(int) 1
```

# Environment search

**Author: Lukas Breitwieser**

In this tutorial we will show how to execute a function for each neighbor of an agent.

Let's start by setting up BioDynaMo notebooks.

```
%jsroot on
gROOT->LoadMacro("${BDMSYS}/etc/rootlogon.C");
```

```
INFO: Created simulation object 'simulation' with UniqueName='simulati
on'.
```

We create three agents in a row along the x-axis with identical y and z values.

```
auto* ctxt = simulation.GetExecutionContext();

auto* a0 = new SphericalAgent({10, 0, 0});
auto* a1 = new SphericalAgent({20, 0, 0});
auto* a2 = new SphericalAgent({30, 0, 0});

a0->SetDiameter(11);
a1->SetDiameter(11);
a2->SetDiameter(11);

ctxt->AddAgent(a0);
ctxt->AddAgent(a1);
ctxt->AddAgent(a2);
```

We finalize the initialization and update the environment so it can be used later. Please not that this is usually done automatically inside `Scheduler::Simulate`.

```
simulation.GetScheduler()->FinalizeInitialization();
simulation.GetEnvironment()->Update();
VisualizeInNotebook();
```

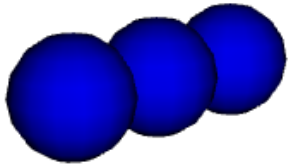

Let's define the function that we want to execute for each neighbor. It prints the unique id of the neighbor and its distance from the querying agent.

In [4]:

```
auto print_id_distance = L2F([](Agent* a, double squared_distance) {
  std::cout << "Neighbor " << a->GetUid() << "  with distance: "
            << std::sqrt(squared_distance) << std::endl;
});
```

The agents have the following ids (in order of increasing x-value) 0-0, 1-0, 2-0

We start by executing print_id_distance for the first agent. We ask for all neighbors within distance 101. Therefore the function should be executed for the agent in the middle with id 1-0

In [5]:

```
ctxt->ForEachNeighbor(print_id_distance, *a0, 101);
```

```
Neighbor 1-0  with distance: 10
```

Let's repeat the experiment for the middle agent. We expect to see two lines for the left and right neighbor.

In [6]:

```
ctxt->ForEachNeighbor(print_id_distance, *a1, 101);
```

```
Neighbor 0-0  with distance: 10
Neighbor 2-0  with distance: 10
```

Lastly, we want to execute the function `print_id_distance` for all neighbors of the righ-most agent. We expect to see one line printing the middle agent as neighbor (1-0)

```
ctxt->ForEachNeighbor(print_id_distance, *a2, 101);
```

```
Neighbor 1-0  with distance: 10
```

# Multi-scale simulations

**Author: Lukas Breitwieser**

In this tutorial we will show how BioDynaMo support multi-scale simulations. Multi-scale simulation means that simulated processes happen in different time-scales---e.g. substance diffusion and neurite growth.

Let's start by setting up BioDynaMo notebooks.

In [1]:

```
%jsroot on
gROOT->LoadMacro("${BDMSYS}/etc/rootlogon.C");
```

INFO: Created simulation object 'simulation' with UniqueName='simulation'.

We define a new [standalone operation (https://biodynamo.org/docs/userguide/operation/)](https://biodynamo.org/docs/userguide/operation/) which only task is to print the current simulation time step if it is executed.

In [2]:

```
struct TestOp : public StandaloneOperationImpl {
  BDM_OP_HEADER(TestOp);
  void operator()() override {
    auto* scheduler = Simulation::GetActive()->GetScheduler();
    auto* param = Simulation::GetActive()->GetParam();
    std::cout << "Processing iteration "
              << scheduler->GetSimulatedSteps()
              << " simulation time "
              << scheduler->GetSimulatedSteps() * param->simulation_time_step
              << std::endl;
  }
};
OperationRegistry::GetInstance()->AddOperationImpl(
    "test_op", OpComputeTarget::kCpu, new TestOp());
```

In [3]:

```
auto set_param = [](Param * param) {
    param->simulation_time_step = 2;
};
Simulation simulation("my-simulation", set_param);
```

Our initial model consists of one agent at origin.

In [4]:

```
auto* ctxt = simulation.GetExecutionContext();
ctxt->AddAgent(new SphericalAgent());
```

Let's create a new instance of our class `TestOp` and add it to the scheduler.

```
auto* op1 = NewOperation("test_op");
auto* scheduler = simulation.GetScheduler();
scheduler->ScheduleOp(op1);
```

Let's simulate 9 steps. We expect that `op1` will be called each time step.

```
scheduler->Simulate(9);
```

```
Processing iteration 0 simulation time 0
Processing iteration 1 simulation time 2
Processing iteration 2 simulation time 4
Processing iteration 3 simulation time 6
Processing iteration 4 simulation time 8
Processing iteration 5 simulation time 10
Processing iteration 6 simulation time 12
Processing iteration 7 simulation time 14
Processing iteration 8 simulation time 16
```

Operations have a frequency attribute which specifies how often it will be executed. An operation with frequency one will be executed at every time step; an operation with frequency two every second, and so on.

```
op1->frequency_ = 3;
scheduler->Simulate(9);
```

```
Processing iteration 9 simulation time 18
Processing iteration 12 simulation time 24
Processing iteration 15 simulation time 30
```

This functionality can be used to set the frequency of different processes in an agent-based model.

# Create a histogram of agent attributes

**Author: Lukas Breitwieser**
In this tutorial we will show how to create a histogram of all agent diameters in the simulation and fit a function to the data.

Let's start by setting up BioDynaMo notebooks.

In [1]:

```
%jsroot on
gROOT->LoadMacro("${BDMSYS}/etc/rootlogon.C");
```

INFO: Created simulation object 'simulation' with UniqueName='simulation'.

We want to define a function that creates a cell at a certain position with diameters drawn from a gaussian distribution with $\mu = 20$ and $\sigma = 5$. The smallest diameter should be larger then $2.0$.

In [2]:

```
simulation.GetResourceManager()->ClearAgents();
auto rng = simulation.GetRandom()->GetGausRng(20, 5);
auto create_cell = [&](const Double3& position) {
  Cell* cell = new Cell(position);
  double diameter = std::max(2.0, rng.Sample());
  cell->SetDiameter(diameter);
  return cell;
};
```

Now that we defined `create_cell` we can use it to create 400 cells on a plane with $z = 0$, $xmin = ymin = -200$, $xmax = ymax = 200$, and spacing = 20 in both dimensions.

```
auto f = [](const double* x, const double* params) { return 0.0; };
ModelInitializer::CreateAgentsOnSurface(f, {}, -200, 200, 20, -200, 200, 20,
                                        create_cell);
simulation.GetScheduler()->FinalizeInitialization();
VisualizeInNotebook(300, 300);
```

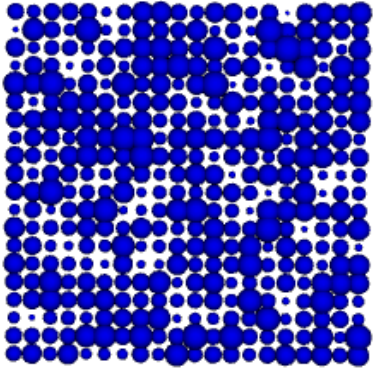

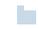

The next step is to create a histogram object with 100 bins in the interval [2, 40].
The second line creates a function which fills the histogram with the diameter of the given agent.
The third line calls the function `fill` for each agent, thus adding all diameters to the histogram.

```
TH1F h("myHisto","Agent Diameter Histogram;Diameter;Count", 100, 2, 40);
auto fill = L2F([&](Agent* a, AgentHandle){ h.Fill(a->GetDiameter()); });
simulation.GetResourceManager()->ForEachAgent(fill);
```

Let's draw the final histogram.
Before we have to create a `TCanvas` object in order to display the result in this notebook.
We also modify the default color and create a grid.

```
TCanvas c("", "", 400, 300);
h.SetFillColor(kBlue - 10);
c.SetGrid();
h.Draw();
c.Draw();
```

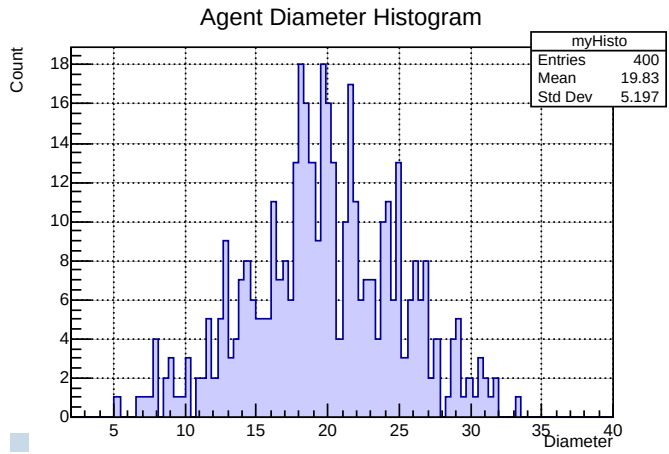

Finally, we can try to fit a function to the data in the histogram.
Since we drew samples from a gaussian random number generator when we created our cells, we expect that a gaussian will fit our data.

```
h.Fit("gaus", "S");
h.Draw();
c.Draw();
```

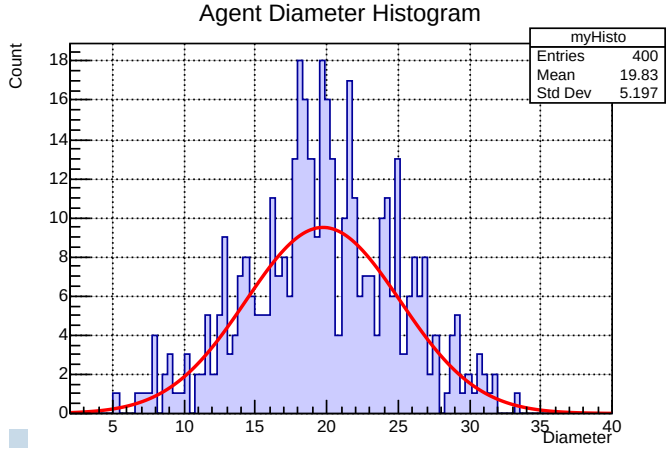

```
 FCN=73.1872 FROM MIGRAD     STATUS=CONVERGED      78 CALLS          79
TOTAL
                        EDM=6.76682e-08    STRATEGY= 1      ERROR MATRIX
ACCURATE
   EXT PARAMETER                                   STEP         FIRST
    NO.   NAME       VALUE          ERROR          SIZE        DERIVATIVE
     1  Constant    9.50544e+00    7.09907e-01    2.21048e-03    4.02840e-
04
     2  Mean        1.97314e+01    3.23457e-01    1.36014e-03   -3.65236e-
04
     3  Sigma       5.40166e+00    3.17423e-01    6.23484e-05    2.34242e-
02
```

# Simulation time series plotting (basics)

**Author: Lukas Breitwieser**

In this tutorial we show how to collect data during the simulation and plot it at the end.
To this extent, we create a simulation where cells divide rapidly leading to exponential growth.

Let's start by setting up BioDynaMo notebooks.

```
%jsroot on
gROOT->LoadMacro("${BDMSYS}/etc/rootlogon.C");
```

INFO: Created simulation object 'simulation' with UniqueName='simulati
on'.

```
using namespace bdm::experimental;
```

```
auto set_param = [](Param* param) {
    param->simulation_time_step = 1.0;
};
Simulation simulation("MySimulation", set_param);
```

Let's create a behavior which divides cells with $10\%$ probability in each time step.
New cells should also get this behavior.
Therefore, we have to call `AlwaysCopyToNew()`.
Otherwise, we would only see linear growth.

```
StatelessBehavior rapid_division([](Agent* agent) {
  if (Simulation::GetActive()->GetRandom()->Uniform() < 0.1) {
    bdm_static_cast<Cell*>(agent)->Divide();
  }
});
rapid_division.AlwaysCopyToNew();
```

Let's create a function that creates a cell at a specific position, with diameter = 10, and the `rapid_division`
behavior.

```
auto create_cell = [](const Double3& position) {
  Cell* cell = new Cell(position);
  cell->SetDiameter(10);
  cell->AddBehavior(rapid_division.NewCopy());
  return cell;
};
```

As starting condition we want to create 100 cells randomly distributed in a cube with $min = 0, max = 200$

```
simulation.GetResourceManager()->ClearAgents();
ModelInitializer::CreateAgentsRandom(0, 200, 100, create_cell);
simulation.GetScheduler()->FinalizeInitialization();
VisualizeInNotebook();
```

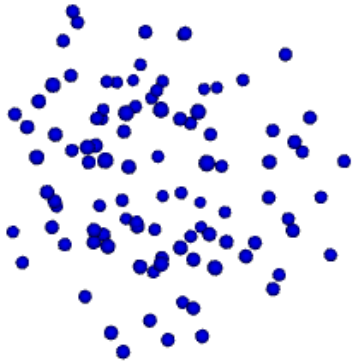

Before we start the simulation, we have to tell BioDynaMo which data to collect.
We can do this with the `TimeSeries::AddCollector` function. In this example we are interested in the number of agents.

In [7]:

```
auto* ts = simulation.GetTimeSeries();
auto get_num_agents = [](Simulation* sim) {
   return static_cast<double>(sim->GetResourceManager()->GetNumAgents());
};
ts->AddCollector("num-agents", get_num_agents);
```

Now let's simulate until there are 4000 agents in the simulation

In [8]:

```
auto exit_condition = [](){
    auto* rm = Simulation::GetActive()->GetResourceManager();
    return rm->GetNumAgents() > 4000;
};
simulation.GetScheduler()->SimulateUntil(exit_condition);
```

Now we can plot how the number of agents (in this case cells) evolved over time.

```
LineGraph g(ts, "My result", "Time", "Number of agents", true, nullptr, 500, 300);
g.Add("num-agents", "Number of Agents");
g.Draw();
```

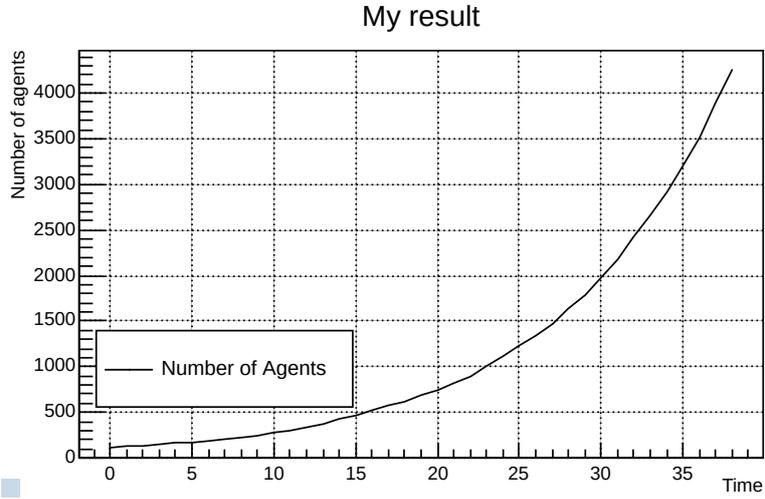

# Simulation time series plotting and analysis

**Author: Lukas Breitwieser**
In this tutorial we show how to collect data during the simulation, and plot and analyse it at the end.
To this extent, we create a simulation where cells divide rapidly leading to exponential growth.

Let's start by setting up BioDynaMo notebooks.

In [1]:

```
%jsroot on
gROOT->LoadMacro("${BDMSYS}/etc/rootlogon.C");
```

INFO: Created simulation object 'simulation' with UniqueName='simulation'.

In [2]:

```
using namespace bdm::experimental;
```

In [3]:

```
auto set_param = [](Param* param) {
    param->simulation_time_step = 1.0;
};
Simulation simulation("MySimulation", set_param);
```

Let's create a behavior which divides cells with $5\%$ probability in each time step.
New cells should also get this behavior.
Therefore, we have to call `AlwaysCopyToNew()`.
Otherwise, we would only see linear growth.

In [4]:

```
StatelessBehavior rapid_division([](Agent* agent) {
  if (Simulation::GetActive()->GetRandom()->Uniform() < 0.05) {
    bdm_static_cast<Cell*>(agent)->Divide();
  }
});
rapid_division.AlwaysCopyToNew();
```

Let's create a function that creates a cell at a specific position, with diameter = 10, and the `rapid_division` behavior.

In [5]:

```
auto create_cell = [](const Double3& position) {
  Cell* cell = new Cell(position);
  cell->SetDiameter(10);
  cell->AddBehavior(rapid_division.NewCopy());
  return cell;
};
```

As starting condition we want to create 100 cells randomly distributed in a cube with $min = 0, max = 200$

```
simulation.GetResourceManager()->ClearAgents();
ModelInitializer::CreateAgentsRandom(0, 200, 100, create_cell);
simulation.GetScheduler()->FinalizeInitialization();
VisualizeInNotebook();
```

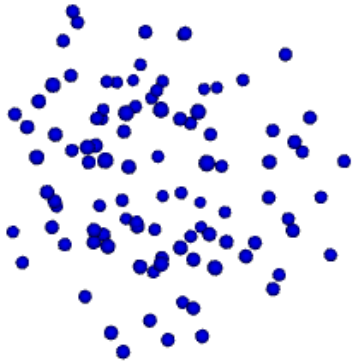

Before we start the simulation, we have to tell BioDynaMo which data to collect.
We can do this with the TimeSeries::AddCollector function.
In this example we are interested in the number of agents ...

```
auto* ts = simulation.GetTimeSeries();
auto get_num_agents = [](Simulation* sim) {
  return static_cast<double>(sim->GetResourceManager()->GetNumAgents());
};
ts->AddCollector("num-agents", get_num_agents);
```

... and the number agents with $diameter < 5$.
We create a condition `cond` and pass it to the function `Count` which returns the number of agents for which
`cond(agent)` evaluates to true.

```
auto* ts = simulation.GetTimeSeries();
auto agents_lt_5 = [](Simulation* sim) {
  auto cond = L2F([](Agent* a){ return a->GetDiameter() < 5; });
  return static_cast<double>(bdm::experimental::Count(sim, cond));
};
ts->AddCollector("agents_lt_5", agents_lt_5);
```

Now let's simulate 40 iterations

```
simulation.GetScheduler()->Simulate(40);
```

Now we can plot how the number of agents (in this case cells) and the number of agents with $diameter < 5$ evolved over time.

```
LineGraph g(ts, "my result", "Time", "Number of agents",
                                 true, nullptr, 500, 300);
g.Add("num-agents", "Number of Agents", "L", kBlue);
g.Add("agents_lt_5", "Number of Agents diam < 5", "L", kGreen);
g.Draw();
```

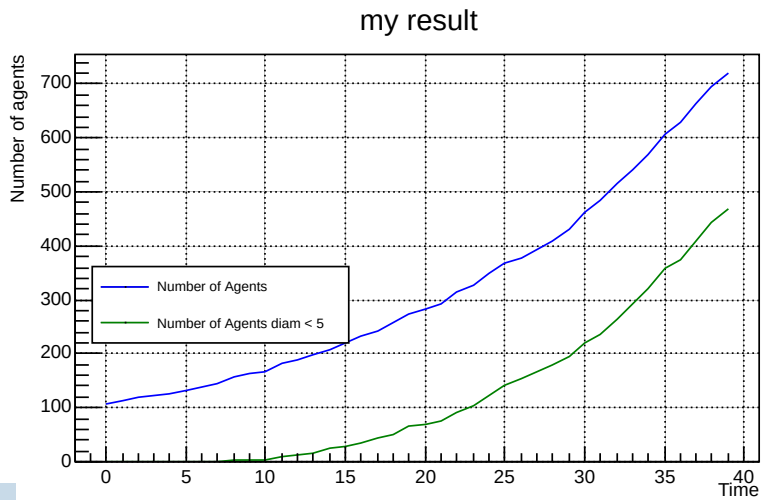

Let's try to fit an exponential function to verify our assumption that the cells grew exponentially.
Please visit the [ROOT user-guide (https://root.cern.ch/root/htmldoc/guides/users-guide/FittingHistograms.html)](https://root.cern.ch/root/htmldoc/guides/users-guide/FittingHistograms.html)
for more information regarding fitting

```
auto fitresult = g.GetTGraphs("num-agents")[0]->Fit("expo", "S");
g.Draw();
```

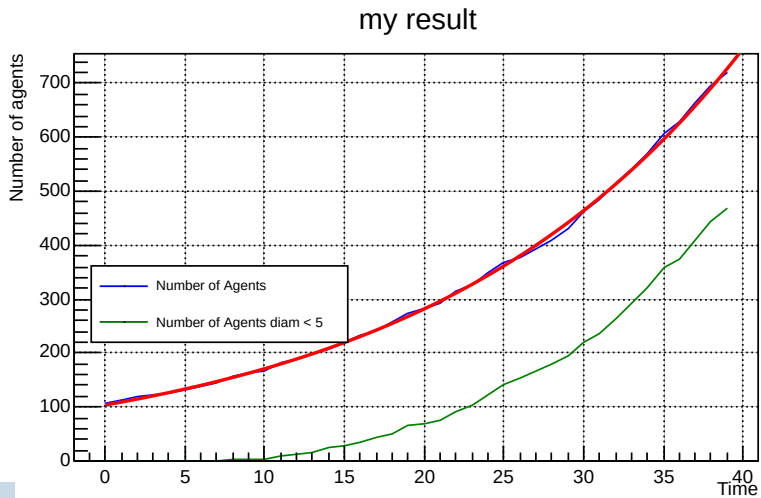

```
****************************************
Minimizer is Minuit / Migrad
Chi2                    =        793.088
NDf                     =             38
Edm                     =    1.27707e-08
NCalls                  =             49
Constant                =        4.64437    +/-   0.00734877
Slope                   =       0.0498693   +/-   0.000234423
```

Indeed, the number of agents follow an exponential function

$$y = \exp(slope * x + constant)$$

with constant = 4.6 and slope = 0.049 This corresponds to the division probability of $0.05$

This is how to change the color after the creation of `g`.
Also the position of the legend can be optimized.

```
g.GetTGraphs("num-agents")[0]->SetLineColor(kBlack);
g.SetLegendPos(1, 500, 20, 700);
g.Draw();
```

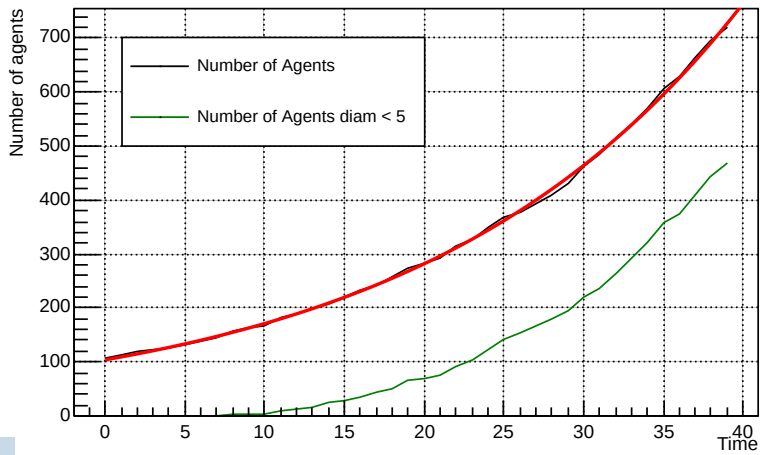

Let's save these results in multiple formats

```
g.SaveAs(Concat(simulation.GetOutputDir(), "/line-graph"),
         {".root", ".svg", ".png", ".C"});
```

# Multiple experiments and statistical analysis

**Author: Lukas Breitwieser**
In this tutorial we show how to collect and analyse data from multiple experiments.
To this extent, we create a simulation where cells divide rapidly leading to exponential growth.

Let's start by setting up BioDynaMo notebooks.

In [1]:

```
%jsroot on
gROOT->LoadMacro("${BDMSYS}/etc/rootlogon.C");
```

INFO: Created simulation object 'simulation' with UniqueName='simulati
on'.

In [2]:

```
using namespace bdm::experimental;
```

In [3]:

```
double gDivProb = 0.05;
```

We use the same simulation as in `ST09-timeseries-plotting-basic` . It is a simulation were agents divide with a specific division probability in each time step leading to exponential growth. We collect the number of agents in each time step. Have a look at `ST09-timeseries-plotting-basic` for more information.

We wrap the required simulation code in a function called `Experiment` which takes two parameters:

- the collected result data from a single invocation (output param)
- the division probability parameter

```cpp
void Experiment(TimeSeries* result, double division_probability) {
    gDivProb = division_probability;

    auto set_param = [](Param* param) {
        param->simulation_time_step = 1.0;
    };
    Simulation simulation("MySimulation", set_param);

    StatelessBehavior rapid_division([](Agent* agent) {
        if (Simulation::GetActive()->GetRandom()->Uniform() < gDivProb) {
            bdm_static_cast<Cell*>(agent)->Divide();
        }
    });
    rapid_division.AlwaysCopyToNew();

    auto create_cell = [&](const Double3& position) {
        Cell* cell = new Cell(position);
        cell->SetDiameter(10);
        cell->AddBehavior(rapid_division.NewCopy());
        return cell;
    };

    simulation.GetResourceManager()->ClearAgents();
    ModelInitializer::CreateAgentsRandom(0, 200, 100, create_cell);
    simulation.GetScheduler()->FinalizeInitialization();

    auto* ts = simulation.GetTimeSeries();
    auto get_num_agents = [](Simulation* sim) {
        return static_cast<double>(sim->GetResourceManager()->GetNumAgents());
    };
    ts->AddCollector("num-agents", get_num_agents);

    simulation.GetScheduler()->Simulate(40);

    // move collected time series data from simulation to object result
    *result = std::move(*simulation.GetTimeSeries());
}
```

We want to run our experiment for 10 times with a different division probability parameter. We choose the division probability randomly between 0.04 and 0.06

```cpp
std::vector<TimeSeries> individual_results(10);
Random rnd;
for(auto& ir : individual_results) {
    Experiment(&ir, rnd.Uniform(0.04, 0.06));
}
```

In the next step we want to combine the individual results. Therefore we calculate the mean, and min (error low), and max (error high) and store it in a merged `TimeSeries` object.

```
TimeSeries merged;
auto merger = [](const std::vector<double>& all_ys,
                 double* y, double* eh, double* el) {
    *y = TMath::Mean(all_ys.begin(), all_ys.end());
    *el = *y - *TMath::LocMin(all_ys.begin(), all_ys.end());
    *eh = *TMath::LocMax(all_ys.begin(), all_ys.end()) - *y;
};
TimeSeries::Merge(&merged, individual_results, merger);
```

Now we can print the merged results and see how the simulations evolved over time, and how they differed from each other.

```
LineGraph g(&merged, "My result", "Time", "Number of agents", false, nullptr, 500,
g.Add("num-agents", "Number of Agents", "LP", kBlue);
g.Draw();
```

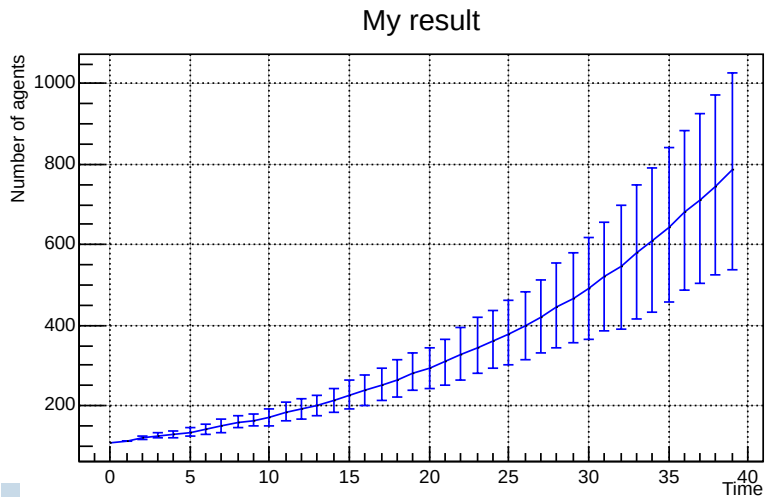

Finally, let's fit an exponential function to the data.

```
g.GetTMultiGraph()->Fit("expo", "S");
g.Draw()
```

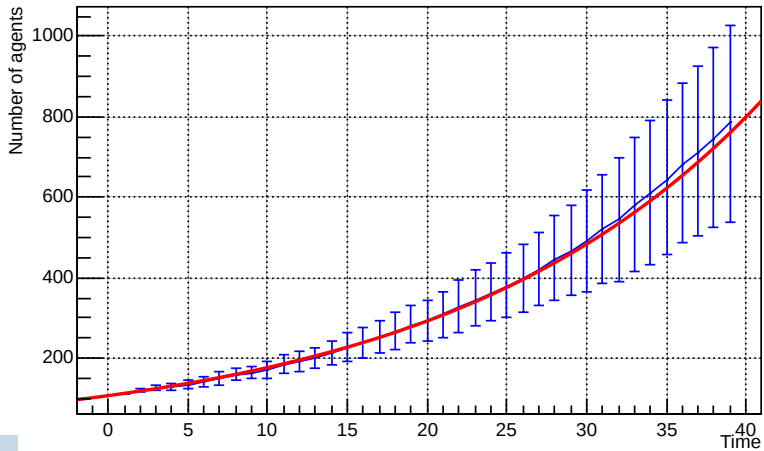

```
 FCN=0.850945 FROM MIGRAD    STATUS=CONVERGED     39 CALLS        4
0 TOTAL
                     EDM=6.73798e-11   STRATEGY= 1      ERROR MATRIX
ACCURATE
  EXT PARAMETER                                STEP         FIRST
  NO.   NAME        VALUE          ERROR       SIZE      DERIVATIVE
   1  Constant    4.67133e+00   1.58086e-02  8.77643e-06   8.20004e-
04
   2  Slope       5.03276e-02   1.84650e-03  1.02514e-06   6.11443e-
03
```

# Hierarchical model support

**Author: Lukas Breitwieser**

Some models require to update certain agents before others. In this tutorial we show how to execute operations first for large agents and afterwards for small ones. Lastly, we demonstrate how to run a different set of operations for large and for small agents.

Let's start by setting up BioDynaMo notebooks.

```
%jsroot on
gROOT->LoadMacro("${BDMSYS}/etc/rootlogon.C");
```

```
INFO: Created simulation object 'simulation' with UniqueName='simulati
on'.
```

To make this demo easier to understand, we turn off multi-threading and load balancing.

```
omp_set_num_threads(1);
ThreadInfo::GetInstance()->Renew();
auto* scheduler = simulation.GetScheduler();
scheduler->UnscheduleOp(scheduler->GetOps("load balancing")[0]);
```

We create a new agent operation which prints out its name and the diameter of the agent it is processing

```
struct TestOp : public AgentOperationImpl {
  BDM_OP_HEADER(TestOp);
  void operator()(Agent* agent) override {
    std::cout << name << " processing agent with diameter "
              << agent->GetDiameter() << endl;
  }
  std::string name = "";
};
OperationRegistry::GetInstance()->AddOperationImpl(
    "test_op", OpComputeTarget::kCpu, new TestOp());
```

We create four agents with diameter {20, 10, 20, 10}

```
auto* ctxt = simulation.GetExecutionContext();
for (int i = 0; i < 4; ++i) {
    double diameter = i % 2 == 0 ? 20 : 10;
    ctxt->AddAgent(new SphericalAgent(diameter));
}
```

We add the new operation to the simulation

```
auto* op1 = NewOperation("test_op");
scheduler->ScheduleOp(op1);
```

Let's simulate one time step and observe the default behavior of BioDynaMo. We expect that the agents are processed in the order they were added ( {20, 10, 20, 10} )

In [6]:

```
scheduler->Simulate(1);
```

```
 processing agent with diameter 20
 processing agent with diameter 10
 processing agent with diameter 20
 processing agent with diameter 10
```

Now we want to define the group of large and small agents and tell BioDynaMo that large agents should be processed before small ones.

This can be done with the following three lines of code.

In [7]:

```
auto small_filter = L2F([](Agent* a) { return a->GetDiameter() < 15; });
auto large_filter = L2F([](Agent* a) { return a->GetDiameter() >= 15; });
scheduler->SetAgentFilters({&large_filter, &small_filter});
```

Let's observe if the output has changed. We expect to see first the large agents {20, 20} , followed by the small ones {10, 10} .

In [8]:

```
scheduler->Simulate(1);
```

```
 processing agent with diameter 20
 processing agent with diameter 20
 processing agent with diameter 10
 processing agent with diameter 10
```

Let's create two more instances of our `TestOp` . We define that:

- `op1` should be run for all agents (large and small).
- `op2` only for small agents
- `op3` only for large agents

```cpp
auto* op2 = NewOperation("test_op");
auto* op3 = NewOperation("test_op");

op1->GetImplementation<TestOp>()->name = "OpAll       ";
op2->GetImplementation<TestOp>()->name = "OpOnlySmall";
op3->GetImplementation<TestOp>()->name = "OpOnlyLarge";

op2->SetExcludeFilters({&large_filter});
op3->SetExcludeFilters({&small_filter});

scheduler->ScheduleOp(op2);
scheduler->ScheduleOp(op3);
```

Now we want to execute another time step with the updated model. We expect that for each agent two operations will be executed.

For large agents `OpAll` and `OpOnlyLarge` and for small agents `OpAll` and `OpOnlySmall`. As before, we expect that first all large agents are executed, followed by all small agents.

```cpp
scheduler->Simulate(1);
```

```
OpAll       processing agent with diameter 20
OpOnlyLarge processing agent with diameter 20
OpAll       processing agent with diameter 20
OpOnlyLarge processing agent with diameter 20
OpAll       processing agent with diameter 10
OpOnlySmall processing agent with diameter 10
OpAll       processing agent with diameter 10
OpOnlySmall processing agent with diameter 10
```

# Dynamic scheduling

**Author: Lukas Breitwieser**

This tutorial demonstrates that behaviors and operations can be added and removed during the simulation. This feature provides maximum flexibility to control which functions will be executed during the lifetime of a simulation.

Let's start by setting up BioDynaMo notebooks.

```
%jsroot on
gROOT->LoadMacro("${BDMSYS}/etc/rootlogon.C");
```

```
INFO: Created simulation object 'simulation' with UniqueName='simulati
on'.
```

```
auto* ctxt = simulation.GetExecutionContext();
auto* scheduler = simulation.GetScheduler();
```

Define a helper variable

```
int test_op_id = 0;
```

We define a standalone operation `TestOp` which prints out that it got executed and which removes itself from the list of scheduled operations afterwards. The same principles apply also for agent operations.

```
struct TestOp : public StandaloneOperationImpl {
  BDM_OP_HEADER(TestOp);
  void operator()() override {
    auto* scheduler = Simulation::GetActive()->GetScheduler();
    std::cout << name << " processing iteration "
              << scheduler->GetSimulatedSteps()
              << std::endl;

    auto* op = scheduler->GetOps("test_op")[test_op_id++];
    scheduler->UnscheduleOp(op);

    std::cout << "  " << name
              << " removed itself from the simulation " << std::endl;
  }
  std::string name = "";
};
OperationRegistry::GetInstance()->AddOperationImpl(
    "test_op", OpComputeTarget::kCpu, new TestOp());
```

Let's define a little helper function which creates a new instance of `TestOp` and adds it to the list of scheduled operations.

```
void AddNewTestOpToSim(const std::string& name) {
    auto* op = NewOperation("test_op");
    op->GetImplementation<TestOp>()->name = name;
    scheduler->ScheduleOp(op);
}
```

Let's define a new behavior `b2` which prints out when it gets executed and which adds a new operation with name `OP2` to the simulation if a condition is met.

In this scenario the condition is defined as `simulation time step == 1`.

In [6]:

```
StatelessBehavior b2([](Agent* agent) {
    std::cout << "B2 " << agent->GetUid() << std::endl;
    if (simulation.GetScheduler()->GetSimulatedSteps() == 1) {
        AddNewTestOpToSim("OP2");
        std::cout << "  B2 added OP2 to the simulation" << std::endl;
    }
});
```

We define another behavior `b1` which prints out when it gets executed, removes itself from the agent, and which adds behavior `b2` to the agent.

In [7]:

```
StatelessBehavior b1([](Agent* agent) {
    std::cout << "B1 " << agent->GetUid() << std::endl;
    agent->RemoveBehavior(agent->GetAllBehaviors()[0]);
    std::cout << "  B1 removed itself from agent " << agent->GetUid() << std::endl;
    agent->AddBehavior(b2.NewCopy());
    std::cout << "  B1 added B2 to agent " << agent->GetUid() << std::endl;
});
```

Now all required building blocks are ready. Let's define the initial model: a single agent with behavior `b1`.

In [8]:

```
auto* agent = new SphericalAgent();
agent->AddBehavior(b1.NewCopy());
ctxt->AddAgent(agent);
```

We also add a new operation to the simulation.

In [9]:

```
AddNewTestOpToSim("OP1");
```

Let's simulate one iteration and think about the expected output.

- Since we initialized our only agent with behavior `b1`, we expect to see a line `B1 0-0`
- Furthermore, `b1` will print a line to inform us that it removed itself from the agent, and that it added behavior `b2` to the agent.

- Because changes are applied immediately (using the default `InPlaceExecCtxt`) also `B2` will be executed. However the condition inside `b2` is not met.
- Next we expect an output from `OP1` telling us that it got executed.
- Lastly, we expect an output from `OP1` to tell is that it removed itself from the simulation.

In [10]:

```
scheduler->Simulate(1);
```

```
B1 0-0
  B1 removed itself from agent 0-0
  B1 added B2 to agent 0-0
B2 0-0
OP1 processing iteration 0
  OP1 removed itself from the simulation
```

Let's simulate another iteration.
This time we only expect output from `B2`. Remember that `B1` and `OP1` have been removed in the last iteration.

This time the condition in `B2` is met and we expect to see an output line to tell us that a new instance of `TestOp` with name `OP2` has been added to the simulation.

In [11]:

```
scheduler->Simulate(1);
```

```
B2 0-0
  B2 added OP2 to the simulation
```

Let's simulate another iteration. This time we expect an output from `B2` whose condition is not met in this iterations, and from `OP2` that it got executed and removed from the simulation.

In [12]:

```
scheduler->Simulate(1);
```

```
B2 0-0
OP2 processing iteration 2
  OP2 removed itself from the simulation
```

Let's simulate one last iteration. `OP2` removed itself in the last iteration. Therefore, only `B2` should be left. The condition of `B2` is not met.

In [13]:

```
scheduler->Simulate(1);
```

```
B2 0-0
```

In summary: We initialized the simulation with `B1` and `OP1`.

In iteration:

0. B1 removed, B2 added, OP1 removed
1. OP2 added

2. OP2 removed

# Randomize iteration order

**Author: Lukas Breitwieser**

In this tutorial we show how to randomize the order that BioDynaMo uses in each iteration to process the agents.

Let's start by setting up BioDynaMo notebooks.

In [1]:

```
%jsroot on
gROOT->LoadMacro("${BDMSYS}/etc/rootlogon.C");
```

INFO: Created simulation object 'simulation' with UniqueName='simulation'.

Let's create two helper functions:

- `AddAgents` to add four agents to the simulation
- `print_uid` which prints the uid of the given agent

In [2]:

```
void AddAgents(ResourceManager* rm) {
    for (int i = 0; i < 4; ++i) {
        rm->AddAgent(new SphericalAgent());
    }
}
auto print_uid = [](Agent* a) {
  std::cout << a->GetUid() << std::endl;
};
```

We define an experiment which

1. takes a simulation object as input
2. adds four agents
3. calls `print_uid` for each agent
4. print a seperator so we can distinguish the output of the two different time steps
5. advances to the next time step
6. calls `print_uid` for each agent again

In [3]:

```
void Experiment(Simulation* sim) {
    auto* rm = sim->GetResourceManager();
    AddAgents(rm);

    rm->ForEachAgent(print_uid);
    rm->EndOfIteration();
    std::cout << "-----------------" << std::endl;
    rm->ForEachAgent(print_uid);
}
```

The default behavior of BioDynaMo is to iterate over the agents in the order they were added (not taking multi-threading and load balancing into account). Therefore, we expect to see the same order twice.

```
Experiment(&simulation)
```

```
0-0
1-0
2-0
3-0
----------------
0-0
1-0
2-0
3-0
```

BioDynaMo also provides a wrapper called `RandomizedRm`, which, as the name suggests, randomizes the iteration order after each iteration. It just takes two lines to add this functionality to the simulation.

```
Simulation simulation("my-sim");
auto* rand_rm = new RandomizedRm<ResourceManager>();
simulation.SetResourceManager(rand_rm);
```

Let's run our experiment again. This time with the simulation which has a randomized resource manager. We expect two different orders.

```
Experiment(&simulation)
```

```
0-0
1-0
2-0
3-0
----------------
2-0
3-0
0-0
1-0
```

# Replace mechanical interaction force

**Author: Lukas Breitwieser**

This tutorial demonstrates how to replace BioDynaMo's default interaction force with a user-defined one. The interaction force is used to calculate forces between agent pairs that are in physical contact with each other.

Let's start by setting up BioDynaMo notebooks.

In [1]:

```
%jsroot on
gROOT->LoadMacro("${BDMSYS}/etc/rootlogon.C");
```

INFO: Created simulation object 'simulation' with UniqueName='simulati
on'.

In [2]:

```
#include "core/operation/mechanical_forces_op.h"
```

We modify the `simulation_max_displacement` parameter to better visualize the difference of the user-defined force that we will add.

In [3]:

```
auto set_param = [](Param* p) {
    p->simulation_max_displacement = 50;
};
Simulation simulation("my-simulation", set_param);
```

In our experiment we create two overlapping cells and visualize the starting condition.

In [4]:

```
void Experiment() {
    simulation.GetResourceManager()->ClearAgents();
    auto* ctxt = simulation.GetExecutionContext();
    auto* scheduler = simulation.GetScheduler();

    auto* cell1 = new Cell({0, 0, 0});
    auto* cell2 = new Cell({10, 0, 0});
    cell1->SetDiameter(20);
    cell2->SetDiameter(20);
    cell1->SetMass(0.1);
    cell2->SetMass(0.1);

    ctxt->AddAgent(cell1);
    ctxt->AddAgent(cell2);

    scheduler->FinalizeInitialization();
    VisualizeInNotebook();
}
```

Let's run our experiment and have a look at the visualization.

In [5]:

```
Experiment();
```

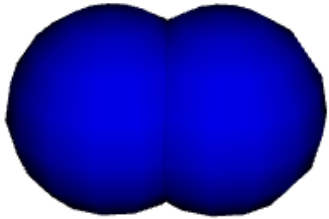

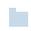

We continue by simulating 10 iterations and observe how the mechanical force pushed the two cells away from each other, until they don't overlap anymore.

In [6]:

```
auto* scheduler = simulation.GetScheduler();
scheduler->Simulate(10);
VisualizeInNotebook();
```

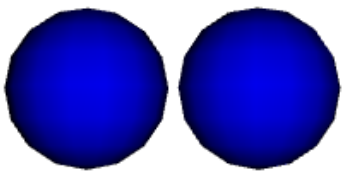

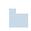

Now we want to add our user-defined force implementation. First, we have to subclass `InteractionForce` and implement our force. In this case, it is an extremely simple (and unrealistic) implementation.

```cpp
class MyInteractionForce : public InteractionForce {
  public:
    MyInteractionForce() {}
    virtual ~MyInteractionForce() {}

    Double4 Calculate(const Agent* lhs, const Agent* rhs) const override {
        if (lhs < rhs) {
            return {100, 0, 0, 0};
        } else {
            return {-100, 0, 0, 0};
        }
    }

    InteractionForce* NewCopy() const override { return new MyInteractionForce(); }
};
```

With the following three lines we instruct BioDynaMo to use our new `MyInteractionForce` instead of the default implementation.

```cpp
auto* myforce = new MyInteractionForce();
auto* op = scheduler->GetOps("mechanical forces")[0];
op->GetImplementation<MechanicalForcesOp>()->SetInteractionForce(myforce);
```

We create the same starting condition as before.

```cpp
Experiment();
```

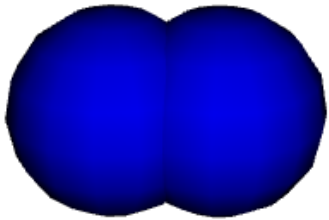

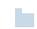

Because `myforce` is so strong, it is sufficient to simulate only one iteration to clearly see its impact.

```
auto* scheduler = simulation.GetScheduler();
scheduler->Simulate(1);
VisualizeInNotebook();
```

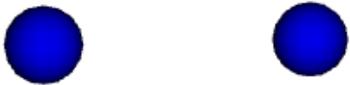

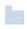



 \clearpage{}

\bibliographystyle{ACM-Reference-Format}

\end{document}